\documentclass[12pt]{article}

\usepackage{epsfig}
\usepackage{graphicx}
\usepackage{psfrag}
\usepackage{amssymb}
\usepackage{amsmath}

\newcommand{\rf}[1]{(\ref{#1})}
\newcommand{\beq}{\begin{equation}}
\newcommand{\beql}[1]{\beq\label{#1}}
\newcommand{\eeq}{\end{equation}}
\newcommand{\bea}{\begin{eqnarray}}
\newcommand{\eea}{\end{eqnarray}}

\newcommand{\e}{\mbox{e}}
\renewcommand{\d}{\mbox{d}}
\newcommand{\g}{\gamma}

\newcommand{\lam}{\lambda}
\newcommand{\La}{\Lambda}
\renewcommand{\b}{\beta}
\renewcommand{\a}{\alpha}
\newcommand{\n}{\nu}
\newcommand{\m}{\mu}

\renewcommand{\th}{\theta}
\newcommand{\ep}{\varepsilon}

\newcommand{\om}{\omega}
\newcommand{\Om}{\Omega}
\newcommand{\del}{\delta}
\newcommand{\Del}{\Delta}
\newcommand{\sg}{\sigma}
\newcommand{\kp}{\kappa}

\newcommand{\oh}{\frac{1}{2}}

\newcommand{\tr}{\mathrm{tr}\,}

\newcommand{\ra}{\rangle}
\newcommand{\la}{\langle}
\newcommand{\lan}{\left\la}
\newcommand{\ran}{\right\ra}
\newcommand{\ket}{\ran}
\newcommand{\bra}{\lan}
\newcommand{\prt}{\partial}
\newcommand{\mi}{\!-\!}
\newcommand{\equ}{\!=\!}
\newcommand{\plu}{\!+\!}

\newcommand{\cD}{{\cal D}}

\newcommand{\cT}{{\cal T}}

\newcommand{\cN}{{\cal N}}

\newcommand{\cO}{{\cal O}}

\newcommand{\cV}{{\cal V}}

\newcommand{\ta}{{\tilde{\a}}}
\newcommand{\ts}{{\tilde{s}}}
\newcommand{\tC}{{\tilde{C}}}

\newcommand{\tV}{{\tilde{V}}}

\newcommand{\tc}{{\tilde{c}}}
\newcommand{\tk}{{\tilde{k}}}

\newcommand{\hG}{{\hat{G}}}

\newcommand{\hH}{{\hat{H}}}

\newcommand{\hA}{{\hat{A}}}
\newcommand{\hC}{{\hat{C}}}
\newcommand{\hX}{{\hat{X}}}
\newcommand{\hP}{{\hat{P}}}
\newcommand{\hY}{{\hat{Y}}}
\newcommand{\hI}{{\hat{I}}}
\newcommand{\hD}{{\hat{D}}}
\newcommand{\hPk}{\hP^{kin}}
\newcommand{\hPp}{\hP^{pot}}

\newcommand{\bN}{{\bar{N}}}

\newcommand{\SL}{\sqrt{\La}}
\newcommand{\SLT}{\sqrt{\La}T}
\newcommand{\R}{{\rm I\!R}}

\begin{document}

\vspace{-36pt}

\begin{center}

{ \large \bf Nonperturbative Quantum Gravity}

\vspace{36pt}

{\sl J. Ambj\o rn}$\,^{a}$,
{\sl A. G\"{o}rlich}$\,^{b}$,
{\sl J. Jurkiewicz}$\,^{b}$ and
{\sl R. Loll}$\,^{c,d}$

\vspace{24pt}

{\footnotesize

$^a$~The Niels Bohr Institute, Copenhagen University\\
Blegdamsvej 17, DK-2100 Copenhagen \O , Denmark.\\
{email: ambjorn@nbi.dk}\\

\vspace{6pt}

$^b$~Institute of Physics, Jagellonian University,\\
Reymonta 4, PL 30-059 Krakow, Poland.\\
{ email: atg@th.if.uj.edu.pl, jurkiewicz@th.if.uj.edu.pl}\\

\vspace{6pt}

$^c$~Institute for Theoretical Physics, Utrecht University, \\
Leuvenlaan 4, NL-3584 CE Utrecht, The Netherlands.\\
{email: r.loll@uu.nl}\\

\vspace{6pt}

$^d$~Perimeter Institute for Theoretical Physics, \\
31 Caroline St. N., Waterloo, Ontario, Canada N2L 2Y5. \\
{email: rloll@perimeterinstitute.ca}

}
\vspace{24pt}
\end{center}

\begin{center}
{\bf Abstract}
\end{center}

Asymptotic safety describes a scenario in which general relativity
can be quantized as a conventional field theory, despite being
nonrenormalizable
when expanding it around a fixed background geometry.
It is formulated in the framework of the Wilsonian renormalization
group and relies crucially on the existence of an ultraviolet fixed point,
for which evidence has been found using renormalization group equations
in the continuum.

``Causal Dynamical Triangulations'' (CDT) is a concrete research
program to obtain a nonperturbative
quantum field theory of gravity via a lattice regularization, and
represented as a sum over spacetime histories.
In the Wilsonian spirit one can use this formulation
to try to locate fixed points of the lattice theory and
thereby provide independent, nonperturbative evidence for the
existence of a UV fixed point.

We describe the formalism of CDT, its phase diagram, possible fixed points
and the ``quantum geometries'' which emerge in the different phases.
We also argue that the formalism may be able to describe a more
general class of Ho\v rava-Lifshitz gravitational models.

\newpage

\section{Introduction}
\label{intro}

An unsolved problem in theoretical physics is how to reconcile
the classical theory of general relativity with quantum mechanics.
Consider the gravitational theory defined by
the Einstein-Hilbert action, plus possible matter
terms. Trying to quantize the fluctuations around a
given solution to the classical
equations of motion one discovers that the
corresponding quantum field theory is perturbatively nonrenormalizable.
Part of the problem is that in spacetime dimension four
the mass dimension of the gravitational coupling constant $G$ is $-2$
in units where $\hbar =1$ and $c=1$.
As a result, conventional perturbative quantum field theory is expected
to be applicable only for energies
\beq\label{i1}
E^2 \ll 1/{G}.
\eeq
Despite being a perfectly good assumption in all experimental situations
we can imagine in the laboratory, this relation can be taken as an
indication that
something ``new'' has to happen at sufficiently large energies
or, equivalently, short distances. If one believes in a fundamental
quantum theory of gravity, one would usually read
the breakdown of perturbation theory
when \rf{i1} is no longer satisfied as signaling the appearance of new degrees of
freedom as part of a different theory, which is valid at higher energies.
A well-known example of this is the electroweak theory, which
was described originally by a four-fermion
interaction. The latter is not renormalizable and perturbation
theory breaks down at sufficiently high energy, namely,
when the energy $E$ no longer satisfies \rf{i1}, with
the gravitational coupling constant $G$ replaced by
the coupling $G_F$ of the four-Fermi interaction,
which also has mass dimension $-2$. The breakdown coincides with the
appearance of new degrees of freedom, the $W$- and $Z$-particles.
At the same time,
the four-Fermi interaction becomes just an approximation to
the process where a fermion interacts via $W$ and $Z$ particles with
other fermions. The corresponding electroweak theory {\it is} renormalizable.

Similarly, in the 1960s a model for the scattering
of low-energy pions was proposed, the so-called non-linear
sigma model.
It is again nonrenormalizable, with a coupling constant of mass
dimension $-2$ when the model is formulated in four
(one time and three space) dimensions. Also in this case
the model did not describe adequately the scattering data at high energy.
Nowadays we understand that this happened because the pions cannot be viewed as
elementary particles, but
are made of quarks and anti-quarks. Again, the correct underlying theory
of these quarks, anti-quarks and gluons {\it is} a renormalizable quantum
field theory.

\subsection{What to do about gravity?}

For the case of gravity there seems to be no simple way of extending it
to a renormalizable quantum field theory by either adding
new fields, like in the electroweak theory, or by introducing
new fields in terms of which the theory becomes renormalizable, as
in the case of the nonlinear sigma model. It may
be possible that this can be done for gravity too, but
so far we have not discovered how.

There have been alternative proposals which share some of the
flavour of the above
``resolutions'' of nonrenormalizable theories. String theory
is an example of a framework which tries to get around the problem of
gravity as a nonrenormalizable quantum field theory by adding new
degrees of freedom, albeit infinitely many. The ambition of string theory
in the 1980s was that of a ``theory
of everything'', unifying gravity and all matter fields
in a single theoretical framework.
One problem with this is that one got much more than
was asked for, including many unobserved particles, symmetries
and spatial dimensions.
Another problem is that it has never been entirely clear what kind of theory
one is dealing with. Best understood is the perturbative expansion around
flat ten-dimensional spacetime, but a genuinely nonperturbative definition
of string theory is still missing. This renders
the role of (a possibly emergent notion of) space, not to mention time,
in string theory somewhat unclear. The world we observe today is not a
simple, obvious consequence of the dynamics of string theory.
With our present understanding of string
theory one has to work hard to extract from it something which even
vaguely resembles the world we can observe. The incompleteness of
this understanding prevents us moreover from making any {\it predictions}
for our universe.
String theory clearly is a versatile and fascinating theoretical framework,
but it is hard to tell whether it is the right one for describing
the real world (including quantum gravity),
or merely a modern incarnation of epicycles, with sufficiently many
free (moduli) parameters to describe anything purely kinematically,
but providing no insights into the dynamics which governs nature.

Loop quantum gravity represents another bold attempt to circumvent
the nonrenormalizability of quantum gravity. It does so by adopting
a non\-standard procedure of quantization where the Hilbert space
of states is non\-separable and the holonomies of connections, viewed
as quantum objects, are finite. It is perhaps too early to tell whether this program
will be successful in its attempts to quantize
four-dimensional gravity, solve all UV problems
and provide us with a semiclassical limit which coincides with Einstein gravity
in the limit $\hbar\to 0$.

\subsection{Searching for fixed points}

A much more mundane approach to quantum gravity, going back
to S.\ Weinberg and known as ``asymptotic safety" \cite{weinberg}
is inspired by the Wilsonian renormalization group.
The key idea is that while a perturbative
expansion around a fixed background geometry leads to a nonrenormalizable
theory, this merely reflects the infrared end of a
renormalization group flow, which originates from a genuinely nonperturbative
UV fixed point governing the short-distance physics of quantum gravity.
Asymptotic safety refers to the assumption that such an ultraviolet
fixed point exists and in its neighbourhood
the co-dimension of the critical surface
associated with it is finite. As a consequence one only
has to adjust a finite number of coupling constants
to reach the critical surface, where the physics is identical
to that of the UV fixed point. In the abstract coupling-constant
space where all possible interactions are allowed,
the couplings which need to be adjusted to reach
the critical surface are called relevant couplings.
In this sense the concept of asymptotic
safety is a generalization of the concept of a renormalizable
field theory. For renormalizable four-dimensional
field theories the fixed point is Gaussian. This implies that the scaling
dimension of the fields when approaching the fixed point is just
the canonical dimension of the field as it appears in the classical
Lagrangian. The above-mentioned finite co-dimension of the
critical surface associated with such a Gaussian fixed point is
equal to the number of independent
polynomials one can form in terms of the fields and their derivatives,
such that the coupling constants of the corresponding
terms in an action have negative (canonical) mass dimension
(in units where $c=\hbar=1$). For marginal couplings, where the
canonical dimension is zero, one needs to calculate the
corrections to the canonical dimensions near the fixed point to decide
whether they are relevant or irrelevant (or stay marginal).
The difficulty one faces in the asymptotic safety scenario is that
the fixed point is not Gaussian and standard perturbation theory
may tell us little or nothing about its existence, let alone the physics
associated with the fixed point if it exists.

Do we have reliable methods to check for the existence of a
non-Gaussian fixed point? To start with,
not too many non-Gaussian fixed points are known outside of two
spacetime dimensions (which is very special) and
involving only bosonic fields\footnote{The situation looks somewhat better
when one considers supersymmetric
quantum field theories. Both in
three and four dimensions there exist a number of conformal
supersymmetric theories which are candidates for fixed points of a
larger class of supersymmetric field theories. The cancellation
of the leading UV divergences imposed by supersymmetry seems
to bring the situation closer to the two-dimensional case
where quantum fluctuations are not necessarily breaking the
conformal invariance one expects at the fixed point. Recently there
has been significant progress in implementing supersymmetric theories
on the lattice \cite{supersym}, and one may
eventually be able to study the detailed approach to
the conformal supersymmetric
fixed points in the same detail as one can currently study bosonic
lattice theories. In the present review we will be mainly
interested in purely bosonic theories.}, with one exception, the
Wilson-Fisher fixed point of three-dimensional
Euclidean scalar field theory.
It plays an important role in nature since it
governs the critical behaviour of many materials. It is nonperturbative,
and its existence and associated properties have been analyzed in
the 4-$\ep$ expansion, the 2+$\ep$ expansion, by an ``exact renormalization
group'' analysis and by other methods, leading to a general agreement
on the values
of its critical exponents. This means that we {\it can} analyze
meaningfully a nonperturbative
fixed point in quantum field theory in spacetime dimensions larger than
two. Unfortunately, the fixed point needed in four-dimensional quantum gravity
is different from the Wilson-Fisher one. First,
it is a UV fixed point, while the Wilson-Fisher fixed point is in the infrared.
Second, and maybe more importantly, the Wilson-Fisher fixed point is
located entirely within the set of renormalizable three-dimensional
quantum field theories in the sense that it can be viewed as a non-trivial
fixed point of a $\phi^4$-field theory. There is a renormalization
group flow from a {\it Gaussian} UV fixed point to
the Wilson-Fisher IR fixed point, while there is no other known
ultraviolte fixed point in the general class of three-dimensional scalar field theories.

The major issue one has to address when applying the 2+$\ep$ expansion
or the exact renormalization
group equations to four-dimensional quantum gravity is how reliable these
methods are, since both rely on truncations. When originally
proposing asymptotic safety, Weinberg referred to the 2+$\ep$ expansion
as an example of an expansion which suggested that there could be
a nontrivial fixed point. This was corroborated later by Kawai and
collaborators \cite{kawai}, but the problem obviously remains that in
spcaetime dimension four the expansion parameter $\ep= 2$ is no
longer small. More recently, starting with the seminal paper
of Reuter \cite{reuter}, there has been major progress in applying
exact renormalization group techniques to quantum gravity \cite{reuteretc}.
The calculations reported so far point to the existence of a
nontrivial fixed point, with additional evidence that one only has a finite
number (three) of relevant coupling constants associated with it,
as required by the asymptotic safety scenario. However,
we will not attempt here to evaluate how reliable this evidence for
asymptotic safety is.

\subsection{Putting gravity on the lattice (correctly)}

In this review article we will present another attempt
to construct a theory of quantum gravity nonperturbatively, namely, by
defining a nonperturbative quantum field theory of gravity as a
sum over spacetime geometries. To make the sum well defined
we introduce a UV cutoff via a spacetime lattice. Under the assumption
that the asymptotic safety scenario is valid, our task
is to search systematically for a fixed point by changing the bare coupling
constants in the lattice action. If such a fixed point is found, we must
investigate whether it qualifies as a UV fixed point for quantum gravity.
This should enable us to make contact with the asymptotic
safety scenario and exact renormalization group calculations.

However, the approach we will be pursuing is more general.
In case the asymptotic safety scenario
is not realized, but something like a string theory is needed to describe
(generalized) spacetime at distances shorter than the Planck scale, the
lattice theory may still provide a good description of physics
down to a few Planck lengths. Independent of what the underlying
fundamental theory turns out to be, there will be an effective
quantum gravity theory, obtained by integrating out all
degrees of freedom except for the spin-two field, although we will
not know a priori at what distances it will cease to be applicable.
The lattice theory we are about to construct may give a nontrivial
description of this near-Planckian regime.

An explicit example of what we have in mind by such an ``effective''
lattice theory is given by the Georgi-Glashow model,
a three-dimensional Higgs model with gauge group $SO(3)$. Rotated
to Euclidean space it contains ``instantons'',
that is, monopole-like configurations. The Georgi-Glashow model was the
first (and for purists still is the only) nonabelian gauge theory
where confinement was proved and monopoles were essential
in the proof. Polyakov, who was the first to provide these
arguments \cite{polyakov},
understood that a compact abelian
$U(1)$-{\it lattice} gauge theory would realize perfectly
the confinement mechanism of the nonabelian theory, despite
the fact that it has only a trivial Gaussian (free photon) UV limit.
The reason is that the lattice theory also contains monopoles.
In the continuum such abelian monopoles would be singular, but for
any finite lattice spacing the singular behaviour is regularized
and the monopole mass proportional to the inverse lattice spacing
$a$. In the case of the Georgi-Glashow model, the Higgs field causes
a spontaneous breaking of  the $SO(3)$-gauge group
to $U(1)$ and the monopole mass is proportional to $m_w$, the
mass of the massive vector particle. The nonabelian massive
vector particles and the Higgs field enter nontrivially only in
the monopole core which is of size $1/m_w$.
When distances are
longer than $1/m_w$, the Georgi-Glashow model behaves essentially
like a $U(1)$-gauge theory, but with monopoles, i.e.\ like the compact
$U(1)$-lattice gauge theory. In this way the lattice theory
provides a perfect long-distance model for the Euclidean Georgi-Glashow
model, including a description of nonperturbative physics like
confinement.

Are there any problems in principle with adopting
a lattice regularization of quantum gravity?
Na\"ively comparing with the classical continuum description one could
be worried that a lattice regularization ``somehow" breaks
the diffeomorphism invariance present in the continuum theory.
Apart from the possibility that symmetries broken by a lattice may be
restored in an appropriate continuum limit, the analogy with the standard
continuum description of gravity may be misleading on this point.
Given a $d$-dimensional topological
manifold one can introduce a piecewise linear geometry on it by choosing a
triangulation, assigning lengths to its links (the 1-simplices), and
insisting that the interior of each $d$-simplex is
flat (this works for both Euclidean and Minkowskian spacetime signature).
This setting allows for geodesic curves between any two points, whose
lengths are well defined, as are the angles between intersecting geodesics.
The key observation is that in this way we equip the manifold with
a continuous geometry {\it without having to introduce any coordinate
system}. Keeping the data which characterize the piecewise linear geometry,
namely, the triangulation together with the link length assignments,
the coordinate gauge redundancy of the continuum theory is no longer
present.

\subsection{Gravitational path integral via triangulation}

When using the path integral
to formulate a nonperturbative quantum field theory of gravity,
one must sum over all geometries
(with certain chosen boundary conditions).
A natural choice for the domain of the path integral is the set of all
continuous geometries, of which the piecewise linear geometries are
a subclass, which presumably will be dense when
the space of continuous geometries is equipped with a suitable distance
measure.

We can change a given linear geometry in two
ways, by changing either the length assignments of its links or the
abstract triangulation itself (which will usually require an accompanying
change of link lengths too).
The domain of the gravitational path integral we are going to
consider consists of distinct triangulations of a given topological manifold
$M$, with the additional specification that all link lengths are fixed
to the same value $a$, which will play the role of a UV lattice cutoff.
This implies that each abstract triangulation is associated with a
piecewise flat geometry (a ``simplicial manifold"), which is unique up to graph
automorphisms. Performing the path integral amounts to
summing over the set of abstract triangulations of $M$.
To the extent we can associate a suitable gravitational action $S[T]$
to each piecewise linear geometry $T$,
we can now approximate the sum over all
geometries on $M$
by summing over the chosen set of abstract triangulations,
\beq\label{0.1}
Z_a =  \sum_T \frac{1}{C_T}\;\e^{iS[T]},
\eeq
where $a$ refers to the cutoff introduced above. Note that the lattice
spacing $a$ is a physical, and not a coordinate length.
The appearance in (\ref{0.1}) of $C_T$, the order of the
automorphism group of $T$, implies that the (very rare)
triangulations which possess such symmetries have a smaller weight
in the path integral.

The partition function $Z_a$ can be understood as a regularization of the
formal continuum path integral
\beq\label{0.2}
Z = \int \cD [g_{\m\n}] \; \e^{i S[g_{\m\n}]},
\eeq
where the integration is nominally over all geometries (diffeomorphism
equivalence classes $[g_{\mu\nu}]$ of smooth four-metrics $g_{\mu\nu}$)
of the given manifold $M$.\footnote{When doing computations
in standard continuum general relativity, one has little choice but to adopt
a concrete coordinate system and work with metric and curvature tensors
which explicitly depend on these coordinates. In the path integral one would
like to get rid of this coordinate freedom again, by ``factoring out"
the four-dimensional diffeomorphism group. The formal notation $\int\cD [g_{\m\n}]$
for the diffeomorphism-invariant integration in the continuum path integral
(\ref{0.2}) leaves of course
entirely open how the quotient of metrics modulo diffeomorphisms is
to be realized in practice (and the measure chosen).}
By contrast, the sum \rf{0.1} does not in any way refer to
diffeomorphisms, since we are summing directly over geometries.
Despite summing over ``lattices'' of spacetime, we are
therefore not breaking diffeomorphism-invariance; what is more, the
diffeomorphism group does not act on the triangulation data.

Using triangulations
(and piecewise linear geometry) in the way just described
goes by the name of
{\it dynamical triangulations} (DT) and was first introduced
in two dimensions \cite{adf,david,kazakov}, mainly as a
regularization of string theory, and later
in three- \cite{av,am3,krzywicki} and four-dimensional gravity
\cite{aj,am4}.

In the case of four-dimensional quantum gravity,
an expression like \rf{0.2} is of course entirely formal.
One way to try to make some sense of it in a nonperturbative
context is
by introducing a cutoff like in \rf{0.1} and investigating whether
the limit $a \to 0$ exists. Existence in this context means
that one can use the partition function $Z_a$ to calculate the
expectation values of certain observables $\cO$ as
\beq\label{0.3}
\la \cO \ra_a = \frac{1}{Z_a} \sum_{T}\frac{1}{C_T} \e^{iS[T]} \cO[T],
\eeq
and relate these regularized observables to observables defined in
the continuum according to standard scaling relations of the type
\beq\label{0.4}
 \la \cO \ra_a =  a^{-\Del} \la \cO \ra_{cont}+ O(a^{-\Del+1}).
\eeq
For an arbitrary choice of bare coupling constants
in the lattice action $S[T]$ such a scaling will in general not be possible,
but close to a fixed point, if it exists, the correlation lengths
of certain correlators may diverge when expressed in terms of
the number of lattice spacings, and a scaling like \rf{0.4} may be
present. Recent work exploring the asymptotic safety scenario in a
continuum setting
provides encouraging evidence of the existence of a fixed
point. If it exists one should be able to
locate it using a lattice approach.

\subsection{The Wilsonian point of view}

To locate the fixed points of a lattice theory,
one first has to choose a lattice action, which
will depend on a set of {\it bare} coupling constants. By varying
these one may encounter phase transitions, which
divide the coupling-constant space
into regions characterized
by different expectation values of certain observables
acting as order parameters. An archetypal system of this type
in statistical mechanics
is a spin model on a lattice, whose order parameter is
the magnetization. The change of coupling constants
in the action (the Hamiltonian of the
statistical system) is in this case
implemented by changing the temperature, which
appears as an overall factor multiplying the various couplings associated
with different pieces of the Hamiltonian.
The system may have a phase transition
of first or higher order at
a certain temperature. The latter is usually associated
with a divergent spin-spin correlation length, when measured in lattice units.
The fact that the
lattice becomes irrelevant when we look at the long-distance properties
of the spin system explains
why we may be able to associate a continuum (Euclidean)
quantum field theory with the spin system at such a critical point.

Finally, the coupling constants which need to be renormalized in the
ensuing continuum quantum field theory are closely related to what we called
the relevant couplings of the lattice theory.
These are the coupling constants which
need to be fine-tuned in order to reach the critical surface.
The ``renormalized'' coupling
constants of the continuum theory are usually not determined
by the values of the relevant lattice coupling constants
{\it at} the critical surface, but rather by the way the relevant
coupling constants {\it approach} their value on the critical surface
(where the long-distance physics is identical to the physics at
the critical point).

Let us give a concrete illustration of how this procedure works.
Consider an observable $\cO(x_n)$, where $x_n$ denotes a lattice point,
with $x_n = a\, n$ and $n$ measuring the position in integer lattice
spacings. The correlation length $\xi(g_0)$ in
lattice units is determined from the leading behaviour of the correlator,
\beq\label{0.5}
-\log \la  \cO(x_n) \cO(y_m) \ra \sim |n-m|/ \xi(g_0) + o(|n-m|).
\eeq
We now approach the critical surface by fine-tuning the
relevant bare coupling constant $g_0$ to its critical value
$g_0^c$ such that the correlation length becomes infinite.
The way in which $\xi(g_0)$
diverges for $g_0 \to g_0^c$ determines how the lattice spacing
$a$ should be taken to zero as a function of the coupling constants, namely,
like
\beq\label{0.6}
\xi(g_0)\propto \frac{1}{|g_0-g_0^c|^\n},~~~~~a(g_0)\propto |g_0-g_0^c|^\n.
\eeq
This particular scaling of the lattice spacing ensures that one can
define a physical mass $m_{ph}$ by
\beq\label{0.7}
m_{ph} a(g_0) = 1/\xi(g_0),
\eeq
such that the correlator
$\la  \cO(x_n) \cO(y_m) \ra$ falls off exponentially like
$\e^{-m_{ph} |x_n-y_m|}$ for $g_0 \to g_0^c$ when $|x_n-y_m|$, but not
$|n-m|$, is kept fixed in the limit $g_0\to g_0^c$.

In this way we obtain a picture
where the underlying lattice spacing goes to zero while the physical mass
(or the correlation length measured in physical length units, not in
lattice spacings) is kept fixed when we approach the
critical point. The mass is thus defined by the approach to the critical
point, and not {\it at} the critical point, where the correlation length is infinite.
This is the standard Wilsonian scenario
for obtaining the continuum (Euclidean) quantum
field theory associated with the critical point $g_0^c$ of a
{\it second-order} phase transition.
Although we obtain a continuum quantum field theory in this way,
one should keep in mind that it could be trivial, in the sense of being
a free field theory. For example, it is generally believed
that for spacetime dimension larger than or equal to four the continuum
field theories corresponding to spin systems are all trivial.
By contrast, the above-mentioned Wilson-Fisher
fixed point in three dimensions {\it is} related
to a nontrivial quantum field theory.
Another thing to note is that -- in the spirit of the asymptotic safety scenario --
the co-dimension of the critical surface
is finite-dimensional and we therefore have only a finite number of relevant
directions in coupling-constant space (just one in the example above).
If this was not the case we would have no chance of finding
the critical surface by varying a few of the coupling constants ``by hand''.

\subsection{Applying Wilsonian ideas to gravity}

Is there any chance of implementing the above construction in the case of
quantum gravity? The answer is yes, if we take some gravity-specific
aspects into account.
A basic assumption underlying the Wilsonian description of critical
phenomena is that of a divergent correlation length when we approach the
critical surface.
However, in the absence of any preferred background metric it is not
immediately clear what will play the role of a correlation length
in quantum gravity.
To address this issue, let us consider first a theory of gravity
coupled to scalar fields. This makes it easier to discuss correlators, since we
can simply use the scalar field $\phi$ as the observable $\cO$ in \rf{0.7}.
Because of the diffeomorphism-invariance of the continuum description of
the theory, it makes little sense to talk about the behaviour of
$\la \phi(x) \phi(y)\ra$
as a function of the {\it coordinate} distance between $x$ and $y$, because
this does not have a diffeomorphism-invariant meaning. The geometrically
appropriate notion is to use the geodesic distance $d_{g_{\m\n}}(x,y)$ instead.
Using it, we arrive at the following invariant continuum definition of the
field correlator in matter-coupled quantum gravity, namely,
\bea\label{0.8}
\lefteqn{\la \phi \phi (R) \ra \equiv} \\
&&\!\!\!\!\! \int \cD [g_{\m\n}]\, e^{i S[g_{\m\n}]} \int\!\!\!\!
\int dx\ dy\sqrt{\mi g(x)} \sqrt{\mi g(y)}
\;\la \phi(x) \phi(y) \ra_{matter}^{[g_{\m\n}]}
\;\del(R \mi d_{g_{\m\n}}(x,y)),
\nonumber
\eea
where the term $\la \phi(x) \phi(y)\ra_{matter}^{[g_{\m\n}]}$
denotes the correlator of the matter
fields calculated for a fixed geometry $[g_{\m\n}(x)]$.
It depends on the specific action chosen for the matter field,
which in turn will depend on the geometry of the manifold.
The $\del$-function ensures that the geodesic distance between points
labeled
by $x$ and $y$ is fixed to $R$, and the double integral implements an
averaging over all pairs of spacetime points. Characteristically, the
definition (\ref{0.8}) is nonlocal; we presently do not know of
a suitable local definition of a correlation function in the full quantum theory.
This aspect is reminiscent of a general feature
of observables in (quantum) gravity. If we insist that metric-dependent
continuum observables should be invariant under diffeomorphisms,
there exist no such quantities which are local.

What we are looking for in the quantum theory are quantities
whose ensemble average is physically meaningful. Even in a formulation
which is purely geometric, like the CDT quantum gravity to be described below,
this is a nontrivial requirement. The point is that
in constructing a two-point function, say, we cannot
mark any two specific points in a way that is meaningful in the
{\it ensemble} of geometries constituting the domain of the path integral.
We can pick two points in a given geometry, but there is no canonical
way of picking ``the same two points" in any other geometry. The best
we can do is to sum over all pairs of points (with mutual distance $R$,
say) for a given geometry, and then repeat the process for all other
geometries. This is precisely how the continuum expression
(\ref{0.8}) was conceived. Despite its nonlocal nature, we can ask physical
questions about $\la \phi \phi (R) \ra $
as a function of $R$, and discuss its short- and long-distance behaviour.
In this way we can realize
a Wilsonian scenario of matter fields coupled to gravity, whose
correlation length diverges when the lattice cutoff is removed.
In two-dimensional toy models of fluctuating geometries coupled
to matter the existence of such divergent correlation lengths has
been verified explicitly \cite{correlationlength}.\footnote{As an
aside note that even in ordinary lattice field theories, in order
to get optimal statistics in computer simulations when measuring
correlators, one uses a nonlocal definition like (\ref{0.8}) (without
the integration of metrics),
taking advantage of the (lattice) translational and rotational
invariance.}

It is less obvious how to use a definition like \rf{0.8} in pure
gravity and how to think about a correlation length in that
case. In the absence of a well-defined classical background,
trying to identify a graviton as the limit of
a massive lattice graviton may not be the most appropriate thing
to do. One can imagine various scenarios here, which will have
to be verified or falsified by explicitly analyzing the lattice 
gravity theory in question. On the one hand,
one may encounter a phenomenon like the Coulomb phase  
in abelian gauge theories in four-dimensional lattice theories,
where the gauge field excitations cannot be considered massive,
despite the fact that they live on a finite lattice. On the other hand,  
the three-dimensional compact $U(1)$-lattice theory mentioned
above is an example where lattice artifacts (the lattice monopoles)
in spite of gauge invariance create a mass gap in the theory,
but one which vanishes when the lattice spacing is taken to zero.

While it may not be immediately useful as a graviton
propagator, an expression like \rf{0.8} can still give us nontrivial
information about the nature of quantum spacetime.
For example, dropping the scalar field in \rf{0.8} by putting
$\la \phi(x) \phi(y) \ra_{matter} =1$, we still have a
diffeomorphism-invariant expression, which tests the average volume
of a geodesic ball of radius $R$. For small $R$, it will determine
the fractal dimension of spacetime, which in a theory of quantum gravity
may be different from the canonical dimension put in by hand
at the outset (when fixing the dimension of the fundamental building blocks).
This fractal dimension has been studied
in two-dimensional quantum gravity, both
analytically and by computer simulations \cite{fractal2d,bowick}, as well as
numerically in four-dimensional quantum gravity models \cite{fractal4d,more4d}.

\subsection{Lorentzian versus Euclidean}\label{wicki}

An expression like \rf{0.8} raises the question of the signature
of spacetime in a theory of gravity. 
Although we introduced this quantity in physical, Lorentzian signature,
using it to extract a fractal dimension (via geodesic balls)
implicitly assumed a rotation to Euclidean signature; also the two-dimensional
measurements of this correlator we mentioned earlier were obtained 
in the context of Euclidean ``gravity".
In equations \rf{0.1}-\rf{0.3}
we kept the ``i'' in front of the action, signaling that we
were dealing with ``real'' quantum field theory, where spacetime has
a Lorentzian signature, and where ``correlation functions'' are
quantum amplitudes, and thus inherently complex. On the other hand, we
have been referring to a ``Wilsonian scenario", which is primarily
valid for statistical
systems. In such a Euclidean context, ``spacetime'' is also Euclidean.

Of course, as
long as we are dealing with quantum field theory in Minkow\-skian {\it flat}
spacetime, the Osterwalder-Schrader axioms guarantee the existence of
a well-defined correspondence between the correlators calculated using
the Euclidean path integral with a Euclidean action $S_E$,
and the correlators calculated via the path integral using a
related Minkowskian action $S_M$ \cite{osterwalder}.
This is why second-order phase transitions
in statistical systems are relevant for quantum field theory. When it comes
to gravitational theories it is not directly clear what it means to
``associate" a Euclidean path integral with
a given Lorentzian one, which for the case of a scalar field on Minkowski space simply
takes the form of an analytic continuation $t\mapsto\tau =it$ in time
with associated map
\bea\label{0.9}
Z_M=\int \cD \phi_M \;\e^{iS_M[\phi_M]} \mapsto
Z_E = \int \cD \phi_E \;\e^{-S_E[\phi_E]}.
\eea
One problem in the continuum theory is that we do not know of a general 
map between
Lorentzian metrics $g_{\mu\nu}^{(M)}$ and real positive-definite metrics
$g_{\mu\nu}^{(E)}$ such that their associated Einstein-Hilbert actions
satisfy
\beq\label{0.10}
iS_M =  i \int \sqrt{-g^{(M)}} \;R(g^{(M)}_{\m\n})\;\; \to\;\;
-\int \sqrt{g^{(E)}}\; R(g^{(E)}_{\m\n}) \equiv -S_E.
\eeq
The reason is that the Wick rotation in the form of an analytic continuation 
in time does not generalize to arbitrary curved spacetimes. Beyond the very special
case of static and stationary metrics, there is no distinguished notion
of time with respect to which we could Wick-rotate, and Wick rotation does
not commute with the action of the diffeomorphism group. Besides, Wick-rotating
will in general produce complex and not real metrics, defeating the purpose
of rendering the path integral real. 
Even if we started from the set of all real Euclidean
metrics, ad hoc declaring it as a fundamental input, we would still need an inverse
Wick rotation and also be faced with the problem that
the Euclidean gravitational action $S_E$ is unbounded from below.
This unboundedness is
caused by the conformal mode of the metric, whose kinetic term enters
the kinetic term of (both the Lorentzian and Euclidean action) with the
``wrong" sign. As a consequence,
rapid variations of the conformal mode can make $S_E$ in
eq.\ \rf{0.10} arbitrarily negative.

There are different routes to try to avoid that this unboundedness causes a
problem in the Euclidean path integral. One possibility is to start from
a different classical action, involving higher-derivative terms
which stabilize the negative, second-order kinetic term of
the conformal mode \cite{Rsquare}. Alternatively, one can take special care of the
conformal mode when defining the Euclidean path integral, for example,
by analytically continuing it differently from the other modes of the
metric \cite{hawking,mottola}. Apart from making the
Euclidean action well defined, the inclusion of higher-derivative terms
may also help to cure gravity's nonrenormalizability,
since its propagator can then contain fourth powers of the
momentum in the denominators. Unfortunately, this prescription is generally
believed to spoil the unitarity of the corresponding Lorentzian theory
by introducing spurious poles in the propagator.

In an asymptotic safety scenario
there is nothing unnatural about a UV fixed point at which higher-derivative
terms play a dominant role, but also in this case it remains a major challenge
to show that a sensible, unitary theory emerges at the end of the day.
It {\it is} possible to avoid the problem of
unitarity violation even when higher-derivative terms are involved,
provided they are not ``generic''. For instance,
special symmetry can prevent unitarity-violating processes.
Arguments of this kind have been given in a version of conformal gravity
\cite{mannheim}. Also, other scale-invariant versions of gravity
theories where the higher-derivative terms come from integrating out
matter fields are likely to be free from this problem \cite{misha}.
Finally, to avoid
nonunitarity but still keep the improved renormalizability associated with
higher-derivative terms, P. Ho\v rava has recently suggested a
new class of gravitational theories. They are asymmetric in space and time
in the sense that higher-derivative terms appear, but only in the form of
{\it spatial} derivatives \cite{horava}.
If this formulation is to lead to a viable theory of quantum gravity,
one will need to show that the spacetime asymmetry is not in contradiction
with observed physics like, for example, the Lorentz invariance of
an (approximately) Minkowskian solution at sufficiently
large distance scales.

\subsection{Causal Dynamical Triangulations 101}

Our attempt to formulate a nonperturbative theory of quantum gravity
has to address the issues raised above. We will
do this here in summary fashion, in order to not get lost in the details.
The theory is defined on a manifold $M$ of topology $\Sigma\times [0,1]$,
with $\Sigma$ a three-dimensional manifold, usually chosen to have
the topology of a three-sphere $S^3$,
but in principle any other fixed topology could be used.
This implies that $M$ has
two boundaries $\Sigma(0)$ and $\Sigma(1)$.\footnote{As we will
see below, this specific choice of boundaries
will not play a significant role in our computer simulations of
four-dimensional quantum gravity.} The four-geometries we are
going to sum over in the path integral are such that the induced
geometry on these boundaries is spatial and that $\Sigma(1)$ is separated
by a proper-time distance $\tau$ from $\Sigma(0)$.
Translated to a continuum language the closest description
for such geometries would
be in terms of 3+1 ADM-decomposed metrics, with infinitesimal line element
given by
\beq\label{0.11}
ds^2 = -N^2(x^i,t) dt^2 + h_{ij}(x^i,t)(dx^i+N^i(x^i,t)dt)(dx^j+N^j(x^i,t)dt),
\eeq
where $N(x,i)$ is the lapse and $N_i(x,t)$ the shift function,
with the additional restriction that the boundaries be separated by
a fixed proper time. The latter restriction is made for the convenience
of the lattice set-up and will allow us in principle to define a transfer matrix
between adjacent spatial slices and from it a
quantum Hamiltonian.\footnote{The continuum analogy (\ref{0.11}) has to be treated with
some care, since the geometries in the path integral will {\it not} be
described in terms of coordinate systems, and moreover will certainly not
be smooth, unlike what is usually assumed in the classical continuum theory.}
The presence of a proper-time slicing brings this path integral over geometries
to a form closely related to the canonical formulation of (quantum) gravity
(see, for example, \cite{canonical}).

Our approach to quantum gravity, dubbed ``Causal Dynamical Triangulations (CDT)",
provides an explicit lattice formulation of quantum gravity,
where the spacetime geometries have Lorentzian signature, and we have identified a
notion of proper time on each of them.\footnote{Previous reviews of CDT quantum gravity 
can be found in \cite{cdtreviews}, sets of lecture notes in \cite{cdtlectures},
and nontechnical accounts in \cite{cdtpopular}. Reference \cite{livrev} is 
a comprehensive review covering lattice approaches
to quantum gravity prior to CDT.} It can be thought of as a lattice
implementation of a continuum proper-time path integral advocated by
Teitelboim \cite{teitelboim}. It turns out that
the CDT formulation allows us to
rotate each lattice geometry to a lattice geometry
with Euclidean signature. More precisely, the lattice construction contains
spacelike links of length-squared $a^2$ and timelike links
of length-squared $-\a a^2$. An analytic continuation
in the lower-half complex plane from positive to negative $\a$
changes the geometry from Lorentzian to Euclidean.
This resembles the standard
analytic continuation $t\mapsto\tau=it$
of the time coordinate in ordinary flat spacetime
when moving from Lorentzian to Euclidean signature. In the case at hand,
it is an analytic continuation of the piecewise linear geometry.
Once we have geometries with Euclidean signature, we can view the
model as a standard statistical model, taking the form of a sum over a class of
Euclidean geometries with positive weights. However, each of these
Euclidean geometries
has a causal, Lorentzian origin and is therefore not generic from
a Euclidean point of view. This makes the CDT path integral -- after
Wick-rotating -- distinct from a sum over all Euclidean geometries.
Whatever expression we obtain after summing over this subclass of
Euclidean geometries, will generically carry an $\alpha$-dependence,
whose role in any
eventual continuum theory has to be understood. Furthermore, a physical
interpretation of results will usually require a suitable ``inverse rotation"
back to Lorentzian signature. For four-dimensional results this
is not straightforward, since they are currently all based on computer
simulations. Their Lorentzian interpretation will typically require
care and further analysis.

The above analytic continuation in $\a$ also leads to a specific analytic
continuation of the Einstein-Hilbert action. As shown by Regge, and
as will be discussed in detail below, the Einstein-Hilbert
action has a natural implementation on piecewise linear geometries.
It is compatible with the analytic continuation in $\a$
in the sense that
\beq\label{0.12}
iS_L(\a) \mapsto -S_E(-\a),
\eeq
where $S_L(\a)$ is the Lorentzian Einstein-Hilbert action for a
given value of $\a$ and $S_E(-\a)$ the Euclidean Einstein-Hilbert
action for the negative value of $\a$. Thus our analytic continuation of
geometries satisfies \rf{0.10} and \rf{0.9}, with the difference that
it is not an ad hoc mapping between the two path integrals, but a
mapping between individual geometries. -- What has happened to the unboundedness of
the Euclidean Einstein-Hilbert action in this formulation? It is regularized by a finite
lattice spacing $a$, but will resurface in the limit $a \to 0$.\footnote{The fact that the
behaviour of the conformal factor is captured correctly, even in the context of
dynamical lattices, is nicely illustrated in two dimensions
\cite{simon-emil}. The formalism of dynamical triangulations (DT), which
provides a regularization of two-dimensional Euclidean Einstein-Hilbert
quantum gravity coupled to conformal matter,
can be used to calculate the conformal factor for
each configuration in the (discretized) path integral. The result
can be compared to the continuum theory (quantum Liouville theory), and
good agreement is found. As described in \cite{simon-emil}, one even
has an analogy to the entropy-driven
Kosterlitz-Thouless phase transition described below, namely, the
condensation of spikes at the $c=1$ barrier of Euclidean 
two-dimensional quantum gravity
coupled to conformal field theory.}
However, inside the path integral it can happen that configurations
with unbounded action are suppressed entropically and thus
play no role in the continuum limit. We will discuss this mechanism shortly.

As long as we stay with Euclidean signature we can use tools and techniques
from statistical field theory and the theory of critical phenomena
when searching for fixed points where the lattice formulation may
have a continuum interpretation. Suppose now that we have a phase diagram
with an order parameter related to geometry. (This describes the actual situation
in four dimensions, as we will see later.) We can then classify
transition points, lines or surfaces according to the
order of the transitions, which can be established
by measuring the behaviour of the order parameter as we approach
the transitions. In line with standard folklore from
the theory of critical phenomena, a second-order transition should be
associated with the existence of a continuum (quantum) field theory.
This answers a question raised earlier: also in theories of quantum gravity the
standard Wilsonian machinery appears to remain at our disposal, although
some of its aspects may have to be adapted.
As mentioned above, the concept of a
divergent correlation length as one approaches the critical
surface plays a key role in the Wilsonian picture, but its construction
and interpretation in quantum gravity
is less straightforward. Fortunately,
there exist definitions like \rf{0.8} which are diffeomorphism-invariant
and have been shown to work in toy models of two-dimensional quantum
gravity (see \cite{correlationlength}).

This brings us to the next point. If we succeed in identifying a potential
UV fixed point, it cannot simply be Gaussian, since
gravity is not renormalizable by conventional power counting (which
would be applicable at a Gaussian fixed point). In addition,
the nature of the fixed point had better be such that the unboundedness
of the Euclidean action does not dominate when the cutoff is taken to
zero. How is this possible without including explicit terms in the bare
action to curb this unboundedness, for example, in the form of
higher-derivative terms, which in turn may create unitarity problems? We hinted
above at the possibility that the path integral configurations leading
to an unbounded action upon removal of the
cutoff could be suppressed for entropical reasons.
Nonperturbatively, the effective Euclidean action contains an explicit
``entropic'' term, coming from the number of geometries that share a
given classical action. Such a term cannot play any role
in a semiclassical expansion, where one can choose
$\hbar$ arbitrarily small and thus the weight of the classical action
for some configurations with negative $S_E$ arbitrarily large.
Since the entropy term is not associated with any adjustable
coupling constant, the action
weight can always be made to dominate over it.
By contrast, a nonperturbative UV fixed point does not necessarily allow
anything like a semiclassical expansion, and the values of
the bare coupling constants close to it may lie in a region where
the Boltzmann weight of the action is comparable with the entropy term.

\subsection{Entropic {\it  QUANTUM} gravity}

An example from lattice field theory, the famous Kosterlitz-Thouless
transition in the two-dimensional XY-model, can
serve as an illustration of the situation.
The XY-model is a lattice spin model, whose ``spins"
are two-dimensional vectors of unit length. In two spatial dimensions, this model
has vortex configurations, with an energy per vortex of approximately
\beq\label{0.13}
E= \kp \ln (R/a),
\eeq
where $\kp$ is a coupling constant, $R$ a measure of the linear size of the system and
$a$ the lattice spacing. Ignoring boundary effects, the centre of the vortex can be
placed at any one of the $(R/a)^2$ lattice points. Saturating the path integral (the
partition function) $Z$ by single-vortex configurations, we obtain\footnote{Our
present discussion is merely qualitative and meant to highlight the competition between
entropy and Boltzmann weights; exact treatments of the Kosterlitz-Thouless transition are given
in many textbooks, see, e.g. \cite{zinnjustin}.}
\beq\label{0.14}
Z \equiv \e^{-F/k_B T} =
\!\!\!\!\sum_{{\rm spin~configurations}}\!\!\!\! \e^{-E[{\rm spin}]/k_B T}
\approx \left(\frac{R}{a}\right)^2 \; \e^{-[\kp \ln(R/a)]/k_B T}.
\eeq
We note that the factor $(R/a)^2 $ is entirely entropic, simply arising from counting
the possible single-vortex configurations, and is independent of any ``bare" coupling
constants (the spin coupling $\kappa$ and temperature $T$). Since the corresponding
entropy $S= k_B \ln ({\rm number~of~configurations})$ has the same functional form as the
vortex energy, we can express the free energy as
\beq\label{0.15}
F= E-ST = (\kp -2k_B T) \ln (R/a).
\eeq
The Kosterlitz-Thouless transition between a low-temperature phase (where vortices
play no role) and a high-temperature phase (where vortices are important)
occurs when $F = 0$, i.e. when the entropy factor is comparable to the Boltzmann
weight of the classical energy. At this point we are far away from the na\"ive
weak-coupling limit of the lattice spin theory, which is just a Gaussian free
field. Instead, the continuum field theory associated with the transition is the
sine-Gordon field theory at the coupling constant value where it changes from a
super-renormalizable to a renormalizable theory.

Are the fixed points of CDT quantum gravity ``entropic" in the sense just described?
The answer is
yes. In fact, it is remarkable that thanks to the geometric
nature of the curvature term, the Regge action in our lattice set-up
assumes a very simple form. We will see later (eq.\ \rf{ny13} below) that as a result
the lattice partition function
corresponding to the Euclidean path integral \rf{0.9} becomes
essentially the generating function for the number of triangulations,
that is, of geometries.
We conclude that in this rather precise sense our quantum
gravity theory {\it is}, quite literally, an entropic theory.

Once a candidate fixed point has been located, one can try
to make contact with the exact renormalization group by following the flow of
the coupling constants when approaching the fixed point. The procedure for
doing this will be discussed in detail below. A potential problem in a UV
fixed point scenario is unitarity. Here the CDT lattice formulation comes
with an additional bonus, since it allows us to
formulate a simple, sufficient criterion for the unitarity of the theory
in terms of properties of its transfer matrix, namely,
{\it reflection positivity}. We will show that the CDT lattice model both
has a transfer matrix and obeys reflection positivity, strongly
supporting the conjecture that any continuum theory -- if it can be
shown to exist -- will be unitary.

Having a handle on unitarity is one
reason for not using a gravitational action different from
the (Regge version of the) Einstein-Hilbert action. In
a Wilsonian renormalization group context it would be natural to
consider more general actions, involving various higher-order curvature
terms. However, for these we would generally not be able to prove generalized
reflection positivity.
In addition, as we will describe below, it appears that we already have
an interesting phase diagram without the explicit inclusion of
higher-order curvature terms. Of course, the effective action
we eventually hope to construct at the phase transition point
will in all likelihood contain such terms, but then presumably of a more
benign nature with respect to unitarity. This is a common situation
in quantum field theory: integrating out one type of field
will usually result in a nonlocal action in the remaining fields, and
the derivative expansion of this action will contain (infinitely many)
higher-derivative terms.

A lattice field theory breaks translational and rotational
symmetry explicitly, which is only restored in the continuum limit.
In the ADM-formulation of general relativity space and time appear
on a different footing, but this only implies a breaking of manifest,
and not of intrinsic diffeomorphism-invariance.
Our lattice formulation also has a built-in asymmetry between
space and time, which persists after rotation to Euclidean
signature.
This may open the possibility that
for some values of the bare coupling constants the theory possesses
a continuum limit in which space and time scale differently. If realized,
it would mean that the framework of CDT quantum gravity is sufficiently
general to allow also for the description and investigation gravitational
theories of Ho\v rava-Lifshitz type.
Although we are not putting in any asymmetric action by hand as
Ho\v rava did, it is possible that
the effective quantum action does contain such an
asymmetry because of the special role
played by time in the lattice construction. In this way the lattice phase
diagram for the theory may {\it a priori} have both
Ho\v rava-Lifshitz and ``isotropic'' fixed points, depending
on the choice of the bare coupling constants.

\subsection{Overview of what is to come}

The rest of this review article is organized as follows: in Sec.\ \ref{latticemodel}
we describe the construction of the CDT lattice model of quantum gravity
in two, three and four dimensions. Sec.\ \ref{trans} deals with the transfer matrix, and
Sec.\ \ref{2d} contains the analytic
solution of the two-dimensional model, as well as a discussion of its relation to
other 2d gravity models. Sec.\ \ref{g2d} discusses various generalizations
of the this model. The higher-dimensional CDT lattice gravity models cannot be
solved analytically. One way to proceed
is via Monte Carlo simulations.
In Sec.\ \ref{MC} we describe the idea behind the Monte Carlo updating
algorithms used. In Sec.\ \ref{phase-diagram} we report on the
Monte Carlo simulations of the four-dimensional CDT model.
The phase diagram is presented, and its resemblance with a
Lifshitz phase diagram emphasized, the order parameter being,
in a loose sense, the ``geometry".  Next, we describe in Sec.\ \ref{S4a}
in detail the geometry observed in the so-called phase C and
argue that one can view it as a metric four-sphere with small
quantum fluctuations superimposed. Emphasis is put on explaining
the entropic and emergent nature of this result,
along the lines sketched above for the XY-model. In Secs.\ \ref{effective}
and \ref{fluctuations}
we analyze the quantum fluctuations around the background geometry of phase C
and show that the observed scale factor of the (quantum) universe
as well as the fluctuations of the scale factor are described well by
a minisuperspace model assuming homogeneity and isotropy of the universe,
and going back to Hartle and Hawking. Under a few assumptions we can
determine the physical size of our quantum universe in Planck units.
This is discussed in Sec.\ \ref{size}.
In Sec.\ \ref{spectraldim}
we define the so-called spectral dimension and describe its
measurement, using a covariant diffusion equation.
The quantitative evaluation of the spectral dimension is one
concrete piece of evidence that the nontrivial UV properties of our
universe are compatible with
{\it both} the asymptotic safety scenario and Ho\v rava-Lifshitz gravity.
These approaches provide independent arguments that the UV-spectral
dimension should be two, which within measuring accuracy is
in agreement with what is found in CDT quantum gravity.
The construction of an effective
action allows us in principle to follow the flow of the coupling constants
entering the effective action as a function of the bare couplings
and the cutoff. In this way we can in principle discuss
the renormalization group flow of the coupling constants
and make contact with exact renormalization group calculations.
We will discuss the general procedure in Sec.\ \ref{renormalization}.
Finally, Sec.\ \ref{discussion} contains a short summary.

\section{The lattice CDT construction}
\label{latticemodel}

\subsection{Discrete Lorentzian spacetimes}\label{discrete}

Our first task will be to define the class of  discrete Lorentzian
geometries $T$ which we will use in  the path integral
(\ref{0.1}). We will mostly follow the treatment of \cite{3d4d}.

Briefly, they can be characterized as ``globally hyperbolic''
$d$-dimensional simplicial manifolds with a sliced structure,
where $(d\mi 1)$-dimensional ``spatial hypersurfaces''
of fixed topology are connected by suitable sets of
$d$-dimensional simplices. The $(d\mi 1)$-dimensional spatial hypersurfaces
are themselves simplicial manifolds, defined to be equilaterally
triangulated manifolds. As a concession to causality, we do not
allow the spatial slices to change topology as a function of time. 
There is a preferred
notion of a discrete ``time'', namely, the parameter
labeling successive spatial slices.
Note, as already emphasized in the introduction,
that this has nothing to do with a gauge choice, since we
are not using coordinates in the first place. This ``proper time''
is simply part of the invariant geometric data common to each of
the Lorentzian geometries.

We choose a particular set of elementary simplicial {\it building blocks}.
All spatial (squared) link lengths are fixed to $a^2$,
and all timelike links to have a squared length
$-\alpha a^2$, $\alpha >0$. Keeping $\alpha$ variable allows
for a relative scaling of space- and timelike lengths and is
convenient when discussing the Wick rotation later.
The simplices are taken to be pieces of flat
Minkowski space, and a simplicial manifold acquires nontrivial
curvature through the way the individual building blocks are
glued together.

As usual in the study of critical phenomena, we expect the final
continuum theory (if it exists) to be largely independent of
the details of the chosen discretization. The virtue of our choice of building
blocks is its simplicity and the availability of a straightforward Wick rotation.

In principle we allow any topology
of the $(d\mi 1)$-dimensional space,
but for simplicity and definiteness Äwe will fix the topology to be that of
$S^{d-1}$. By assumption we have a foliation of spacetime, where ``time''
is taken to mean proper time. Each time-slice, with the
topology of $S^{d-1}$, is
represented by a $(d\mi 1)$-dimensional triangulation.
Each abstract triangulation
of $S^{d-1}$ can be viewed as constructed by gluing together
$(d\mi 1)$-simplices whose
links are all of (spatial) length $a_s=a$, in this way
defining a $(d\mi 1)$-dimensional piecewise linear geometry
on $S^{d-1}$ with Euclidean signature.

We now connect two neighbouring
$S^{d-1}$-triangulations $T_{d-1}(1)$ and $T_{d-1}(2)$, associated with
two consecutive discrete proper times labeled 1 and 2, and create a
$d$-dimensional, piecewise linear geometry,
such that the corresponding $d$-dimensional
``slab'' consists of $d$-simplices, has the topology of $[0,1]\times S^{d-1}$,
and has $T_{d-1}(1)$ and $T_{d-1}(2)$ as its $(d\mi 1)$-dimensional boundaries.
The spatial links (and subsimplices) contained in these
$d$-dimensional simplices lie in either $T_{d-1}(1)$ or $T_{d-1}(2)$,
and the remaining links are declared
timelike with proper length squared $a_t^2 = -\a a^2$, $\a >0$.
Subsimplices which contain at least
one timelike link we will call ``timelike".
In discrete units, we can say that $T_{d-1}(1)$ and $T_{d-1}(2)$
are separated by a
single ``step'' in time direction, corresponding to a timelike
distance $\sqrt{\a} a$ in the sense that each link in the slab
which connects the two boundaries has a squared proper length $-\a a^2$.
It does {\it not} imply that
all points on the piecewise linear manifold defined by $T_d(1)$ have
a proper distance squared $-\a a^2$ to the piecewise linear manifold defined
by $T_d(2)$ in the piecewise Minkowskian metric of the triangulation, so
when we sometimes say that the time-slices $T_d(1)$ and $T_d(2)$ are separated
by a proper-time $a_t$, it is meant in the above sense.

Thus, our slabs or ``sandwiches" are assembled from
$d$-dimensional simplicial
building blocks of $d$ kinds, which are labeled according to the
number of vertices they share with the two adjacent spatial slices
of (discretized) proper time which we labeled 1 and 2.
A $(d,1)$-simplex has one $(d\mi 1)$-simplex (and
consequently $d$ vertices) in common with $T_{d-1}(1)$, and only one
vertex in common with $T_{d-1}(2)$. It has $d$
timelike links, connecting each of the $d$ vertices in $T_{d-1}(1)$
to the vertex belonging to $T_{d-1}(2)$. The next kind of $d$-simplex
shares  a $(d\mi 2)$-dimensional spatial {\it subsimplex} with  $T_{d-1}(1)$
and a one-dimensional spatial {\it subsimplex} (i.e.\ a link)
with  $T_{d-1}(2)$,
and is labeled a $(d\mi 1,2)$-simplex, where the label again
reflects the number of vertices it shares with $T_{d-1}(1)$ and  $T_{d-1}(2)$.
It has $2 (d\mi 2)$ timelike links. This continues
all the way to a $(1,d)$-simplex.  We can view the $(d\mi k,k\plu 1)$ simplex
as the ``time-reversal'' of the $(k\plu 1,d\mi k)$-simplex.
Gluing together the $d$-simplices such
that they form a slab means that we identify some of timelike
$(d\mi 1)$-dimensional subsimplices belonging to different $d$-simplices.
It is only possible to
glue a $(k\plu 1,d\mi k)$-simplex to a simplex of the same type or
to simplices of types $(k,d\plu 1\mi k)$ and $(k\plu 2,d \mi k\mi 1)$.
An allowed $d$-dimensional triangulation  of the
slab has topology $[0,1]\times S^{d-1}$,
is a simplicial manifold with boundaries,
and is constructed according to the recipe above.

\begin{figure}[t]
\centerline{\scalebox{0.3}{\includegraphics{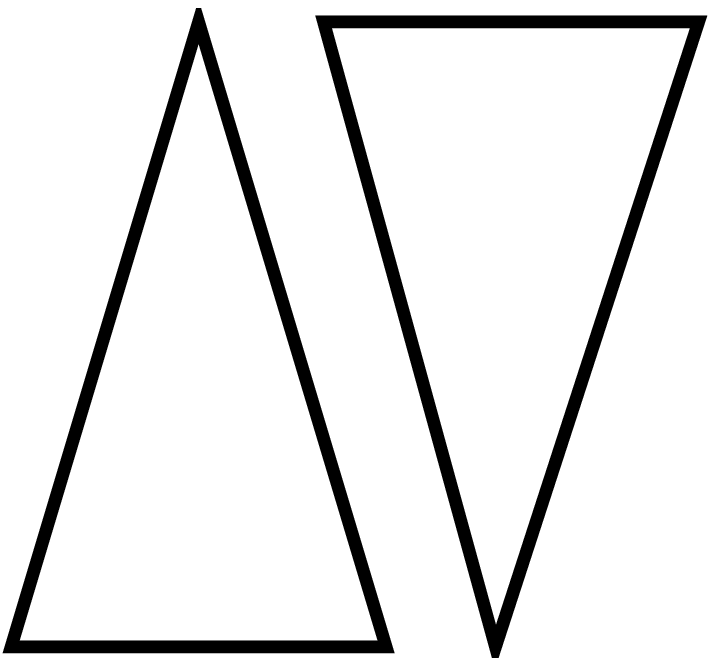}}~~~~~~~~\scalebox{0.4}{\includegraphics{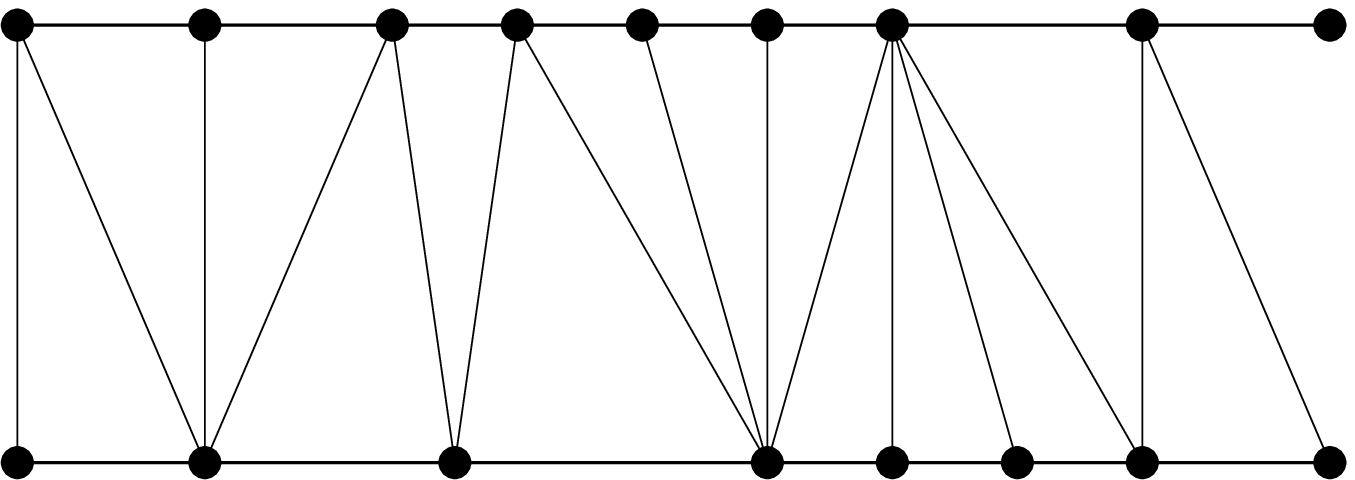}}}
\centerline{\scalebox{0.5}{\rotatebox{0}
{\includegraphics{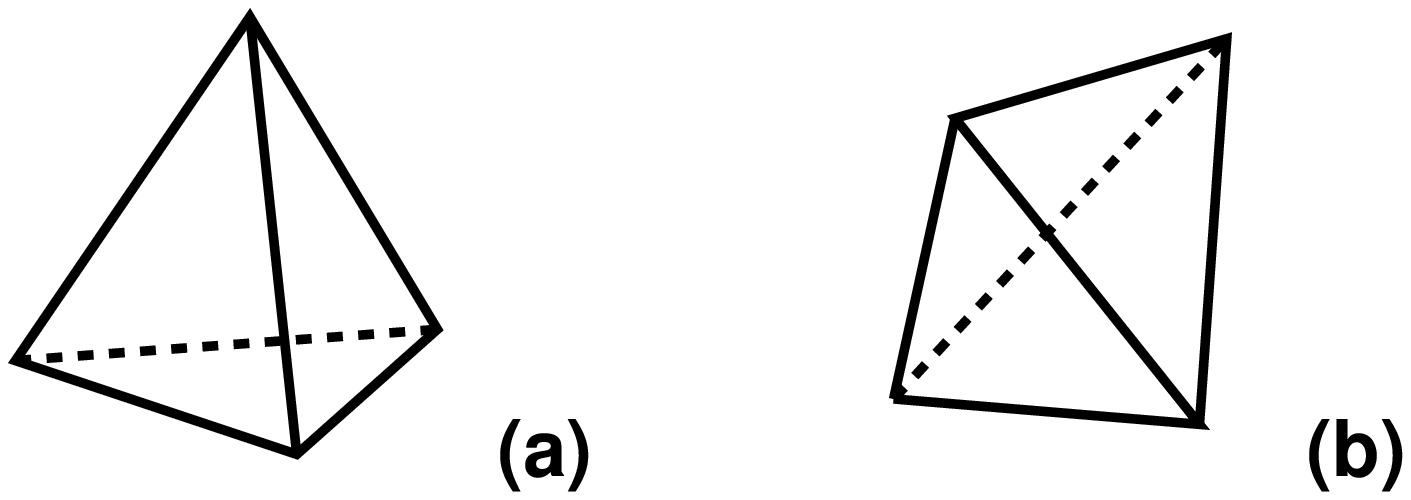}}}}
\centerline{\scalebox{0.5}{\rotatebox{0}
{\includegraphics{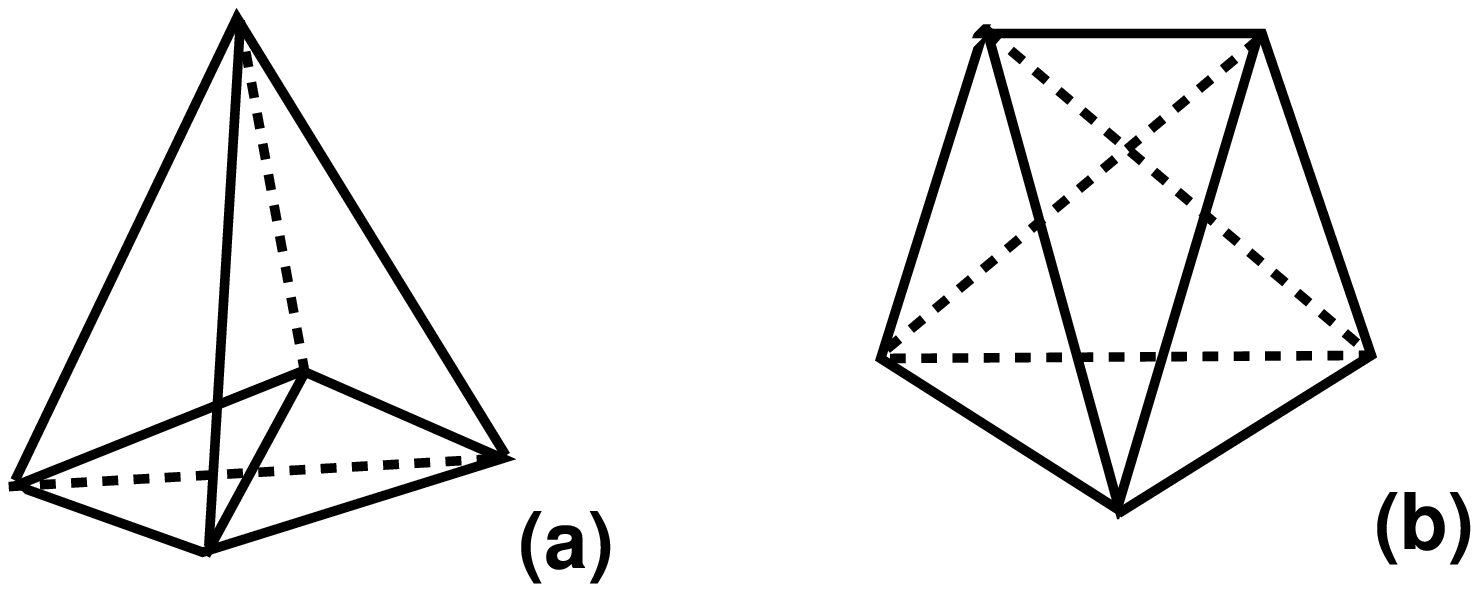}}}}
\caption[3dsimplex]{{\footnotesize 
Top figure: a (2,1)-simplex and a (1,2)-simplex and the
way they are glued together to form a ``slab'' (strip).
The ends of the strip should be joined to form a band with topology
$S^1 \times [0,1]$. Middle figure: a (3,1)-tetrahedron (fig.\ (a)) 
and a (2,2)-tetrahedron (fig. (b))in three
dimensions. Bottom figure: a (4,1)-simplex (fig. (a)) and a 
(3,2)-simplex (fig. (b)) in four dimensions.}}
\label{connect}
\end{figure}

The allowed simplices (up to time reversal for d=3,4), are
shown in Fig.\ \ref{connect} for d=2,3,4.

\subsection{The action associated with piecewise linear geometries}

A path in the gravitational path integral consists of a
sequence of triangulations of $S^{d-1}$, denoted
by $T_{d-1}(k)$, $k=0,\ldots,n$, where the spacetime between each pair
$T_{d-1}(k)$ and $T_{d-1}(k+1)$ has been filled in by a layer of
$d$-simplices as just described.
In the path integral we sum over all possible sequences
$\{T_{d-1}(k)\}$ and all possible
ways of triangulating the slabs in between $T_{d-1}(k)$ and $T_{d-1}(k+1)$.
The weight assigned to each geometry depends on the Einstein-Hilbert action
associated with the geometry. Let us discuss how the Einstein-Hilbert action
can be defined on piecewise linear geometry in an entirely geometric way.
This description goes back to Regge \cite{regge}.

Let us consider a piecewise linear $d$-dimensional
geometry with Euclidean signature, constructed by gluing  $d$-dimensional
simplices together such that they form a triangulation.
We change for a moment to Euclidean signature just to make the
geometric discussion more intuitive. The $d$-simplices are building blocks for our
piecewise linear geometries. The curvature of such a piecewise linear
geometry is located at the $(d\mi 2)$-dimensional subsimplices.
A number of $d$-simplices $\sg_i^d$ will share a $(d\mi 2)$-subsimplex
$\sg^{d-2}_0$.
Each of the $d$-simplices $\sg^d_i$
has a {\it dihedral angle}\footnote{Given a $d$-simplex
and all its subsimplices, any of the $(d\mi 2)$-subsimplices will be
the intersection (the common face) of precisely two $(d\mi 1)$-subsimplices.
The angle (and it {\it is} an angle for any $d\geq 2$)
between these $(d \mi 1)$-subsimplices is called the dihedral
angle of the $(d\mi 2)$-subsimplex in the given $d$-simplex.}
$\th(\sg^d_i,\sg_0^{d-2})$ associated with the given
$(d\mi 2)$-dimensional subsimplex $\sg_0^{d-2}$.  The sum
of dihedral angles of the $d$-simplices
sharing the subsimplex would add up to $2\pi$
if space  around that subsimplex is flat.
If the dihedral angles add up to something
different it signals that the piecewise linear space is not flat.
The difference to $2\pi$ is called the deficit angle
$\ep_{\sg_0^{d-2}}$ associated with
the subsimplex $\sg_0^{d-2}$,
\beq\label{2.0}
\ep_{\sg^{d-2}} = \Big(2\pi-
\sum_{\sg^d \ni \sg^{d-2}} \th(\sg^d,\sg^{d-2})\Big)\;\e^{i \phi(\sg^{d-2})}.
\eeq
The phase $\phi(\sg^{d-2})$ is 0 if $\sg^{d-2}$ is timelike
and $-\pi/2$ if $\sg^{d-2}$ is spacelike.
The reason a phase factor appears in the definition is that
in a geometry with Lorentzian geometry the
dihedral angles can be complex (see later for explicit expressions, and
Appendix 1 for a short discussion of Lorentzian angles).
For  a subsimplex $\sg^{d-2}$ which is entirely spacelike
the real part of the sum of the dihedral angles is $2\pi $ and
the phase factor ensures that $\ep_{\sg^{d-2}}$ is real. For
a timelike subsimplex $\sg^{d-2}$ the sum of the dihedral angles
is real. If we have a triangulation of flat Minkowskian spacetime
the deficit angles are zero.

We can view the piecewise linear geometry as
flat except when we cross the $(d\mi 2)$-subsimplices.
Regge proved (see also \cite{sorkin} for a detailed discussion)
that  one should associate the  curvature
\beq\label{2.1}
2 \ep_{\sg^{d-2}} V_{\sg^{d-2}}
\eeq
with a $(d\mi 2)$-subsimplex $\sg^{d-2}$, where
$V_{\sg^{d-2}}$ denotes the
volume of the subsimplex $\sg^{d-2}$. Let
us define
\beq\label{2.2}
 V_d(\sg^{d-2}) = \frac{2}{d(d+1)}
\sum_{\sg\, \ni \,\sg^{d-2}} V_{\sg^d},
\eeq
where $V_{\sg^d}$ is the volume of the $d$-simplex $\sg^d$ and where
the factor $2/d(d\plu 1)$ distributes the volume of the $d$-simplices equally
between its $d(d\plu 1)/2$ $(d\mi 2)$-subsimplices. If $T_{d-2}$ denotes
the $(d\mi 2)$-subsimplices in the triangulation $T$, the total volume and
total curvature of $T$ are now
\beq\label{2.3}
V_d(T) = \sum_{\sg^{d-2}\in \, T_{d-2}} V_d(\sg^{d-2})
\eeq
and
\beq\label{2.4}
R_{tot}(T)=
2 \sum_{\sg^{d-2}\in \, T_{d-2}} V_d(\sg^{d-2})\;
\ep_{\sg^{d-2}} \;\frac{V_{\sg^{d-2}}}{V_d(\sg^{d-2})}.
\eeq
One can think of \rf{2.3} and \rf{2.4} as the total volume and total
curvature of the piecewise linear geometry associated with $T$,
the counterparts of the integrals of the volume
and curvature densities on a smooth manifold $M$ with
geometry defined by a metric $g_{\m\n}(\xi)$, that is,
\beq\label{2.5}
V_d(g)=\int_{M} \d^d \xi \sqrt{-g(\xi)}~~~~{\rm and}~~~~
R_{tot}(g) =\int_{M} \d^d\xi \sqrt{-g(\xi)}\; R(\xi).
\eeq
More generally, since we are considering manifolds with boundaries,
when the $(d\mi 2)$-simplex is a boundary subsimplex, the deficit
angle is defined as in eq.\ \rf{2.0} except that in this case $2\pi$ is replaced
by $\pi$.
THe continuum expression corresponding to \rf{2.4} is
\beq\label{2.5a}
R_{tot}(g) =\int_{M} \d^d\xi \sqrt{-g(\xi)}\; R(\xi) +
\int_{\partial M} \d^{d-1} \xi \sqrt{h(\xi)} \;K(\xi),
\eeq
where the last integral is over the boundary of $M$ with
the induced metric and $K(\xi)$ denotes the trace of second
fundamental form.

\subsection{Volumes and dihedral angles}

We will now go on to compute the volumes and dihedral angles of
the $d$-dimen\-sio\-nal Minkowskian simplices, because they are
needed in the gravitational Regge action in dimension $d$.
All (Lorentzian) volumes in any dimension we will be using
are by definition real and positive. Formulas for Euclidean volumes
and dihedral angles can be derived from elementary geometric arguments
and may be found in many places in the literature \cite{angles}.
They may be continued to Lorentzian geometries by taking suitable
care of factors of $i$ and $-1$. We will follow Sorkin's
treatment and conventions for the Lorentzian case \cite{sorkin}.
(Some basic facts about Lorentzian angles are summarized in Appendix 1).
The dihedral angles $\Theta$ are chosen such that $0\leq {\rm Re}
\Theta \leq \pi$, so giving $\sin\Theta$ and $\cos \Theta$ fixes them
uniquely. The angles are in general complex, but everything can be
arranged so that the action comes out real in the end,
as we shall see.

We now list systematically the geometric data we will need
for simplices of various dimensions $d > 0$. As above we denote
a simplex $(k,d+1-k)$. We also allow $k\equ 0$, which means that
all vertices of the simplex belong to a spatial hyper-plane $t\equ
constant$. These have no $\a$-dependence.
We also only list the geometric properties of
the simplices with $k\geq d\plu 1 \mi k$ since the rest of the  simplices
can be obtained by time-reversal.

\subsubsection{d=0}
In this case the simplex is a point and we have by convention
Vol(point)$= 1$.

\subsubsection{d=1}
A link can be spacelike or timelike in accordance with the
definitions given above, and a spacelike link has by definition
the length $a$, with no $\a$-dependence. For the timelike link we have
\begin{equation}\label{2.6}
\mbox{ Vol(1,1)$\ =
\sqrt{\alpha }\;a$.}
\end{equation}

\subsubsection{d=2}
Also for the triangles, we must distinguish between
space- and timelike. The former lie entirely in planes $t\equ const$
and have no $\a$-dependence,
whereas the latter extrapolate between two such slices. Their
respective volumes are
\begin{equation}\label{2.7}
\mbox{ Vol(3,0)$\ =
\frac{\sqrt{3}}{4} a^2$ \hspace{.4cm} and \hspace{.4cm}
Vol(2,1 )$\ =\frac{1}{4} \sqrt{4 \alpha +1} \;a^2$.}
\end{equation}
We do not need the dihedral angles for the two-dimensional
simplices since they only enter when calculating the total curvature
of a two-dimensional simplicial manifold, and it is a topological
invariant, $2\pi \chi$, where $\chi$ is the Euler characteristic
of the manifold.

\subsubsection{d=3}

We have three types of three-simplices (up to time-reflection), but need
only the dihedral angles of the timelike three-simplices. Their volumes are

\beq\label{2.8}
{\rm Vol}(4,0)= \frac{\sqrt{2}}{12}a^3,~~~
{\rm Vol}(3,1)=\frac{\sqrt{3\alpha +1}}{12}a^3,~~~
{\rm Vol}(2,2)=\frac{\sqrt{2\alpha +1}}{6 \sqrt{2}}a^3.
\end{equation}.

A timelike three-simplex has dihedral angles around its timelike
links (TL) and its spacelike links (SL) and we find
\begin{alignat}{2}
\cos\Theta_{(3,1)}^{SL}& = -\frac{i}{\sqrt{3} \sqrt{4 \alpha+ 1}}&
\qquad
\sin\Theta_{(3,1)}^{SL} &= \frac{2 \sqrt{3 \alpha
+1}}{\sqrt{3} \sqrt{4\alpha+1}}
\label{2.9}\\
\cos\Theta_{(3,1)}^{TL}& = \frac{2\alpha +1}{4 \alpha +1} &\qquad
\sin\Theta_{(3,1)}^{TL} &= \frac{2 \sqrt{\alpha} \sqrt{3\alpha +1}}{4\alpha +1}
\label{2.10}\\
\cos\Theta_{(2,2)}^{SL} &= \frac{4 \alpha +3}{4 \alpha +1} &\qquad
\sin\Theta_{(2,2)}^{SL} &= -i\ \frac{2\sqrt{2}\sqrt{2\alpha +1}}{4\alpha+1}
\label{2.11}\\
\cos\Theta_{(2,2)}^{TL} &= -\frac{1}{4 \alpha+1} &\qquad
\sin\Theta_{(2,2)}^{TL} &=
\frac{2 \sqrt{2 \alpha} \sqrt{2\alpha +1}}{4\alpha +1}
\label{2.12}
\end{alignat}

\subsubsection{d=4}

In $d\equ 4$ there are up to reflection symmetry two types of four-simplices,
(4,1) and (3,2). Their volumes are given by
\begin{equation}
{\rm Vol}(4,1)= \frac{1}{96} \sqrt{8\alpha +3}\, a^4,\qquad
{\rm Vol}(3,2)=\frac{1}{96} \sqrt{12\alpha +7} \,a^4.
\end{equation}

For the four-dimensional simplices the dihedral angles are located
at the triangles, which can be spacelike (SL) or timelike (TL).
For the (4,1)-simplices there are four SL-triangles and six TL-triangles.
For the (3,2)-simplices there is one SL-triangle. However, there are
now two kinds of TL-triangles. For one type (TL1) the dihedral angle
is between two (2,2)-tetrahedra belonging to the four-simplex,
while for the other type (TL2) the dihedral angle is between a
(3,1)- and a (2,2)-tetrahedron. Explicitly, the list of angles is 

\begin{alignat}{2}
\cos\Theta_{(4,1)}^{SL} &= -\frac{i}{2\sqrt{2} \sqrt{3\alpha+1}} &\quad
\sin\Theta_{(4,1)}^{SL} &= \sqrt{ \frac{3(8\alpha +3)}{8(3\alpha+1)}}
\label{angle41s}\\
\cos\Theta_{(4,1)}^{TL} &= \frac{2\alpha+1}{2 (3 \alpha+1)}&\quad
\sin\Theta_{(4,1)}^{TL} &= \frac{ \sqrt{4\alpha+1} \sqrt{8\alpha +3}}
{2(3\alpha +1)}
\label{angle41t}\\
\cos\Theta_{(3,2)}^{SL}& = \frac{6 \alpha +5}{2 (3\alpha +1)}&\quad
\sin\Theta_{(3,2)}^{SL}& = -i\ \frac{\sqrt{3}\sqrt{12 \alpha +7}}
{2 (3 \alpha+1)}
\label{angle32s}\\
\cos\Theta_{(3,2)}^{TL1}& = \frac{4\alpha +3}{4 (2\alpha +1)}&\quad
\sin\Theta_{(3,2)}^{TL1}& = \frac{\sqrt{(4\alpha+1) (12\alpha +7)}}
{4 (2 \alpha +1)}
\label{angle32t1}\\
\cos\Theta_{(3,2)}^{TL2}& =
\frac{-1}{2\sqrt{2(2\alpha +1) (3\alpha +1)}}&\quad
\sin\Theta_{(3,2)} &= \frac{\sqrt{(4\alpha+1)
(12\alpha +7)}}{2\sqrt{2(2 \alpha +1)(3\alpha +1)}}.
\label{angle32t2}
\end{alignat}

\subsection{Topological identities for Lorentzian
triangulations}\label{topo}

In this section we derive some important linear relations
among the ``bulk'' variables $N_{i}$, $i=0,\ldots,d$ which count the
numbers of $i$-dimensional simplices in a given
$d$-dimensional Lorentzian triangulation.
Such identities are familiar from Euclidean dynamically
triangulated manifolds (see, for example, \cite{aj,book1,mauroetal}).
The best-known of them is the Euler identity
\begin{equation}
\chi = N_{0}-N_{1}+N_{2}-N_{3}+\ldots,
\end{equation}
for the Euler characteristic $\chi$ of a simplicial manifold with or
without boundary.
For our purposes, we will need refined versions
where the simplices are distinguished by their Lorentzian properties.
The origin of these relations lies in the simplicial {\it mani\-fold}
structure. They can be derived in a systematic way by establishing
relations among simplicial building blocks in local neighbourhoods and
by summing them over the entire triangulation. Our notation for the
numbers $N_{i}$ is
\begin{eqnarray}
&&N_{0} = \mbox{ number of vertices}\nonumber\\
&&N_{1}^{\rm TL} = \mbox{ number of timelike links}\nonumber\\
&&N_{1}^{\rm SL} = \mbox{ number of spacelike links}\nonumber\\
&&N_{2}^{\rm TL} = \mbox{ number of timelike triangles}\label{nnumbers}\\
&&N_{2}^{\rm SL} = \mbox{ number of spacelike triangles}\nonumber\\
&&N_{3}^{\rm TL_{1}}\equiv N_{3}^{(3,1)} =
\mbox{ number of timelike (3,1)- and (1,3)-tetrahedra}\nonumber\\
&&N_{3}^{\rm TL_{2}}\equiv N_{3}^{(2,2)} =
\mbox{ number of timelike (2,2)-tetrahedra}\nonumber\\
&&N_{3}^{\rm SL} = \mbox{ number of spacelike tetrahedra}\nonumber\\
&&N_{4}^{\rm TL_{1}}\equiv N_{4}^{(4,1)} = \mbox{ number of timelike
(4,1)- and (1,4)-simplices}\nonumber\\
&&N_{4}^{\rm TL_{2}}\equiv N_{4}^{(3,2)} = \mbox{ number of timelike
(3,2)- and (2,3)-simplices}.
\nonumber
\end{eqnarray}

\subsubsection{Identities in 2+1 dimensions}

We will be considering compact spatial slices ${}^{(2)}\Sigma$,
and either open or periodic boundary conditions in time-direction.
The relevant spacetime topologies are therefore $I\times {}^{(2)}\Sigma$
(with an initial and a final spatial surface) and
$S^{1}\times {}^{(2)}\Sigma$. Since the latter results in
a closed three-manifold, its Euler characteristic vanishes. From this we
derive immediately that
\begin{equation}
\chi (I\times {}^{(2)}\Sigma)= \chi ({}^{(2)}\Sigma).
\end{equation}
(Recall also that for closed two-manifolds with $g$ handles, we have
$\chi = 2- 2g$, for example, $\chi (S^{2}) =2$ for the two-sphere.)

Let us for simplicity consider the case of periodic boundary
conditions.
A three-dimensional closed triangulation is characterized by the seven
numbers $N_{0}$, $N_{1}^{\rm SL}$, $N_{1}^{\rm TL}$, $N_{2}^{\rm SL}$,
$N_{2}^{\rm TL}$, $N_{3}^{(3,1)}$ and $N_{3}^{(2,2)}$. Two relations
among them are directly inherited from the Euclidean case, namely,
\begin{eqnarray}
&& N_{0} -N_{1}^{\rm SL} -N_{1}^{\rm TL} +N_{2}^{\rm SL}+
N_{2}^{\rm TL} -N_{3}^{(3,1)} -N_{3}^{(2,2)} =0,\label{cons1}\\
&& N_{2}^{\rm SL}+ N_{2}^{\rm TL} = 2 (N_{3}^{(3,1)}+N_{3}^{(2,2)}).
\label{cons2}
\end{eqnarray}
Next, since each spacelike triangle is shared by two
(3,1)-tetrahedra, we have
\begin{equation}
N_{3}^{(3,1)} =\frac{4}{3} N_{1}^{\rm SL}.
\label{31id}
\end{equation}
Lastly, from identities satisfied by the two-dimensional spatial
slices, one derives
\begin{eqnarray}
&& N_{1}^{\rm SL} =\frac{3}{2} N_{2}^{\rm SL},\\
&& N_{0} =  \chi ({}^{(2)}\Sigma) t+\frac{1}{2} N_{2}^{\rm SL},
\end{eqnarray}
where we have introduced the notation $t$ for the number of
time-slices in the triangulation.

We therefore have five linearly independent conditions on the seven
variables $N_{i}$, leaving us with two ``bulk'' degrees of freedom,
a situation identical to the case of Euclidean dynamical triangulations.
(The variable $t$ does not have the same status as the
$N_{i}$, since it scales (canonically) only like a length, and not like a
volume.)

\subsubsection{Identities in 3+1 dimensions}

Here we are interested in four-manifolds which are of the form of a
product of a compact three-manifold ${}^{(3)}\Sigma$ with either an
open interval or a circle, that is,
$I\times {}^{(3)}\Sigma$ or $S^{1}\times {}^{(3)}\Sigma$. (Note that
because of $\chi ({}^{(3)}\Sigma)=0$, we have
$\chi (I\times {}^{(3)}\Sigma) =\chi (S^{1}\times {}^{(3)}\Sigma)$.
An example is $\chi (S^{1}\times T^{3})\equiv \chi (T^{4})=0$.)
In four dimensions, we need the entire set (\ref{nnumbers}) of ten
bulk variables $N_{i}$. Let us again discuss the linear constraints
among them for the case of periodic boundary conditions in time.

There are three constraints which are inherited from the
Dehn-Sommerville conditions for general four-dimensional triangulations
\cite{aj,book1,mauroetal},
\begin{eqnarray}
\!\!\!\!\!\!\!\!\!\!\!\!
&& \!\!\!\!\!\!\!\!\!
N_{0} -N_{1}^{\rm SL} -N_{1}^{\rm TL} +N_{2}^{\rm SL} +N_{2}^{\rm TL}
- N_{3}^{\rm SL}- N_{3}^{\rm TL_{1}} -
N_{3}^{\rm TL_{2}}+ N_{4}^{\rm TL_{1}}+ N_{4}^{\rm TL_{2}}\! =\! \chi,
\nonumber\\
&&\!\!\!\!\!\!\!\!\!
2 (N_{1}^{\rm SL}\plu N_{1}^{\rm TL}) -\! 3 (N_{2}^{\rm SL}\plu N_{2}^{\rm
TL}) +\! 4(N_{3}^{\rm SL}\plu N_{3}^{\rm TL_{1}}\plu N_{3}^{\rm TL_{2}}) -
\! 5 (N_{4}^{\rm TL_{1}}\plu N_{4}^{\rm TL_{2}}) \!=\! 0,\nonumber\\
&&\!\!\!\!\!\!\!\!\!
 5 (N_{4}^{\rm TL_{1}}+ N_{4}^{\rm TL_{2}}) = 2
(N_{3}^{\rm SL}+ N_{3}^{\rm TL_{1}} +N_{3}^{\rm TL_{2}}).
\end{eqnarray}
The remaining constraints are special to the sliced, Lorentzian
spacetimes we are using. There are two which arise from conditions on
the spacelike geometries alone (cf. (\ref{cons1}), (\ref{cons2})),
\begin{eqnarray}
&&N_{0}- N_{1}^{\rm SL} +N_{2}^{\rm SL} -N_{3}^{\rm SL} =0,\nonumber\\
&&N_{2}^{\rm SL}=2 N_{3}^{\rm SL}.
\end{eqnarray}
Furthermore, since each spacelike tetrahedron is shared by a pair of a
(4,1)- and a (1,4)-simplex,
\begin{equation}
2 N_{3}^{\rm SL} = N_{4}^{(4,1)},
\end{equation}
and since each timelike tetrahedron of type (2,2) is shared by a
pair of (3,2)-simplices, we have
\begin{equation}
2 N_{3}^{\rm TL_{2}} = 3 N_{4}^{(3,2)}.
\end{equation}
In total, these are seven constraints for ten variables.

\subsection{The Einstein-Hilbert action}\label{actions}

We are now ready to construct the gravitational
actions of Lorentzian dynami\-cal triangulations explicitly.
The Einstein-Hilbert action is
\beq\label{3.1}
S_M(\La,G)= \frac{1}{16\pi G}
\int_{M} \d^d\xi \sqrt{-g(\xi)}\; ( R(\xi) -2\La),
\eeq
where $G$ is the gravitational coupling constant and $\La$ the
cosmological constant. We can now formulate
\rf{3.1} on a piecewise linear manifold using \rf{2.3}-\rf{2.5}:
\beq\label{3.2}
S_M(\La,G) = \frac{1}{16\pi G a^2}
\sum_{\sg^{d-2}\in \, T_{d-2}} \Big(  2\ep_{\sg^{d-2}}\;V_{\sg^{d-2}} -
2 \La \; V(\sg^{d-2})\Big),
\eeq
and by introducing the dimensionless quantities
\beq\label{3.3}
\cV(\sg^{d-2}) = a^{-d} V(\sg^{d-2}),~~~\cV_{\sg^{d-2}}a^{2-d},~~~
\kp=\frac{a^{d-2}}{16\pi G},~~~
\lam = \frac{2 \La a^d}{16\pi G},
\eeq
we can write
\beq\label{3.4}
S_M(\lam,\kp;T)=
\sum_{\sg^{d-2}\in \, T_{d-2}}  \Big( \kp \;2\ep_{\sg^{d-2}}\; \cV_{\sg^{d-2}}
-\lam \; \cV(\sg^{d-2})\Big)
\eeq
In each dimension ($d\equ 2,3,4$) we have only a finite number
of building blocks and for each building block we have explicit
expressions for the volume and the deficit angles. Thus we
can find an expression for the action which can be expressed
as a sum over the number of $(d\mi 2)$- and $d$-simplices of the various
kinds times some coefficients. Using the topological
relations between the different simplices we can replace the number of
some of the $(d-2)$-simplices with the number of vertices.
In this way the action becomes simple, depending only on the
global number of simplices and $(d\mi 2)$-subsimplices.
We will now give the explicit
expressions in $d=2,3$ and 4 spacetime dimensions.

\subsection{The action in 2d}

Two-dimensional spacetime is special because we have the
discretized version of the Gauss-Bonnet theorem,
\beq\label{4.1}
\sum_{\sg^{d-2}\in \, T_{d-2}}  \cV_{\sg^{d-2}} \;2\ep_{\sg^{d-2}} =
2\pi \chi,
\eeq
where $\chi$ is the Euler characteristic of the manifold
(including boundaries). Thus
the Einstein term is trivial (the same for all metric configurations) 
when we do not allow the spacetime
topology to change. We will therefore consider only the
cosmological term
\beq\label{4.2}
S_L(\lam;T) =  - \lam \frac{\sqrt{4\a+1}}{4} \;N_2(T),
\eeq
where $N_2(T)$ denotes the number of triangles in the triangulation $T$.

\subsection{The action in three dimensions}

The discretized action in three dimensions becomes
(c.f. \cite{sorkin})
\begin{eqnarray}
S^{(3)}&=&\kp\sum_{\stackrel{\rm spacelike}{l}}\!\!{\rm Vol}(l)\ \frac{1}{i}
\Big(2\pi \mi \!\!\!\sum_{\stackrel{\rm tetrahedra}{{\rm at}\ l}}\Theta\Big)+
\kp\sum_{\stackrel{\rm timelike}{l}}\!\!{\rm Vol}(l)\
\Big(2\pi \mi \!\!\!
\sum_{\stackrel{\rm tetrahedra}{{\rm at}\ l}}\Theta\Big)\nonumber\\
&& - \lambda\;
\Big(\sum_{\stackrel{\rm (3,1)\& (1,3)-}{\rm tetrahedra}}\!\!{\rm Vol}(3,1)
+\sum_{\stackrel{\rm (2,2)-}{\rm tetrahedra}}\!\!{\rm Vol}(2,2)\Big).
\label{act3dis}
\end{eqnarray}
Performing the sums, and taking into account how many tetrahedra meet
at the individual links, one can re-express the action as a function
of the bulk variables $N_{1}$ and $N_{3}$, namely,
\begin{eqnarray}
\lefteqn{S^{(3)}\!\! =\!\! \kp \frac{2\pi}{i} N_{1}^{\rm SL}+}
\label{act3disN}\\
&& -\kp\Big(\frac{2}{i}
N_{3}^{(2,2)} \arcsin\frac{-i\ 2\sqrt{2}\sqrt{2\alpha +1}}{4\alpha
+1}+\frac{3}{i}
N_{3}^{(3,1)} \arccos\frac{-i}{\sqrt{3}\sqrt{4\alpha +1}}\Big)\nonumber\\
&&+\kp\sqrt{\alpha} \Big( 2\pi N_{1}^{\rm TL}-4 N_{3}^{(2,2)}
\arccos \frac{-1}{4\alpha +1} -3 N_{3}^{(3,1)} \arccos
\frac{2\alpha+1}{4\alpha+1} \Big)\nonumber\\
&&-\lambda \Big( N_{3}^{(2,2)}\ \frac{1}{12} \sqrt{ 4\alpha
+2}+N_{3}^{(3,1)}\ \frac{1}{12} \sqrt{3\alpha +1} \Big).
\nonumber
\end{eqnarray}
Our choice for the inverse trigonometric functions with imaginary
argument avoids
branch-cut ambiguities for real, positive $\alpha$.
Despite its appearance, the action (\ref{act3disN})
is {\it real} in the relevant
range $\alpha >0$, as can be seen by applying elementary
trigonometric identities and the relation (\ref{31id}).
The final result for the Lorentzian action can be written as a
function of three bulk variables (c.f. Sec.\ \ref{topo}), for example,
$N_{1}^{\rm TL}$, $N_{3}^{(3,1)}$ and $N_{3}^{(2,2)}$, as
\begin{eqnarray}
S^{(3)}&=& 2\pi \kp\sqrt{\alpha} N_{1}^{\rm TL} \nonumber\\
&-&3 \kp\, N_{3}^{(3,1)}
\Bigl(  {\rm arcsinh}\ \frac{1}{\sqrt{3}\sqrt{4\alpha +1}}
+ \sqrt{\alpha} \arccos \frac{2\alpha+1}{4\alpha +1} \Bigr)\nonumber \\
&+& 2\kp \,N_{3}^{(2,2)}
\Bigl(  {\rm arcsinh}\ \frac{2\sqrt{2}\sqrt{2\alpha +
1}}{4\alpha +1}
-2  \sqrt{\alpha} \arccos \frac{-1}{4\alpha+1} \Bigr)\nonumber\\
&-& \frac{\lambda}{12} \Bigl( N_{3}^{(2,2)} \sqrt{ 4\alpha +2}
 +N_{3}^{(3,1)} \sqrt{3\alpha +1}\Bigr). \nonumber \\
\label{3dloract}
\end{eqnarray}

\subsection{The action in four dimensions}

The form of the discrete action in four dimensions is completely
analogous to (\ref{act3dis}), that is,
\begin{eqnarray}
S^{(4)}&=&k\sum_{\stackrel{\rm spacelike}{ \triangle}}{\rm Vol}(\triangle)
\ \frac{1}{i}
\Big(2\pi \mi \!\!\!\!\!
\sum_{\stackrel{\rm 4-simplices}{{\rm at}\ \triangle}}\!\!\!\Theta\Big)+
k\sum_{\stackrel{\rm timelike}{\triangle}}{\rm Vol}(\triangle)\
\Big(2\pi \mi \!\!\!\!\!
\sum_{\stackrel{\rm 4-simplices}{{\rm at}\ \triangle}}
\!\!\!\Theta\Big)\nonumber\\
&&-\lambda  \sum_{\stackrel{\rm (4,1)\& (1,4)-}{\rm tetrahedra}}
{\rm Vol}(4,1) -
\lambda \sum_{\stackrel{\rm (3,2) \& (2,3)-}{\rm tetrahedra}}{\rm Vol}(3,2).
\label{act4dis1}
\end{eqnarray}
Expressed in terms of the bulk variables $N_{2}$ and $N_{4}$, the
action reads
\begin{eqnarray}
\lefteqn{S^{(4)}=\kp \biggl( \frac{2\pi}{i} \frac{\sqrt{3}}{4} N_{2}^{\rm SL} -
\frac{\sqrt{3}}{4i} N_{4}^{(3,2)} \arcsin\frac{-i\ \sqrt{3}\sqrt{12 \alpha +7}}
{2 (3 \alpha+1)}} \label{act4disN}\\
&& - \frac{\sqrt{3}}{i} N_{4}^{(4,1)}
\arccos\frac{-i}{2\sqrt{2}\sqrt{3\alpha +1}} \biggr)
+ \frac{\kp}{4} \sqrt{4\alpha +1}\
\Biggl( 2\pi N_{2}^{\rm TL} -\nonumber\\
&&-N_{4}^{(3,2)} \Bigl(
6\arccos\frac{-1}{2\sqrt{2}\sqrt{2\alpha +1}\sqrt{3\alpha +1}} +
3\arccos\frac{4\alpha +3}{4 (2\alpha +1)} \Bigr) \nonumber\\
&&-6 N_{4}^{(4,1)} \arccos\frac{2\alpha +1}{2 (3\alpha +1)}\Biggr)
-\lambda \Bigl( N_{4}^{(4,1)}\ \frac{\sqrt{8\alpha +3}}{96} +
N_{4}^{(3,2)}\ \frac{\sqrt{12\alpha +7}}{96} \Bigr) .
\nonumber
\end{eqnarray}
We have again taken care in choosing the inverse functions of the
Lorentzian angles in (\ref{angle41s}) and (\ref{angle32s}) that
make the expression (\ref{act4disN}) unambiguous. Using the manifold
identities for four-dimensional simplicial Lorentzian triangulations
derived in Sec.\ \ref{topo},
the action can be rewritten as a function of the three bulk
variables $N_{2}^{\rm TL}$, $N_{4}^{(3,2)}$ and $N_{4}^{(4,1)}$, in a
way that makes its real nature explicit,
\begin{eqnarray}
\lefteqn{S^{(4)} = \kp \sqrt{4\alpha +1}
\Biggl[\frac{\pi}{2} N_{2}^{\rm TL}+}   \label{4dloract}\\
&& N_{4}^{(4,1)}
\Bigl( -\frac{\sqrt{3}}{\sqrt{4\alpha +1}}
{\rm arcsinh} \frac{1}{2\sqrt{2}\sqrt{3\alpha +1}}
-\frac{3 }{2} \arccos\frac{2\alpha +1}{2 (3\alpha +1)}
\Bigr)+\nonumber \\
&&N_{4}^{(3,2)}
\Biggl( \frac{\sqrt{3}}{4\sqrt{4\alpha +1}}
{\rm arcsinh}\frac{\sqrt{3}\sqrt{12 \alpha +7}}
{2 (3 \alpha+1)} - \nonumber\\
&&~~~~~~~\frac{3 }{4}
\biggl( 2\arccos\frac{-1}{2\sqrt{2}\sqrt{2\alpha +1}\sqrt{3\alpha +1}} +
\arccos\frac{4\alpha +3}{4 (2\alpha +1)}\biggr) \Biggr)\nonumber\\
&&-\lambda\;\Big( N_{4}^{(4,1)}\frac{\sqrt{8\alpha +3}}{96}+
 N_{4}^{(3,2)} \frac{\sqrt{12\alpha +7}}{96}\Big).
\nonumber
\end{eqnarray}
It is straightforward to verify that this action is real for real
$\alpha \geq -\frac{1}{4}$, and purely imaginary for $\alpha\in\R$,
$\alpha \leq -\frac{7}{12}$. Note that this implies that we could
in the Lorentzian case choose to work with building blocks possessing
lightlike (null) edges ($\alpha =0$) instead of timelike edges, or even
work entirely with building blocks whose edges are all spacelike.

\subsection{Rotation to Euclidean signature}

The standard rotation from Lorentzian to Euclidean signature
in quantum field theory is given by
\beq\label{5.1}
t_L \mapsto -i t_E,~~~~t>0,~t_E >0,~~~~({\rm or}~~t<0,~t_E <0)
\eeq
in the complex lower half-plane (or upper half-plane).
The rotation \rf{5.1} implies
\bea\label{5.0}
S_L(\La,G) &=& \frac{1}{16\pi G}
\int_{M} \d^d\xi \sqrt{-g(\xi)}\; ( R_L(\xi) -2\La) \nonumber\\
 \mapsto iS_E(\La,G) &=& \frac{i}{16\pi G}
\int_{M} \d^d\xi \sqrt{g(\xi)}\; ( -R_E(\xi) +2\La)
\eea
for the Einstein-Hilbert action.
This translation to Euclidean signature is of course formal.
As already explained in Sec.\ \ref{wicki} above, for a given metric $g_{\m\n}(x_i,t)$ 
there is in general no suitable analytic continuation of this kind. 
However, it turns out that
our particular geometries {\it do} allow for a continuation
from a piecewise linear geometry with Lorentzian signature to
one with Euclidean signature and such
that \rf{5.0} is satisfied. The form this prescription takes is the one
given in eq.\ \rf{0.12} for $\a$ not too small, as we will now show.

In our geometric notation where we have a proper time
\beq\label{5.2}
dt = \sqrt{\a} dl,~~~~dt >0~~~\a > 0,~~~\sqrt{\a} >0,
\eeq
we see that \rf{5.1} corresponds to a rotation $\a \mapsto -\a$
in the complex lower-half plane, such that $\sqrt{-\a} = -i\sqrt{\a}$.
Treating the square roots in this way we can now perform the
analytic continuation to Euclidean signature, resulting in
the length assignments 
\beq\label{5.3}
a_t^2 = -\a a^2 \mapsto \ta a^2,~~~a_s=a,~~~~\ta>0
\eeq
to ``timelike'' and spacelike links. 
After this rotation all invariant length assignments are positive,
contrary to the Lorentzian situation where we made the explicit
choice for some of the edges to be timelike with 
$a_t^2 < 0$.\footnote{This was not strictly necessary - 
in Lorentzian signature one can also have building blocks all of
whose links are space- or timelike.}
However, while all $\a >0$ are allowed in the
Lorentzian case, we would like the rotated Euclidean simplices to
be realizable in flat Euclidean $\R^d$, i.e.\ we want the triangle inequalities
to be satisfied for all simplices. For a triangle in two dimensions this implies that
$\ta > 1/4$ since else the total length $2a_t \equ 2\sqrt{\ta} a$ of the two
``timelike'' sides of the triangle is less than the length $a_s \equ a$
of the spacelike side. Similar constraints exist in three and four
dimensions, namely,
\bea
d=2:~~~\ta &>& \frac{1}{4} \nonumber\\
d=3:~~~\ta &>& \frac{1}{2} \nonumber\\
d=4:~~~\ta &>& \sqrt{\frac{7}{12}}.\nonumber\\
\eea
Assuming these constraints on the values of $\a,\ta$ to be satisfied, 
we can now rotate the Lorentzian actions to Euclidean signature.

\subsubsection{The Euclidean action in 2d}

Using the prescription given above, the Lorentzian action \rf{4.2} is 
readily rotated to Euclidean signature, resulting in
\beq\label{6.1}
S_E(T) = \lam \;\frac{\sqrt{4\ta-1}}{4} N_2(T).
\eeq
We have explicitly
\beq\label{6.1a}
S_L(T;\a) \to S_L(T,-\a) = iS_E(T;\ta),~~~~\a=\ta >0.
\eeq

\subsubsection{The Euclidean action in three dimensions}

The analytic continuation of \rf{3dloract} becomes
\begin{eqnarray}
S^{(3)}_E(\ta) &=& -2\pi \kp\sqrt{\ta} N_{1}^{\rm TL} \nonumber\\
&+&N_{3}^{(3,1)}
\Bigl( -3 \kp\ {\rm arcsin}\ \frac{1}{\sqrt{3}\sqrt{4\ta -1}}
+3 k \sqrt{\alpha} \arccos \frac{2\ta-1}{4\ta -1}\Bigr)\nonumber \\
&+&N_{3}^{(2,2)}
\Bigl( 2 \kp\ {\rm arcsin}\ \frac{2\sqrt{2}\sqrt{2\ta -1
1}}{4\ta-1}
+4 \kp \sqrt{\ta} \arccos \frac{1}{4\ta-1} \Bigr) \nonumber \\
&+&\frac{\lambda}{12} \Bigl( N_{3}^{(2,2)}\sqrt{ 4\ta -2} +
\frac{\lambda}{12} N_{3}^{(3,1)}\sqrt{3\ta-1}\Bigr).
\label{3dact}
\end{eqnarray}
The terms which are multiplied by the coupling constant $\kp$
constitute the Einstein term while the terms multiplied by
$\lam$ make up the cosmological term.
Again one has explicitly
\beq\label{7.0}
S_L(T;\a) \to S_L(T,-\a) = iS_E(T;\ta),~~~~\a=\ta >0.
\eeq

The expression \rf{3dact} simplifies considerably
when $\ta\equ 1$, in which case all three-simplices 
(tetrahedra) are identical and equilateral, yielding
\beq\label{7.1}
S^{(3)}_E(\ta=1)=
 -2 \pi  \kp N_1 +N_3 (6  \kp \arccos \frac{1}{3} +
\frac{\sqrt{2}}{12}\lambda ).
\eeq
One recognizes $\arccos \frac{1}{3}$ as the dihedral angle of an equilateral
tetrahedron, and the term $2\pi N_1$ as coming from the $2\pi$ which enters
in the definition of the deficit angle associated with a link. One
can replace the number of links by the number of vertices using \rf{cons1}
and \rf{cons2}, obtaining
\beq\label{7.2}
N_0-N_1+N_3=0.
\eeq

\subsubsection{The Euclidean action in four dimensions}\label{subsecd4act}

Finally, the analytic continuation of \rf{4dloract} becomes
\begin{eqnarray}
\lefteqn{ S_E=-\kp\sqrt{4\tilde\alpha -1}\Biggl[\pi \Big(N_0-\chi
+\oh N_{4}^{(4,1)} +N_{4}^{(3,2)}\Big)+ }\label{act4dis}\\
&&  N_{4}^{(4,1)}\biggl( -\frac{\sqrt{3}}{\sqrt{4\tilde\alpha -1}}
\arcsin\frac{1}{2\sqrt{2}\sqrt{3\tilde\alpha -1}} +
\frac{3}{2} \arccos\frac{2\tilde\alpha -1}{6\tilde \alpha -2}\biggr)+
\nonumber\\
&&N_{4}^{(3,2)} \biggl(
+\frac{\sqrt{3}}{4\sqrt{4\tilde\alpha -1}}\arccos
\frac{6\tilde\alpha -5}{6\tilde \alpha -2}
+\frac{3}{4} \arccos\frac{4\tilde\alpha -3}{8\tilde \alpha -4}+
\nonumber\\
&&\frac{3}{2}
\arccos\frac{1}{2\sqrt{2}\sqrt{2\tilde\alpha \mi 1}
\sqrt{3\tilde\alpha \mi 1}}\biggr)
\Biggr]
 \plu\lambda\Big( N_{4}^{(3,2)}\frac{\sqrt{12\tilde\alpha \mi 7}}{96}
\plu N_{4}^{(4,1)}\frac{\sqrt{8\tilde\alpha \mi 3}}{96}\Big).
\nonumber
\end{eqnarray}
In this expression, $\chi$ is the Euler characteristic of the piecewise flat
four-manifold and $\tilde\alpha \equiv -\alpha$ denotes the positive
ratio between the two types of squared edge lengths after the
Euclideanization. In order to satisfy the triangle inequalities, we need
$\tilde\alpha >7/12$ as noted above. For simplicity,
we have assumed that the manifold
is compact without boundaries. In the presence of boundaries, appropriate
boundary terms must be added to the action.

\begin{figure}[t]
\centering
\includegraphics[width=0.6\textwidth]{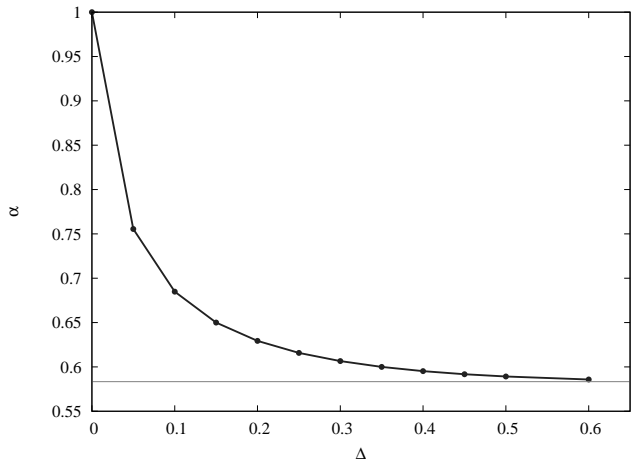}
\caption{\label{figalfa} {\footnotesize The asymmetry factor $\alpha$,
plotted as a function of $\Del$, for $\kappa_0=2.2$.
The horizontal line is $\ta=7/12$, the lowest kinematically allowed value
of $\ta$ (see \rf{5.3}), where
the (3,2)-simplices degenerate because of a saturation of a triangle
inequality.}
}
\end{figure}

For the simulations, a convenient alternative parametrization of the action
is given by
\begin{equation}
S_E=-(\kappa_0+6\Delta) N_0+\kappa_4 (N_{4}^{(4,1)}+N_{4}^{(3,2)})+
\Delta (2 N_{4}^{(4,1)}+N_{4}^{(3,2)}),
\label{actshort}
\end{equation}
where the functional dependence of the $\kappa_i$ and $\Delta$ on the
bare inverse Newton constant $\kp$, the bare cosmological constant
$\lambda$ and $\tilde\alpha$ can be computed from \rf{act4dis}. We have
dropped the constant term proportional to $\chi$, because it will be irrelevant
for the quantum dynamics. Note that $\Delta\equ 0$ corresponds to
$\tilde\alpha\equ 1$, and $\Delta$ is therefore a measure of the asymmetry
between the lengths of the spatial and timelike edges of the simplicial
geometry if we {\it insist} that \rf{actshort} represents
the Regge version of the Einstein-Hilbert
action on the piecewise linear geometry.
Given $\kp_0,\kp_4$ and $\Del$ one can go from \rf{actshort} to \rf{act4dis}
and find $\ta$ (and $\kp$ and $\lam$).
In Fig.\ \ref{figalfa} we have shown such an inversion. It
corresponds to values of $\kp_0,\kp_4$ and $\Del$ actually used in
computer simulations which we will discuss later. This is why the inversion
is only performed for a finite number of points which are connected
by linear interpolation. More precisely, they correspond to
a given value of $\kp_0$ ($\kp_0\equ 2.2$) and various values 
of $\Delta$. Given $\kp_0$ and $\Delta$, the value of $\kp_4$ used is
the so-called critical value $\kp_4(\kp_0,\Del)$, where the
statistical system becomes critical, as will be discussed in detail below.

\section{The transfer matrix}\label{trans}

We now have all the necessary prerequisites to study the
amplitude or (after rotation to Euclidean signature) 
the partition function for pure gravity for our CDT model, that is, 
\begin{equation}
Z~=~~\sum_{ T} ~\frac{{\rm e}^{iS(T)}}{C(T)}~~~
\to ~~~\sum_{T} ~\frac{{\rm e}^{-S(T)}}{C(T)}.\label{part}
\end{equation}
The class of triangulations appearing in the sums are as described
above corresponding to our CDT model.
The weight factor is the standard phase factor given by the classical
action (or the corresponding Boltzmann factor after rotation to Euclidean signature),
except when the triangulation has a special symmetry. If the symmetry group
(also called the automorphism group) of the triangulation $T$ has $C(T)$
elements we divide by $C(T)$.
One may think of this factor as the remnant of the division by the volume
of the diffeomorphism group Diff($M$) that would occur in a formal
gauge-fixed continuum expression for $Z$. Its effect is to suppress geometries
possessing special symmetries. This is analogous to what
happens in the continuum where the diffeomorphism orbits through
metrics with special isometries are smaller than the ``typical'' orbits,
and are therefore of measure zero in the quotient space
Metrics$(M)$/Diff$(M)$. 

One way to see how the nontrivial measure factor $C(T)^{-1}$ arises is as
follows. The triangulations we have been discussing so far can be associated
with  {\it unlabeled} graphs. Working with {\it labeled}
triangulations is the discrete counterpart of
introducing coordinate systems in the continuum, and is often convenient
from a practical point of view (for example, when storing geometric data in
a computer). One possibility is to label the vertices.
A link is then identified by a pair
of labels, a triangle by three labels, etc. If we have a triangulation
$T$ with $N_0(T)$ vertices, there are $N_0(T)!$ possible ways to label
the same triangulation. Two labeled triangulations are said to
represent the same abstract triangulation if there is a one-to-one map between
the labels, such that also links are mapped to links, triangles to
triangles, etc. If we work with labeled triangulations,
we only want to count physically distinct ones once and therefore 
want to divide by
$N_0(T)!$. This can be thought of as the discrete analogue of dividing by the
volume of the diffeomorphism group Diff($M$), since also in the
standard continuum formulation we only want to count geometries, not the number of
parametrizations of the same geometry. We can write
\beq\label{diff}
Z = \sum_{T_l} \frac{1}{N_0(T_l)!} \; \e^{-S(T_l)} =
\sum_{ T} \frac{1}{C(T)}\;{\rm e}^{-S(T)},
\eeq
where the first summation is over labeled triangulations, while
the second sum is over abstract triangulations.
The reason for the presence of the factor $C(T)$ in the expression on the right
is that any symmetries
of the graph $T$ make the cancellation of the factor
$N_0(T)!$ incomplete. 

As an example, consider the simplest
triangulation of the two-sphere, as the boundary of a single tetrahedron, which
contains four triangles, six links and four vertices. Let us
label the vertices 1 to 4. The triangulation is defined assigning
(unordered) labels $(i,j)$, $i\neq j$, to the links,
and (unordered) labels $(i,j,k)$, $i \neq j \neq k$,
to the triangles, $i,j,k$ taking values 1 to 4. We can perform
4! permutations of the labeling of the vertices.
However, no matter which of the 4! permutations  we consider
we end up with the same labeled triangulation since the list of
vertices, links and triangles will be identical except for a possible
ordering. Thus $C(T) \equ 24$.

The most natural boundary conditions for $Z$ in our discretized
model are given by specifying the spatial piecewise linear
geometries  at given initial and final proper times, which
we take as 0 and $t$. We assume, as we have done above, that
the slices at constant proper time $t$ are by construction spacelike and of fixed
topology, and defined by (unlabeled)
$(d\mi 1)$-dimensional triangulations.
Given an initial and a final geometry, in the form of triangulations
$T_{d-1}(1)$ and $T_{d-1}(2)$,
we may think of the corresponding amplitude $Z$ as the matrix element
of the quantum propagator of the system, evaluated between two
states $|T_{d-1}(1)\rangle$ and $|T_{d-1}(2)\rangle$. Since we regard
distinct spatial triangulations $T_{d-1}(i)$ as physically
inequivalent, a natural scalar product is given by
\beq\label{norm1}
\la T_{d-1}(1) | T_{d-1}(2) \ra = \frac{1}{C(T_{d-1}(1))}\;
\delta_{T_{d-1}(1),T_{d-1}(2)},~~~~
\eeq
\beq
\sum_{T_{d-1}} C(T_{d-1}) \; |T_{d-1}\ra \la T_{d-1} | = \hat{1}.
\label{norm}
\end{equation}
In line with our previous reasoning, we have included a symmetry
factor $C(T_{d-1})$ for the spatial triangulations.

In the regularized context, it is natural to have a cutoff on the
allowed size of the spatial slices so that their volume is
$\leq N$. The spatial discrete volume Vol($T_{d-1}$) is simply the number of
$(d\mi 1)$-simplices in a slice of constant integer $t$, $N_{d-1}(t)$
(times a factor proportional to $a^{d-1}$ which we will not write).
We define the finite-dimensional Hilbert space
$H^{(N)}$ as the space spanned by
the vectors $\{ |T_{d-1}\rangle,\ N_{\rm min}\leq N_{d-1}(T) \leq N \}$,
endowed with the scalar product (\ref{norm}). The lower bound
$N_{\rm min}$ is the minimal size of a spatial triangulation of
the given topology satisfying the simplicial manifold conditions.
It is believed that
the number of states in $H^{(N)}$ is exponentially bounded as a
function of $N$ \cite{expobound}.

For given volume-cutoff $N$, we can now associate
with each time-step  $\Delta t \equ 1$ a transfer matrix
${\hat T}_N$ describing the evolution of the system from $t$ to $t\plu 1$,
with matrix elements
\beq\label{transfer}
\langle T_{d-1}(2)|\hat T_N(\alpha ) |T_{d-1}(1)\rangle
=\sum_{T_d:\ T_{d-1}(1) \rightarrow T_{d-1}(2)}  \frac{1}{C(T_d)}\
\e^{-\Delta S_\alpha (T_d)}.
\eeq
The sum is taken over all distinct interpolating $d$-dimensional
triangulations $T_d$ from $T_{d-1}(1)\rightarrow T_{d-1}(2)$
of the ``sandwich" with boundary geometries $T_{d-1}(1)$
and $T_{d-1}(2)$, contributing $\Delta S$
to the action, according to (\ref{3dact}),
(\ref{act4dis}).
The ``propagator'' $G_N(T_{d-1}(1),T_{d-1}(2);t)$ for arbitrary time intervals
$t$ is obtained by iterating the transfer matrix $t$ times,
\beq\label{prop}
G_N(T_{d-1}(1),T_{d-1}(2);t)=\langle T_{d-1}(2) |\hat T_N^t|T_{d-1}(1)\rangle,
\eeq
and satisfies the semigroup property
\bea\label{iter}
\lefteqn{G_N(T_{d-1}(1),T_{d-1}(2);t_1+t_2)=}\\
&&
\sum_{T_{d-1}} C(T)\
G_N(T_{d-1}(1),T_{d-1};t_1)\ G_N(T_{d-1},T_{d-1}(2);t_2),
\nonumber
\eea
where the sum is over all spatial geometries (of bounded volume) at
some intermediate time.
Because of the appearance of {\it different} symmetry factors in
(\ref{transfer}) and (\ref{iter}) it is at first not obvious why
the composition property (\ref{iter}) should hold.
In order to understand that it does, one has to realize that
by continuity and by virtue of the manifold property there are no
nontrivial automorphisms of a sandwich
$T = T_{d-1}(1)\rightarrow T_{d-1}(2)$ that
leave its boundary invariant. It immediately follows that the (finite)
automorphism group of $T$ must be a subgroup
of the automorphism groups of both of the boundaries, and that
therefore $C(T)$ must be a divisor of both $C(T_{d-1}(1))$ and
$C(T_{d-1}(2))$.
It is then straightforward to verify that the factor $C(T)$
appearing in the composition law (\ref{iter}) ensures that
the resulting geometries appear exactly with the symmetry
factor they should have according to (\ref{transfer}).

If $d\equ 2$ one can solve analytically for the transfer matrix and
the propagator as we will discuss in detail below.

\subsection{Properties of the transfer matrix}\label{proper}

In this section we will establish some desirable properties of
the transfer matrix $\hat T_{N}(\alpha)$, defined in (\ref{transfer}),
which will enable us to define a self-adjoint Hamiltonian operator.
In statistical mechanical models the Hamilton operator $\hat{h}$ is
usually defined by
\beq\label{transmatrix}
\hat T={\rm e}^{-a\hat h},~~~{\rm i.e.}~~~\hat{h} = -\frac{1}{a} \ln \hat T ,
\eeq
where $a$ denotes the lattice spacing in
time-direction.
A set of sufficient conditions on $\hat T_{N}$
guaranteeing the existence of a
well-defined quantum Hamiltonian $\hat h$ are that the transfer
matrix should satisfy
\begin{itemize}
\item[(a)] {\it symmetry}, that is, $\hat T_{N}^{\dagger}=\hat T_{N}$.
This is the same as self-adjointness since the Hilbert space $H^{(N)}$
which $\hat T_{N}$ acts on is finite-dimensional. It is necessary if the
Hamiltonian is to have real eigenvalues.
\item[(b)] {\it Strict positivity} is required in addition to (a),
that is, all eigenvalues must be greater than zero; otherwise,
$\hat h_{N}=-a^{-1} \log \hat T_{N}$ does not exist.
\item[(c)] {\it Boundedness}, that is, in addition to (a) and (b),
$\hat T_{N}$ should be bounded above
to ensure that the eigenvalues of the Hamiltonian are bounded below,
thus defining a stable physical system.
\end{itemize}
Establishing (a)-(c) suffices to show that our discretized systems are
well-defined as regularized statistical models. This of course does
not imply that they will possess interesting continuum limits and that
these properties will necessarily persist in the limit (as would be desirable).
On the other hand, it is difficult to imagine how the continuum limit
could have such properties unless also the limiting sequence of
regularized models did.

All of the above properties are indeed satisfied for the Lorentzian
model in $d\equ 2$  where moreover the quantum Hamiltonian and
its complete spectrum in the continuum limit are known explicitly
\cite{nakayama,lottietal}, as we will discuss below.
In $d>2$ we are able to prove a slightly weaker statement than
the above, namely, we can verify (a)-(c) for the
two-step transfer matrix $\hat T_N^2$. This is still
sufficient to guarantee the existence of a well-defined Hamiltonian.
Note that self-adjointness of the continuum
Hamiltonian $\hat H$ implies a unitary time evolution operator
${\rm e}^{-i\hat H T}$
if the continuum proper time $T$ is analytically continued.
As was mentioned in the introduction this is an important point
because  we do not at the present stage know the
effective field theory which is associated with a
conjectured UV fixed point. This effective field theory could
contain an arbitrary number of higher-derivative terms and the question
of unitarity becomes an issue, as discussed in Sec.\ \ref{intro}.
The existence of a unitary time evolution operator ensures unitarity.

One verifies the symmetry of the transfer matrix by inspection of
the explicit form of the matrix elements (\ref{transfer}). The
``sandwich actions'' $\Delta S_{\alpha}$ as functions of the boundary
geometries $T_{d-1}(1)$, $T_{d-1}(2)$
in three and four dimensions can be read
off from (\ref{3dloract}) and (\ref{4dloract}).
To make the symmetry explicit, one may simply rewrite these
actions as separate functions of the simplicial building blocks and
their mirror images under time-reflection (in case they are
different). Likewise, the symmetry factor $C(T)$ and the counting
of interpolating geometries in the sum over $T$ are invariant
under exchange of the in- and out-states, $|T_{d-1}(1)\rangle$ and
$|T_{d-1}(2)\rangle$.

Next, we will discuss the
reflection (or Osterwalder-Schrader) positivity
\cite{osterwalder} of our model,
with respect to reflection at planes of constant integer and
half-integer time (see also \cite{momu} and references therein).
These notions
can be defined in a straightforward way in the Lorentzian model
because it possesses a distinguished notion of (discrete proper) time.
Reflection positivity implies the {\it positivity} of the transfer
matrix, $\hat T_{N}\geq 0$.

``Site reflection'' denotes the reflection $\theta_{s}$
with respect to a spatial
hypersurface of constant integer-$t$ (containing ``sites'', i.e.
vertices), for example, $t\equ 0$, where it takes the form
\begin{equation}
\theta_{s}:\ t\rightarrow -t.
\label{sreflect}
\end{equation}
Let us accordingly split any triangulation $T$ along this hypersurface,
so that $T^{-}$ is the triangulation with $t\leq 0$ and $T^{+}$ the one
with $t\geq 0$, and
$T^{-} \cap T^{+} = T_{d-1}(t\equ 0)\equiv T^0_{d-1}$, where
$T^0_{d-1}$ denotes a spatial triangulation at $t\equ 0$.
Consider now functions $F$ that depend only on $T^{+}$ (that is,
on all the connectivity data specifying $T^{+}$ uniquely, in some
parametrization of our choice). Site-reflection positivity means the
positivity of the Euclidean expectation value
\begin{equation}
\langle (\theta_{s}F) F\rangle \geq 0,
\label{fspos}
\end{equation}
for all such functions $F$. The action of $\theta_{s}$ on a function
$F(T^+)$ is defined by anti-linearity,
$(\theta_{s}F)(T^-):= \bar F(\theta_{s}(T^-))$.
By virtue of the composition property (\ref{iter}), we can write
\begin{eqnarray}
\lefteqn{\langle (\theta_{s}F) F\rangle = Z^{-1} \sum_{T} \frac{1}{C(T)}
(\theta_{s}F) F\ {\rm e}^{-S(T)} } \label{scal}\\
&&
=Z^{-1} \sum_{T^0_{d-1}} C(T^0_{d-1}) \!\!\!\!\sum_{
{ T^{-}\atop T^{-}(t=0)=T^0_{d-1} } }
 \!\!\!\!\frac{(\theta_{s}F)(T^{-})}{C(T^{-})}\ {\rm e}^{-S(T^{-})}
 \!\!\!\!\sum_{{ T^{+}\atop T^{+}(t=0)=T^0_{d-1} } }
 \!\!\!\!\frac{F(T^{+})}{C(T^{+})}\ {\rm e}^{-S(T^{+})}\nonumber\\
&&
=Z^{-1} \sum_{T^0_{d-1}} C(T^0_{d-1})
\bar{\cal F}(T^0_{d-1}) {\cal F}(T^0_{d-1})
\geq 0. \nonumber
\end{eqnarray}
The equality in going to the last line
holds because both the action and the symmetry factor $C$
depend on ``outgoing'' and ``incoming''
data in the same way (for example, on (m,n)-simplices in the same way
as on (n,m)-simplices). Note that the standard procedure of
extracting a scalar product and a Hilbert space from (\ref{scal})
is consistent with our earlier definition (\ref{norm}). One
associates (in a many-to-one fashion) functions $F(T^{+})$ with
elements $\Psi_{F}$ of a Hilbert space at fixed time $t$ with
scalar product $\langle\cdot ,\cdot\rangle$, where
$\langle \Psi_{F},\Psi_{G}\rangle =\langle (\theta_{s}F)G\rangle$
\cite{osterwalder}. A set of representative functions which
reproduces the states and orthogonality relations (\ref{norm}) is
given by
\begin{equation}
F_{T_{d-1}}(T^{+})=\left\{ {1/\sqrt{C(T_{d-1})},\;\;\; T^{+}(t=0)=T_{d-1}\atop
0\;\;\;\;\;\;\;\;\;\;\; {\rm otherwise,} } \right.
\end{equation}
which can be verified by an explicit computation of the expectation
values $\langle (\theta_{s}F_{T_{d-1}'})F_{T_{d-1}''}\rangle$.
We have therefore proved site-reflection positivity of our model.
This is already enough to construct a Hamiltonian from
the square of the transfer matrix $\hat T_{N}$ (see eq.\ (\ref{ham2})
below), since it implies the positivity of $\hat T_{N}^{2}$ \cite{momu}.

Proving in addition link-reflection positivity (which would imply
positivity of the ``elementary'' transfer matrix, and not only of
$\hat T_{N}^{2}$) turns out to be more involved.
A ``link reflection'' is the reflection
$\theta_{l}$ at a hypersurface of
constant half-integer-$t$, for example, $t\equ 1/2$,
\begin{equation}
\theta_{l}:\ t\rightarrow 1-t.
\label{lreflect}
\end{equation}
To show link-reflection positivity in our model we would need
to demonstrate that
\begin{equation}
\langle (\theta_{l}F) F\rangle \geq 0,
\label{flpos}
\end{equation}
where $F$ is now any function that depends only on the part
$T^+$ of the triangulation $T$ at times later or equal to 1.
We can again write down the expectation value,
\begin{eqnarray}
\lefteqn{\langle (\theta_{l}F) F\rangle =
Z^{-1} \sum_{T_{d-1}^0} \sum_{T_{d-1}^1}
C(T_{d-1}^0) C(T_{d-1}^1) G_N(T_{d-1}^0,T_{d-1}^1;1)\times}\nonumber\\
&& ~~~~~~~\sum_{ { T^{-}\atop T^{-}(t=0)=T_{d-1}^{0} } }
\!\!\!\!\frac{(\theta_{s}F)(T^{-})}{C(T^{-})}\ {\rm e}^{-S(T^{-})}
\!\!\!\!\sum_{{ T^{+}\atop T^{+}(t=1)=T_{d-1}^{1} } }
\!\!\!\!\frac{F(T^{+})}{C(T^{+})}\ {\rm e}^{-S(T^{+})}.
\label{linkref}
\end{eqnarray}
In order to show that this is positive, one should try to
rewrite the right-hand side as a sum of positive terms.
The proof of this is
straightforward in $d=2$ (see Appendix\ 2 for details),
but it is considerably more
difficult to understand what happens in higher dimensions.
The reason for this is the nontrivial way in which the various
types of simplicial building blocks fit together in between
slices of constant integer-$t$. In a way, this is a desirable
situation since it means that there is a more complicated
``interaction'' among the simplices. It is perfectly possible
that $\hat T_{N}$ itself is {\it not} positive for
$d=3,4$. This may depend both on the values of the bare
couplings and on the detailed choices we have made as part of
our discretization.
(By contrast, it is clear from our proof of site-reflection positivity
that this property is largely independent of the choice of building
blocks.)

Nevertheless, as already mentioned above, site-reflection positivity
is perfectly sufficient for the construction of a well-defined
Hamiltonian. So far, we have only shown that the eigenvalues of the
squared transfer matrix are positive. In order for the
Hamiltonian
\begin{equation}
\hat h'_N := -\frac{1}{2a} \log \hat T_N^2
\label{ham2}
\end{equation}
to exist, we must achieve strict positivity.
We do not expect that the Hilbert space $H^{(N)}$ contains any
zero-eigenvectors of $\hat T_N$
since this would entail a ``hidden''
symmetry of the discretized theory. It is straightforward to
see that none of the basis vectors $|T_{d-1}\rangle$ can
be zero-eigenvectors. However, we cannot in principle
exclude ``accidental'' zero-eigenvectors of the form of linear
combinations
$\sum_i \alpha_i |T_{d-1}(i)\rangle$. In case such vectors exist, we will
simply adopt the standard procedure of
defining our physical Hilbert space as the quotient space
$H^{(N)}_{ph}=H^{(N)}/{\cal N}^{(N)}$, where ${\cal N}^{(N)}$
denotes the span of all zero-eigenvectors.

Lastly, the boundedness of the transfer matrix (and therefore also of
$\hat T_{N}^{2}$) for finite spatial volumes
$N$ follows from the fact that (i) there is only a finite number of
eigenvalues since the Hilbert space $H^{(N)}$ is finite-dimensional,
and (ii) each matrix element
$\langle T_{d-1}(2)|\hat T_{N}|T_{d-1}(1)\rangle$
has a finite value, because it has the form of a {\it finite} sum of
terms ${\rm e}^{-S}/C(T)$. (Note that this need not be
true in general if we abandoned the simplicial manifold restrictions,
because then the number of interpolating geometries for given, fixed boundary
geometries would not necessarily be finite.)

\section{A two-dimensional toy model}
\label{2d}

The two-dimensional CDT model is easy to solve by combinatorial
methods \cite{al,alet}. Since we do not allow topology changes the action is
simply given by the cosmological term, eq.\ \rf{4.1}, or after
rotation to Euclidean signature, eq.\ \rf{6.1}.

In the following we will assume that we have performed the rotation to
Euclidean signature, and we will in addition put $\ta \equ 1$ since
a different $\ta$ clearly can be absorbed in redefinition of the
cosmological constant.

Our manifold has the topology of a cylinder and we want to find
the ``propagator'' defined by eq.\ \rf{prop}. The ``triangulations''
of the boundaries are simply specified by the number of links
constituting the boundaries. This is in agreement with the
continuum description: the geometry of the boundary $S^1$ is
entirely specified by its {\it length}. In our discretized theory
the length of a boundary will be $L(l) = a \cdot l$ where $l$ denotes
the number of links of the boundary. Similarly, with $\ta\equ 1$, the
volume of spacetime (the area of the surface) will be
$V = \sqrt{3} a^2/4 \cdot n$, where $n$ denotes the number of triangles
of the triangulation. The action, i.e.\ the cosmological term,
for a given triangulation $T$ with $n(T)$ triangles, will then be
(see \rf{6.1})
 \beq\label{disact}
S = \La_0 \frac{\sqrt{3}a^2}{4}\;n(T) = \lam \;n(T),~~~~
\lam \equiv  \La_0 \frac{\sqrt{3}a^2}{4},
\eeq
where $\La_0$ is the {\it bare} dimensionful cosmological coupling
constant, while $\lam$ is a convenient dimensionless coupling
constant (which differs by a factor $\sqrt{3}/4$ from the one used
in \rf{6.1}). Shortly, we will define the {\it renormalized} dimensionful
cosmological constant.

As stated above the boundaries will be characterized by
integers $l_1$ and $l_2$, the number of vertices or links at the two
boundaries. The path integral amplitude for the ``propagation'' from
geometry $l_1$ to $l_2$ will be the sum over all interpolating
surfaces of the
kind described above, with a weight given by
\beq\label{1}
\e^{-S(T)} = \e^{-\lam n(T)},
\eeq
according to \rf{disact}.
Let us call the corresponding amplitude $G^{(1)}_\lam(l_1,l_2)$.
Thus we have
\bea
G_\lam^{(1)}(l_1,l_2;t) &=&
\sum_{l=1}^\infty G_\lam^{(1)}(l_1,l;1)\;l\;G_\lam^{(1)}(l,l_2,t-1),\label{4}\\
G_\lam^{(1)}(l_1,l_2;1) &=& \frac{1}{l_1}\sum_{\{k_1,\dots,k_{l_1}\}}
\e^{-\lam  \sum_{i=1}^{l_1} k_i}, \label{5}
\eea
where the factors $l$ and $l_1$ appearing in \rf{4} and \rf{5} are
the factors $C(T)$ appearing in \rf{iter} and \rf{transfer}, while
the $k_i$ is the number of timelike links connected to vertex $i$
on the entrance loop (see Fig.\ \ref{connect}, where it will be clear
that the sum of $k_i$'s is just the total number of triangles in
the ``slab'' interpolating between
the two boundaries of lengths $l_1$ and $l$).
From a combinatorial point of view it is convenient to mark a
vertex on the entrance loop in order to get rid of these factors, that is,
\beq\label{6}
G_\lam (l_1,l_2;t) \equiv l_1 G_\lam^{(1)}(l_1,l_2;t).
\eeq
$G_\lam(l_1,l_2;1)$ is of course precisely the transfer
matrix defined by \rf{transfer} (up to the factor $C(T)$), satisfying
\bea
G_\lam(l_1,l_2,t_1+t_2)
&=& \sum_{l} G_\lam(l_1,l;t_1)\; G_\lam(l,l_2;t_2)\label{7}\\
G_\lam(l_1,l_2;t+1) &=& \sum_{l} G_{\lam}(l_1,l;1)\;G_\lam(l,l_2;t).\label{8}
\eea
It is defined by summing over all piecewise linear geometries of a
``slab'' like the one shown in Fig.\ \ref{connect} with fixed lengths
of the boundaries.

Knowing $G_\lam(l_1,l_2;1)$ allows us to find $G_\lam(l_1,l_2;t)$
by iterating \rf{8} $t$ times. This program is conveniently
carried out by introducing the generating function for the numbers
$G_\lam(l_1,l_2;t)$,
\beq\label{9}
G_\lam(x,y;t)\equiv \sum_{k,l} x^k\,y^l \;G_\lam(k,l;t),
\eeq
which we can use to rewrite \rf{7} as
\beq\label{10}
G_\lam(x,y;t_1+t_2) = \oint \frac{\d z}{2\pi i z} \;
G_\lam(x,z^{-1};t_1) G_\lam(z,y;t_2),
\eeq
where the contour should be chosen to include the singularities
in the complex $z$--plane of $G_\lam(x,z^{-1};t_1)$ but not those
of $G_\lam(z,y;t_2)$.

One can either view the introduction of $G_\lam(x,y;t)$ as a purely
technical device or take $x$ and $y$ as related to boundary cosmological
constants. Let $\lam_i$ and $\lam_f$ denote dimensionless lattice
boundary cosmological constants, such that if the entrance boundary
consists of  $k$ and the exit loop of $l$ links,
the lattice boundary action will be
\beq\label{boundaryact}
S_{i}^{(b)} = \lam_i k = \La_i^0 L_i,~~~~
S_{f}^{(b)} = \lam_f l = \La_f^0 L_f.
\eeq
In \rf{boundaryact} we have introduced also
dimensionful bare lattice boundary cosmological
constants $\La_i^0 \equ \lam_i/a$ and $\La_f^0= \lam_f/a$, as well
as continuum boundary lengths $L_i\equ k\, a$ and $L_f\equ l \, a$.
We now write
\beq\label{10a}
x=\e^{-\lam_i}= \e^{- \La_i a},~~~~y=\e^{-\lam_f}=\e^{-\La_f a},
\eeq
such that $x^k= \e^{-\lam_i \,k}$ becomes
the exponential of the boundary cosmological term,
and similarly for $y^l= \e^{-\lam_f \, l}$.
Let us for notational convenience define
\beq\label{11}
g=\e^{-\lam}.
\eeq
For the technical purpose of counting we view $x,y$ and $g$ as
variables in the complex plane. In general the function
\beq\label{11a}
G(x,y;g;t)\equiv G_\lam(x,y;t)
\eeq
will be analytic in a neighbourhood of $(x,y,g)=(0,0,0)$.

From the definitions \rf{5} and \rf{6} it follows by standard techniques
of generating functions that we may associate a factor $g$ with each
triangle, a factor $x$ with each vertex on the entrance loop and
a factor $y$ with each vertex on the exit loop, leading to
\beq\label{12}
G(x,y;g;1) =\sum_{k=0}^\infty \left( gx \sum_{l=0}^\infty
 (gy)^l \right)^k -
\sum_{k=0}^\infty (gx)^k = \frac{g^2 xy}{(1-gx)(1-gx-gy)}.
\eeq
Formula \rf{12} is simply a book-keeping device for all possible
ways of evol\-ving from an entrance loop of any length in one step to
an exit loop of any length, i.e. a sum over all possible configurations
of the kind shown in Fig.\ \ref{connect}.
The subtraction of the term $1/(1-gx)$ has been performed to
exclude the degenerate cases where either the entrance or the exit
loop is of length zero. The asymmetry between $x$ and $y$ in \rf{12}
is due to the marking of a vertex belonging to the entrance loop. We
can undo the marking by dividing $G_\lam (l_1,l_2;1)$ by $l_1$.
This operation translates to acting with $\int \frac{d x}{x}$ on $G(x,y;g;1)$
leading to the symmetric
\beq\label{12a}
G^{(1)}(x,y;g;1) = -\ln \Big[\frac{1-gx-gy}{(1-gx)(1-gy)}\Big].
\eeq

From \rf{12} and eq.\ \rf{10}, with $t_1=1$, we obtain
\beq\label{13}
G(x,y;g;t) = \frac{gx}{1-gx}\; G(\frac{g}{1-gx},y;g;t-1).
\eeq
This equation can be iterated and the solution written as
\beq\label{14}
G(x,y;g;t) =
\frac{g^2 xy \;F_1^2(x)F_2^2(x)
\cdots F_{t-1}^2(x)}{[1-gF_{t-1}(x)][1-gF_{t-1}(x)-gy]},
\eeq
where $F_t(x)$ is defined iteratively by
\beq\label{15}
F_t(x) = \frac{g}{1-gF_{t-1}(x)},~~~F_0(x)=x.
\eeq
Let $F$ denote the fixed point of this iterative equation. By standard
techniques one readily obtains
\beq\label{16}
F_t(x)= F\ \frac{1-xF +F^{2t-1}(x-F)}{1-xF +F^{2t+1}(x-F)},~~~~
F=\frac{1-\sqrt{1-4g^2}}{2g}.
\eeq
Inserting \rf{16} in eq.\ \rf{14}, we can write
\bea
\lefteqn{G(x,y;g,t)\!\!\! =
\frac{ F^{2t}(1-F^2)^2\; xy}{(A_t-B_tx)(A_t-B_t(x+y)+C_txy)}} \label{17a} \\
& &
= \frac{F^{2t}(1-F^2)^2\;xy}{\Big[(1\!\!-\!xF)\!-\!F^{2t+1}(F\!\!-\!x)\Big]
\Big[(1\!\!-\!xF)(1\!\!-\!yF)\!-\!F^{2t} (F\!\!-\!x)(F\!\!-\!y)\Big]}~,
\nonumber
\eea
where the time-dependent coefficients are given by
\beq\label{18}
A_t =1-F^{2t+2},~~~B_t=F(1-F^{2t}),~~~C_t=F^2(1-F^{2t-2}).
\eeq
The combined region of convergence of the
expansion in powers $g^kx^ly^m$, valid for all $t$ is
\beq\label{18a}
|g| < \oh,~~~~ |x|< 1,~~~~|y|<1.
\eeq
The asymmetry between $x$ and $y$ in the expression \rf{17a}
is due to the marking of the entrance loop. Thus the corresponding
expression for $G^{(1)}_\lam(x,y;t)$ is symmetric.

We can compute $G_\lam(l_1,l_2;t)$ from $G(x,y;g;t)$ by
a (discrete) inverse Laplace transformation
\beq\label{19}
G_\lam(l_1,l_2;t) =\oint \frac{\d x}{2\pi i x} \oint \frac{\d y}{2\pi i y}\;
\frac{1}{x^{l_1}}\,
\frac{1}{y^{l_2}}\; G(x,y;g;t),
\eeq
where the contours should be chosen in the region where $G(x,y;g;t)$ is
analytic,
\beq\label{19aNEW!}
G_\lam (l_1,l_2;t) = l_1
\left(\frac{B_t}{ A_t}\right)^{l_1\plu l_2} \;
\;\sum_{k=0}^{\min (l_1,l_2)}
\frac{(l_1\plu l_2\mi k\mi 1)!}{k!(l_1\mi k)!(l_2\mi k)!}
\left( \mi \frac{A_tC_t}{B_t^2}\right)^k,
\eeq
which, as expected, is symmetric with respect to $l_{1}$ and $l_{2}$
after division by $l_{1}$ ($l_1,l_2 > 0$ are assumed).

In the next section we will give explicit expressions
for $G_\lam(l_1,l_2;t)$ and $G_\lam(x,y;t)$ in a certain continuum limit.

\subsection{The continuum limit}

The path integral formalism we are using here
is very similar to the one used to re\-pre\-sent the free particle as
a sum over paths. Also there one performs a
summation over geometric objects (the paths), and the path integral itself
serves as the propagator. From the particle case it is known that the bare mass
undergoes an additive renormalization (even for the free particle),
and that the bare propagator is subject to a wave-function renormalization
(see \cite{book1} for a review). The same is true here.
The coupling constants
with positive mass dimension, i.e.\ the cosmological constant and the
boundary cosmological constants, undergo
additive renormalizations, while the propagator itself
undergoes a multiplicative wave-function renormalization.
We therefore expect the bare dimensionless
coupling constants $\lam,\lam_i$ and $\lam_f$ to behave as
\beq\label{20a}
\lam = \lam_c + {\La}a^2,~~~~
\lam_i= \lam_{i}^c+X a,~~~
\lam_f = \lam_{f}^c+{Y}a,
\eeq
where $\La,X,Y$ denote the renormalized dimensionful
cosmological and boundary cosmological constants and where we have
absorbed a factor $\sqrt{3}/4$ in the definition of ${\La}$.

If we introduce the notation
\beq\label{20c}
g_c = \e^{-\lam_c},~~~~x_c= \e^{- \lam_{i}^c},~~~~y_c=\e^{-\lam_{f}^c},
\eeq
for critical values of the coupling constants,
it follows from \rf{10a} and \rf{11} that
\beq\label{20b}
g=g_c\,\e^{-a^2\La},~~~~x=x_c\,\e^{-a\,X},~~~~y=y_c\,\e^{-a\,Y}.
\eeq
The wave-function renormalization will appear as a multiplicative
cutoff dependent factor in front of the ``bare''
propagator $G(x,y;g;t)$,
\beq\label{20}
G_\La (X,Y;T) = \lim_{a \to 0} a^{\eta} G(x,y;g;t),
\eeq
where $T=a\, t$, and where the critical exponent $\eta$
should be chosen such that
the right-hand side of eq.\ \rf{20} exists. In general this will only be
possible for particular
choices of $g_c$, $x_c$ and $y_c$ in \rf{20}.

The basic relation \rf{7} can survive the limit \rf{20} only
if $\eta=1$, since we have assumed that the boundary lengths
$L_1$ and $L_2$ have canonical dimensions and satisfy $L_i = a\, l_i$.

From eqs.\ \rf{16}, \rf{17a} and \rf{18} it is clear that we can only
obtain a nontrivial continuum limit if $|F| \to 1$ and the
natural  choice is to take $g_c = 1/2$. With this choice
the continuum propagator \rf{20} is defined by {\it the approach
to the critical point}, and this approach takes place
from within the region of convergence of the power series defining
$G(x,y;g;t)$ as function of $g$.  Corresponding to
$g_c = 1/2$ we have $\lam_c = \ln 2$.

From \rf{17a} it
follows that we can only get macroscopic loops in the limit
$a \to 0$ if we simultaneously take $x,y \to 1$. Thus the
critical values are $x_c\equ y_c\equ 1$, corresponding to
critical points $\lam^c_i= \lam^c_f=0$ and
\beq\label{25c}
\lam_i = X \,a,~~~~\lam_f= Y \,a.
\eeq
Summarizing, we have
\beq\label{25}
g=\oh \e^{-\La a^2},~~~~x=\e^{-Xa},~~~~~~y=\e^{-aY},
\eeq
and with these definitions it is straightforward
to perform the continuum limit of $G(x,y;g,t)$ as $(x,y,g) \to
(x_c,y_c,g_c)=(1,1,1/2)$, yielding
\bea
\lefteqn{G_\La(X,Y;T)= \frac{4\La\ \e^{-2\SLT}}{(\SL+X)+\e^{-2\SLT}(\SL-X)}}
\nonumber\\
&&\times \, \frac{1}{(\SL+X)(\SL+Y)-\e^{-2\SLT}(\SL-X)(\SL-Y)}.
\label{26}
\eea

From $G_\La(X,Y;T)$
we can finally calculate $G_\La(L_1,L_2;T)$, the continuum
amplitude for propagation from a loop of length $L_1$,
with one marked point, at time-slice $T=0$ to a loop of length $L_2$
at time-slice $T$, by an inverse Laplace transformation,
\beq\label{22}
G_\La(L_1,L_2;T) = \int_{-i\infty}^{i\infty} d X \int_{-i\infty}^{i\infty} d Y
\; \e^{X L_1}\;\e^{Y L_2}\; G_\La(X,Y;T).
\eeq
This transformation can be viewed as the limit of \rf{19} for
$a \to 0$. The continuum version of \rf{10} thus reads
\beq\label{22a}
G_\La(X,Y;T_1+T_2) = \int_{-i\infty}^{i\infty}d Z \;
G_\La(X,-Z;T_1) \, \,G_\La(Z,Y;T_2),
\eeq
where it is understood that the complex contour of integration
should be chosen to the left of
singularities of $G_\La(X,-Z;T_{1})$, but to the right of those of
$G_\La(Z,Y,T_{2})$.

By inverse Laplace transform
of formula \rf{26} we obtain
\beq\label{30}
G_\La(L_1,L_2;T) = \frac{\e^{-[\coth \SLT] \SL(L_1+L_2)}}{\sinh \SLT}
\; \frac{\sqrt{\La L_1 L_2}}{L_2}\; \;
I_1\left(\frac{2\sqrt{\La L_1 L_2}}{\sinh \SLT}\right),
\eeq
where $I_1(x)$ is a modified Bessel function of the first kind.
The asymmetry between $L_1$ and $L_2$ arises because the entrance loop
has a marked point, whereas the exit loop has not. The amplitude with
both loops marked is obtained by multiplying with $L_2$, while the
amplitude with no marked loops is obtained after dividing
\rf{30} by $L_1$. The highly nontrivial
expression \rf{30} agrees
with the loop propagator obtained from a bona-fide continuum calculation
in proper-time gauge of pure two-dimensional gravity by Nakayama \cite{nakayama}.

The important point we want to emphasize here is that the additive
renormalization of the cosmological constant is an entropic effect
when calculated after rotation to Euclidean signature.
In fact, we can write the propagator \rf{11a} as
\beq\label{ny1}
G(x,y,g;t)= \sum_{k,l,n}x^ky^lg^n \sum_{T(k,l,n)} \frac{1}{C(T)},
\eeq
where the summation is over all {\it causal} triangulations $T(k,l,n)$
consisting of $n$ triangles and with the two boundaries
made of $k$ and $l$ links.
The critical point is $g_c=1/2$. That can only be the case because the
number of (causal) triangulations constructed from  $n$ triangles
grows exponentially as $e^{n \ln 2}$. The continuum renormalized
cosmological constant, as defined by eq.\ \rf{25}, emerges when taking the
difference between the value of the action for a geometry made of
$n$ triangles and the {\it entropy} of the configurations with a given
action (which in this case is proportional to the number of triangles
$n$). More precisely, let the number of causal triangulations
which can be constructed from
$n$ triangles be
\beq\label{ny2}
\cN(n) = f(n)\, \e^{\lam_c n},~~~~~\lam_c= \ln 2,
\eeq
where $f(n)$ is a pre-factor growing slower than exponentially, and which
can also depend on the boundary cosmological constants $x,y$, a dependence
we will suppress here. We can now write eq.\ \rf{ny1} as
\beq\label{ny3}
G(\lam) = \sum_n f(n)\; \e^{-(\lam - \lam_c)n}, ~~~g\equiv\e^{-\lam}.
\eeq
and using \rf{20a}, written as
\beq\label{ny4}
\lam = \lam_c + \La a^2,
\eeq
and the continuum area  $A=n a^2$ (again
disposing of a factor $\sqrt{3}/4$ for notational simplicity),
eq.\ \rf{ny3} can now be written as
\beq\label{ny5}
G(\La) \propto \int_0^\infty \d A \; f(A/a^2) \;\e^{-\La A},
\eeq
with the continuum action $\La A$ and the nontrivial physics
contained in the function $f(A/a^2)$.

Let us end this subsection by highlighting the entropic interpretation
of \rf{ny1}. This formula tells us that the theory
is nothing {\it but} entropy of geometries -- in the sense of counting
geometric ``microstates". The partition
function (or proper-time propagator) $G(x,y,g;t)$ is simply the
generating function of the number of geometries with fixed
area and boundary length. One could object that this is
not so surprising, since the action in two dimensions
is sufficiently simple to allow for this interpretation. However, the same
is true for DT and CDT in three and four dimensions.
It is again a consequence of the geometric nature of the
Einstein-Hilbert action, together with the simple form it takes on the DT
piecewise linear geometries. It is nevertheless surprising
that even in four dimensions one has an entirely entropic
expression for the partition function of CDT quantum gravity,
namely, eq.\ \rf{ny13} below, which is almost identical to \rf{ny1}.
If one were able to count the number of four-dimensional triangulations
(as we did for the two-dimensional triangulations), one
would have an analytical expression for the four-dimensional
CDT partition function!

\subsection{The transfer matrix in two dimensions}\label{2dtransfer}

If we interpret the propagator $G_\La(L_1,L_2;T)$ as the matrix element
between two boundary states of a Hamiltonian evolution in
``time'' $T$,
\beq\label{ham}
G_\La(L_1,L_2;T)=\langle L_2|\e^{-\hat{H} T}|L_1\rangle
\eeq
the propagator has to satisfy the heat kernel equation
\beq\label{heat}
\frac{\prt}{\prt T} \, G_\La(L_1,L_2;T) = -\hH(L_1) G_\La(L_1,L_2;T).
\eeq
From eq.\ \rf{13} we can in the limit $a\to 0$, using \rf{25},
directly read off the Laplace-transformed eq. \rf{heat},
\beq\label{heat1}
\frac{\prt}{\prt T} \, G_\La(X,Y;T) = -\hat{H}(X) G_\La(X,Y;T);
~~~\hH (X) = \frac{\prt}{\prt X} (X^2 -\La).
\eeq
An inverse Laplace transformation leads to
\beq\label{35b}
\hat{H}(L,\frac{\partial}{\partial L})=
 -L\frac{\partial^2}{\partial L^2}+\La L .
\eeq
However, a little care should be exercised when looking at a
matrix element like \rf{ham}, since we have for combinatorial
convenience used a different measure on the entrance boundary
(where we marked a point) and the exit boundary where no point
was marked. If we want to follow the conventions of Sec.\ \ref{trans}
where we worked with boundaries without markings, the labeling of the
boundary triangulations are simply $|l\ra$, where $l$ is the
number of links, and in the continuum limit the relations
\rf{norm1} and \rf{norm} become
\beq\label{norm2}
\int_0^\infty \d L \;L \; |L\ra\la L| = \hat{1},~~~~
\la L_2|L_1\ra = \frac{1}{L_1} \del(L_1-L_2).
\eeq
This leads to an expansion of functions on $[0,\infty]$ according to
\beq\label{si}
| \phi\ra = \int \d L \; | L\ra \;L \phi(L),~~~~\phi(L) = \la L|\phi\ra,
\eeq
with the measure given by
\beq\label{measure}
\la \psi|\phi\ra = \int \d L \; L \; \psi^*(L) \phi(L).
\eeq
Thus the Hamiltonian \rf{35b} really acts on functions $L \phi(L)$.
Rewriting it as a Hermitian operator on functions $\phi(L)$ with
the measure \rf{measure} leads to the 
Hamiltonian 
\beq\label{ham1}
\hat{H}^{(1)}(L,\frac{\partial}{\partial L})=
-\frac{\partial^2}{\partial L^2} L+\La L,  ~~~~
\la \hH \psi| \phi\ra = \la \psi | \hH \phi\ra
\eeq
for the propagator of unmarked loops.
We thus have
\beq\label{unmarked}
G^{(1)}_\La(L_1,L_2;T) = \la L_2 |\e^{-T \hat{H}^{(1)}}|L_1\ra =
\frac{1}{L_1} \, G_\La(L_1,L_2;T),
\eeq
where the measure used on both entrance and exit loops is given
by  \rf{si} and \rf{measure}.

As discussed in Sec.\ \ref{trans} one obtains
the discretized Hamiltonian from the logarithm of the transfer matrix,
see eq.\ \rf{transmatrix}. We circumvented this by using
relation \rf{12a} to derive the continuum Hamiltonian. However, we would
like to emphasize that one can indeed use the discretized transfer matrix
elements \rf{12} or \rf{12a} to obtain
directly the Hamiltonians \rf{35b} and \rf{ham1}.

Let us use \rf{12} to derive \rf{35b}. We take the limit given by \rf{20b}
to obtain
\bea\label{xx1}
\la y | \hat{T} | x\ra &=& G(x,y;g;1)\\
&\to& \frac{1}{a} \frac{1}{X+Y}
\left(1 +a \frac{\oh (X^2+Y^2)-2\La}{X+Y}-a(2X+Y) + O(a^2)\right).
\nonumber
\eea
The first factor is precisely the factor $a^\eta$ from \rf{20}.
The next factor is the Laplace transform of the kernel
$\del(L_1-L_2)$ of the unit operator coming from the expansion
\beq\label{unit}
\hat{T} = \e^{-a\hat{H}} = \hat{1}-a \hat{H} +O(a^2).
\eeq
Finally, one can show that the matrix element 
of order $a$ in the parentheses is the Laplace transform of the matrix element
$\la L_2| \hat{H} |L_1\ra$ with $\hat H$ given by \rf{35b}.

One can solve for the eigenvalue equation
\beq\label{eigenvalue}
\hH^{(1)} \psi = E \psi,
\eeq
with the constraint that $\psi(L)$ is square-integrable on $[0,\infty]$
with respect to the measure $L dL$. One finds \cite{lottietal,zohren}
\beq\label{eigenvalue1}
\psi_n(L) = P_n(L) \e^{-\SL \, L}, ~~~~E_n = 2\SL (n+1),
\eeq
where $P_n(L)$ is a polynomial of order $n$, $n=1,2,\ldots$.
The ``ground-state'' wave function of $\hH^{(1)}$ is simply
\beq\label{groundstate}
\psi_0(L) = 2\SL \, \e^{-\SL \, L},
\eeq
and the propagator has the spectral decomposition
\beq\label{spectral}
G^{(1)}_\La (L_1,L_2;T) =
\sum_{n=0}^\infty \psi^*_n(L_2)\psi(L_1)\; \e^{-E_n T}.
\eeq

\begin{figure}[t]
\vspace{-1cm}
\centerline{\scalebox{0.35}{\rotatebox{0}{\includegraphics{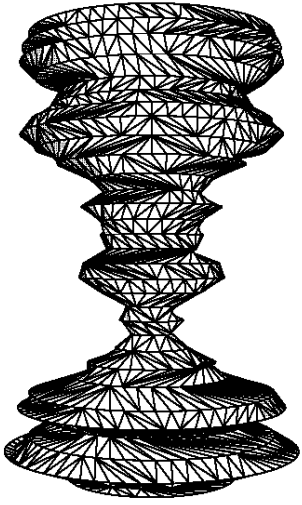}}}}
\vspace{-0.3cm}
\caption{{\footnotesize Piecewise linear spacetime
in (1+1)-dimensional  quantum gravity, generated by computer.
Proper time runs along the vertical direction.}}
\label{fig2d}
\end{figure}

We can view $G^{(1)}_\La (L_1,L_2;T)$ as the partition function
for universes where a spatial boundary of length $L_2$ is separated
by a geodesic distance $T$ from the ``initial'' spatial boundary of
length $L_1$. One can then ask for the probability distribution $P_{T'}(L)$
of the length (the spatial volume) of space at a proper time $T'$ between
zero and $T$. By using (the continuum version of) the decomposition rule
\rf{4} we have
\beq\label{average}
\la L^n\ra_{T'} = \frac{\int L \,\d L \; G_\La^{(1)} (L_1,L;T') \; L^n \;
  G_\La^{(1)} (L,L_2;T-T')}{ G_\La^{(1)} (L_1,L_2;T)}.
\eeq
Using the spectral decomposition it is easy to
show that for  $T \gg T' \gg 1/\SL$ one has
\beq\label{PL}
P_{T'}(L) =L \psi_0^2(L), ~~~~\la L\ra_{T'} = \frac{1}{\SL},
\eeq
up to corrections of order $e^{-2\SL T'}$, $e^{-2\SL (T-T')}$ and
$e^{-2\SL L}$. This is illustrated in Fig.\ \ref{fig2d} which shows
a two-dimensional spacetime configuration generated by computer
simulations, to be described below. The length of the spatial
slices follows the distribution \rf{PL} up to corrections of the
kind mentioned above.

Finally note that we have
\beq\label{hh}
\hH^{(1)} W^{(1)}_\La(L) =0,~~~~W^{(1)}_\La(L) \equiv \frac{\e^{-\SL\, L}}{ L}.
\eeq
Thus $W^{(1)}_\La(L)$ is formally the function
$\psi_{-1}(L)$ in \rf{eigenvalue1},
but is not a normalizable eigenfunction of $\hH^{(1)}$, since
it is not integrable at zero with respect to the measure $L\,dL$.
However, it has another interesting interpretation. Let us calculate
the amplitude for the process that  a spatial universe of length $L$ at a later
time $T$ has length zero. From \rf{30} we obtain
\beq\label{hh1}
G_\La^{(1)}(L,L_2=0;T) =
\La \frac{\e^{-[\coth \SLT] \;\SL L_1}}{\sinh^2 \SLT},
\eeq
and the integral of this amplitude over all time can be viewed
as a Hartle-Hawking amplitude for the universe vanishing
into ``nothing''. We have
\beq\label{hh2}
W^{(1)}_\La(L) = \int_0^\infty \d T \;  G_\La^{(1)}(L,L_2=0;T) = \
\frac{\e^{-\SL\, L}}{ L}.
\eeq
Thus this Hartle-Hawking wave function satisfies
\beq\label{hhw}
\hH^{(1)} W^{(1)}_\La(L) =0,
\eeq
which can be viewed as a special case of the so-called Wheeler-DeWitt equation.

\section{Generalized two-dimensional CDT and ``quantum geometry''}
\label{g2d}

We have dealt in some detail with the two-dimensional CDT model
because it is exactly solvable and illustrates some
of the points mentioned in the earlier sections: the entropic
nature of the renormalization of bare coupling constants, the
use of the transfer matrix to extract a ``quantum gravity Hamiltonian''
and the way the continuum limit is obtained by {\it approaching}
the critical point.
Also, it provided us with an explicit picture of probability
distribution of the spatial slices.

The model can be generalized in a number of ways.
One can use different weights and explore the universality of the model
\cite{charlotte} (see also \cite{ansarimarko}). There also exists a so-called ``string bit''
Hamiltonian formulation \cite{bergfin}. One can add a field variable
corresponding to the ``lapse" of the ADM formulation to the CDT model
\cite{markosmolin}.
Matter can be coupled to the model in a straightforward fashion
\cite{cdtmatter1,cdtmatter2,cdtmatter3}. One can systematically
weaken the causality constraints \cite{cap}, and it is possible to develop
a complete ``string field theory'' for this generalized CDT theory
\cite{cdtSFT}. It is also possible to use matrix models
to describe the CDT-models \cite{cdtmatrix}. In addition, one
can relax the constraint that the geometry is bounded by fine-tuning
the boundary cosmological constant \cite{d-branes}.
The proper time used in the model has an amazing interpretation
as stochastic time, an observation going all the
way back to \cite{stochastic} in the case of
two-dimensional Euclidean quantum gravity. However, the representation
of the model by stochastic quantization is 
realized in a much simpler manner in the CDT framework \cite{cdt-stochastic}.
We refer to the literature for details on most of these topics.

\subsection{Suppression of baby universes}

In this subsection  we will discuss  the
concept of ``quantum geometry'', which is beautiful and simple, and
its relation to ``baby universes''.
Underlying the composition rules \rf{4} or \rf{7} is the assumption that
the spatial geometries form a complete set  of states, as discussed in
Sec.\ \ref{trans}. Assuming that we only want to include purely geometric
quantities like area and length, we are led to essentially
two models: the CDT model described above, or two-dimensional
Euclidean (Liouville) quantum gravity.
The reason for this is essentially entropic.  We are summing over certain
classes of surfaces and want the ensemble to have a well-defined
continuum limit, i.e. a scaling limit where the cutoff $a$, the lattice
length of the elementary links, is taken to zero.

\begin{figure}[t]
\centerline{\scalebox{0.55}{\rotatebox{0}{\includegraphics{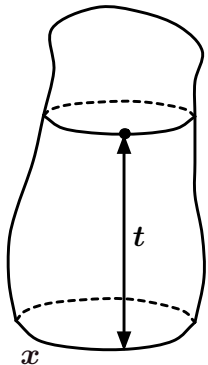}}}}
\caption[figcap]{{\footnotesize
Graphical representation of relation (\ref{2.50}): differentiating the
disc amplitude $W_\La(X)$ (represented by the entire figure) with respect to
the cosmological constant $\La$ corresponds to
marking a point somewhere inside
the disc. This point has a geodesic distance $T$ from the initial loop.
Associated with the point one can identify a connected curve of length $L$,
all of whose points
also have a geodesic distance $T$ to the initial loop. This loop can now be
thought of as the curve along which the lower part of the figure
(corresponding to the loop-loop propagator $G_\La (X,L;T)$) is glued
to the cap, which itself is the disc amplitude $W_\La(L)$.}}
\label{figcap}
\end{figure}

As an example of the constraints imposed by entropy, consider the disc
amplitude $W^{(1)}(L)$, which in the CDT context we met in Sec.\ \ref{2dtransfer}
above.
Let us for combinatorial convenience change to the disc amplitude
$W_\La(L) = L W_\La^{(1)}(L)$, where a point has been marked on the boundary.
In contrast with what we did above, we will in the following argument 
not confine ourselves per se to causal triangulations, but allow for a larger class 
of piecewise linear
surfaces, namely, {\it all} piecewise linear surfaces built out of equilateral
triangles and of disc topology. This will create a larger context within
which both strictly Lorentzian and generalized (acausal, Euclidean) models
can be discussed.
Let us define the
Laplace-transformed disc amplitude as the partition function
\beq\label{ldisk}
W_\La(X) = \int_0^\infty \d L \; \e^{-X \,L} W_\La(L)
\eeq
of two-dimensional
quantum gravity, summed over all two-dimensional spacetimes
with the topology of a disc, whose boundary is characterized
by either the length $L$ or its conjugate boundary cosmological
coupling constant $X$. If we differentiate this partition
function with respect to the cosmological constant $\La$, it corresponds
to multiplying the integrand in the path integral by
the volume factor $V$, since the cosmological term appears
in the integrand as $e^{-\La V}$.
Thus, in a discretized context such a differentiation
is  equivalent to marking a vertex or a triangle, since in this
case the {\it counting}  of different surfaces will increase
by a similar factor. The marking of a vertex gives rise to the
decomposition of the disc amplitude shown in Fig.\ \ref{figcap},
implying the functional relation
\beq\label{2.50}
-\frac{\prt W_\La(X)}{\prt \La} =
 \int_0^\infty \d T \int_{0}^{\infty}
\d L\ G_\La (X,L;T)\; L W_\La(L).
\eeq
It encodes the following geometric situation: each configuration appearing in the
path integral has a unique decomposition into a cylinder of
proper-time extension $T$
(where the proper time is defined as the geodesic distance of the marked point
to the boundary),
and the disc amplitude itself, as summarized in eq.\ \rf{2.50}.

Starting from a regularized theory with a cutoff $a$,
there are two natural solutions to eq.\ \rf{2.50} \cite{al}.
In one of them, the regularized disc amplitude diverges with
the cutoff $a$ and the geodesic distance $T$ scales canonically
with the lattice spacing $a$ according to
\bea\label{2.51}
W_{reg} &\xrightarrow[a\to 0]{}& a^{\eta}\, W_\La(X),~~~~~~~~~~~~\eta < 0, \\
t_{reg} &\xrightarrow[a\to 0]{}&  T/a^\ep,~~~~~~~~~~~~~~~~~~~\ep =1.
\label{2.51a}
\eea
In the other, the scaling goes as
\bea\label{2.52}
W_{reg} &\xrightarrow[a\to 0]{}& {\rm const.} +a^{\eta}\, W_\La(X),
~~~~~~~~~~~~~\eta=3/2\\
t_{reg}& \xrightarrow[a\to 0]{}&  T/a^\ep,
~~~~~~~~~~~~~~~~~~~~~~~~~~~~~~~\ep=1/2,
\label{2.52a}
\eea
where the subscript ``{\it reg}" denotes the
regularized quantities in the discrete lattice formulation.
The first scaling \rf{2.51}-\rf{2.51a}, with $\eta =- 1$,
is encountered in CDT,
while the second scaling \rf{2.52}-\rf{2.52a} is realized in Euclidean
gravity \cite{davidloop,multiloop}.

\begin{figure}[t]
\centerline{\scalebox{0.40}{\rotatebox{0}{\includegraphics{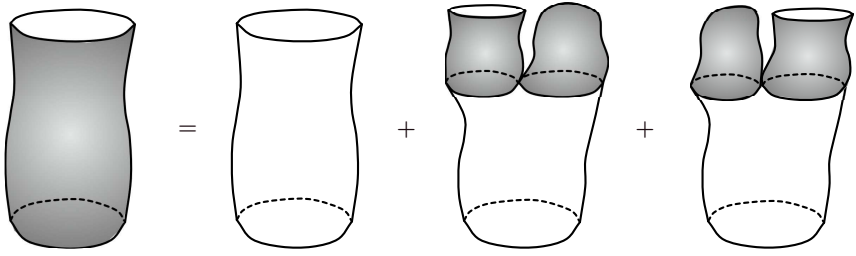}}}}
\caption[figcap2]{{\footnotesize
In all four graphs, the geodesic distance from the final to the initial
loop is given by $T$. Differentiating
with respect to $T$ leads to eq.\ \rf{2.55}. Shaded parts of graphs represent
the full, $g_s$-dependent propagator and disc amplitude, and unshaded
parts the CDT propagator.}}
\label{figcap2}
\end{figure}

Allowing for the creation of baby universes during the ``evolution''
in proper time $T$ (by construction, a process forbidden in CDT) leads
to a generalization of \rf{heat1}, namely,
\bea\label{2.53}
\lefteqn{
a^{\ep}\frac{\prt}{\prt T} G_{\La,g_s}(X,Y;T) =
} \\
&&
- \frac{\prt}{\prt X} \Big[\Big(a(X^2-\La)+
2 g_s\, a^{\eta-1} W_{\La,g_s}(X)\Big) G_{\La,g_s}(X,Y;T)\Big],
\nonumber
\eea
where we have introduced a new coupling constant $g_s$, associated
with the creation of baby universes, and also made the additional dependence
explicit in the amplitudes. The graphic illustration of eq.\ \rf{2.53}
is shown in Fig.\ \ref{figcap2}.
If $g_s=1$, i.e.\ if there is no suppression of baby universes and each
geometry with the same area is allowed with the same weight, one obtains
precisely the equation known from Euclidean quantum gravity \cite{fractal2d}.
This happens because according
to \rf{2.51} and \rf{2.52}, we have either $\eta= -1$, which is
inconsistent with \rf{2.53}, or we have from \rf{2.52} that
$\eta = 3/2$ and thus $\ep = 1/2$, which is consistent with \rf{2.52}.
Thus, in the limit $a \to 0$ the CDT term disappears and on
Fig.\ \ref{figcap2} we are left with the shaded part and eq.\ \rf{2.53}
reduces to
\beq\label{euclidequ}
 \frac{\prt}{\prt T} G_{\La,g_s=1}(X,Y;T) =
- \frac{\prt}{\prt X} \Big[2 W_{\La,g_s=1}(X)
G_{\La,g_s=1}(X,Y;T)\Big].
\eeq
On the other hand, setting $g_s = 0$, thereby forbidding the creation of
baby universes, leads of course back to \rf{heat1}.

In the case of CDT, where $g_s\equ 0$, eq.\ \rf{2.53} does not
contain any reference to the disc amplitude and we can solve
for $G_\La(X,Y;T)$, and subsequently find the disc amplitude,
defined e.g.\ by \rf{hh2}. In Euclidean quantum gravity $g_s\equ 1$ and
the term $a (X^2 -\La)$ vanishes in the limit $a \to 0$. However, now
eq.\ \rf{euclidequ} contains the unknown disc amplitude $W_\La(X)$. The
disc amplitude can be found independently \cite{davidloop}, but it can also
be found by combining \rf{euclidequ} and \rf{2.50}.
Note that due to \rf{2.52} eq.\ \rf{2.50} reads:
\beq\label{quantumg}
-\frac{\prt W_\La(x)}{\prt \La} \propto
 \int_0^\infty \d T \; G_\La (X,L=0;T).
\eeq
If we integrate \rf{euclidequ}
with respect to $T$ from zero to infinity and with respect to $Y$
and use \rf{quantumg} on the right-hand side, we obtain
\beq\label{disk}
-1 \propto \frac{\prt}{\prt X} \left( W_\La (X)
\frac{\prt}{\prt \La} W_\La (X)\right)
\eeq
which has the Euclidean gravity solution
$$
W_\La(X) = \Big(X- \oh \SL\Big)\sqrt{X + \SL}
$$
(see \cite{al} for details).
Thus ``quantum geometry''  considerations alone entirely determine the
disc amplitude and the cylinder amplitude both in CDT and Euclidean
two-dimensional quantum gravity.

From the above discussion it is clear that if we start from the CDT
model where no baby universe creation is allowed, and then drop this
geometric constraint, we end up with Euclidean two-dimensional quantum
gravity. This was already discussed in the original article which introduced
CDT \cite{al}. However, it is interesting that the CDT model can also
be viewed as an ``effective'' model of Euclidean gravity. It is
possible to define in a precise way the concept of ``integrating out'' baby 
universes, at the level of the discretized geometries which appear
in the path integral defining Euclidean two-dimensional quantum gravity. By doing
so one arrives at the CDT model \cite{ackl}.

The CDT and Euclidean models were characterized by being
purely geometric.
It is possible to find an interpolating model, different
from CDT, but with the same scaling, but still allowing
for baby universes during the time evolution like the Euclidean
model. The price is the introduction of a new
coupling constant, the $g_s$ already included in
eq.\ \rf{2.53}. By allowing the coupling constant $g_s$
to be a function of the cutoff it is
possible to make the ``kinetic'' CDT term of the same strength
as the baby universe term if we assume
\beq\label{2.54}
g_s = G_s a^3,
\eeq
where $G_s$ is a coupling constant of mass dimension three, which is
kept constant when $a \to 0$. With this choice, eq.\ \rf{2.53} is turned into
\beq\label{2.55}
 \frac{\prt}{\prt T} G_{\La,G_s}(X,Y;T) =
- \frac{\prt}{\prt T} \Big[\Big((X^2-\La)+2 G_s\;W_{\La,G_s}(X)\Big)
G_{\La,G_s}(X,Y;T)\Big].
\eeq
The graphical representation of eq.\ \rf{2.55} is the one
given in Fig.\ \ref{figcap2}. It is clear that while the creation
of baby universes can take place, they are severely suppressed by
the scaling of $g_s$, compared to the Euclidean case where they
completely dominate over the CDT-like, ordinary propagation.

While the disc amplitude is
at this stage unknown, it is possible to obtain an equation
for $W_{\La,G_s}(X)$ by the same methods as used for the purely
Euclidean case just discussed (see \cite{cap} for details).
The model can also be extended to include higher topologies \cite{cdtSFT},
and it is possible to find a closed expression for the disc
amplitude summed over all genera \cite{alltop}.

\subsection{Coupling to matter}

At a formal level it is easy to couple matter
to CDT quantum gravity in two dimensions as well as in higher
dimensions. In this subsection we describe the results
obtained for two-dimensional CDT, exemplified by the
coupling of the Ising model to CDT-gravity.

The presence of the time slicing makes it natural
to put the Ising spins at the vertices since they are
located at the discrete times defined by the lattice. However, 
this assignment is not important; 
one could just as well have placed the spins at the 
triangle centres. For simplicity we assume that the time is
defined to be periodic. This implies that our two-dimensional lattice
has the topology of a torus where the number of steps
in the time direction is fixed to $t$, while the length
of a spatial slice can fluctuate dynamically.
The partition function can be written as
\beq\label{is1}
G(\lam,\b,t) = \sum_{T \in {\cal{T}}_t} \e^{-\lam N_2(T)} \; Z_{T}(\b),
\eeq
where ${\cal{T}}_t$ is the CDT-class of triangulations
of toroidal topology as described,
$\lam$ is the cosmological constant, $N_2(T)$ the number
of triangles in the triangulation $T$, and $Z_{T}(\b)$ denotes
the Ising model on the lattice defined by the triangulation $T$, that is,
\beq\label{is2}
Z_{T}(\b) = \sum_{\{\sg_i(T)\}} \e^{\b \sum_{\la ij\ra \in T}
\del_{\sg_i\sg_j}}.
\eeq
In eq.\ \rf{is2} we sum over the spin variables $\sg_i$ located
at vertices $i$, while $\la ij \ra$ denotes the link connecting
neighbouring vertices $i$ and $j$. The parameter $\b$ is the spin coupling
constant $J$ divided by the temperature. The Ising spin variable
$\sg_i$ can take values $\pm 1$. The generalization for which the spin variable 
can take the
values $1,\dots,q$ leads to the $q$-state Potts model.
We can write eq.\ \rf{is1} as
\beq\label{is3}
 G(\lam,\b,t) = \sum_{N_2} \e^{-\lam N_2} \; Z_{N_2}(\b),
~~~Z_{N_2}(\b)= \sum_{T \in {\cal{T}}_t(N_2)} Z_{T}(\b),
\eeq
where ${\cal{T}}_t(N_2)$ is the CDT-class of triangulations defined as above,
but constrained to have $N_2$ triangles.

In general one expects a leading behaviour of the form
\beq\label{is4}
Z_{N_2}(\b) \sim \e^{\m(\b) N_2} N_2^{\g(\b)-3}\Big(1+ \cO (1/N_2)\Big),
\eeq
which means that we can view
$\m(\b)$ as the free energy density at $\b$ for the ensemble of
random lattices when the size $N_2$ goes to infinity. If the
spin system defined on this ensemble  
has a phase transition from a low-temperature magnetized phase
to a high-temperature phase with magnetization zero at
a certain value $\b_c$, this fact will
show up as a discontinuity in the derivative of $\m(\b)$ at $\b_c$ at
a certain order, depending on the order of the phase transition.
It is also possible that the exponent $\g(\b)$ can be
discontinuous at $\b_c$.

The same construction works for the DT-class of triangulations.
In this case it is possible to solve the discretized models analytically
using a mapping to matrix models \cite{two-matrix}. The solution reveals
that at a certain critical (inverse) temperature $\b_c$
there is a third-order phase transition from a magnetized phase (high $\b $) to
a phase at low $\b$ where the magnetization is zero. The critical exponents 
associated
with this transition differ from the flat-space Onsager exponents
and agree with the so-called KPZ exponents \cite{KPZ}, which can be
calculated because two-dimensional Euclidean quantum gravity
coupled to conformal field theories with central charge $c < 1$
can be solved analytically by conformal bootstrap \cite{KPZ}.
For $\b \neq \b_c$, when the spin system is not critical,
the geometric properties of the coupled geometry-spin system are
identical to that of two-dimensional Euclidean quantum gravity
without matter. However, at $\b_c$ the geometric properties of the
coupled system change from that of ``pure'' Euclidean gravity to
something else, for example, the entropy exponent $\g(\b)$ changes from
$-1/2$ to $-1/3$ at $\b_c$. This value, first calculated in \cite{two-matrix},
is in agreement with the continuum calculation \cite{KPZ}.

The results for CDT quantum gravity coupled to the Ising model are different
from the KPZ results. At this stage it has not been possible
to solve the combined model analytically in the same way as
in Euclidean two-dimensional gravity coupled to matter. One can resort to strong-coupling
expansions and Monte Carlo simulations, with both methods showing that
there is a phase transition between a magnetized and
an unmagnetized phase \cite{cdtmatter1}. The critical matter
exponents found are the (flat-space) Onsager exponents.
In addition, the entropy exponent $\g(\b)$
in eq.\ \rf{is4} is equal to 1/2 for all values of $\b$, including
$\b_c$, with evidence of similar results for the
three-state Potts model \cite{cdtmatter2}. In two dimensions, the coupling
between geometry and matter therefore 
appears to be weaker for CDT than for
Euclidean gravity. This should not come as a surprise since we have already
seen that one can view CDT in two dimensions as an ``effective'' version of
2d Euclidean quantum gravity, where baby universes have been integrated out.
At the same time there exist persuasive arguments explaining the
difference between the flat-space Onsager and the KPZ
exponents by the presence of baby universes \cite{baby-Ising}.
Following this logic it is then not surprising that two-dimensional CDT gravity
coupled to conformal matter gives rise to flat-spacetime exponents.
It would be very interesting if the result could be
derived from the Euclidean model by integrating out baby universes,
in the same way as pure CDT can be obtained from the pure Euclidean
model. Until now the fact that one has to integrate over the
matter configurations at the boundaries of the baby universes
has prevented us from imitating the pure-gravity construction.

\section{Changing geometries: Monte Carlo moves}
\label{MC}

\subsection{The need for Monte Carlo simulations}\label{need}

We have used the two-dimensional model as an illustration
of how one can obtain a continuum limit of ``quantum geometry", starting from a
discretized lattice theory. This is a nontrivial proof of principle, but 
our main interest is of course the fully fledged 
four-dimensional theory of quantum gravity.
We expect this theory to behave differently, reflecting the presence of
true propagating degrees of freedom (suitable nonperturbative analogues of
the ``gravitons" of the linearized theory) and the fact that
it is a nonrenormalizable field theory. 

From this
point of view, three-dimensional quantum gravity has a place
in between two- and four-dimensional quantum gravity.
It has a nontrivial Einstein-Hilbert action,
and associated with it a formally nonrenormalizable
gravitational coupling constant. However, since
three-dimensional gravity has no propagating degrees of
freedom, it most likely will tell us nothing about
the problem of nonrenormalizability. The fact that physical results
depend only on
a finite number of degrees of freedom (and therefore no {\it field} degrees
of freedom) suggests
that the (quantum) theory could be exactly solvable, like the
two-dimensional quantum theory. There have been a number of attempts
to solve the three-dimensional CDT theory analytically, so far with only limited
success, due to the difficulty of generalizing the exact solution methods
to three dimensions \cite{3dCDTanalytic,hexa,blz}. Thus, even
in three dimensions, to make progress beyond these results, we must turn to
numerical methods \cite{3dajl,joe}, a situation also shared by general, 
non-gravitational lattice models.

Also in four dimensions we have presently no analytic tools for tackling
the CDT path integral directly.
The results reported below for four-dimensional CDT quantum gravity
are based on Monte Carlo simulations,
using the action \rf{actshort}. In the remainder of this section we will 
discuss how to conduct Monte Carlo simulations for causal dynamical
triangulations, which are important even in lower dimensions.
Neither the two-dimensional model with matter coupling nor the
three-dimensional model (with and without matter) have so far been
solved analytically.

\subsection{Monte Carlo simulation with dynamical lattices}\label{DT}

Conventional Monte
Carlo simulations in quantum field theory use a fixed lattice of
(Euclidean) spacetime, and have fields defined on this fixed lattice.
The theory is defined by a partition function
\beq\label{mc-2}
Z = \sum_{\phi} \e^{-S[\phi]},
\eeq
where we sum over all field configurations on the lattice.
Let $\cO(\phi)$ be an observable depending on the fields.
Using the path integral we can in principle calculate the expectation value
\beq\label{10.1}
\la \cO (\phi)\ra = Z^{-1}\sum_{\phi} \cO(\phi)\;
\e^{-S[\phi]}.
\eeq
The purpose of the Monte Carlo simulations is to generate a sequence
of statistically independent field configurations $\phi(n)$,
$n=1,\ldots,N$ with the probability distribution
\beq\label{mc-3}
P(\phi(n)) =  Z^{-1} \e^{-S[\phi(n)]}.
\eeq
Then
\beq\label{10.2}
\la \cO (\phi)\ra_N =  \frac{1}{N}
\sum_{n=1}^N \cO(\phi(n))
\eeq
serves as an {\it estimator} of the expectation value \rf{10.1}
and one has
\beq\label{mc-4}
\la \cO (\phi)\ra_N \to \la \cO(\phi)\ra~~~{\rm for}~~~N\to \infty.
\eeq

One starts out with a field configuration $\phi(0)$ or, more generally,
with a probability distribution $P_0(\phi)$. One
then performs certain changes, starting from $\phi_0$
or, more generally, from a sequence of configurations $\phi$
chosen with probability $P_0(\phi)$. These changes are
accepted or rejected according to certain criteria which ensure
that field configurations after sufficiently many changes
will occur with the correct weight \rf{mc-3},
dictated by the path integral.  In a Monte Carlo simulation
a change $\phi \to \phi'$ of the field configuration is
usually generated by a stochastic process $\cT$,
a Markov chain. The field will perform a random walk  in the space of
field configurations with a transition function, or transition probability,
$\cT(\phi \to \phi')$. Thus, if after a certain number $n$ of steps (changes
of the field configuration) we have arrived at a field configuration $\phi(n)$,
$\cT(\phi(n) \to \phi(n+1))$ is the probability of changing $\phi(n)$ to
$\phi(n + 1)$ in the next step. We have
\beq\label{mc-1}
\sum_{\phi'} \cT(\phi\to \phi') =1~~~\mbox{\rm for all $\phi$}.
\eeq
The transition probability should be chosen such that
\begin{itemize}
\item[(i)] any field configuration $\phi$ can be reached in a finite
number of steps ({\it ergodicity});
\item[(ii)] the probability distribution of field configurations
converges, as the number of steps goes to infinity, to the
Boltzmann distribution \rf{mc-3}.
\end{itemize}
The convergence of the Markov chain is usually ensured by choosing
$\cT$ to satisfy the so-called rule of {\it detailed balance}
\beq\label{detailed}
P(\phi) \; \cT(\phi \to \phi') = P(\phi') \cT(\phi'\to\phi).
\eeq
Starting out with an initial configuration $\phi(0)$ or, more
generally, an initial probability distribution of configurations
$P_0(\phi)$, the stochastic process will after $n$ steps have
generated a new probability distribution $P_n(\phi)$. The 
relation between $P_n(\phi)$ and $P_{n+1}(\phi)$ is
\beq\label{mc-5}
P_{n+1}(\phi) = \sum_{\phi'} P_n(\phi') \;\cT(\phi'\to \phi).
\eeq
Using the identity \rf{mc-1}, the right-hand side 
of this equation can be written as
\beq\label{mc-6}
P_n(\phi) + \sum_{\phi'} \Big(P_n(\phi') \;\cT(\phi'\to \phi)-
P_n(\phi) \;\cT(\phi\to \phi') \Big).
\eeq
According to relations \rf{mc-5} and \rf{mc-6}, the rule \rf{detailed}
of detailed balance ensures that the distribution $P(\phi)$ is a
{\it stationary} distribution under the stochastic process $\cT$.
We refer to standard textbooks for a proof that $P_n$ 
converges to $P$.
Basically, the detailed-balance condition tells us that $P(\phi)$ is
the unique eigenvector of $\cT(\phi \to \phi')$ with eigenvalue one, 
whereas all other eigenvectors have smaller eigenvalues.

A solution to \rf{detailed} is that either
${\cal T}(\phi\to \phi')={\cal T}(\phi'\to \phi)=0$ or
\beq\label{detailed1}
 \frac{{\cal T}(\phi\to\phi')}{{\cal T}(\phi'\to\phi)}=
\frac{P(\phi')}{P(\phi)}.
\eeq
Usually one chooses ${\cal T}(\phi\to \phi')$ to be
nonzero only if some notion
of distance between $\phi$ and $\phi'$ is small,
since else most attempted changes will be rejected.
On the other hand, the class of nonvanishing
transition amplitudes among the configurations
$\phi$ must satisfy
the condition of ergodicity, so it cannot be too restricted.
The choice of nonvanishing transition amplitudes
has a crucial influence
on the efficiency of the algorithm. Using \rf{detailed1}, one
usually decomposes
the transition probability $\cT(\phi \to \phi')$
into a {\it selection probability}
$g(\phi\to \phi')$ and an {\it acceptance ratio} $A(\phi \to \phi')$.
According to eq.\ \rf{detailed1} we then have
\beq\label{mc2}
\frac{P(\phi')}{P(\phi)}= \frac{\cT(\phi\to \phi')}{\cT(\phi'\to \phi)} =
\frac{g(\phi\to \phi') A(\phi\to \phi')}{g(\phi'\to \phi) A(\phi'\to \phi)}.
\eeq
The selection probability $g(\phi\to \phi')$ is now designed to select
the configurations $\phi,\phi'$, where $\cT(\phi \to \phi')$ is different
from zero and assigns a weight of our choice to the
transition $\phi \to \phi'$. The acceptance ratio $A(\phi\to \phi')$
should then be chosen to ensure detailed balance in the form \rf{mc2}.
A general choice, used in many Monte Carlo simulations, is
the so-called {\it Metropolis algorithm}, defined by
 \bea
A(\phi \to  \phi') &=&
\min \Big(1, \frac{g(\phi'\to \phi)}{g(\phi\to \phi')}
\;\frac{P(\phi')}{P(\phi)}\Big), \label{mc5a}\\
A(\phi'\to \phi)  &=&
\min \Big(1, \frac{g(\phi\to \phi')}{g(\phi'\to\phi)}
\frac{P(\phi)}{P(\phi')}\Big).    \label{mc6a}
\eea

In standard lattice field theory the lattice is fixed and not part of the dynamics.
For quantum gravity models based on dynamical triangulations
the situation is different. The lattice is represented
by an abstract triangulation, which itself represents a curved spacetime
geometry. In the path integral we sum over all triangulations,
each associated with an action and thus a
weight. It is the lattice itself which
changes in this type of quantum gravity model.
The Monte Carlo simulation must be set up 
to move us around in the class of abstract CDT triangulations
with the correct weight.

In Euclidean quantum gravity models, where one performs the summation
over all abstract triangulations of a fixed topology, the so-called
{\it Pachner moves} \cite{pachner} are a minimal set of
local topology-preserving changes of
the triangulations which are ergodic. This means that 
by repeated use of these
``moves'' it is possible to get from one triangulation of a certain
topology to any other of the same topology.
 In the computer we work with {\it labeled triangulations}.
Thus the partition function will be
represented in the numerical simulations as
a sum over labeled triangulations, as described in the
discussion surrounding eq.\ \rf{diff} in Sec.\ \ref{trans}.
Speaking in general terms, the computer then keeps a list
of the vertices of the triangulation, for each vertex a list of
all the neighbouring vertices and a list of how this set of points
is organized into simplices and subsimplices. There are many
different ways to organize and store this information. It is not our
intention to go into a detailed discussion of this, but in the next
subsection we will give an example of how this 
can be done in the very simple case of two-dimensional CDT, 
and how the tables are updated when a move is performed.

The Pachner moves used in Euclidean dynamical triangulations
are not directly applicable in CDT since 
they generally do not respect the sliced structure of the
Lorentzian discrete geometries.
Our strategy for constructing suitable sets of moves
is to first select moves that are ergodic {\it within}
the spatial slices $t=const.$ (this is clearly a necessary condition for
ergodicity in the complete set of four-dimensional 
triangulations of fixed topology),
and then supplement them
by moves that act within the sandwiches $\Delta t\equ 1$.
In manipulating triangulations, especially in four dimensions,
it is  useful to know the numbers of simplicial building blocks of
type A contained in a given building block of type B of equal or
larger dimension (Table \ref{ntable1}).

If the moves are used as part of a Monte Carlo updating
algorithm, they will be rejected whenever the resulting
triangulation would violate the simplicial manifold constraints.
The latter are restrictions on the possible gluings (the pairwise
identifications of subsimplices of dimension $d-1$) to make
sure that the boundary of the neighbourhood of each vertex has
the topology of a $S^{d-1}$-sphere \cite{mauroetal}. In dimension
three, ``forbidden'' configurations are conveniently
characterized in terms of the intersection pattern of the
triangulation at some half-integer time.
The latter is given by a two-dimensional tessellation
in terms of squares and triangles or,
alternatively, its dual three- and four-valent graph
(see Appendix 2 in \cite{3dajl} for details).
Also in four dimensions, one may characterize sandwich geometries by
the three-dimensional geometry obtained when cutting the four-dimensional
geometry at a half-integer time between two adjacent spatial
triangulations height. The corresponding three-dimensional geometry
consists of two types of three-dimensional building blocks,
tetrahedra and prisms (see also \cite{countingbhs}).

\begin{table}
	\begin{center}
	\renewcommand{\arraystretch}{1.25}
	\begin{tabular}{ |c||c|c|c|c|c|c|c|c|c|c|}
		\hline
		contains &
		$N_{0}$ & $N_{1}^{\rm TL}$ & $N_{1}^{\rm SL}$ &
		$N_{2}^{\rm TL}$ & $N_{2}^{\rm SL}$ &
 	    $N_{3}^{(3,1)}$ & $N_{3}^{(2,2)}$ & $N_{3}^{\rm SL}$ &
 	    $N_{4}^{(4,1)}$ & $N_{4}^{(3,2)}$ \\
		\hline\hline
		$N_{0}$ & 1 & 2 & 2 & 3 & 3 & 4 & 4 & 4 & 5 & 5  \\
		\hline
		$N_{1}^{\rm TL}$ &   & 1 & 0 & 2 & 0 & 3 & 4 & 0 & 4 & 6  \\
		\hline
		$N_{1}^{\rm SL}$ &   &   & 1 & 1 & 3 & 3 & 2 & 6 & 6 & 4  \\
		\hline
		$N_{2}^{\rm TL}$ &   &   &   & 1 & 0 & 3 & 4 & 0 & 6 & 9  \\
		\hline
		$N_{2}^{\rm SL}$ &   &   &   &   & 1 & 1 & 0 & 4 & 4 & 1  \\
		\hline
		$N_{3}^{(3,1)}$ &   &   &   &   &   & 1 & 0 & 0 & 4 & 2  \\
		\hline
		$N_{3}^{(2,2)}$ &   &   &   &   &   &   & 1 & 0 & 0 & 3  \\
		\hline
		$N_{3}^{\rm SL}$ &   &   &   &   &   &   &   & 1 & 1 & 0  \\
		\hline
		$N_{4}^{(4,1)}$ &   &   &   &   &   &   &   &   & 1 & 0  \\
		\hline
		$N_{4}^{(3,2)}$ &   &   &   &   &   &   &   &   &   & 1  \\
		\hline
	\end{tabular}
	\end{center}	
	\caption{\label{ntable1}{\footnotesize Numbers of simplicial
building blocks contained in simplices of equal or higher dimension.}}
\end{table}

To summarize: the moves define the possible changes we can
make to go from one triangulation to another. Thus they
serve precisely as part of the section function $g(T \to T')$
mentioned above. In this way the moves allow us to define the
selection function $g(T \to T')$ mentioned above. As a next step, we 
choose the acceptance ratio
$A(T\to T')$ such that detailed balance is satisfied for a
particular move and its ``inverse''. In the next subsections we
will describe the CDT moves in two, three and four dimensions explicitly.
In addition, we will discuss how the functions
$g(T\to T')$ and $A(T\to T')$ are determined in two dimensions.

\subsection{Moves in two dimensions}

\begin{figure}[t]
\centerline{\scalebox{0.8}{\rotatebox{0}{\includegraphics{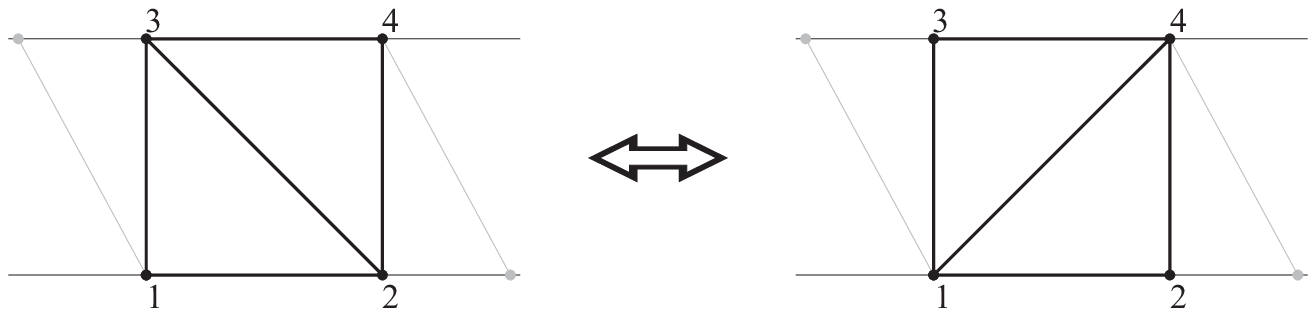}}}}
\vspace{0.5cm}
\centerline{\scalebox{0.8}{\rotatebox{0}{\includegraphics{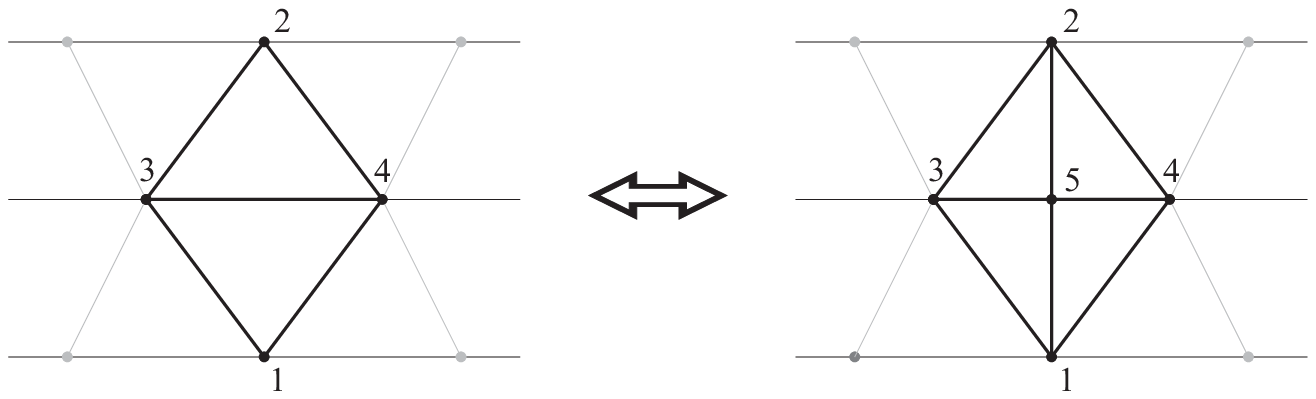}}}}
\caption[move]{{\footnotesize The moves used in the Monte Carlo updating of
the two-dimensional CDT geometry. The top figure shows the (2,2)-move,
and the bottom figure the (2,4)-move.}}
\label{move}
\end{figure}

The two-dimensional case is simple.
Two moves which are ergodic and respect the CDT structure
of spacetime are shown in Fig.\ \ref{move}. We denote the two moves
by (2,2) and (2,4). The (2,2)-move takes two adjacent triangles
positioned as shown in Fig.\ \ref{move}, that is, a (1,2)- and a
(2,1)-triangle sharing a timelike link, and flips the diagonal,
thus not changing the boundary of the subcomplex consisting
of the two triangles. Similarly, the (2,4)-move takes two adjacent
triangles sharing a spacelike link and replaces the two triangles by
four triangles, as shown in Fig.\ \ref{move}. Again, the
boundary of this new subcomplex is identical to the boundary of
the original subcomplex consisting of the two adjacent triangles.
The inverse move, a (4,2)-move, takes a vertex of order four and
deletes it, together with
the two timelike links and one of the spacelike links sharing
the vertex. As a result, two triangles are deleted and one is left with a
subcomplex of two adjacent triangles sharing a spacelike link.
If we use the labeling of vertices shown in Fig.\ \ref{move}, the changes
are encoded by writing for the (2,2)-move
\beq\label{22move}
123+234 \leftrightarrow 134 + 124,
\eeq
and similarly for the (2,4)-move
\beq\label{24movea}
134+234 \leftrightarrow 135+145+235+245.
\eeq
The labeling $abc$ denotes a triangle consisting of vertices with
labels $a$, $b$ and $c$, regardless of their orientation.
Similarly, $ab$ denotes a link. Thus 123+234 stands for a subcomplex 
whose triangles share link 23, while the boundary is made of the links
which are not shared, i.e.\ 12, 13, 24 and 34. This ``algebraic'' representation
is very convenient and can be generalized in a straightforward manner
to higher dimensions. It is the one used in the computer. For example,
in four dimensions two four-simplices 12345 and 23456 will 
share the tetrahedron 2345.

Let us use the two-dimensional case to discuss the
detailed-balance aspect of the Monte Carlo updating, since it is
slightly more complicated than in ordinary lattice Monte Carlo simulations.
Given is the space of configurations, in this case, the abstract
two-dimensional triangulations with the topology of a cylinder,
satisfying the CDT requirements. To simplify the discussion of boundaries
let us compactify the time direction, thereby creating a torus.
In the computer we work with labeled triangulations,
with the action given by \rf{disact}. For a given triangulation
$T$, since the number of triangles $N_2(T)$ is two times the number of
vertices $N_0(T)$, we have $S(T)=2\lam N_0(T)$. Since we work with labeled
triangulations in the computer, according to \rf{diff}
the probability assigned to a given triangulation is given by
\beq\label{update1}
P(T) = \frac{1}{Z} \;\frac{1}{N_0(T)!} \; \e^{-2\lam N_0(T)}.
\eeq
Let us consider the (2,4)-move and its inverse. In order to
have an ergodic set of moves we will also need to invoke the
(2,2)-move, which we will simply do by alternating with a suitable
frequency between (2,4)- and (2,2)-moves.
We require detailed balance for the (2,4)-move and its inverse.
According to \rf{mc2} we write
\beq\label{mc2a}
\frac{P(T')}{P(T)}= \frac{\cT(T\to T')}{\cT(T'\to T)} =
\frac{g(T\to T') A(T\to T')}{g(T'\to T) A(T'\to T)}.
\eeq

Given a labeled triangulation $T_{N_0}$ and performing move (2,4),
we create a new labeled triangulation with $N_0+1$ vertices. If the
old vertices are labeled 1 to $N_0$, we assign the label $N_0+1$
to the new vertex. (New links and triangles are defined by
pairs and triples of vertex labels, the triples also defining the correct
orientation.) Starting from a labeled triangulation $T_{N_0}$ we can 
construct $N_0$ labeled triangulations $T_{N_0+1}$ by choosing different vertices
and performing the (2,4)-move. We define the selection probability
$g(T_{N_0} \to T_{N_0+1})$ to be the same for all triangulations $T_{N_0+1}$
that can be reached in this way and zero for all other labeled
$T_{N_0+1}$ triangulations. For the labeled triangulations
which can be reached in this way we have
\beq\label{mc3}
g(T_{N_0}\to T_{N_0+1}) = \frac{1}{N_0}.
\eeq
We implement this in the computer program by choosing randomly
with uniform probability a vertex in $T_{N_0}$.

Given a labeled triangulation $T_{N_0+1}$ we perform the (4,2)-move
as follows. Select the vertex labeled $N_0+1$, and
assume for the moment it is of order four. Delete it from
the list, in this way creating a labeled triangulation $T_{N_0}$.
If the vertex labeled $N_0+1$ is not of order four do not perform
the move.\footnote{One is free to avoid this situation
by keeping a list of vertices of order four and choosing at random one
of these vertices, rather than the one with label $N_0+1$. This can be viewed
as redefining the labeling, interchanging the label of the
chosen vertex with the vertex labeled $N_0+1$, and reassigning 
the labeling of links and triangles correspondingly. Of course, we do not really
have to make the reassignment in the computer since all labelings
are equivalent.} Thus, for the triangulations $T_{N_0+1}$ where
the move can be performed we can only reach one triangulation
$T_{N_0}$ and the selection probability
$g(T_{N_0+1}\to T_{N_0})$ defined by this procedure is one.
Finally, choose the acceptance ratios $A(T\to T')$ in accordance
with the Metropolis algorithm, \rf{mc5a}-\rf{mc6a} as
\bea
A(T_{N_0} \to  T_{N_0+1}) &=&
\min \Big(1, \frac{N_0}{N_0+1} \;\e^{-2\lam}\Big), \label{mc5}\\
A(T_{N_0+1} \to  T_{N_0}) &=&
\min \Big(1, \frac{N_0+1}{N_0} \;\e^{2\lam}\Big).    \label{mc6}
\eea

The line of argument given here for the (2,4)-move applies
to all moves discussed below, namely, one first chooses a
suitable selection probability $g(T \to T')$ and then adjusts
the acceptance ratio accordingly, such that the detailed-balance
equation is satisfied.

\subsection{Moves in three dimensions}

For the three-dimensional CDT triangulations there are five basic moves
(counting inverse moves as separate) \cite{3dajl,ajl4d}. They map
one three-dimensional CDT triangulation
into another, while preserving the constant-time slice structure, as
well as the total proper time $t$.
We label the moves by how they affect the number of simplices of
top-dimension, i.e. $d\equ 3$.
A $(m,n)$-move is one that operates on
a local subcomplex of $m$ tetrahedra and replaces it by a different
one with $n$ tetrahedra. In all cases the two-dimensional boundaries
of the two subcomplexes are identical\footnote{The notation of a
$(m,n)$-move can be generalized to any dimension $d$. We have already
used it in the case $d=2$ for the (2,2)-move and the (2,4)-move.
We have two subcomplexes consisting of $m$ and $n$ $d$-simplices,
but with identical $(d-1)$-dimensional boundaries.}.
The tetrahedra themselves are characterized
by 4-tuples of vertex labels. Throughout, we will not distinguish
moves that are mirror images of each other under time reflection.
In detail, the moves are
\begin{itemize}
\item[(2,6):]
this move operates on a pair of a (1,3)- and a (3,1)-tetrahedron
(with vertex labels 1345 and 2345) sharing
a spatial triangle (with vertex labels 345). A vertex (with label 6)
is then inserted at the centre of the triangle and connected by new
edges to the vertices 1, 2, 3, 4 and 5. The final configuration
consists of six tetrahedra, three below and three above the spatial
slice containing the triangles (Fig.\ \ref{26m}).
This operation may be encoded by writing
\begin{equation}
1{345} +2{345} \rightarrow
1{346} +2{346} +
1{356} +2{356} +
1{456} +2{456},
\label{26move}
\end{equation}
where the shared triangles
on the right-hand side of the arrow have labels 346, 356 and 456.
The inverse move (6,2) corresponds to a reversal of the arrow in
(\ref{26move}). Obviously, it can only be performed if
the triangulation contains a suitable subcomplex of six tetrahedra.
\begin{figure}[t]
\centerline{\scalebox{0.5}{\rotatebox{0}{\includegraphics{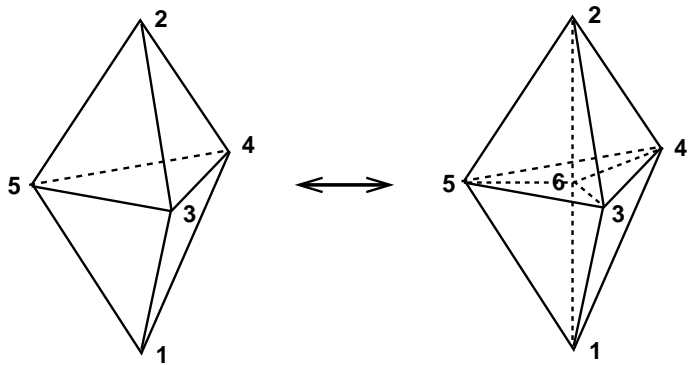}}}}
\caption[26m]{{\footnotesize The (2,6)-move in three dimensions.}}
\label{26m}
\end{figure}
\item[(4,4):]
This move can be performed on a subcomplex of two (1,3)- and
two (3,1)-tetrahedra
forming a ``diamond'' (see Fig.\ \ref{44m}),
with one neighbouring pair each above and below a spatial slice.
The move is then
\begin{equation}
1{235}+{235}6+1{345}+{345}6
\rightarrow 1{234} +{234}6+1{245}
+{245}6.
\label{44move}
\end{equation}
From the point of view of the spatial ``square'' (double triangle) 2345,
the move (\ref{44move}) corresponds to a flip of its diagonal.
It is accompanied by a corresponding
reassignment of the tetrahedra constituting
the diamond.
\begin{figure}[t]
\centerline{\scalebox{0.5}{\rotatebox{0}{\includegraphics{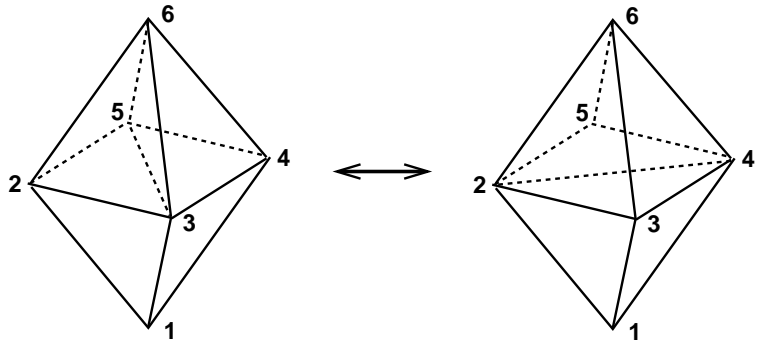}}}}
\caption[44m]{{\footnotesize The (4,4)-move in three dimensions.}}
\label{44m}
\end{figure}
The (2,6)- and (6,2)-moves, together with the (4,4)-move (which is its
own inverse) induce moves within the spatial slices that are known
to be ergodic for two-dimensional triangulations.
\item[(2,3):]
The last move, together with its inverse, affects the sandwich
geometry without changing the spatial slices at integer-$t$.
It is performed on a pair of a (3,1)- and a (2,2)-tetrahedron which
share a triangle 345 in common (see Fig.\ \ref{23m}),
and consists in substituting this
triangle by the one-dimensional edge 12 dual to it,
\begin{equation}
1{345}+2{345}\rightarrow
{12}34+{12}35+{12}45.
\label{23move}
\end{equation}
The resulting configuration consists of one (3,1)- and two
(2,2)-tetrahe\-dra, sharing the link 12. Again, there is an obvious
inverse.
\begin{figure}[t]
\centerline{\scalebox{0.5}{\rotatebox{0}{\includegraphics{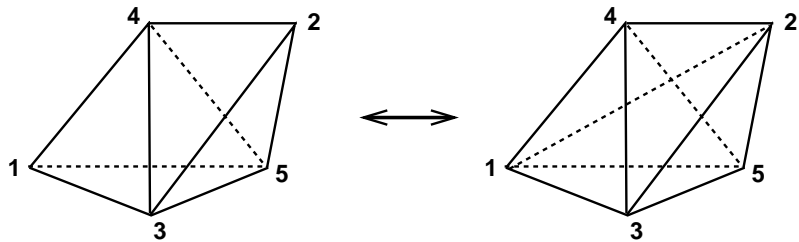}}}}
\caption[23m]{{\footnotesize The (2,3)-move in three dimensions.}}
\label{23m}
\end{figure}
\end{itemize}

\subsection{Moves in four dimensions}

If we distinguish between the space- and timelike character of
all of the four-dimensional moves, there is a total of ten moves
(including inverses). We will again characterize simplices in terms of
their vertex labels. The first two types of moves, (2,8) and (4,6),
reproduce a set of ergodic moves in three dimensions when restricted
to spatial slices. We will describe each of the moves in turn \cite{ajl4d}.
\begin{itemize}
\item[(2,8):] The initial configuration for this move is a pair of
a (1,4)- and a (4,1)-simplex,
sharing a purely spacelike tetrahedron 3456 (Fig.\ \ref{28m}).
\begin{figure}[t]
\centerline{\scalebox{0.5}{\rotatebox{0}
{\includegraphics{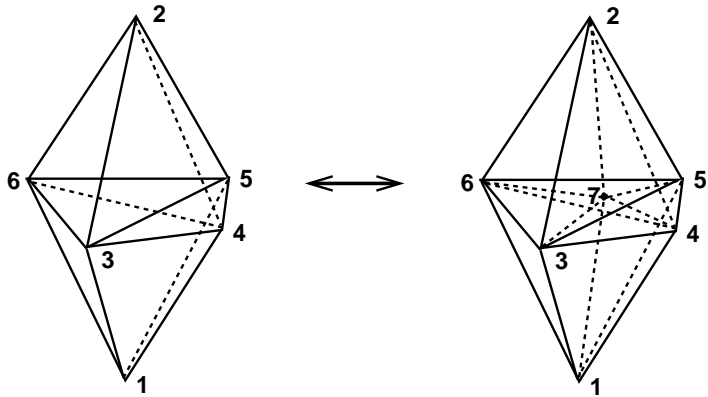}}}}
\caption[28m]{{\footnotesize The (2,8)-move in four dimensions.}}
\label{28m}
\end{figure}
The move consists in inserting an additional vertex 7 at the center
of this tetrahedron and subdividing the entire subcomplex so as to
obtain eight four-simplices,
\bea
\lefteqn{1{3456}+2{3456}\rightarrow}\label{28move}\\
&&1{3457}+2{3457}+
1{3467}+2{3467}+
1{3567}+2{3567}+
1{4567}+2{4567},
\nonumber
\eea
with an obvious inverse.
\item[(4,6):] In this configuration, we start from a pair (2345,3456)
of spatial tetrahedra sharing a common triangle 345, which are
connected to a vertex 1 at time $t\mi 1$ and another vertex 7 at time
$t\plu 1$, forming together a subcomplex of size four (Fig.\ \ref{46m}).
The move
consists in swapping the triangle 345 with its (spatially) dual edge 26,
during which the two spatial tetrahedra are substituted by three,
and the geometry above and below the spatial slice at time $t$
changes accordingly. In our by now familiar notation, this amounts to
\bea\label{46move}
\lefteqn{1{2345} +{2345}7 +
1{3456} +{3456}7 \rightarrow}\\
&&1{2346} +{2346}7 +
1{2356} +{2356}7 +
1{2456} +{2456}7,
\nonumber
\eea
with the arrow reversed for the corresponding inverse move.
\begin{figure}[t]
\centerline{\scalebox{0.5}{\rotatebox{0}
{\includegraphics{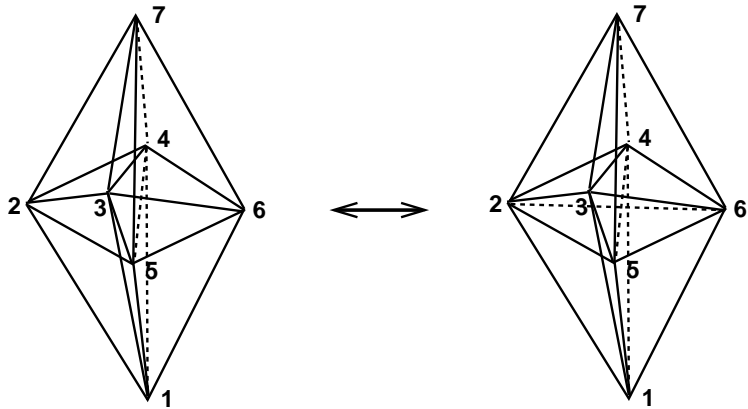}}}}
\caption[46m]{{\footnotesize The (4,6)-move in four dimensions.}}
\label{46m}
\end{figure}
\item[(2,4):] This type of move comes in two varieties.
Its general structure is as follows:
the initial configuration is a pair of four-simplices with
a common tetrahedron. During the move, this tetrahedron
is ``deleted'' and substituted by its dual (in a four-dimensional
sense) edge. The end result is a subcomplex consisting of four
4-simplices. From a Lorentzian point of view, there are two
situations where the application of this move does not interfere with
the slice-structure or the manifold constraints. In the first one,
a (4,1)- and a (3,2)-tetrahedron from the same sandwich share
a (3,1)-tetrahedron 3456 (Fig.\ \ref{24ma}).
The dual edge 12 is timelike and
shared in the final configuration by one (4,1)- and three
(3,2)-simplices. The second possibility is that of two
(3,2)-simplices sharing a (2,2)-tetrahedron. One of the
(3,2)-simplices is ``upside-down'', such that the entire subcomplex
has spatial triangles in both the slices at $t$ and at $t\plu 1$
(134 and 256, see Fig.\ \ref{24mb}). After the move, the total number of
(3,2)-simplices has again increased by two.
Both types of moves are described by the relation
\begin{equation}
1{3456}+2{3456}\rightarrow {12}345 +
{12}346 +{12}456 +{12}356,
\label{24move}
\end{equation}
and their inverses by the converse relation.
\begin{figure}[t]
\centerline{\scalebox{0.5}{\rotatebox{0}
{\includegraphics{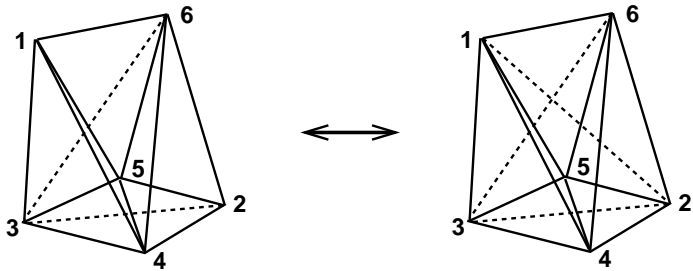}}}}
\caption[24ma]{{\footnotesize 
The (2,4)-move in four dimensions, first version.}}
\label{24ma}
\end{figure}
\begin{figure}[t]
\centerline{\scalebox{0.5}{\rotatebox{0}
{\includegraphics{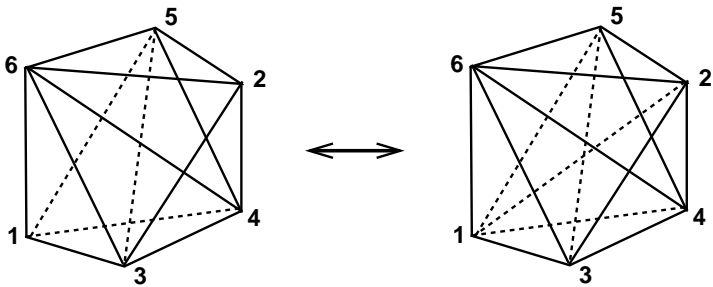}}}}
\caption[24mb]{{\footnotesize 
The (2,4)-move in four dimensions, second version.}}
\label{24mb}
\end{figure}
\item[(3,3):] The initial subcomplex in this type of move is made up
of three 4-simplices which share a triangle in common. In the course
of the move, this triangle is ``deleted'' and substituted by its
dual (in a four-dimensional sense), which is again a triangle.
It is straightforward to verify that this move can only occur
in Lorentzian gravity if both of the triangles involved are timelike.
Again, there are two allowed variations of the move. In the first
one, both the initial and final subcomplex consist of one (4,1)- and
two (3,2)-simplices, and the spacelike edge of each of the triangles
123 and 456 lies in the same slice $t\equ const.$ (Fig.\ \ref{33ma}).
The four-simplices are rearranged according to
\begin{equation}
12{456}+13{456}+23{456}
\rightarrow {123}45+{123}46+{123}56.
\label{33move}
\end{equation}
\begin{figure}[t]
\centerline{\scalebox{0.5}{\rotatebox{0}{\includegraphics{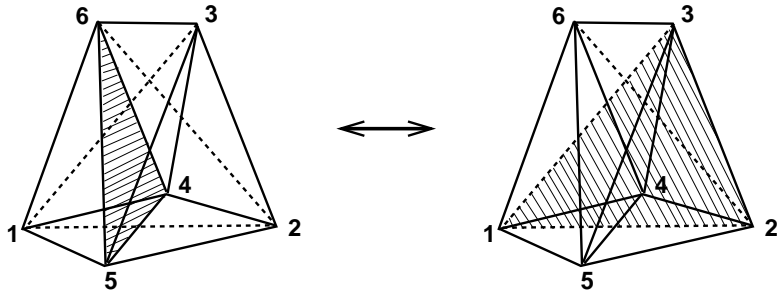}}}}
\caption[33ma]{{\footnotesize 
The (3,3)-move in four dimensions, first version.}}
\label{33ma}
\end{figure}
\begin{figure}[t]
\centerline{\scalebox{0.5}{\rotatebox{0}{\includegraphics{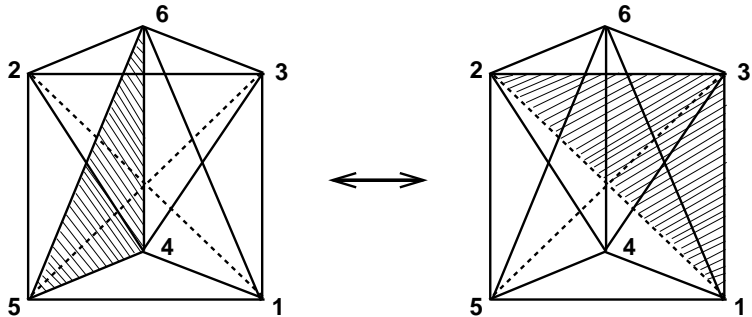}}}}
\caption[33mb]{{\footnotesize 
The (3,3)-move in four dimensions, second version.}}
\label{33mb}
\end{figure}
The other initial configuration for which (\ref{33move}) can be
performed involves only (3,2)-simplices of both orientations, as
illustrated in Fig.\ \ref{33mb}.
The ``swapped'' triangle 123 now has its spacelike
edge in the opposite spatial slice from that of the original triangle
456. As in the (4,4)-move in three dimensions, this type of move
is its own inverse.

\end{itemize}

\section{The entropic nature of CDT and the phase diagram}
\label{phase-diagram}

We are now ready to set up Monte Carlo simulations of
the four-dimensio\-nal CDT quantum gravity model. We store the
simplicial geometry as a set of lists, which consist
of sequences of labels
for simplices of dimension $n$, $0\leq n\leq 4$, together with their
position and orientation with respect to the time direction.
Additional list data include information about nearest neighbours, i.e. how the
triangulation ``hangs together", and other discrete data
(for example, how many four-simplices meet at a given edge)
which help improve the acceptance rate of Monte Carlo moves.

The simulation is set up to generate a random walk in
the ensemble of causal geometries of a fixed time extension $t$.
We are of course interested in conducting the simulations with
triangulations that are as large as possible. However, since the moves
used to change the triangulation are local, the time it takes
to change a given triangulation into a new ``independent'' triangulation
will increase with the size of the triangulation. Thus one has
to find a compromise between (computer) time and size. Until now
we have used triangulations which consists of up to 362.000 four-simplices.
To get an idea of the corresponding linear size, 
suppose we organized the four-simplices such that they approximately
formed a regular four-dimensional lattice. The size of this lattice
turns out to be up to $9.5^4$. While we are working with our
dynamically generated universe in the computer, hoping to find
measurable observables,
it is worth keeping in mind that we are
trying to squeeze the universe into a $10^4$ lattice! 
Not unlike in lattice QCD, this obviously limits 
the kind of questions we can ask. Clearly,
if we added matter, it would be difficult to study the formation
of galaxies, even if we could manage to find a way to address
this question after rotating to Euclidean signature.

\subsection{The entropic nature of CDT and the continuum limit}\label{entro}

The action we are using is \rf{actshort} is a compact rewriting
of the Einstein-Hilbert action \rf{act4dis}. There are three
coupling constants, $\kp_0$, $\kp_4$ and $\Del$, where $\kp_4$ is related
to the cosmological coupling constant as described in
Sec.\ \ref{subsecd4act}
and dictates the average number of $N_4$. 
The cosmological-constant term contributes at the same leading order
as the entropy (the number of triangulations with a
given number $N_4$ of four-simplices), which also grows
exponentially with $N_4$ \cite{expobound}.
We encountered a similar phenomenon in the two-dimensional model which
could be solved analytically (see \rf{ny2}).\footnote{Contrary to the situation
in two dimensions there is no analytic proof of the exponential
bound in three- or four-dimensional DT. At some stage, there 
were claims (based on Monte Carlo simulations) that the number
grows much faster \cite{expobound1}.
However, it is now generally believed
that the growth is only exponential as a function of $N_4$.} 
As discussed in connection
with \rf{ny2}, the exponential growth defines a critical point, to which
one has to fine-tune the bare cosmological constant. If the bare
cosmological constant is smaller than the critical value, the partition
function is simply ill defined. If it is larger than the critical value
the expectation value $\la N_4 \ra$ will remain finite even as 
$N_4\rightarrow\infty$. We test arbitrarily large
values of $N_4$ when fine-tuning the bare
cosmological constant to its critical value, precisely as in the
two-dimensional case. In addition, the renormalized, physical
cosmological constant is defined by this approach to the critical
value, again like in two dimensions. However, the subleading correction to the
exponential growth of the number of triangulations, i.e.\ the function
$f(n)$ in \rf{ny2} in the two-dimensional case, can be different.

Let us make the discussion more explicit to illustrate the analogy
with the two-dimensional case and to emphasize the role of ``entropy'',
i.e.\ the number of configurations. We can write the partition function as
\beq\label{ny11}
Z(\kp_0,\kp_4,\Del) = \sum_{N_4,N_4^{(4,1)},N_0} \e^{-(\kp_4+\Del) N_4} \;
\e^{-\Del N_4^{(4,1)}} \;\e^{(\kp_0+6\Del)N_0}\!\!\!\!\!
\sum_{T(N_4,N_4^{(4,1)},N_0)}  \frac{1}{C_T} .
\eeq
Introducing
\beq\label{ny12}
x=\e^{-(\kp_4+\Del)},~~~~y=\e^{-\Del},~~~~z=\e^{(\kp_0+6\Del)} ,
\eeq
this can be rewritten as 
\beq\label{ny13}
\tilde Z(x,y,z) = \sum_{N_4,N_4^{(4,1)},N_0} x^{N_4}\;y^{N_4^{(4,1)}}\;z^{N_0}
\; \cN(N_4,N_4^{(4,1)},N_0),
\eeq
where $\cN(N_4,N_4^{(4,1)},N_0)$ denotes the number of CDT configurations
with $N_4$ four-simplices, $N_4^{(4,1)}$ of which are of type $(4,1)$ or
$(1,4)$, and with $N_0$ vertices, including symmetry factors.
We conclude that the calculation of the partition function is in principle a
combinatorial problem, just as in two dimensions where
we could solve the problem explicitly.
Our rationale for calling the model entropic is that
{\it the partition function is entirely determined by the number of geometries
in the simplest possible way, in the sense of being the generating function
for these numbers.} The logarithm of the
number of geometric ``microscopic" configurations
for given numbers $(N_4,N_4^{(4,1)},N_0)$ is proportional to their entropy
in the sense of statistical models.\footnote{For a statistical system, like a
spin system, the entropy is defined as (proportional to) the logarithm
of the number of states which appear
in the definition of the partition function as a state sum. When talking
about ``entropy", it is by analogy with such statistical systems.
However, while a ``state'' in the classical partition function for spins
can be considered as physical, a ``history''
contributing to the path integral in quantum field theory (or in quantum
mechanics) is {\it not} physical. After rotation to
Euclidean time, the path integral can be viewed as partition function
of a (virtual) statistical system. It is only in this sense that we talk about
the entropy of geometries.}
Unlike in two dimensions, it has until now not been possible to solve this
counting problem analytically. As mentioned above, one
cannot even prove the exponential bound. This means that
we will in the first place probe the properties of \rf{ny11} with numerical methods.

Let us now turn to the renormalization of the cosmological constant
by fine-tuning $\kp_4$.
We write the partition function as
\beq\label{ny14}
Z(\kp_0,\kp_4,\Del) = \sum_{N_4} \e^{-(\kp_4+\Del) N_4} \; Z_{N_4}(\kp_0,\Del),
\eeq
where $Z_{N_4}(\kp_0,\Del)$ is the partition function for a fixed number
$N_4$ of four-simplices, namely,
\beq\label{ny15}
Z_{N_4}(\kp_0,\Del)= \sum_{T_{N_4}}  \frac{1}{C_T}
\e^{-\Del N_4^{(4,1)}(T_{N_4})} \;\e^{(\kp_0+6\Del)N_0(T_{N_4})}.
\eeq
As already mentioned, there is numerical evidence that $Z_{N_4}(\kp_0,\Del)$
is exponentially bounded as a function of $N_4$ \cite{expobound},
\beq\label{ny16}
Z_{N_4}(\kp_0,\Del) \leq \e^{(\kp_4^c+\Del)N_4}f(N_4,\kp_0,\Del),
\eeq
where $f(N_4)$ grows slower than exponentially. We call $\kp_4^c$ the
{\it critical} value of $\kp_4$. It is a function of $\Del$ and $\kp_0$
and plays the same role as $\lam_c$ in the two-dimensional model discussed
above: the partition function is only defined for $\kp_4 > \kp_4^c$ and
the ``infinite-volume'' limit, where $\la N_4\ra \to \infty$ can only
be achieved for $\kp_4 \to \kp_4^c$. We are interested in sending the
lattice spacings $a=a_s,a_t$ to zero while keeping the physical
four-volume, which is roughly $N_4 a^4$, fixed. Thus
we want to consider the limit $N_4 \to \infty$, and fine-tune
$\kp_4$ to $\kp_4^c$ for fixed $\kp_0,\Del$. This fine-tuning is similar
to the fine-tuning $\lam \to \lam_c$ in the two-dimensional model.
Like there, we expect the {\it physical} cosmological constant
$\Lambda$ to be defined
by the {\it approach} to the critical point according to
\beq\label{ny17}
\kp_4 = \kp_4^c + \frac{\La}{16\pi G} a^4,
\eeq
which is the counterpart to \rf{ny4}. It ensures that the term
\beq\label{ny17a}
(\kp_4-\kp_4^c)\;N_4 =  \frac{\La}{16\pi G} V_4,~~~~V_4=N_4a^4,
\eeq
gives rise to the standard cosmological term in the Einstein-Hilbert action.
Thus the fine-tuning of $\kp_4 \to \kp_4^c$ brings us to
the limit of infinite four-volume. It
does {\it not necessarily} imply a continuum limit too.
The situation here may be different from that of the
two-dimensional model, where
approaching $\lam_c$ automatically implied a continuum limit.
Two-dimensional quantum gravity is of course a very simple model with
no propagating degrees of freedom, which we {\it do} expect to be present 
in four-dimensional quantum gravity. Thus the situation
is more like in ordinary Euclidean lattice field theory/critical phenomena,
where ``infinite volume'' does not
necessarily mean ``continuum limit''.

A good example of what one might
expect is the Ising model on a finite lattice. To obtain a phase
transition for this model one has to take the lattice volume
to infinity, since there are no genuine phase transitions for finite
systems. However, just taking the lattice volume to infinity is not
sufficient to ensure critical behaviour of the Ising model. We also
have to tune the coupling constant to its critical value, at which
point the spin-spin correlation length diverges.
Similarly, in CDT, having adjusted $\kp_4(\kp_0,\Del)$
we have to search the
coupling constant space of
$\Del$ and $\kp_0$ in order to find regions where we do not only
have an infinite four-volume limit but also an interesting
limit from the point of view of continuum physics.

How can one imagine obtaining an interesting continuum behaviour
as a function of $\kp_0$? For purposes of illustration we ignore
$\Del$ and
assume that the subleading correction $f(N_4,k_0)$ has the form
\beq\label{ny17b}
f(N_4,\kp_0) = \e^{k (\kp_0) \sqrt{N_4}}.
\eeq
(We will later on check numerically that such a term
is indeed present.) The partition function now becomes
\beq\label{ny17x}
Z(\kp_4,\kp_0) = \sum_{N_4} \e^{-(\kp_4-\kp_4^c)N_4 + k(\kp_0) \sqrt{N_4}}.
\eeq
For dimensional reasons we expect the classical Einstein term in the action
to scale like
\beq\label{ny17c}
\frac{1}{16 \pi G} \int \d^4 \xi \, \sqrt{g(\xi)}\;\;R(\xi)
\;\;\;\propto \;\;\;\frac{\sqrt{V_4}}{G},
\eeq
motivating the search for a value $\kp_0^c$ with
$k(\kp_0^c) =0$, with the approach to this point governed by
\beq\label{ny17d}
k(\kp_0)\propto \frac{a^2}{ G},~~~~{\rm i.e.}~~~~~
k(\kp_0)\sqrt{N_4}\;\propto \;\frac{\sqrt{V_4}}{G}.
\eeq
With such a choice we can identify a continuum limit where
$\la N_4\ra$, calculated from \rf{ny17x}
(by a trivial saddle point calculation), goes to infinity while $a \to 0$,
\beq\label{ny17e}
\la N_4\ra = \frac{ \sum_{N_4} N_4 \;
\e^{-(\kp_4-\kp_4^c)N_4 + k(\kp_0) \sqrt{N_4}}}{\sum_{N_4}
\e^{-(\kp_4-\kp_4^c)N_4 + k(\kp_0) \sqrt{N_4}}} \approx
\frac{k^2(\kp_0)}{4(\kp_4-\kp_4^c)^2} \propto \frac{1}{\La^2 a^4}.
\eeq
Thus we find
\beq\label{ny17f}
\la V_4\ra \propto \frac{1}{\La^2},~~~~Z(\kp_4,\kp_0) \approx
\exp\Big(\frac{k^2(\kp_0)}{4(\kp_4-\kp_4^c)}\Big) =
\exp\Big(\frac{c}{G\La}\Big),
\eeq
as one would na\"ively expect from Einstein's equations, with the partition
function being dominated by a typical instanton contribution, for a
suitable constant $c>0$.

\subsection{Set-up of computer simulations}

The actual set-up for the computer simulations is slightly
different from the theoretical framework discussed above, in that we choose
to work with a fixed number of four-simplices $N_4$,
rather than fine-tuning $\kp_4$ to its critical
value. We can perform computer simulations for various $N_4$
(and fixed $\kp_0,\Del$) and in this way check scaling with respect
to $N_4$. This is an alternative to fine-tuning $\kp_4$,
and much more convenient from a computational point of view. For large
$N_4$ we can then check whether there are any finite-size effects or
whether effectively we have already reached the infinite-volume limit.
At a formal level we can perform a Laplace transformation
of the partition function,
\beq\label{constantV1}
Z(\La) = \int_0^\infty \d V \; \e^{-\La \, V}\; Z(V),
\eeq
where
\beq\label{constantV2}
Z(V) = \int \cD [g_{\m\n}] \; \e^{-S_{Ein}[g_{\m\n}]} \;
\del\!\left( \int \sqrt{g} -V\right),
\eeq
is the partition function for a constant four-volume and
$S_{Ein}[g_{\m\n}]$ the gravitational action without
the cosmological term.
The expectation values we are calculating are thus
\beq\label{constantV3}
\la \cO(g_{\m\n})\ra_V = \frac{1}{Z(V)} \int \cD [g_{\m\n}] \;
\cO(g_{\m\n})\; \e^{-S[g_{\m\n}]} \;
\del\!\left( \int \sqrt{g} -V\right).
\eeq
 It is in principle possible to reconstruct  the expectation
values of observables for a fixed value of the cosmological constant
$\La$ by measuring the $\la \cO(g_{\m\n})\ra_V$ for all $V$ and constructing
$Z(V)$ (which cannot be measured as an observable in a Monte Carlo
simulation). This reconstruction is somewhat cumbersome but feasible, since
the action only depends on the ``global'' variables $N_0$, $N_4^{(1,4)}$
and $N_4^{(2,3)}$. However, we will restrict ourselves to the measurement
of observables for fixed $V$.

\begin{figure}[t]
\vspace{-2cm}
\psfrag{V}{\bf{\Large $N_4^{(3,2)}$}}
\psfrag{P(V)}{\Large\bf $P(N_4^{(3,2)})$}
\centerline{\scalebox{0.45}{\rotatebox{-90}{\includegraphics{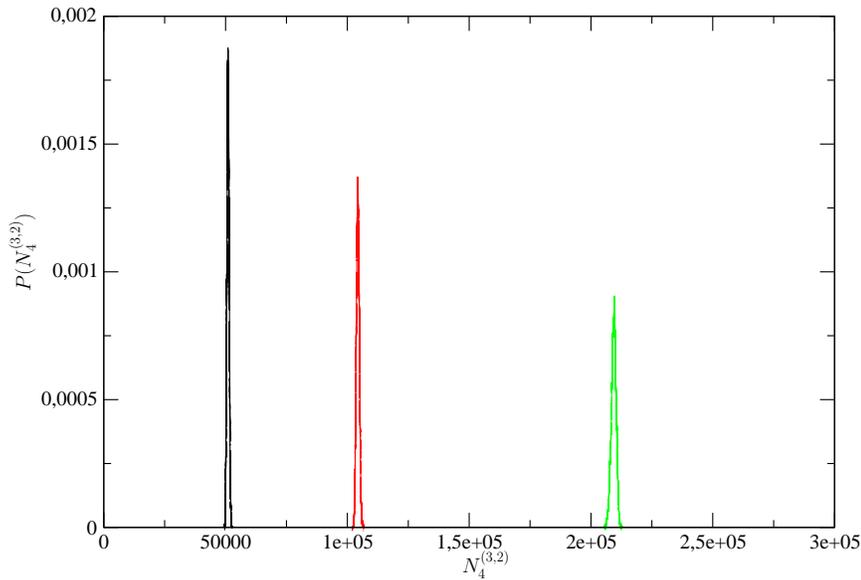}}}}
\vspace{-0.0cm}
\caption[phased]{{\footnotesize Unnormalized distribution of the number
$N_4^{(3,2)}$ of four-dimensional
simplices of type (3,2), at fixed numbers $N_4^{(4,1)}=$ 40, 80
and 160k (left to right) of four-simplices of
type (4,1), at $\kappa_0\equ 2.2$ and $\Delta\equ 0.6$.
}}
\label{volumes}
\end{figure}

The local moves do not in general preserve the numbers
$N_4^{(4,1)}$ and $N_4^{(3,2)}$ of four-simplices, or their
sum. We deal with this in a standard way which was
developed for dynamically triangulated models in
dimensions three and four \cite{av,aj}
to ensure that the system volume is peaked at a prescribed value,
with a well-defined range of fluctuations.
Adapting it to the current
situation of causal triangulations with nonvanishing asymmetry
$\Delta$, we
implement an approximate four-volume constraint by
adding a term
\beq
\delta S = \epsilon |N_4^{(4,1)}- \tilde N_4|,
\label{fixvolume}
\eeq
to the Euclidean action \rf{actshort},
with typical values of $\epsilon$ lying in
the range of 0.01 to 0.02, except during thermalization where we
set $\epsilon\equ 0.05$. The reason for fixing $N_4^{(4,1)}$
instead of $N_4\equ N_4^{(4,1)}\plu N_4^{(3,2)}$ in eq.\ \rf{fixvolume}
is mere technical convenience.
We have checked in the phase space
region relevant to four-dimensional quantum gravity
(phase C, see below) that for $N_4^{(4,1)}$ fixed according
to \rf{fixvolume}, the number $N_4^{(3,2)}$ of four-simplices
of type (3,2) is likewise peaked sharply, see Fig.\ \ref{volumes}.
\begin{table}
\begin{center}
\renewcommand{\arraystretch}{1.4}
\begin{tabular}{ |c||c|c|c|c|c|}
\hline
``four-volume" $\tilde N_4= N_4^{(4,1)}$  &   10 &   20 &  40 & 80 & 160 \\
\hline
actual four-volume $N_4= N_4^{(4,1)}\plu N_4^{(3,2)}$  & 22.25  &  45.5
& 91 & 181 & 362   \\
\hline
\end{tabular}
\end{center}
\caption{\label{ntab} {\footnotesize 
Translation table between the two types of discrete
four-volume, $\tilde N_4$ and $N_4$, listing the values at which most of
the numerical simulations have been performed so far,
in units of 1000 building blocks. The table is valid for the
coupling constants $(\kp_0,\Del) = (2.2,0.6)$, which is well inside phase C.
}}
\end{table}
The ``four-volumes" $\tilde N_4$ and the corresponding ``true"
discrete four-volumes $N_4$ used in the simulations are
listed in Table \ref{ntab} for $(\kp_0,\Del)= (2.2,0.6)$ well inside phase
C, at the choice of coupling constants where most of our simulations
are performed.
In order to stabilize the total volume after thermalization,
$\kappa_4$ has to be fine-tuned to its pseudo-critical value
(which depends weakly on the volume) with accuracy smaller
than $\epsilon$, in practice to about $0.2\epsilon$.
The measurements reported in this paper were taken
at $\tilde N_4\equ$ 10, 20, 40, 80 and 160k, and the runs
were performed on individual PCs or a PC farm for the smaller
systems and a cluster of work stations for the larger systems.

Before measurements can be performed, one needs  a well
thermalized configuration of a given volume. In order to
double-check the quality of the thermalization, we used two
different methods
to produce starting configurations for the measurement runs.
In the first method, we evolved from an
initial minimal four-dimensional triangulation of prescribed topology
and given time extension $t$, obtained by
repeated gluing of a particular triangulated spacetime slice
(a ``slab'') of $\Delta t\equ 1$ (one lattice step in the
time direction) and topology $[0,1]\times S^3$,
which consists of 30 four-simplices. The spatial in- and out-geometries
of the slice are minimal spheres $S^3$, made of five tetrahedra.
The two types of spatial boundary conditions used are (i) periodic
identification of the geometries at initial and final integer times,
and (ii) free boundary conditions, where all vertices contained
in the initial slice at time $t_0$ are
connected by timelike edges to a single vertex at time $t_0\mi 1$,
and similarly for the vertices contained in the final spatial slice.
From this initial configuration, the geometry evolves to its
target volume $\tilde N_4$ as specified in $\delta S$. During the
evolution the volume-volume correlator (to be defined below)
was monitored and used to gauge the auto-correlation of configurations.
The number of sweeps to reach the
thermalized configuration changes linearly with $\tilde N_4$ and
ranged from $10^5$ to $10^8$ sweeps for the largest volumes,
the latter of which took several weeks on a work station.

In the second method, we started instead from a
thermalized configuration of smaller volume, which we let
evolve towards the target volume. In this case the final
volume-volume distribution is reached from a narrower
distribution, namely, that of the smaller volume. During
thermalization, this width grows very slowly. The timing of
the entire process is similar to that of the first method.

\subsection{Phase structure of the model}\label{phase}

\begin{figure}[t]
\center
\scalebox{1.0}{\rotatebox{0}{\includegraphics{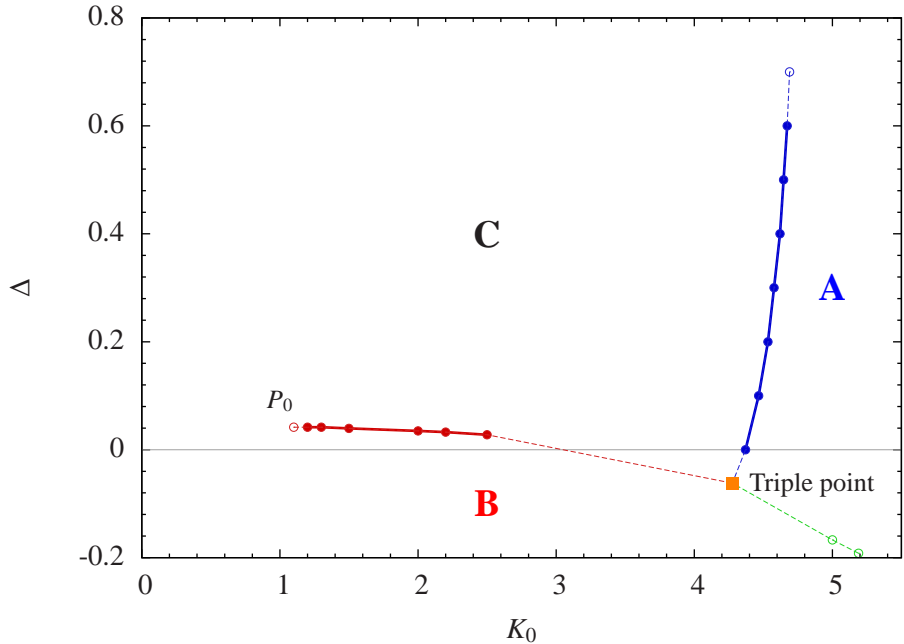}}}
\caption{{\footnotesize 
The phase diagram of four-dimensional quantum gravity, defined
in terms of causal dynamical triangulations, parametrized by the inverse
bare gravitational coupling $\kappa_0$ and the asymmetry parameter $\Delta$.}}
\label{phasediagram}
\end{figure}

As described above, the bare cosmological
constant $\kappa_4$ is tuned to its
(pseudo-)critical value in the
simulations, tantamount to approaching the infinite-volume limit.
Depending on the values of the two remaining parameters $\kappa_0$
and $\Delta$ in the discretized action \rf{actshort}, we have identified
three different phases, A, B and C, mutually separated by phase transition
lines \cite{emerge,blp}. Fig.\ \ref{phasediagram} shows the phase diagram,
based on computer simulations with $N_4^{(4,1)}=80.000$ \cite{horava-simu}.
Because there are residual finite-size effects for universes of this size,
one can still expect minor changes in the location of the transition lines
as $N_4 \to \infty$. The dotted lines in Fig.\ \ref{phasediagram}
represent mere extrapolations,
and lie in a region of phase space which is difficult to access due
to inefficiencies of our computer algorithms.
Thus the phase diagram is only tentative where the lines
are dotted, and in particular the
triple point is our invention. However, once we are in the
``bulk'' of phase A, B and C there are no problems with the simulations.
Thus the phase diagram in Fig.\ \ref{phasediagram} is essentially
correct in the sense that we have only observed these three phases.
Also, moving away from the putative triple point there is
eventually no problem in identifying a transition between phases A and B.
We have not done so except for the single point shown on Fig.\
\ref{phasediagram} because this phase transition presently does not
have our main interest (since we do not know of an interesting continuum interpretation).
In summary, the most likely scenario, even without
having identified part of the transition lines in detail, is the one
presented in Fig.\ \ref{phasediagram}.

We describe each of the three phases in turn:
\begin{figure}[ht]
\centerline{\scalebox{0.5}{\rotatebox{0}{\includegraphics{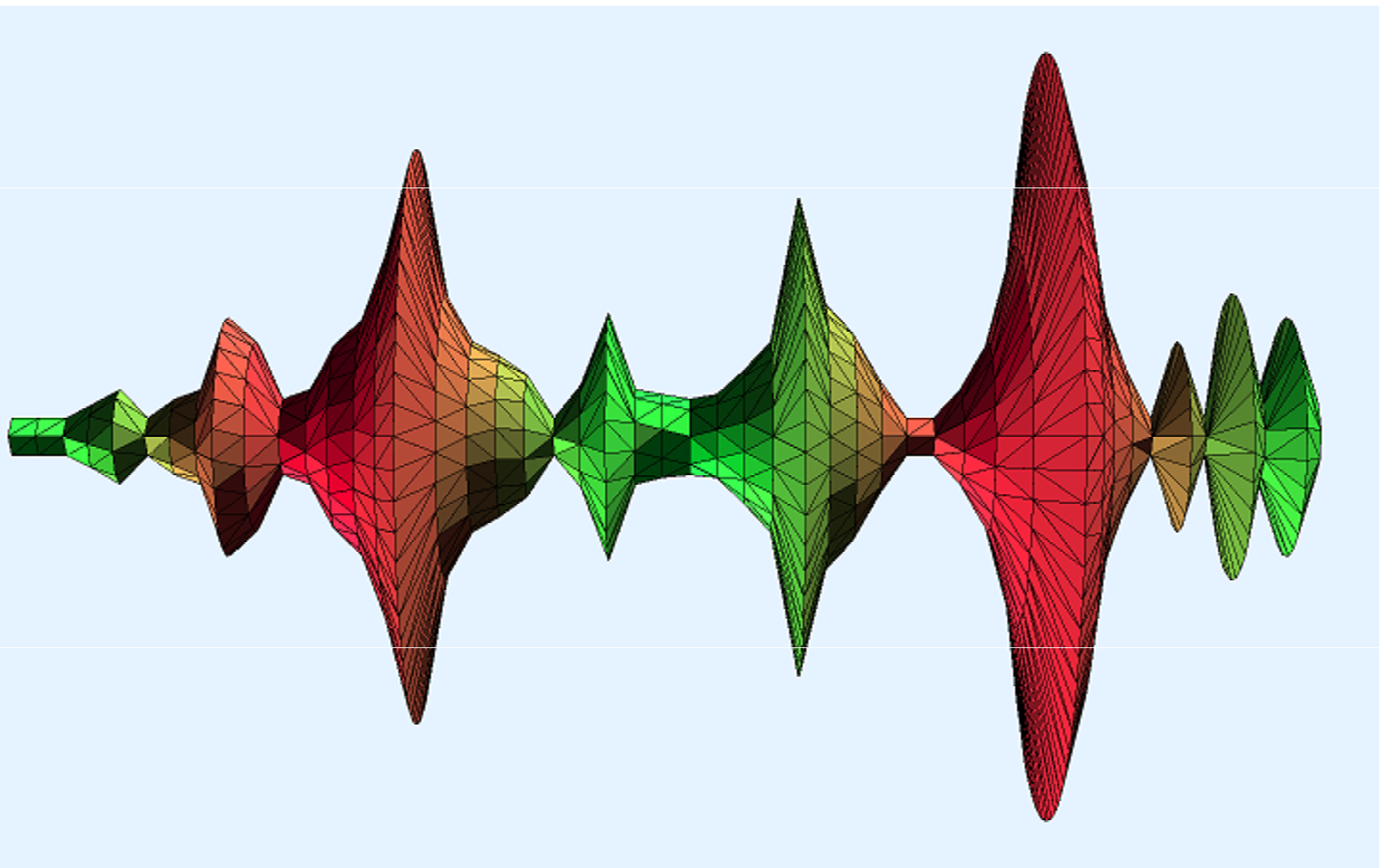}}}}
\caption[phased]{{\footnotesize
Monte Carlo snapshot of a typical universe in phase A ($\kappa_0\equ 5.0$,
$\Delta\equ 0$),
of discrete volume $N_4\equ$ 45.5k
and total time extent (horizontal direction) $t_{total}\equ 50$.
In this and the following two figures, the circumference at
integer proper time $t_n$ is chosen proportional to the spatial three-volume
$N_3(t_n)$. The surface represents an interpolation between
adjacent spatial volumes, without capturing the
actual four-dimensional connectivity between neighbouring spatial slices.}}
\label{uni5p0}
\end{figure}
\begin{itemize}
\item[(A)] This phase prevails for sufficiently large $\kappa_0$
(recall $\kappa_0$ is proportional to
the inverse of the bare Newton's constant).
When plotting the volume of the spatial slices of constant $t_n$ as a function
of $t_n$, we observe an irregular sequence of maxima and minima,
where the minimal size is of the order of the cutoff, and the sizes of
the maxima vary, see Fig.\ \ref{uni5p0}. The time intervals during which
the spacetime has a macroscopic spatial extension are small and
of the order of $\Delta n\equ 3-4$, $n$ being the subscript in $t_n$.
\item[(B)] This phase occurs for sufficiently small $\kappa_0$ and
for small asymmetry $\Delta$. In it,
spacetime undergoes a ``spontaneous dimensional reduction"
in the sense that all four-simplices are concentrated in a slice of
minimal time extension between $t_n$ and $t_{n+1}$ and $t_n$ and $t_{n-1}$,
such that the three-volumes $N_3(t_k)$ at times $t_k$ is
only large at time $t_n$ and remain close to their kinematic minimum
everywhere else (Fig.\ \ref{uni1p6}).
\end{itemize}
\begin{figure}[t]
\centerline{\scalebox{0.5}{\rotatebox{0}{\includegraphics{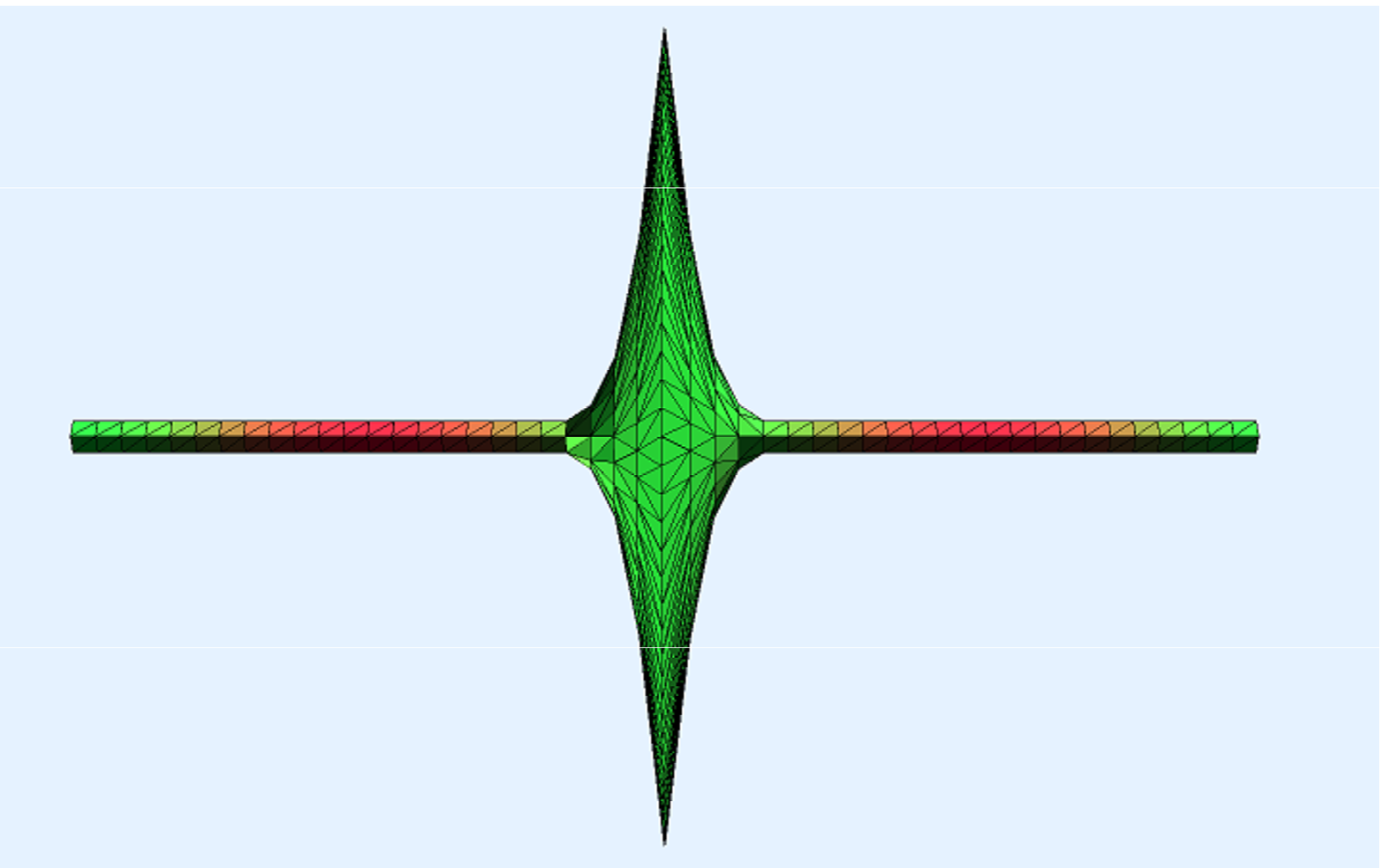}}}}
\caption[phased]{{\footnotesize
Monte Carlo snapshot of a typical universe in phase B ($\kappa_0\equ 1.6$,
$\Del=0$),
of discrete volume $N_4\equ$ 22.25k
and total time extent $t_{total}\equ 50$.
The entire universe has collapsed into a
slice of minimal time extension.}}
\label{uni1p6}
\end{figure}
However, there is yet another phase where the dynamics is not
reduced to just the spatial directions but is genuinely
four-dimensional.
\begin{itemize}
\item[(C)] This phase occurs for sufficiently small $\kappa_0$ and
nonvanishing asymmetry $\Delta$. In this phase,
where $\Delta >0$ and $\tilde{\alpha}<1$ ($\tilde{\alpha}$ was defined
above, eq.\ \rf{act4dis}),
there seems to be a sufficiently strong coupling between successive
spatial slices to induce a change in the spatial
structure itself. We will discuss
this new geometrical structure in detail below in Sec.\ \ref{S4a}.
Stated shortly, we
observe a genuine four-dimensional universe, {\it the blob} shown
in Fig.\ \ref{unipink}. This blob is four-dimensional in the sense that
as a function of the  four-volume $N_4$,
the time extent scales as $N^{1/4}_4$ and the spatial volume
as $N_4^{3/4}$ (see Fig.\ \ref{unipink}), and the blob itself
is independent of the total time extent $t_{total}$ chosen, as long as
$t_{total}$ is larger than the time extent of the blob.
\end{itemize}
\begin{figure}[ht]
\centerline{\scalebox{0.5}{\rotatebox{0}{\includegraphics{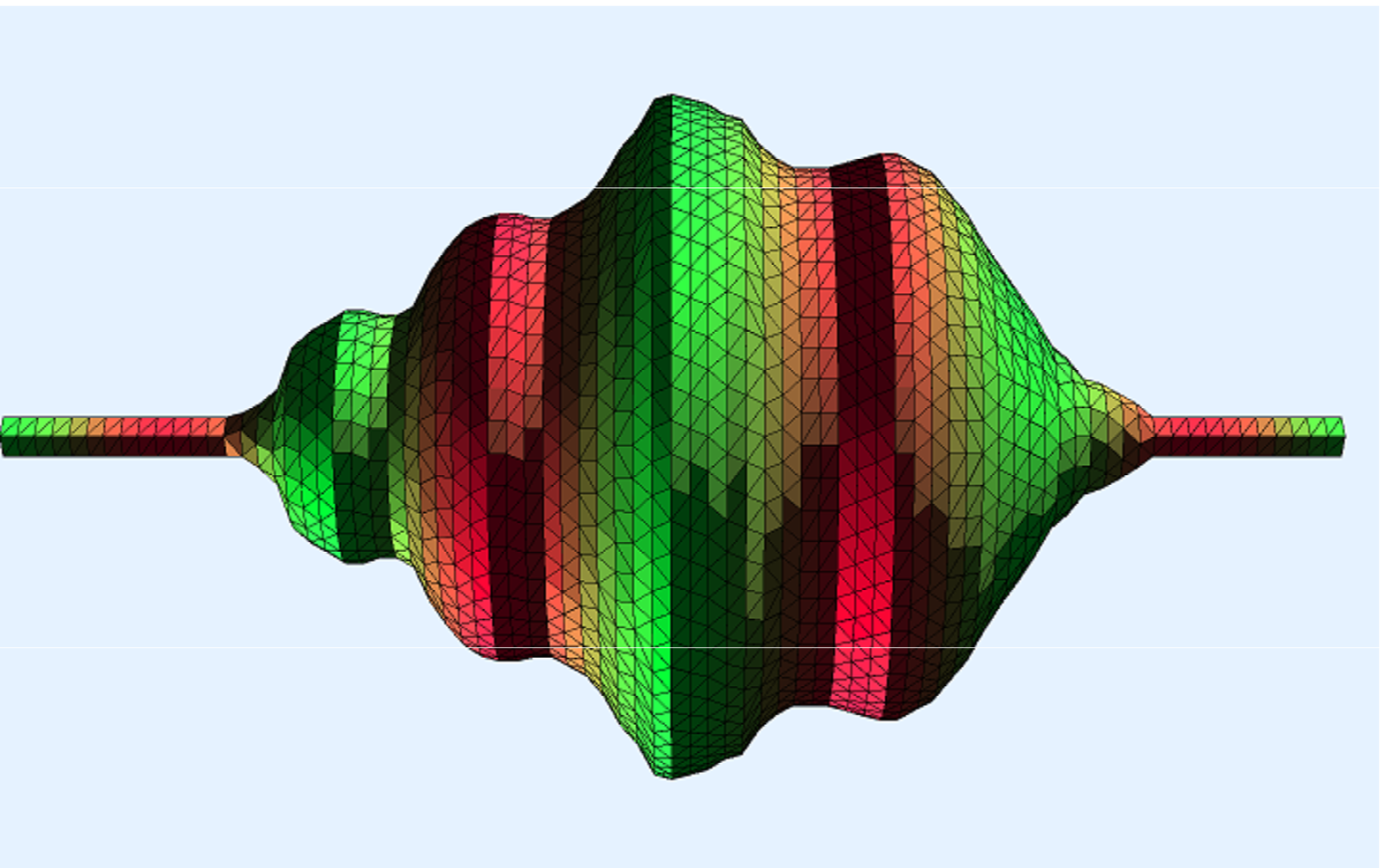}}}}
\caption[phased]{{\footnotesize
Monte Carlo snapshot of a typical universe in phase C ($\kappa_0\equ 2.2$,
$\Del=0.6$)
of discrete volume $N_4\equ$ 91.1k
and total time extent $t_{total}\equ 80$.}}
\label{unipink}
\end{figure}

We have performed measurements to determine the order of the transitions in the CDT
phase diagram. Based on our numerical investigation so far, the
A-C transition appears to be
a first-order transition \cite{horava-simu},
while the B-C transition is second order \cite{second-order}.
Below we report on some details of these measurements.

\subsubsection{The A-C transition}

\begin{figure}[t]
\center
\scalebox{1.05}{\rotatebox{0}{\includegraphics{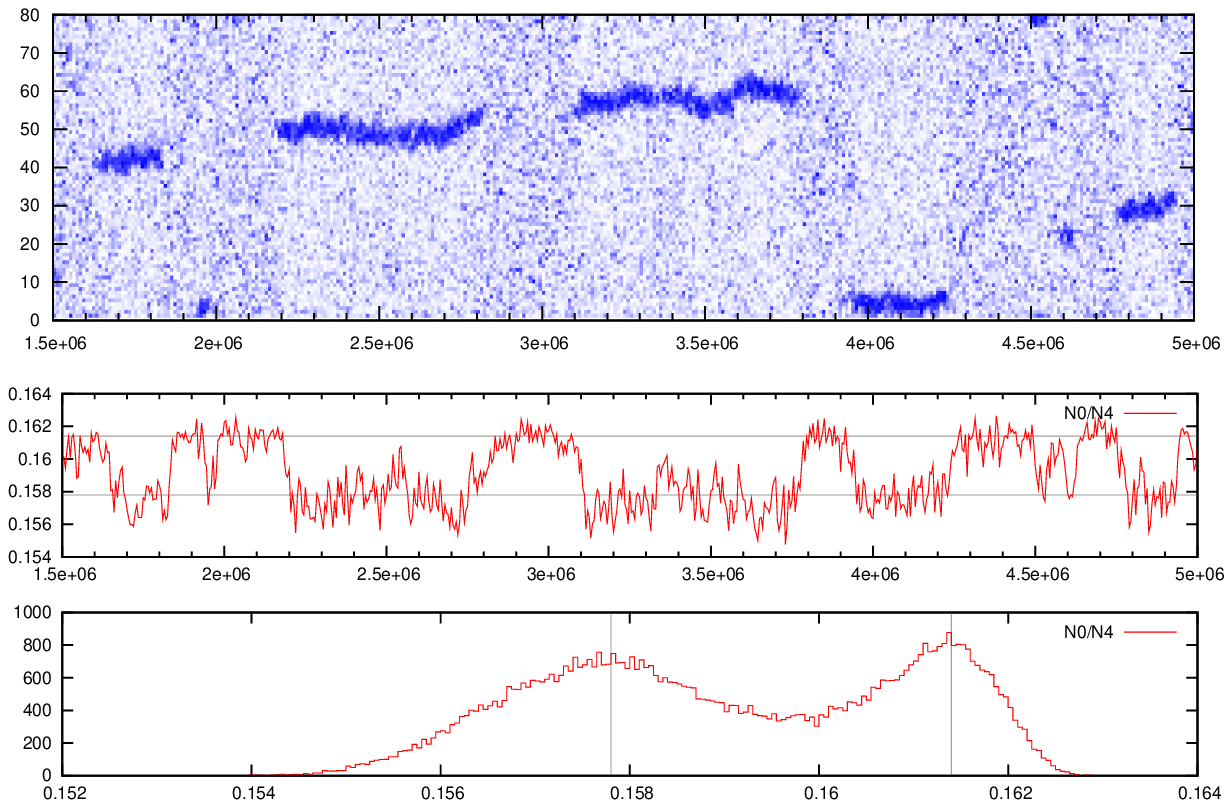}}}
\caption{{\footnotesize 
Transition between phases $C$ (smaller $\kp_0$) and $A$ (larger $\kp_0$)
at $\kp_0 = 4.711$ and $\Delta = 0.6$ for $N_4 = 120k $.
We observe that configurations \emph{jump} between two regimes,
which is a strong evidence of first-order transition.
Top: Density plot of the spatial volume as a function
of Monte Carlo simulation time (horizontal axis) and
slicing time $t_n$ (vertical axis).
Darker colours mean larger spatial volumes $N_3(t_n)$.
Middle: Order parameter $N_0 / N_4$, conjugate to $\kp_0$,
as a function of Monte Carlo time.
Bottom: Distribution of the values
taken by the order parameter $N_0 / N_4$, exhibiting a double-peak structure.}}
\label{phtr-order}
\end{figure}

The two graphs at the bottom of Fig.\ \ref{phtr-order} illustrate the behaviour
of $N_0/N_4$ at the A-C phase transition line.
Since we can approach this line by changing the coupling constant $\kp_0$
while keeping $\Del$ fixed, the quantity conjugate to $\kappa_0$
($N_4$ is fixed), namely, the ratio $N_0/N_4$, is a natural
candidate for an order parameter. The graph at the
centre of Fig.\ \ref{phtr-order}
shows $N_0/N_4$ as a function of Monte Carlo time.
One sees clearly that it jumps between two values, corresponding to
the distinct nature of geometry in phases A and C. We have checked that
the geometry indeed ``jumps" in the sense that no
smoothly interpolating typical configurations have been found. Lastly,
we have also established that the jump becomes more and more
pronounced as the four-volume $N_4$ of the universe increases, further
underlining the archetypical first-order behaviour at this transition line.

The top graph in Fig.\ \ref{phtr-order} shows the location of the universe
along the vertical ``proper-time
axis" ($t_n\in[ 0, 80]$, and to be periodically identified) as a function of
Monte Carlo time, plotted along the horizontal axis.
The value of the spatial three-volume $N_3(t_n)$ in the slice labeled by
$t_n$ is colour-coded; the darker, the bigger the volume at time $t_n$.
We can distinguish two types of behaviour as a function of Monte Carlo time,
(i) presence of an extended universe centred at and fluctuating weakly around
some location on the proper-time axis; (ii) absence of such a universe with
a well-defined ``centre-of-volume". The former is associated with the presence
of a distinct dark band in the figure, which disappears abruptly as a function
of Monte Carlo time, only to reappear at some different location $t_n$ later
on in the simulation. Comparing with the middle graph, it is clear that
these abrupt changes in geometry correlate perfectly with
the changes of the order parameter $N_0/N_4$.
When $N_0/N_4$ is small, we witness the extended universe of phase C,
whose ``equator" coincides with the dark blue/red line of the colour plot.
Conversely, at the larger values of $N_0/N_4$ characteristic of phase A
this structure disappears, to be replaced by an array of universes
too small to be individually identifiable on the plot. When jumping back
to phase C the centre-of-volume of the single, extended universe
reappears at a different location in time.
Finally, the bottom graph in Fig.\ \ref{phtr-order}
illustrates the double-peak
structure of the distribution of the values taken by the order
parameter $N_0/N_4$.

\subsection{The B-C transition}

\begin{figure}[t]
\center
\scalebox{1.05}{\rotatebox{0}{\includegraphics{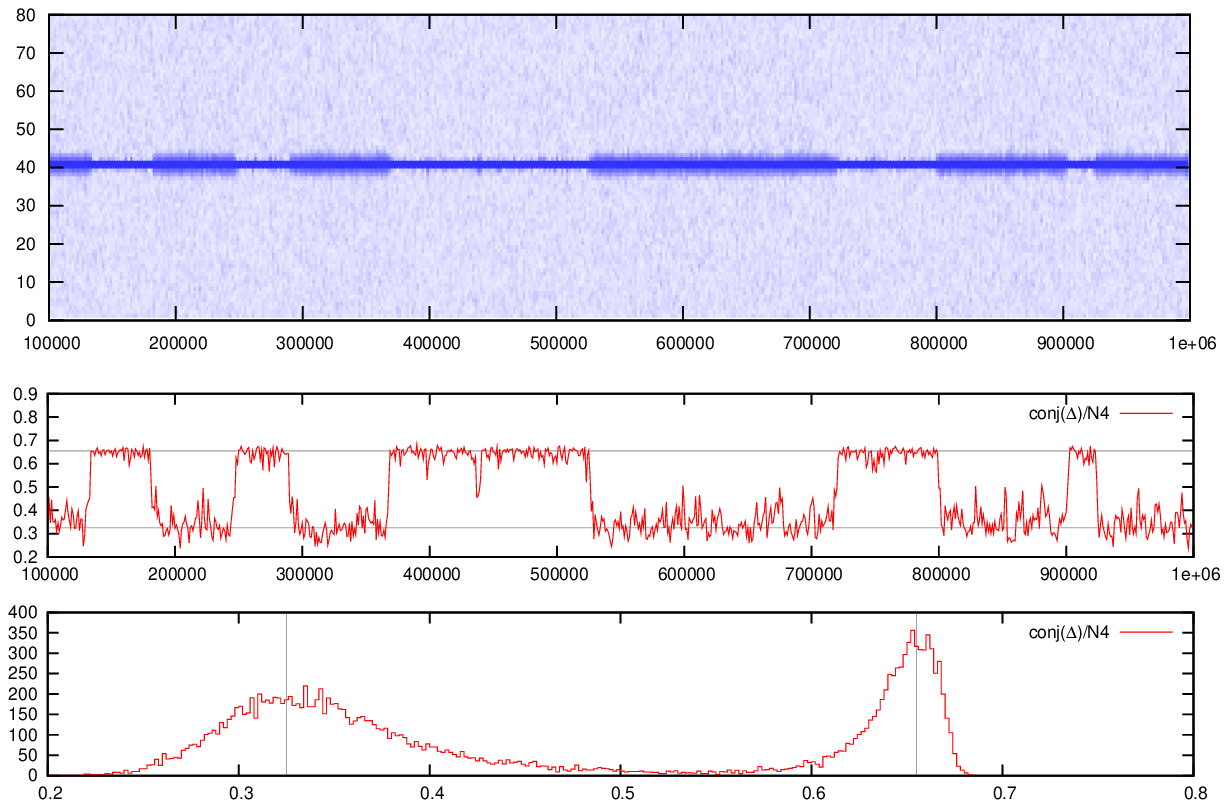}}}
\caption{{\footnotesize
Transition between phases $C$ (larger $\Delta$) and $B$ (smaller $\Delta$)
at $\kp_0 = 2.2$ and $\Delta = 0.022$ for $N_4 = 40k$.
Although the configurations \emph{jump} between two regimes,
the effect gets weaker with increasing total volume $N_4$.
Thus the jump cannot be taken as evidence of a first-order
transition.
Top: Density plot of the spatial volume as a function
of Monte Carlo simulation time (horizontal axis)
and slicing time $i$ (vertical axis).
Darker colours mean larger spatial volumes $N_3(t_n)$.
Middle: Order parameter $(N_{41} - 6 N_0) / N_4$, conjugate to $\Delta$,
as a function of Monte Carlo time.
Bottom: Distribution of the values taken
by the order parameter, again exhibiting a double-peak structure.}}
\label{phtr-bc}
\end{figure}

Our measurements to determine the character of the B-C transition are
depicted in an analogous manner in Fig.\ \ref{phtr-bc}.
Since we are varying $\Delta$ to reach this transition from inside phase C,
we have chosen the variable conjugate to
$\Delta$ in the action \rf{actshort} (up
to a constant normalization $N_4$),
$\mathrm{conj}(\Delta)= (-6N_0 +N_4^{(4,1)})/N_4$, as our order parameter.
Looking at the graph at the centre,
we see that this parameter exhibits the
jumping behaviour as a function of Monte Carlo time that is characteristic of a
first-order transition. Small values of the parameter indicate the system is
in phase C, while large values correspond to phase B.
The time extent of the universe diminishes
as one approaches the phase transition line from phase C,
and is close to zero when we are at the transition line.
It {\it is} zero when we cross the line.
Some indication of this behaviour is given by the colour-coded three-volume
profile $N_3(t)$ as a function of the Monte Carlo time in the top graph of
Fig.\ \ref{phtr-bc}. In phase B,
only one lattice time $t_n$ has a number of
tetrahedra $N_3(t_n)$ much larger than zero. The ``universe" is
concentrated on a single
time slice, while well inside phase C it has a nontrivial time extension.
The bottom graph in Fig.\ \ref{phtr-bc} again
shows the double-peak structure of the order parameter.

Looking at Fig.\ \ref{phtr-bc} and comparing it
with the previous Fig.\ \ref{phtr-order},
the evidence for a first-order transition at the
B-C phase boundary seems even more clear-cut than in the case of the
A-C transition.

\begin{figure}[t]
\centerline{\scalebox{0.8}{\includegraphics{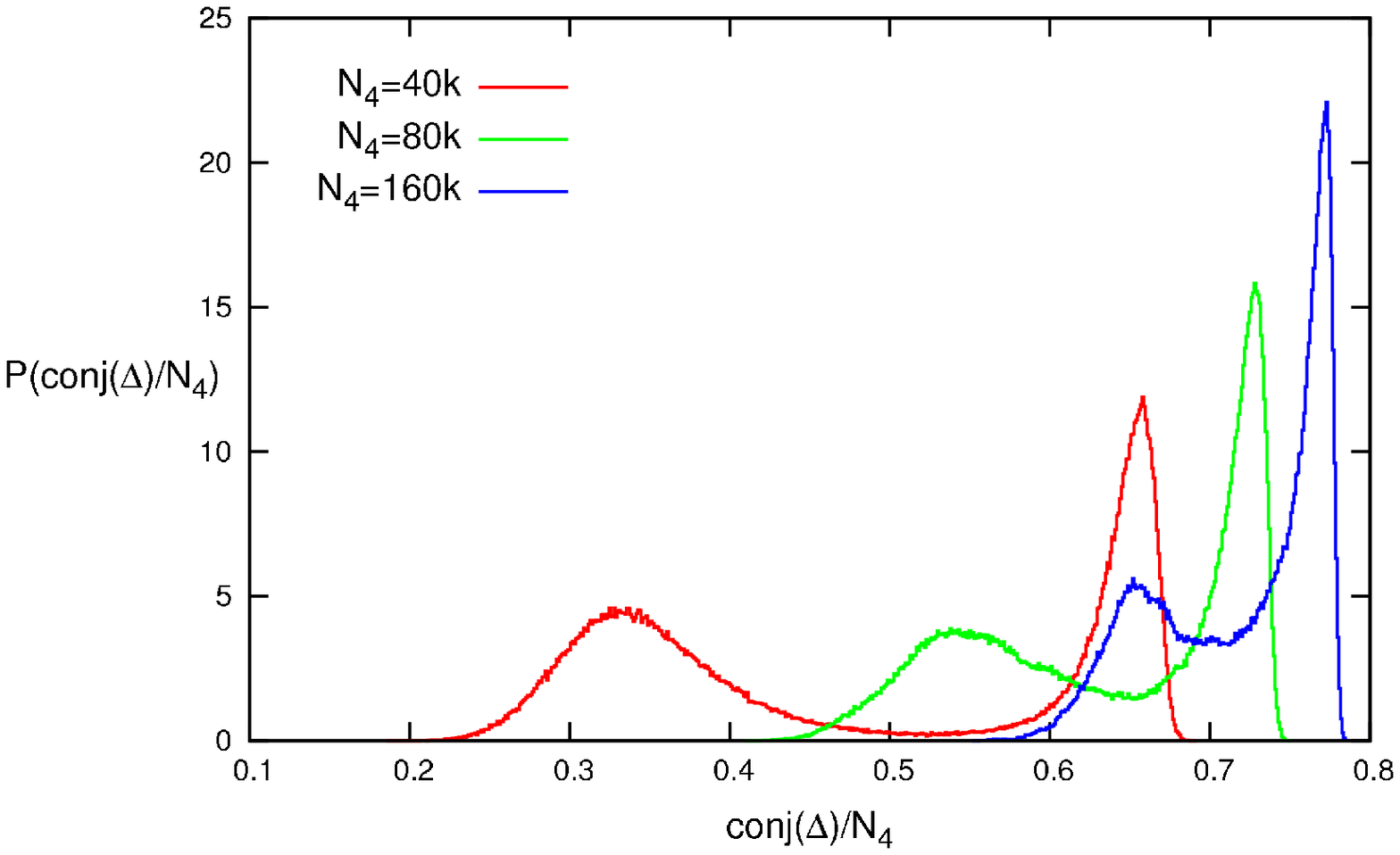}}}
\caption{{\footnotesize
Histograms of $\mathrm{conj}(\Delta)$ at the B-C transition
for three different system sizes (40K, 80K and 160K).
The histograms are normalized to probability distributions.
By fine-tuning $\Del$, one can achieve that the two peaks
have the same height. We have not done this because it is
quite time-consuming.}}
\label{peaks}
\end{figure}

However, a double peak for a given $N_4$ does not
tell us whether the transition is first or second order. One
has to establish how the double peak behaves as a function of $N_4$.
If the size of the peaks increases and the separation of the peaks
does not diminish when $N_4$ increases it is a strong signal that
the transition is first order. This appears to be the situation
for the A-C transition. For the B-C transition the opposite
happens: the peaks approach each other, approximately as $1/N_4$, and
the height of the peaks relative to the minimum between the peaks
also diminishes, again approximately as $1/N_4$. This is shown
in Fig.\ \ref{peaks}. Thus the peak structure does not have
the characteristics of a first-order transition.

In order to establish that we are really dealing with a higher-order transition we
have looked at various critical exponents. They all point to
a higher-order transition. We will limit the discussion here to
the so-called shift exponent, $\tilde{\n}$ (for a detailed
discussion we refer to \cite{second-order}), defined
as follows. For fixed $\kp_0$,
we vary the coupling constant $\Del$ to reach a critical
point on the B-C phase transition line. For a system of finite volume,
the precise definition of the critical $\Del^c$ is
ambiguous, since there is strictly speaking only a phase transition
for $N_4 = \infty$. We choose to define $\Del^c(N_4)$
using the location of the maximum of the susceptibility
$\chi_{\mathrm{conj}(\Delta)}=
\left<\mathrm{conj}(\Delta)^2\right>-\left<\mathrm{conj}(\Delta)\right>^2$.
One expects a $N_4$-dependence of the form
\begin{equation}
\Delta^c(N_4)=\Delta^c(\infty)-C N_4^{-1/\tilde{\nu}},
\label{shiftexpo}
\end{equation}
with a first-order transition characterized by $\tilde{\n}=1$. Thus
observing a $\tilde{\n}$ different from 1 is a signal of a higher
order transition.

\begin{figure}[t]
\centerline{\scalebox{1.1}{\includegraphics{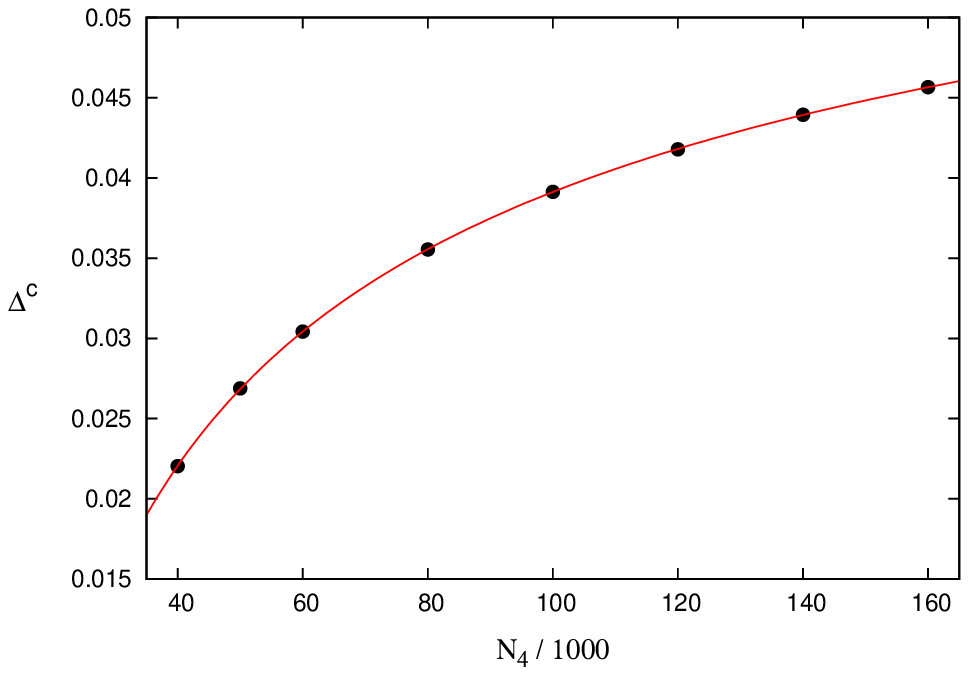}}}
\caption{{\footnotesize 
This plot shows measurements of B-C transition points at
$\kappa_0=2.2$ for different system sizes,
which allow to determine the shift exponent $\tilde{\nu}$.}}
\label{shiftdata}
\end{figure}

Figure \ref{shiftdata} displays the data and the best fit
for $\kp_0=2.2$. We have not included error bars
because they turned out to be smaller than the data point dots.
The best fit through all
data points yields $\tilde{\nu}=2.40\pm 0.03$, which is
far from the value $\tilde{\n}=1$ characterizing a first-order
transition. We can perform similar measurements for
the other values of $\kp_0$ along the B-C transition line
which are shown in Fig.\ \ref{phasediagram} and establish that for
these points the transition seems to be of second (or higher) order.

It would be interesting to perform the
same measurements moving along the B-C transition line
towards the conjectured triple point. However we need more
a more efficient algorithm\footnote{As described above,
we are basically using the simplest Metropolis algorithm for
updating geometry, using the local moves.
This local algorithm might be
very inefficient when it comes to  changing certain types of geometries.
In four-dimensional Euclidean DT gravity, one type of geometry
which could be created was a so-called baby universe,
only connected to the rest of the universe with a thin ``neck''.
These baby universes were both a blessing and a curse. By counting
them one could measure the so-called susceptibility exponent of
the partition function \cite{baby}, but the local moves had a hard
time getting rid of them, once they were created. This issue could be
resolved by
inventing an algorithm that cut away the baby universes in one place 
and glued them
back onto the triangulation in another \cite{baby-surgery}.
This ``baby-universe surgery'' decreased the auto-correlation time
in the simulations by several orders of magnitude. We have been searching
for a similar algorithm in the CDT case, but so far without success.}. It is possible that
a multicanonical algorithm can deal better with the
double-peak structure which seems to get more pronounced
when we move towards the conjectured triple point.

We have not studied the A-B transition in any detail since it seems
not interesting from a gravitational point of view, where we want to start
out in a phase with an extended, quasi-classical four-dimensional universe,
i.e.\ in phase C.

\subsection{Relation to Ho\v rava-Lifshitz gravity}

We can now give the following qualitative characterization of the three
phases in terms of what we will provisionally call ``average geometry".
As we will show below, the universe of phase C
exhibits a classical four-dimensional background
geometry on large scales. We may rephrase this by saying 
that $\la {\it geometry}\ra \neq 0$.
One may even argue that  $\la {\it geometry}\ra = const.$,
in view of the fact that according to the minisuperspace analysis of
\cite{agjl,bigs4,semi,rounds4} -- to be presented below --
the universe can be identified with a round
$S^4$, the maximally symmetric de Sitter space of constant scalar curvature.
By contrast, in phase B the universe presumably has no extension or
trace of classicality, corresponding to ``$\la {\it geometry}\ra = 0$".
Lastly, in phase A, the geometry of the universe appears to be
``oscillating'' in the time direction.
The three phases are separated by three phase transition lines which
we have conjectured to meet in a triple point as
illustrated in Fig.\ \ref{phasediagram}.

We have chosen this particular qualitative description to match precisely
that of a Lifshitz phase diagram \cite{lifshitz}.
In an effective Lifshitz theory, the Landau free energy density $F(x)$
as function of an order parameter $\phi(x)$
takes the form\footnote{see, for example, \cite{gold} for an introduction to
the content and scope of ``Landau theory"}
\bea\label{2.f2}
F(x) &=& a_2 \phi(x)^2 + a_4 \phi(x)^4 +a_6\phi(x)^6 + \cdots\\
&&+c_2(\prt_\a \phi)^2 +d_2 (\prt_\b \phi)^2
+ e_2 (\prt_\b^2 \phi)^2 +\cdots ,
\nonumber\eea
where for a $d$-dimensional system $\a =m+1,\ldots,d$, $\b=1,\ldots,m$.
Distinguishing between ``$\alpha$"- and ``$\beta$"-directions allows
one to take anisotropic behaviour into account.
For a usual system, $m=0$ and a phase transition can occur when
$a_2$ passes through zero (say, as a function of temperature).
For $a_2> 0$ we have $\phi=0$, while for $a_2 <0$ we have $|\phi| >0$
(always assuming $a_4 >0$).  However, one also has a transition
when anisotropy is present ($m>0$) and
$d_2$ passes through zero. For negative $d_2$
one can then have an oscillating behaviour of $\phi$ in the $m$
``$\beta$"-directions.
Depending on the sign of $a_2$, the transition to this
so-called modulated or helical phase can occur either from the
phase where $\phi=0$, or from the phase where $|\phi| >0$.
We conclude that the phases C, B, and A
of CDT quantum gravity depicted in
Fig.\ \ref{phasediagram} can be put into a one-to-one correspondence
with the ferromagnetic, paramagnetic and helical phases
of the Lifshitz phase diagram\footnote{For
definiteness, we are using here a ``magnetic'' language for the Lifshitz
diagram. However, the Lifshitz diagram can
also describe a variety of other systems, for instance, liquid crystals.}.
The triple point
where the three phases meet is the so-called Lifshitz point.

The critical dimension beyond which the mean-field
Lifshitz theory alluded to above is believed to be valid is
$d_c = 4+m/2$. In lower dimensions, the fluctuations play an
important role and so does the number of components of the field $\phi$.
This does not necessarily affect the general structure of the phase
diagram, but can alter the order of the transitions.
Without entering into the details of the rather complex general situation,
let us just mention that for $m=1$ fluctuations will often turn the
transition along the A-C phase boundary into a first-order transition.
Likewise, most often the transition between phases B and C is
of second order.

As mentioned in the Introduction P.\ Ho\v rava has suggested recently
a new class of quantum gravity theories \cite{horava}.
The basic idea is that in a theory of quantum gravity
one should insist on manifest unitarity as well as renormalizability.
These two requirements clash when we use the Einstein-Hilbert action.
It contains only second-order time derivatives and in this respect
has a chance to be unitary, but it is not renormalizable as we have discussed
in detail. In order to  obtain a
renormalizable theory one can add higher-curvature terms like $R^2$
to the action. Expanding the action around flat spacetime, 
higher-derivative terms will appear in the quadratic part, making the
propagators better behaved at large momenta. As also
mentioned in Sec.\ \ref{intro} one can prove that a number of such theories
is renormalizable \cite{Rsquare}. However,
it is believed that in general they will not be unitary, the unitarity
being spoiled by the higher-order time derivatives
which appear in the propagators.
Covariance implies that higher-order derivatives result in higher-order
time derivatives, which usually lead to acausal propagation. This conundrum
can be avoided if we give up covariance and allow space and time to
behave differently. Having higher-derivative spatial terms  but only
second-order time derivatives in the equations of motion makes it
possible to construct a unitary and renormalizable theory.
The price one pays is that local Lorentz invariance is broken. Since we
know that Lorentz invariance is valid to a good approximation at
large distances, the terms which break Lorentz invariance have to be
suitably small, such that they only play an important role at high
(spatial) momenta. Ho\v rava used the Lifshitz free energy \rf{2.f2}
as an example where higher-derivative terms play an important role and
where there is asymmetry between the various (spatial)
directions\footnote{In the Lifshitz free energy density \rf{2.f2} all
directions are of course spatial directions, but the derivative terms
break the symmetry. In the Ho\v rava-Lifshitz model, the asymmetry is between
space directions and time. After a formal rotation to Euclidean
signature the Ho\v rava Lagrangian would be a Lifshitz-like
free energy density.}.

From the above discussion it should be clear that our CDT program
effectively has a lot in common with the Ho\v rava program, although
our primary intention is to provide a regularization of a covariant theory of
gravity. Our starting point is different too, being a sum over {\it geometries} on spacetimes
with Lorentzian signature. In this sense we are not breaking any symmetry
between space and time. However, after rotating to Euclidean signature
geometries it is clear that in the path-integral histories 
we effectively treat one direction
(the former time direction) differently from the space directions.
In addition, as we have explained in the Introduction, the effective
action corresponding to a nontrivial fixed point will
have only a vague memory of the bare action we put in by hand, since
entropic contributions are going to be important. It is possible therefore
that this effective action has some Lifshitz-like features.
We have already noted that by using geometry in a loose sense as an order
parameter, Fig.\ \ref{phasediagram} resembles closely a Lifshitz phase diagram.
Furthermore, as described above, CDT quantum gravity has a reflection-positive
transfer matrix, pointing to a unitary theory, again in line with the
philosophy of Ho\v rava.

Let us finally try to understand in more detail the reason
for the phase transitions in geometry observed in
Fig.\ \ref{phasediagram}. Let us recall some results from
the earlier attempt to formulate a fully Euclidean theory of quantum gravity
using dynamical triangulations (DT).
This gravity theory was defined by summing over {\it all}
Euclidean geometries of a fixed topology, approximating the sum
by the piecewise linear geometries obtained by using DT
\cite{aj,am4,more4d}. The
weight of a geometry was given by the Regge action, as described above,
with an even simpler action because all four-simplices are identical.
As for CDT we have a bare gravitational  coupling  constant $1/\kp_0$ and
a bare cosmological coupling constant $\kp_4$ (which is traded in the
computer simulations for a fixed number of four-simplices).

One observed two phases as function of $\kp_0$ \cite{aj,fractal4d,am4,more4d}.
One phase, called the ``crumpled phase" is characterized by having
essentially no extension, no matter how large one chooses $N_4$.
This happens because a few links and their vertices acquire a very high order,
such that they are shared by almost all four-simplices \cite{singular}.
This phase occurs for small values of $\kp_0$,
in particular, for $\kp_0\equ 0$. For $\kp_0 \equ 0$,
when the number $N_4$ of simplices is fixed,
there is no action, but only the entropy of the configurations. 
These ``crumpled'' triangulations
are the generic, entropically preferred triangulations.
However, when $\kp_0$ is increased, i.e.\ the bare gravitational
coupling constant is decreased, a first-order phase transition
takes place.\footnote{First it was believed that the transition
was second order, which generated some excitement since a scenario
like the one outlined by eqs.\ \rf{ny17}-\rf{ny17f} would then
provide the basis for a continuum theory. However, when larger four-volumes
were used it was understood that the transition was most likely
a first-order transition \cite{firstorder}.}
Instead of having no extension, the generic triangulation in
this other phase has a maximal linear extension, i.e.\ a branched-polymer
structure\footnote{A branched polymer is a tree graph. Of course the
triangulation is really $d$-dimensional, but the $d\mi 1$ dimensions
have little extension.}. These polymer-like configurations are
important even in $d\equ 2$ where they become dominant at the
so-called $c\equ 1$ barrier of noncritical string theory
\cite{ad,dfj,david1}.
This barrier reflects the appearance of the tachyon of bosonic
string theory, which means that the effective Euclidean action becomes
unbounded from below. A related unboundedness is present in Euclidean
quantum gravity due to the conformal factor. The phase transition
between the crumpled and the branched-polymer phase
can be viewed as a transition between a phase where entropy dominates
and a phase where the conformal factor dominates. Since the entropy
of the configurations is independent of the bare coupling constant
$\kp_0$, it is natural that the bare action will start to dominate for
large values of $\kp_0$, i.e.\ small values of the bare gravitational
coupling constant, a phenomenon we already discussed in the
Introduction.

Given this scenario it is tempting to view the A-C
and A-B phase transitions, which occur when changing $\kp_0$,
as the CDT manifestation of the dominance of branched polymers for large
values of $\kp_0$. Although we have only constructed an effective
action well inside phase C in Fig.\ \ref{phasediagram} (see below for details),
the Lifshitz picture fits well with the above remarks about instability
of the conformal factor in Euclidean quantum gravity. As described
below, well inside phase C we have a positive kinetic term  for
the scale factor. This term comes from entropic contributions to
the effective action since the scale factor coming from
the bare action will contribute with a negative kinetic term
(leading to the unboundedness of the Euclidean bare action in the
continuum limit).\footnote{We have emphasized
the nonperturbative nature of this cure for the
gravitational ``conformal sickness" inside phase C since the early days 
of CDT research \cite{cure1}. How an analogous mechanism may play
out in a continuum path-integral formulation through Faddeev-Popov 
contributions to the measure has been discussed in \cite{cure2}.}
As we increase $\kp_0$, the coefficient of the
kinetic term will decrease since the bare action part will become more
and more dominant, and eventually go through zero. In the Lifshitz
theory this is exactly the point where the coefficient $d_2$ in
the Landau free energy density \rf{2.f2} becomes zero and we
(usually) have a first-order transition to the oscillatory phase, with
negative $d_2$. This is precisely what we observe! We have
not attempted to construct the effective action in the
``oscillatory'' phase A, but the qualitative description seems quite
consistent. It would have been even more interesting if
the A-C phase transition had been second order, but the situation
here seems to be rather like in Euclidean DT in that the transition seems to be 
first order, and thus not suited for taking a
continuum limit of the lattice theory.

In this respect, the transition from phase C to B seems to be more interesting, 
since it appears to be of
second order. In the Lifshitz theory it is an ``ordinary''
phase transition from an ordered phase C to a disordered phase B.
Above we used the somewhat vague label of ``average geometry"
as order parameter in the gravitational case. In Fig.\ \ref{unipink} we depict
a configuration in phase C as it is generated by the computer. As stated
there (and to be discussed in detail below), this ``blob''  scales with
$N_4$, as if it were a real, four-dimensional universe. The parameter $\Del$
we have introduced is in some ways a trivial parameter. If we
take the microscopic picture seriously, it just redefines
the length ratio $a_t/a_s$ between time- and spacelike links,
while the Regge action used is still the Einstein-Hilbert action
of the corresponding geometry. If entropic terms were unimportant,
one would simply expect to see the same ``physical'' universes on average.
Looking at Fig.\ \ref{figalfa} we see that large values of $\Del$ correspond
to small $\ta$. As a consequence, if we plot the universe
using a graphic representation where the temporal and spatial lattice
steps are chosen identical, the universes should appear increasingly elongated
for increasing $\Del$. This is precisely
what we observe (see Fig.\ \ref{cos3_b}), contrary to the
situation when we change $\kp_0$ (see Fig.\ \ref{cos3_a}). In other words, 
deep inside phase C
we have at least qualitative evidence for the independence of
the geometric universe of the parameter $\Del$. Quantitatively, this is
harder to nail down, and at any rate becomes increasingly
inaccurate when we approach the B-C phase transition line by decreasing
$\Del$. One possible explanation is that entropic contributions mix with the bare action
terms and one should not put too much emphasis on the microscopic
interpretation of the various terms in the action. When we approach
the phase transition line, the time extent of the universe will shrink
to a few lattice spacings as described above
(for the sizes of universe we have studied). After
crossing the transition line the universe has collapsed,
at least on the part of the B-C line we have been able to investigate.
This signals that an asymmetry between space and time has developed.
It is natural to conjecture that the transition is related to the
breaking of this symmetry. 

Of course, in Ho\v rava's scenario
one would exactly expect such an asymmetry since space and time
in the UV regime are assigned different
dimensions. How would one observe such a difference
in the present lattice approach? Consider a universe of time
extent $T$, spatial extension $L$ and total four-volume $V_4(T,L)$.
By measuring $T$ and $L$ we can establish the mutual relations
\beql{x5.1}
T \propto V_4^{1/d_t},~~~~L\propto \Big(V_4^{1-1/d_t}\Big)^{1/d_s}\propto
T^{(d_t-1)/d_s}.
\eeq
Well inside phase $C$ we find $d_t=4$ and $d_s=3$, in agreement
with what is expected for an ordinary four-dimensional spacetime.
If the dimension [T] of time was $z$ times the dimension [L] of length
we would have
\beql{x5.2}
z= \frac{d_s}{d_t-1}.
\eeq
Well inside phase B both $d_s$ and $d_t$ must
be large, if not infinite. Since the B-C phase transition
appears to be of second order, it is possible
that $z$ goes to a value different from 1 when we approach
the transition line. One could even imagine
that $z$ is a function of the position on the B-C line.
Ho\v rava suggested that $z\equ 3$ for four-dimensional quantum gravity.
By direct measurement we find that for $\Del >0.3$ one
obtains convincingly $d_t \approx4$ and
$d_s \approx 3$ and thus $z\approx 1$, but for smaller $\Del$
the quality of our results does not allow for any definite statements.
Auto-correlation times seem to become very long and there may be large
finite-volume effects, which obscure the measurements which are
precisely based on finite-size scaling.

A situation where $z$ is a function of the position on the B-C transition
line would be interesting. It highlights one aspect of the phase diagram of
Fig.\ \ref{phasediagram} which {\it is} different from the standard
Lifshitz diagram, namely, that the B-C line ends for sufficiently
small $\kp_0$. One can envisage several intriguing scenarios where $z$
changes as one moves along the phase transition line. For example,
$z$ may change
from the Ho\v rava value 3 at the tricritical point where the A, B and
C phases meet to 1 (corresponding to isotropy between space
and time) at the left end point of the B-C transition line. Or perhaps 
the opposite is true, namely,
that the conjectured triple point, and not the end point, is the
point of isotropy. Such a scenario is also appealing, since it
singles out the isotropy point more distinctly and naturally from a
renormalization group point of view. However, it would fit less
well into the Ho\v rava-Lifshitz picture.
Presently neither of these is supported by the numerical data,
since the numerical problems mentioned above make precise statements
difficult. Much work remains to be done to understand which of these
scenarios is realized. Primarily, we need better algorithms
to beat the critical slowing down when approaching the B-C line. --
In the remainder of this article we will discuss the physics
one can extract from the universes we observe well inside phase C.

\section{The macroscopic de Sitter universe\label{S4a}}

The Monte Carlo simulations referred to above will generate a sequence of
spacetime histories. An individual spacetime history is not an
observable, in the same way as a path $x(t)$ of a particle in the
quantum-mechanical path integral is not. However,
it is perfectly legitimate to talk about
the {\it expectation value} $\la x(t) \ra$ as
well as the {\it fluctuations around} $\la x(t) \ra$. Both of these quantities
are in principle calculable in quantum mechanics.

\begin{figure}[t]
\psfrag{a}{{\bf{\LARGE $t_i$}}}
\psfrag{b}{{\bf{\LARGE $t_f$}}}
\psfrag{c}{{\hspace{-20pt}\bf{\LARGE $x_i$}}}
\psfrag{d}{{\hspace{-20pt}\bf{\LARGE $x_f$}}}
\psfrag{t}{{\bf{\LARGE $t$}}}
\psfrag{x}{{\bf{\LARGE $x$}}}
\psfrag{u}{{\bf{\LARGE $a$}}}
\centerline{\scalebox{1.0}{\rotatebox{0}{\includegraphics{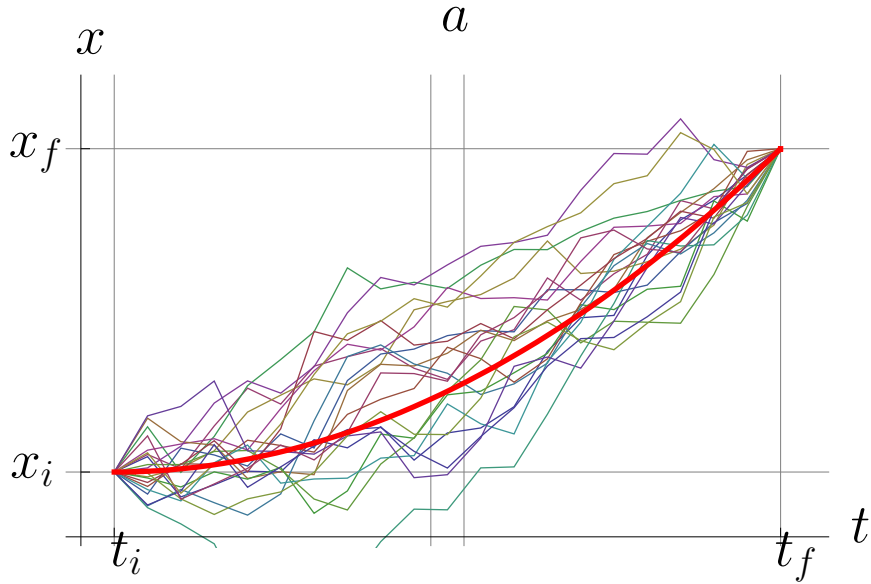}}}}
\caption{{\footnotesize Piecewise linear spacetime histories
in quantum mechanics. The paths shown are from Monte Carlo simulations
of a harmonic oscillator, rotated to Euclidean time.
We show a number of  such histories as well
as the average path (thick line).}}
\label{fig0}
\end{figure}

Let us make a slight digression and discuss the situation in some more 
detail, since
it illustrates well the picture we also hope to emerge in a theory
of quantum gravity. Consider the particle shown in Fig.\ \ref{fig0},
moving from point $x_i$ at time $t_i$ to point $x_f$ at time $t_f$.
In general there will be a classical particle trajectory satisfying
these boundary conditions (which we will assume for simplicity). If $\hbar$
can be considered small compared to the other parameters entering into
the description of the system, the classical path will be a good
approximation to $\la x(t) \ra$ according to Ehrenfest's theorem.
The smooth curve in Fig.\ \ref{fig0} represents the average path $\la x(t) \ra$.
In the path integral we sum over all continuous paths from $(x_i,t_i)$ to
$(x_f,t_f)$, as also illustrated by Fig.\ \ref{fig0}. However, when all
other parameters in the problem are large compared to $\hbar$ we expect
a ``typical'' path to be close to $\la x(t) \ra$, which will also be
close to the classical trajectory. Let us make this explicit in the simple
case of the harmonic oscillator. Let $x_{cl}(t)$ denote the solution to
the classical equations of motion such that $x_{cl}(t_i)=x_i$ and
$x_{cl}(t_f)=x_f$. For the harmonic oscillator the decomposition
\begin{equation}
x(t) = x_{cl}(t) + y(t),~~~~ y(t_i)=y(t_f)=0
\end{equation}
leads to an exact factorization of the path integral thanks
to the quadratic nature of the action.
The part involving $x_{cl}(t)$
gives precisely the classical action and the part involving $y(t)$
the contributions from the fluctuations,
independent of the classical part.
Again for a quadratic action one has the
exact equality $\la x(t)\ra = x_{cl}(t)$,
which however is not really important for the arguments that follow.
Taking the classical path to be macroscopic gives a picture of
a macroscopic path dressed with small
quantum fluctuations\index{quantum fluctuations}, small
because they are independent of the classical motion. Explicitly,
we have for the fluctuations (in a Euclidean-signature calculation)
\begin{equation}
\bra y^2(t)\ket =
\frac{\hbar}{2m\omega} \;
\left(\frac{\cosh \om (t_f-t_i) -
\cosh \om(t_f+t_i-2t)}{\sinh \om (t_f-t_i)}\right).
\end{equation}
This is a simple illustration of what we hope for in quantum gravity.
Once we set the system size (by hand) to be macroscopic -- in the present
example by choosing a macroscopic $x_{cl}(t)$ -- the quantum fluctuations
around it are small and of the order
\begin{equation}
\la |y|\ra \approx \sqrt{\frac{\hbar}{2 m \omega}}.
\end{equation}
We hope that the dynamics of the universe is amenable to a similar logic:
its macroscopic size is determined by the (inverse) cosmological constant in any
Euclidean description\footnote{This is trivial to
show in the model by simply differentiating the partition
function with respect to the cosmological constant; 
in the simulations it is therefore put in by hand.}, and the small
quantum fluctuations\index{quantum fluctuations} around it are dictated by the
other coupling constant, namely, Newton's constant.

Obviously, there are many more dynamical variables in quantum gravity than
there are in the particle case.
We can still imitate the quantum-mechanical situation
by picking out a particular one, for example,
the spatial three-volume $V_3(t)$ at proper time $t$. We
can measure both its expectation value $\la V_3(t)\ra $ as well as fluctuations
around it. The former gives us information about the large-scale
``shape'' of the universe we have created in the computer. In this
section, we will describe the measurements of $\la V_3(t)\ra $, keeping a more
detailed discussion of the fluctuations to Sec.\ \ref{fluctuations} below.

A ``measurement'' of $V_3(t)$ consists of a table $N_3(i)$, where
$i=1,\ldots,N$ labels the time slices. Recall from
eq.\ \rf{fixvolume}
that the sum over slices $\sum_{i=1}^N N_3(i)$ is basically
kept constant ($N_4^{(4,1)} = 2 N_3$). The
time axis has a total length of $N$ time steps, where $N=80$ in the
actual simulations, and we have cyclically identified time slice $N+1$ with
time slice 1. As is clear from Fig.\ \ref{unipink} this identification
has no consequences for the physics of the ``blob'', as long as the number
of time steps $N$ is sufficiently large compared to the time extension
of the ``blob''.

What we observe in the simulations \cite{emerge,blp}
is that for the range of discrete volumes
$N_4$ under study the universe does {\it not} extend
(i.e. has appreciable three-volume) over the entire time axis, but rather is
localized in a region much shorter than 80 time slices.
Outside this region the spatial extension $N_3(i)$ will be minimal,
consisting of the minimal number (five) of tetrahedra needed to
form a three-sphere $S^3$, plus occasionally a few more
tetrahedra.\footnote{This
kinematical constraint ensures that the triangulation remains a {\it simplicial
manifold} in which, for example,
two $d$-simplices are not allowed to have more than
one $(d-1)$-simplex in common.}
This thin ``stalk" therefore
carries little four-volume and in a given simulation
we can for most practical purposes
consider the total four-volume of the remainder,
the extended universe, as fixed.

In order to perform a meaningful average over geometries
which explicitly refers to the extended part of the universe,
we have to remove the translational zero mode which is present
\cite{agjl,bigs4}.
During the Monte Carlo simulations the extended universe will
fluctuate in shape and its {\it centre of mass}
(or, more to the point, its {\it centre of volume})
will perform a slow random walk along the time axis.
Since we are dealing with a circle (the compactified time axis),
the centre of volume is not uniquely defined
(it is clearly arbitrary for a constant volume distribution), and we must
first define what we mean by such a concept.
Here we take advantage of the empirical fact that our dynamically generated
universes decompose into an extended piece and a stalk,
with the latter containing less than one per cent of the total volume.
We are clearly interested in a definition such that the
centre of volume of a given configuration lies in the
centre of the extended region. One also expects that any sensible definition
will be unique up to contributions related
to the stalk and to the discreteness of the time steps. In total
this amounts to an ambiguity of the centre of volume of one
lattice step in the time direction.

In analyzing the computer data we have chosen one specific definition
which is in accordance with the discussion above\footnote{Explicitly,
we consider the quantity
\begin{equation}
CV(i') = \left| \sum_{i=-N/2}^{N/2-1} (i+0.5) N_3(1 + {\rm mod}(i' + i - 1, N) ) \right|
\end{equation}
where {\it mod(i,N)} is defined as the remainder, on division of $i$ by $N$,
and find the value of $i' \in \{1,\dots,N \}$ for which $CV(i')$ is smallest.
We denote this $i'$ by $i_{cv}$. If there is
more than one minimum, we choose the value which has the largest
three-volume $N_3(i')$. Let us stress that
this is just one of many definitions of $i_{cv}$. All other sensible
definitions will for the type of configurations considered here
agree to within one lattice spacing.}.
Maybe surprisingly, it turns out that the inherent
ambiguity in the choice of a definition
of the centre of volume -- even if it is only of the order of one lattice
spacing -- will play a role later on in our
analysis of the quantum fluctuations.
For each universe used in the
measurements (a ``path" in the gravitational path integral)
we will denote the centre-of-volume time coordinate calculated
by our algorithm by $i_{cv}$. From now on,
when comparing different universes, i.e.\ when performing ensemble
averages, we will redefine the temporal coordinates according to
\beq\label{cm}
N_3^{new} (i) =  N_3(1 + {\rm mod}(i + i_{cv} - 1, N)),
\eeq
such that the centre of volume is located at 0.

\begin{figure}[t]
\centerline{\scalebox{1.0}{\rotatebox{0}{\includegraphics{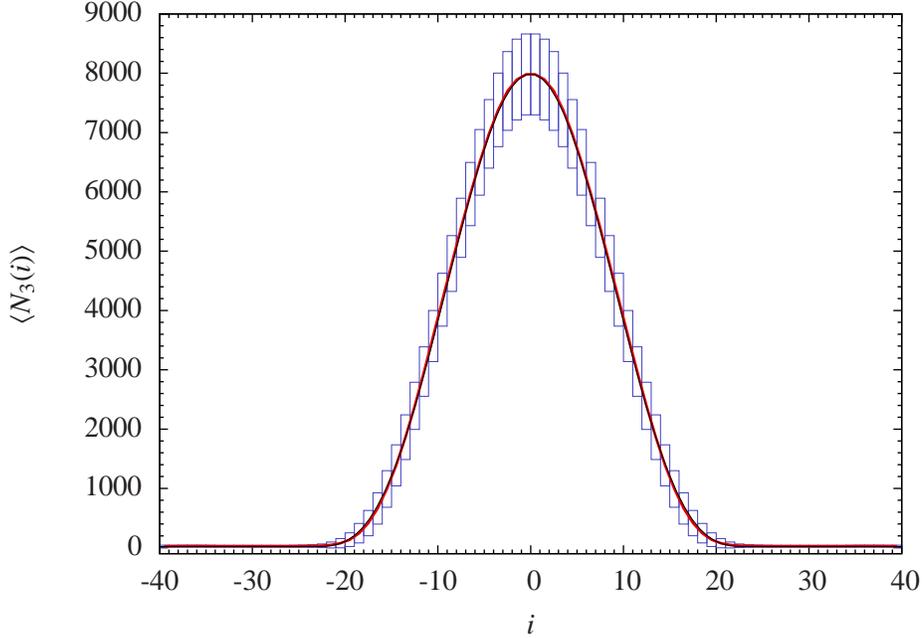}}}}
\caption{\label{cos3} {\footnotesize 
Background geometry $\langle N_3(i)\rangle$:
MC measurements for fixed $N_4^{(4,1)}= 160.000$ ($N_4=362.000$)
and best fit \rf{n1} yield indistinguishable curves at given plot resolution.
The bars indicate the average size of quantum fluctuations.}}
\end{figure}

Having defined in this manner the centre of volume along the
time direction of our spacetime configurations we can now
perform superpositions of such configurations and
define the average $\la N_3(i)\ra $ as a function of the discrete time $i$.
The results of measuring the average discrete
spatial size of the universe at various
discrete times $i$ are illustrated
in Fig.\ \ref{cos3} and can be succinctly
summarized by the formula \cite{agjl,bigs4}
\beq\label{n1}
N_3^{cl}(i):= \la N_3(i)\ra  = \frac{N_4}{2(1+\xi)}\;
\frac{3}{4} \frac{1}{s_0 N_4^{1/4}}
\cos^3 \left(\frac{i}{s_0 N_4^{1/4}}\right),
\eeq
where $N_3(i)$ denotes the number of three-simplices in the spatial slice
at discretized time $i$ and $N_4$ the
total number of four-simplices in the entire universe.
The constants $\xi$ and $s_0$ depend on $\kp_0$ and $\Del$, and
$\xi$ is defined as follows.
According to the formulas for four-volumes for a given $\Del$ in
\rf{actshort}, and thus in principle a given asymmetry parameter $\ta$,
the continuum Euclidean four-volume of the universe is given by
\beq\label{vol1}
V_4 = C_4\, a^4\;\Big(\frac{\sqrt{8\tilde\a -3}}{\sqrt{5}}\; N_4^{(4,1)} +
\frac{\sqrt{12\tilde\a -7}}{\sqrt{5}}\; N_4^{(3,2)} \Big),
\eeq
where $C_4 = \sqrt{5}/96$ is the four-volume of an equilateral four-simplex
with edge length $a=1$.
It is convenient to rewrite expression \rf{vol1} as
\beq\label{vol2}
V_4 = \tilde{C}_4(\xi)\, a^4\, N_4^{(4,1)} =
\tilde{C}_4(\xi)\, a^4 \, N_4/(1+\xi),
\eeq
where $\xi$ is the ratio
\beq\label{vol3}
\xi = N_4^{(3,2)}/N_4^{(4,1)},
\eeq
and $\tilde{C}_4(\xi)\, a^4$ is a measure of the ``effective four-volume''
of an ``average'' four-simplex. Since we are
keeping $N_4^{(4,1)}$ fixed in the simulations and since $\xi$ changes
with the choice of bare coupling constants, it is sometimes
convenient to rewrite \rf{n1} as
\beq\label{n1x}
N_3^{cl}(i) = \oh N_4^{(4,1)}\;\frac{3}{4}
\frac{1}{\tilde{s}_0 (N_4^{(4,1)})^{1/4}}
\cos^3 \left(\frac{i}{\tilde{s}_0 (N_4^{(4,1)})^{1/4}}\right),
\eeq
where $\tilde{s}_0$ is defined by
$\tilde{s}_0 (N_4^{(4,1)})^{1/4} = s_0 N_4^{1/4}$.
Of course, formula \rf{n1} is only valid in the extended part of the universe
where the spatial three-volumes are larger than the minimal cutoff size.

The data shown in Fig.\ \ref{cos3} have been
collected at the particular values
$(\kp_0,\Del) = (2.2,0.6)$ of the bare coupling constants and for
$N_4= 362.000$ (corresponding to $N_4^{(4,1)} = 160.000$).
For these values of $(\kp_0,\Del)$ we find $s_0 = 0.59$ and
have verified relation \rf{n1} for $N_4$ ranging from 45.500 to 362.000
building blocks (45.500, 91.000, 181.000 and 362.000).
After rescaling the time and volume variables by suitable powers of $N_4$
according to relation (\ref{n1}), and plotting them in the same way as
in Fig.\ \ref{cos3}, one finds almost total
agreement between the curves for different spacetime volumes.\footnote{By
contrast, the quantum fluctuations indicated in
Fig.\ \ref{cos3} as vertical bars
{\it are} volume-dependent. After a rescaling which
puts universes of different size on the same curve, the quantum
fluctuations will be the larger the
smaller the total four-volume, see Sec.\ \ref{fluctuations} below for details.}
Eq.\ \rf{n1} shows that
spatial volumes scale according to $N_4^{3/4}$ and time intervals
according to $N_4^{1/4}$, as one would expect for
a genuinely {\it four}-dimensional spacetime. This strongly suggests
a translation of relation \rf{n1} to a continuum notation.
The most natural identification is via a proper-time gauge choice
\beq\label{propertime}
ds^2 = g_{tt} dt^2 + g_{ij} dx^idx^j,
\eeq
suggested by the CDT setup.  With this choice we have
\beq\label{n2}
\sqrt{g_{tt}}\; V_3^{cl}(t) = V_4 \;
\frac{3}{4 B} \cos^3 \left(\frac{t}{B} \right),
\eeq
where we have made the identifications
\beq\label{n3}
\frac{t_i}{B} = \frac{i}{s_0 N_4^{1/4}}, ~~~~
\Del t_i \sqrt{g_{tt}}\;V_3(t_i) = 2 \tC_4 N_3(i) a^4,
\eeq
such that
\beq\label{n3z}
\int dt \sqrt{g_{tt}} \; V_3(t) = V_4.
\eeq
In the second of relations \rf{n3}, $\sqrt{g_{tt}}$ is the constant proportionality
factor between the time
$t$ and genuine continuum proper time $\tau$, $\tau=\sqrt{g_{tt}}\; t$.
(The combination $\Del t_i\sqrt{g_{tt}}V_3$ contains $\tC_4$, related to the
four-volume of a four-simplex rather than the three-volume corresponding
to a tetrahedron, because its time integral must equal $V_4$.)
Writing $V_4 =8\pi^2 r^4/{3}$, and $\sqrt{g_{tt}}=r/B$,
eq.\ \rf{n2} is seen to describe
a Euclidean {\it de Sitter universe}
(a four-sphere, the maximally symmetric space for
positive cosmological constant)
as our searched-for, dynamically generated background geometry!
In the parametrization of \rf{n2}
this is the classical solution to the action
\beq\label{n5}
S= \frac{1}{24\pi G} \int d t \sqrt{g_{tt}}
\left( \frac{ g^{tt}\dot{V_3}^2(t)}{V_3(t)}+k_2 V_3^{1/3}(t)
-\lam V_3(t)\right),
\eeq
where $k_2= 9(2\pi^2)^{2/3}$ and  $\lam$ is a Lagrange multiplier,
fixed by requiring that the total four-volume
be $V_4$, $\int d t \sqrt{g_{tt}} \;V_3(t) = V_4$.
Up to an overall sign, this is precisely
the Einstein-Hilbert action for the scale
factor $a(t)$ of a homogeneous, isotropic universe \cite{hawking}
\beq\label{a-action}
S = -\frac{3\pi}{2G} \int d t \sqrt{g_{tt}}
\left( g^{tt} a(t)\dot{a}^2(t)+a(t)
-\frac{\lam}{9} a^3(t)\right),
\eeq
rewritten in terms of the spatial three-volume $V_3(t) =2\pi^2 a(t)^3$,
although we of course never put any such
simplifying symmetry assumptions into the CDT model.

For a fixed, finite four-volume $V_4$ and when applying scaling arguments
it can be convenient to rewrite eq.\ \rf{n5} in terms of dimensionless units by
introducing $s=t/V_4^{1/4}$ and $V_3(t) = V_4^{3/4} v_3(s)$, in which case
\rf{n5} becomes
\beq\label{n5a}
S =  \frac{1}{24\pi}\; \frac{\sqrt{V_4}}{G} \int d s \sqrt{g_{ss}}
\left( \frac{ g^{ss}\dot{v_3}^2(s)}{v_3(s)}+k_2 v_3^{1/3}(s) \right),
\eeq
now assuming that $\int ds \sqrt{g_{ss}}\; v_3(s) = 1$,
and with $g_{ss}\equiv g_{tt}$.
A discretized, dimensionless version of \rf{n5} is
\beq\label{n7b}
S_{discr} =
k_1 \sum_i \left(\frac{(N_3(i+1)-N_3(i))^2}{N_3(i)}+
\tilde{k}_2 N_3^{1/3}(i)\right),
\eeq
where $\tilde{k}_2\propto k_2$.
This can be seen by applying the scaling \rf{n1}, namely,
$N_3(i) = N_4^{3/4} n_3(s_i)$
and $s_i = i/N_4^{1/4}$. With this scaling, the action \rf{n7b} becomes
\beq\label{n7bb}
 S_{discr} =
k_1 \sqrt{N_4} \sum_i  \Delta s  \left(\frac{1}{n_3(s_i)}
\left(\frac{n_3(s_{i+1})-n_3(s_i)}{\Delta s}\right)^2+
\tilde{k}_2 n_3^{1/3}(s_i)\right),
\eeq
where $\Delta s = 1/N^{1/4}$, and therefore has the same form as \rf{n5a}.
This enables us to finally conclude that the identifications
\rf{n3} when used in the action \rf{n7b} lead
na\"ively to the continuum expression \rf{n5} under the identification
\beq\label{n7c}
G = \frac{a^2}{k_1} \frac{\sqrt{\tC_4}\; \ts_0^2}{3\sqrt{6} }.
\eeq
Comparing the kinetic terms in \rf{n7b} and  \rf{n5} in more detail,
\beq\label{new1}
\frac{1}{24\pi G} \sum_i \frac{(V_3(t_i+\Del t_i)- V_3(t_i))^2}{\Del t_i
\sqrt{g_{t_it_i}} \;V_3(t_i)} = k_1 \sum_i \frac{(N_3(i+1)-N_3(i))^2}{N_3(i)},
\eeq
and using eq.\ \rf{n3} leads to
\beq\label{new2}
G = \frac{a^4}{k_1}\; \frac{2\sqrt{\tC_4}}{24 \pi \; g_{t_it_i} (\Del t_i)^2}.
\eeq
Eq.\ \rf{n7c} now follows from the equations
\beq\label{new3}
(\Del t_i)^2 = \frac{B^2}{s_0^2\sqrt{N_4}},~~~\ts_0^2=s_0^2\,\sqrt{1+\xi},
\eeq
\beq\label{new4}
V_4 =\frac{8\pi^2}{3}\, r^4 = \frac{\tC_4}{1+\xi} \;N_4 a^4,~~~~
g_{t_it_i}= \frac{r^2}{B^2}.
\eeq
\begin{figure}[t]
\centerline{\scalebox{1.00}{\rotatebox{0}{\includegraphics{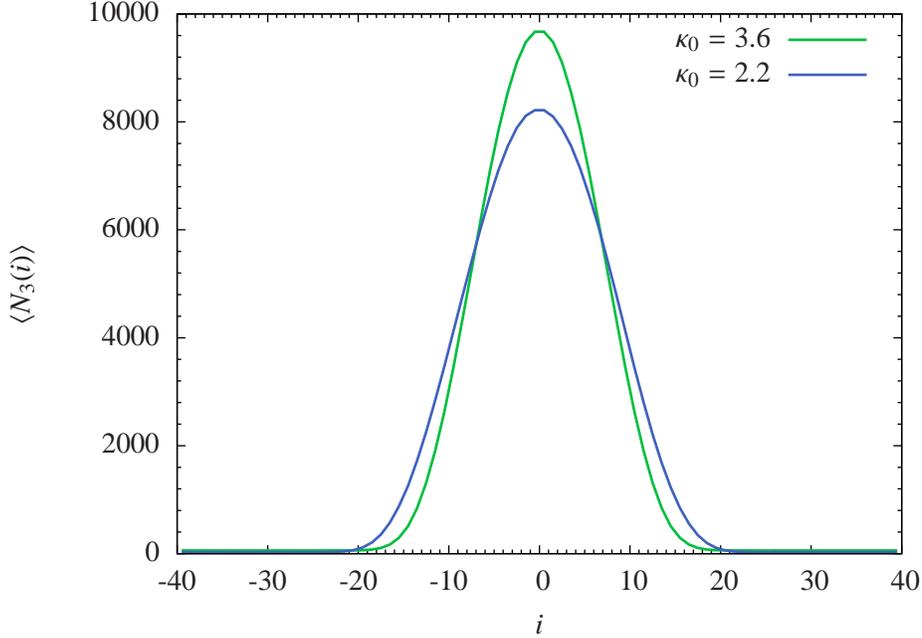}}}}
\caption{\label{cos3_a} {\footnotesize 
The measured average shape $\la N_3(i)\ra $
of the quantum universe at $\Del = 0.6$, for $\kp_0$
= 2.2 (broader distribution) and $\kp_0$ = 3.6 (narrower distribution), taken
at $N_4^{(4,1)}=160.000.$}}
\end{figure}

Next, let us comment on the universality of these results.
First, we have checked that they are not
dependent on the particular definition of time slicing we have
been using, in the following sense. By construction of the
piecewise linear CDT geometries
we have at each integer time step $t_i= i\, a_t $
a spatial surface consisting of $N_3(i)$ tetrahedra. Alternatively, one
can choose as reference slices for the measurements of the spatial volume
non-integer values of time, for example,
all time slices at discrete times $i-1/2$,
$i=1,2,...$. In this case the ``triangulation" of the
spatial three-spheres consists of
tetrahedra -- from cutting a (4,1)- or a (1,4)-simplex half-way --
and ``boxes" (prisms), obtained by cutting a (2,3)- or (3,2)-simplex
(as discussed in Sec.\ \ref{MC}). We again find a relation
like \rf{n1} if we use the total number of spatial building blocks
in the intermediate slices (tetrahedra+boxes) instead of just the
tetrahedra.

Second, we have repeated the measurements for other values of the bare
coupling constants. As long as we stay in the phase C well away
from the boundaries, a relation like \rf{n1} remains
valid. In addition, the value of $s_0$, defined in eq.\ \rf{n1},
is almost unchanged until we get close to the phase transition
lines beyond which the extended universe disappears.
Fig.\ \ref{cos3_a} shows the average shape
$\la N_3(i)\ra $ for $\Del = 0.6$ and for $\kp_0$
equal to 2.2 and 3.6.
Only for the values of $\kp_0$ around 3.6 and larger will the
measured $\la N_3(i)\ra $ differ significantly from the value at 2.2.
For values larger than 3.8 (at $\Del = 0.6$),
we cross the phase transition line between the A and the C phase
and the universe will disintegrate into a number of small
components distributed along the time axis (as described in detail above),
and one can no longer fit the distribution $\la N_3(i)\ra $
to the formula \rf{n1}.

\begin{figure}[t]
\centerline{\scalebox{1.00}{\rotatebox{0}{\includegraphics{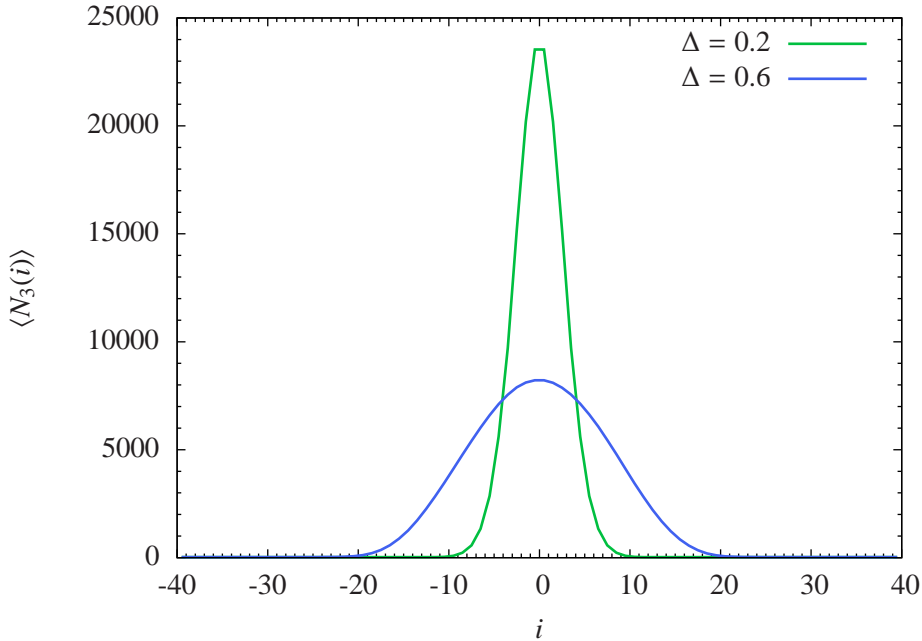}}}}
\caption{\label{cos3_b}{\footnotesize The measured average shape
$\la N_3(i)\ra $ of the quantum
universe at $\kp_0 = 2.2$, for $\Del$
= 0.6 (broad distribution) and $\Del$= 0.2 (narrow distribution),
both taken at $N_4^{(4,1)}=160.000$.}}
\end{figure}

Fig.\ \ref{cos3_b} shows the average shape
$\la N_3(i)\ra $ for $\kp_0= 2.2$ and
$\Del $ equal to 0.2 and 0.6. Here the value $\Del = 0.2$ is close to the
phase transition where the extended universe will flatten out to a
universe with a time extension of a few lattice spacings only.
Later we will show that while $s_0$ is almost unchanged, the constant
$k_1$ in \rf{n7b}, which governs the quantum fluctuations around the
mean value $\la N_3(i)\ra $, is more sensitive to a change
of the bare coupling constants, in particular in the case where we
change $\kp_0$ (while leaving $\Del$ fixed).

\section{Constructive evidence for the effective action\label{effective}}

After the first ``observations'' of the ``blob'' and measurements
which showed that its large-scale behaviour was four-dimensional
it was realized that the data were well described by the minisuperspace
action \rf{n5} (or \rf{a-action}) \cite{semi}. However, only after
it was understood how to fix the translational mode did we obtain
high-quality data which could provide us with a serious test of
the prediction coming from the action \rf{n5}.
While the functional form \rf{n1} for the three-volume fits the data
perfectly and the corresponding continuum effective action \rf{n5} reproduces
the continuum version \rf{n2} of \rf{n1}, it is still of interest to
check to what extent one can reconstruct the discretized
version \rf{n7b} of the continuum action \rf{n5} from the data explicitly.
Stated differently, we would like to understand whether there are
other effective actions which reproduce the data equally well.
As we will demonstrate by explicit construction in this section,
there is good evidence for the uniqueness of the action \rf{n7b}.

We have measured two types of data,  
the three-volume $N_3(i)$ at the discrete time step $i$,
and the three-volume correlator $\la N_3(i) N_3(j)\ra$.
Having created $K$ statistically independent configurations
$N_3^{(k)} (i)$ by Monte Carlo simulation allows us to construct the average
\beq\label{5.s1}
\bar{N}_3(i) := \la N_3(i) \ra \cong \frac{1}{K} \sum_k N_3^{(k)} (i),
\eeq
where the superscript in $(\cdot)^{(k)}$ denotes the result
of the k'th configuration sampled,
as well as the covariance matrix
\beq\label{5.s2}
C(i,j) \cong \frac{1}{K} \sum_k
(N^{(k)}_3 (i) -\bar{N}_3(i))(N^{(k)}_3 (j) -\bar{N}_3(j)).
\eeq
Since we have fixed the sum $\sum_{i=1}^{N} N_3(i)$
(recall that $N$ denotes the
fixed number of time steps in a given simulation), the covariance
matrix has a zero mode, namely, the constant vector $e^{(0)}_i$,
\beq\label{5.s3}
\sum_j C(i,j)e^{(0)}_j = 0,~~~~e^{(0)}_i = 1/\sqrt{N}\quad \forall i.
\eeq
A spectral decomposition of the symmetric covariance matrix gives
\beq\label{5.4}
\hat{C} = \sum_{a=1}^{N-1} \lam_a | e^{(a)}\ra \la e^{(a)}|,
\eeq
where we assume the $N-1$ other eigenvalues of the covariance matrix
$\hat{C}_{ij}$ are different from zero. We now define the ``propagator"
$\hat{P}$ as
the inverse of $\hat{C}$ on the subspace orthogonal to the zero mode
$e^{(0)}$ , that is,
\beq\label{5.5}
\hat{P} = \sum_{a=1}^{N-1} \frac{1}{\lam_a} | e^{(a)}\ra \la e^{(a)}| =
(\hat{C} +\hat{A})^{-1}-\hat{A},~~~\hat{A}=| e^{(0)}\ra \la e^{(0)}|.
\eeq

We now assume we have a discretized action which can
be expanded around the expectation value $\bar{N}_3(i)$
according to
\beq\label{5.6}
 S_{discr}[\bar{N}+n] = S_{discr}[\bar{N}] +
\oh \sum_{i,j} n_i \hat{P}_{ij} n_j  +O(n^3).
 \eeq
If the quadratic approximation describes
the quantum fluctuations around the expectation value $\bar{N}$ well,
the inverse of $\hat{P}$ will be a good
approximation to the covariance matrix.
Conversely, still assuming the quadratic approximation gives
a good description of the
fluctuations, the $\hat{P}$ constructed from the covariance matrix will to a
good approximation allow us to reconstruct the action via \rf{5.6}.

Simply by looking at the inverse $\hat P$ of the measured covariance matrix,
defined as described
above, we observe that it is to a very good approximation
small and constant except on the
diagonal and the entries next to the diagonal.
We can then decompose it into a ``kinetic'' and a ``potential'' term.
The kinetic part $\hPk$ is defined as the matrix with non-zero
elements on the diagonal and in the
neighbouring entries, such that the sum of the elements
in a row or a column is always zero,
\beq\label{5.6a}
\hPk = \sum_{i=1}^N p_i \hX^{(i)},
\eeq
where the matrix $\hX^{(i)}$ is given by
\beq\label{5.7}
\hX^{(i)}_{jk}= \del_{ij}\del_{ik}+\del_{(i+1)j}\del_{(i+1)k}-
\del_{(i+1)j}\del_{ik}-\del_{ij}\del_{(i+1)k}.
\eeq
Note that the range of $\hPk$ lies by definition in the
subspace orthogonal to the zero mode.
Similarly, we define the potential term
as the projection of a diagonal matrix $\hat D$ on the subspace
orthogonal to the zero mode
\beq\label{5.8}
\hPp = (\hI -\hA) \hD (\hI -\hA) = \sum_{i=1}^{N} u_i \hY^{(i)}.
\eeq
The diagonal matrix $\hD$ and the matrices $\hY^{(i)}$ are defined by
\beq\label{5.9}
\hD_{jk} = u_j \del_{jk},~~~\hY^{(i)}_{jk} = \del_{ij}\del_{ik} -\frac{\del_{ij}+\del_{ik}}{N}+\frac{1}{N^2},
\eeq
and $\hI$ denotes the $N\times N$ unit matrix.

The matrix $\hP$ is obtained from the numerical data by inverting
the covariance matrix $\hC$ after subtracting the zero mode,
as described above.
We can now try to find the best values of
the $p_i$ and $u_i$ by a least-$\chi^2$ fit\footnote{A $\chi^2$-fit
of the form \rf{5.10} gives the same weight to each
three-volume $N_3(i)$. One might argue that more weight
should be given to the larger $N_3(i)$ in a configuration
since we are interested in the continuum physics and
not in what happens in the stalk where
$N_3(i)$ is very small. We have tried various $\chi^2$-fits
with reasonable weights associated with
the three-volumes $N_3(i)$. The kinetic term, which is the
dominant term, is insensitive to any (reasonable) weight
associated with $N_3(i)$. The potential term, which will be analyzed
below, is more sensitive to the choice of the weight. However,
the general power-law dependence reported below is again unaffected
by this choice.} to
\beq\label{5.10}
\tr \left(\hP - (\hPk+\hPp)\right)^2.
\eeq

\begin{figure}[ht]
\centerline{\scalebox{1.00}{\rotatebox{0}{\includegraphics{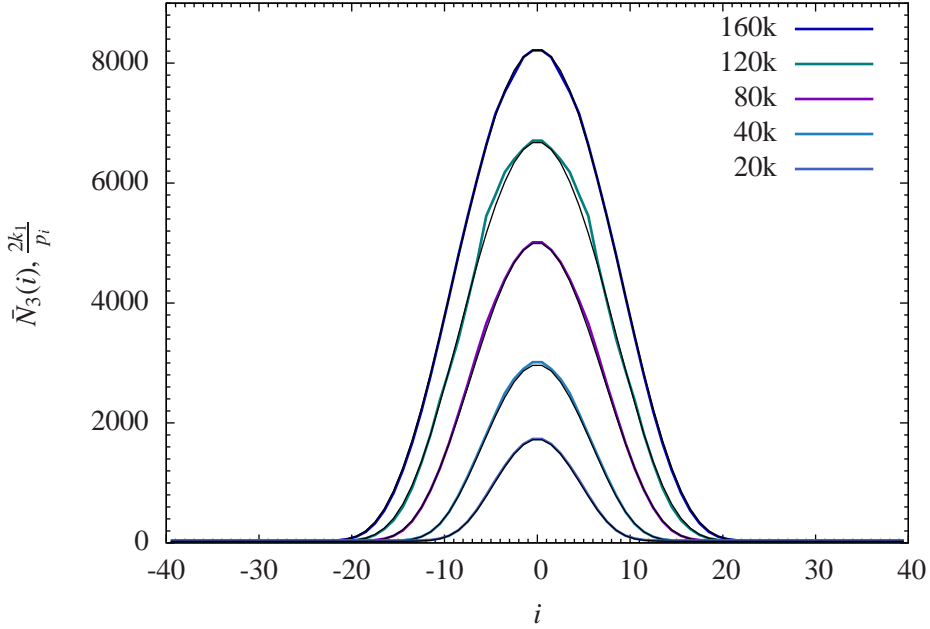}}}}
\caption{\label{reconstruct} {\footnotesize 
The measured expectation values
$\bar{N}_3(i)$ (thick lines), compared to the averages $\bar{N}_3(i)$
reconstructed from the measured covariance matrix $\hat C$ (thin black lines),
for $\kp_0=2.2 $ and $\Del
= 0.6$, at various fixed volumes $N_4^{(4,1)}$.
The two-fold symmetry of the interpolated
curves around the central symmetry axis
results from an explicit symmetrization of the
collected data.}}
\end{figure}

Let us look at the discretized minisuperspace action
\rf{n7b} which obviously has served as an
inspiration for the definitions of $\hPk$ and $\hPp$.
Expanding $N_3(i)$ to second order
around $\bN_3(i)$ one obtains the identifications
\beq\label{5.11}
\bN_3(i) = \frac{2 k_1}{p_i},~~~~~~U''(\bN_3(i)) = u_i,
\eeq
where $U(N_3(i))=k_1\tilde k_2 N_3^{1/3}(i)$ denotes the potential term in \rf{n7b}.
We use the fitted coefficients $p_i$
to reconstruct $\bN_3(i)$ and then compare these reconstructed
values with actually measured averages $\bN_3(i)$.
Similarly, we can use the measured $u_i$'s to
reconstruct the second derivatives $U''(\bN_3(i))$ and
compare them to the form $\bN^{-5/3}_3(i)$ coming from \rf{n7b}.

The reconstruction of $\bN_3(i)$ is illustrated in
Fig.\ \ref{reconstruct} for a variety
of four-volumes $N_4$ and compared with the  measured
expectation values $\bN_3(i)$. One observes
that the reconstruction works very well and,
{\it most importantly},
the coupling constant $k_1$, which in this
way is determined independently for each
four-volume $N_4$ {\it is}  really independent of
$N_4$ in the range of $N_4$ considered, as it should.

\begin{figure}[t]
\centerline{\scalebox{1.00}{\rotatebox{0}{\includegraphics{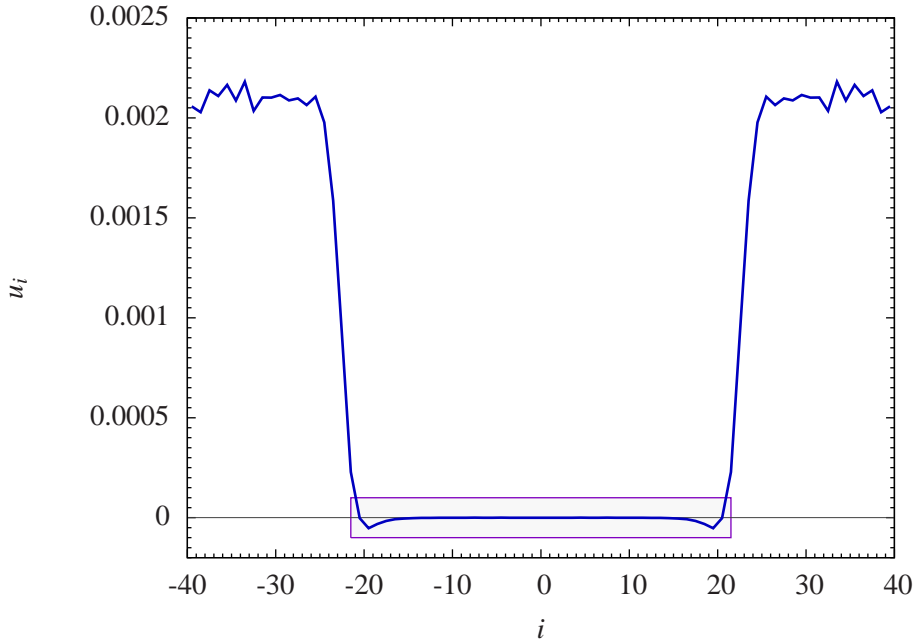}}}}
\caption{\label{figuu} {\footnotesize Reconstructing the second
derivative $U''(\bar{N}_3(i))$ from
the coefficients $u_i$, for $\kp_0=2.2$
and $\Del= 0.6$ and $N_4^{(4,1)}=160.000$. The ``blob'' is located
at $|i| \leq 21$.}}
\end{figure}

We will now try to extract
the potential $U''(\bN_3(i))$ from the
information contained in the matrix $\hPp$. The determination of
$U''(\bN_3(i))$ is not an
easy task as can be understood from Fig.\ \ref{figuu},
which shows the measured coefficients $u_i$ extracted
from the matrix $\hPp$. We  consider this figure
somewhat remarkable. The interpolated curve makes an
abrupt jump by two orders of magnitude going from the
extended part of the universe (stretching over roughly 40 time steps)
to the stalk. The occurrence of this jump is entirely dynamical,
no distinction has ever been made by hand between stalk and bulk.

\begin{figure}[t]
\centerline{\scalebox{1.00}{\rotatebox{0}{\includegraphics{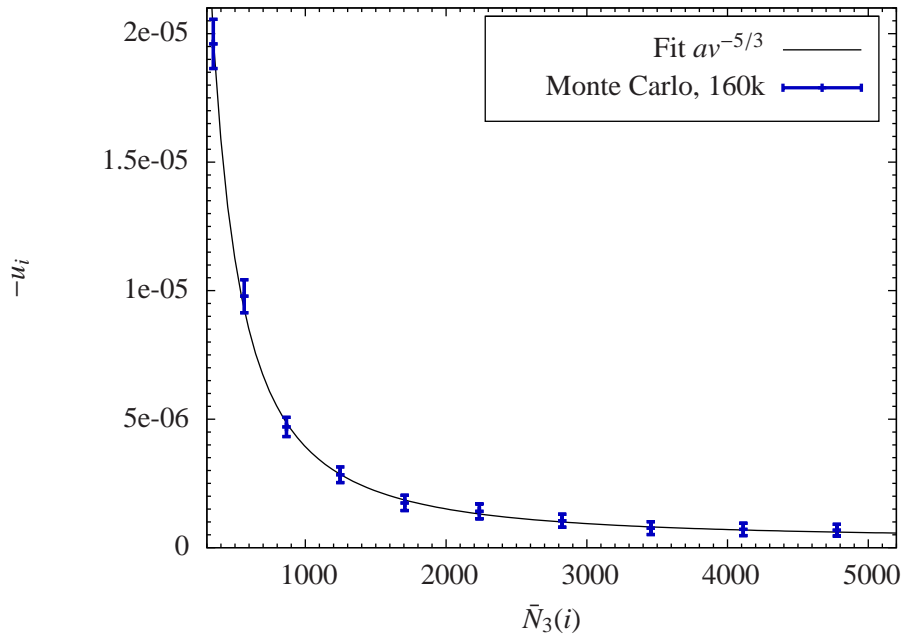}}}}
\caption{\label{figuu_a}{\footnotesize The second derivative
$-U''(N_3)$ as measured for $N_4^{(4,1)}= 160.000$,
$\kp_0 = 2.2$ and $\Del = 0.6$.}}
\end{figure}

There are at least two reasons for why it is difficult
to determine the potential numerically. Firstly, the results
are ``contaminated" by the presence of the stalk.
Since it is of cutoff size, its dynamics is dominated
by fluctuations which likewise are of cutoff size.
They will take the form of short-time subdominant contributions
in the correlator matrix $\hC$.
Unfortunately, when we invert $\hC$ to obtain the propagator $\hP$, the
same excitations will correspond
to the largest eigenvalues and give a very large contribution.
Although the stalk contribution in the matrix $\hC$ is located
away from the bulk-diagonal, it can be seen from the appearance of
the $1/N^2$-term in eqs.\ \rf{5.8} and \rf{5.9} that after
the projection orthogonal to the zero mode the contributions
from the stalk will also
affect the remainder of the geometry in the form of fluctuations
around a small constant value. In deriving Fig.\ \ref{figuu_a} we have
subtracted this constant value as best possible.
However, the {\it fluctuations} of the stalk cannot be subtracted and
only good statistics can eventually eliminate their effect on the
behaviour of the extended part of the quantum universe.
The second (and less serious) reason is that from a
numerical point of view the potential term is always subdominant
to the kinetic term for the individual spacetime histories in
the path integral. For instance, consider the simple example of the
harmonic oscillator. Its discretized action reads
\beq\label{k1}
S = \sum_{i=1}^N  \Del t
\left[\Big(\frac{x_{i+1}-x_i}{\Del t}\Big)^2 + \om^2 x_i^2\right],
\eeq
from which we deduce that the ratio between the kinetic and potential
terms will be of order $1/\Del t$ as $\Del t$ tends to zero.
This reflects the well-known fact
that the kinetic term will dominate and
go to infinity in the limit as $\Del t \to 0$, with
a typical path $x(t)$ being nowhere differentiable, but closer to
a random walk. The same will be true when
dealing with a more general action like \rf{n5} and its discretized version
\rf{n7b}, where $\Del t$ scales like $\Del t \sim 1/N_4^{1/4}$. Of course,
a {\it classical} solution
will behave differently: there the kinetic term will be
comparable to the potential term.
However, when extracting the potential term directly from the data, as we
are doing, one is confronted with this issue.

The range of the discrete three-volumes $N_3(i)$
in the extended universe is from several
thousand down to five, the kinematically allowed minimum.
However, the behaviour
for the very small values of $N_3(i)$ near the edge of the extended universe
is likely to be mixed in with discretization effects, to be discussed shortly.
In order to test whether one really has a $N_3^{1/3}(i)$-term
in the action one should therefore only use values of $N_3(i)$ somewhat larger
than five. This has been done in
Fig.\ \ref{figuu_a}, where we have
converted the coefficients $u_i$ from functions of the discrete time steps $i$
into functions of the background spatial three-volume $\bN_3(i)$
using the identification in \rf{5.11}
(the conversion factor can be read off the relevant curve
in Fig.\ \ref{reconstruct}).
It should be emphasized that Fig.\ \ref{figuu_a} is based
on data from the extended part of the spacetime only;
the variation comes entirely
from the central region between times -20 and +20 in Fig.\ \ref{figuu},
which explains why it has been numerically demanding to extract a good signal.
The data presented in Fig.\ \ref{figuu_a} were taken at a discrete volume
$N_4^{(4,1)}= 160.000$, and fit well the form $N_3^{-5/3}$,
corresponding to a potential $\tk_2 N_3^{1/3}$. There is a very small
residual constant term present in this fit, which presumably is due to the
projection onto the space orthogonal to the zero mode,
as already discussed earlier. In view of the fact that its value is quite
close to the noise level with our present statistics,
we have simply chosen to ignore it in the remaining discussion.

\begin{figure}
\centering
\includegraphics[width=0.85\textwidth]{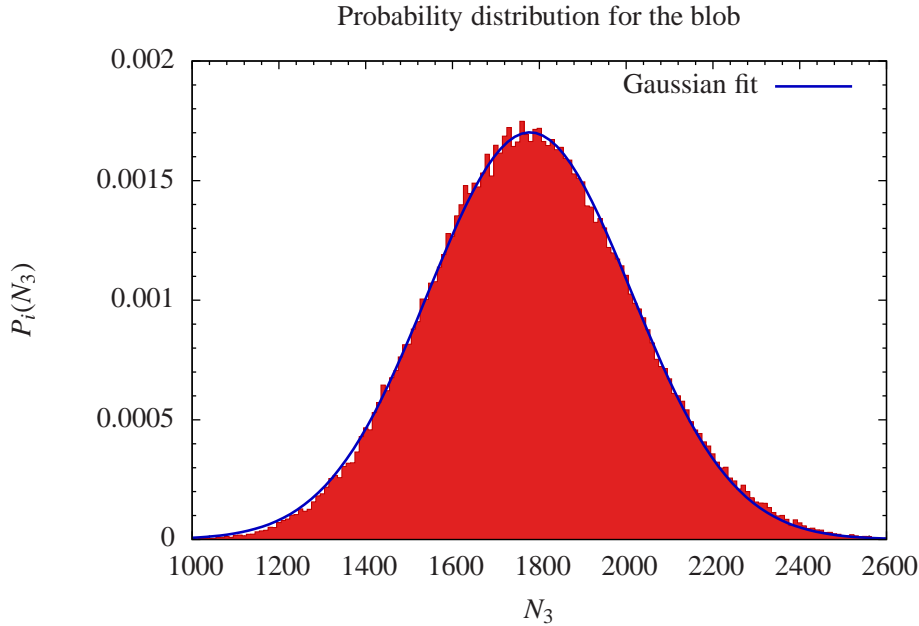}\\
\caption{{\footnotesize Probability distribution ${P}_i(N_3)$ of three-volumes
at fixed time $t_i$ well inside the blob ($i=11$). }}
\label{gauss}
\end{figure}

Let us finally address the question of how small values of $N_3(i)$
we can use without encountering finite-size effects \cite{agjl-etal}.
Let us consider a universe of size $N_4^{(4,1)}= 160.000$ as above.
If we are well inside the bulk (more than
6 time steps away from the stalk) we observe a nice Gaussian
distribution $P_i(N_3)$, where $N_3$ denotes the number of tetrahedra
associated to the time-slice at $t_i$ (see Fig. \ref{cos3} in order
to relate the time index $i$ to the position of the blob and the
position of the stalk).
It is peaked around the average value $\la N_3(i)\ra$, where
the average value with high precision is given by formula \rf{n1}.
A typical measured distribution and a Gaussian fit is shown in Fig.\
\ref{gauss}.

\begin{figure}
\centering
\includegraphics[width=0.80\textwidth]{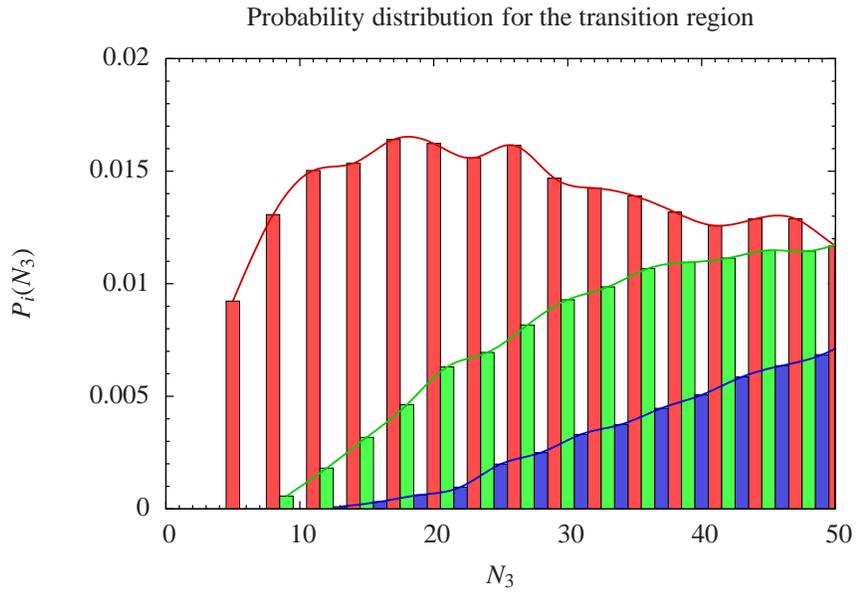}\\
\includegraphics[width=0.80\textwidth]{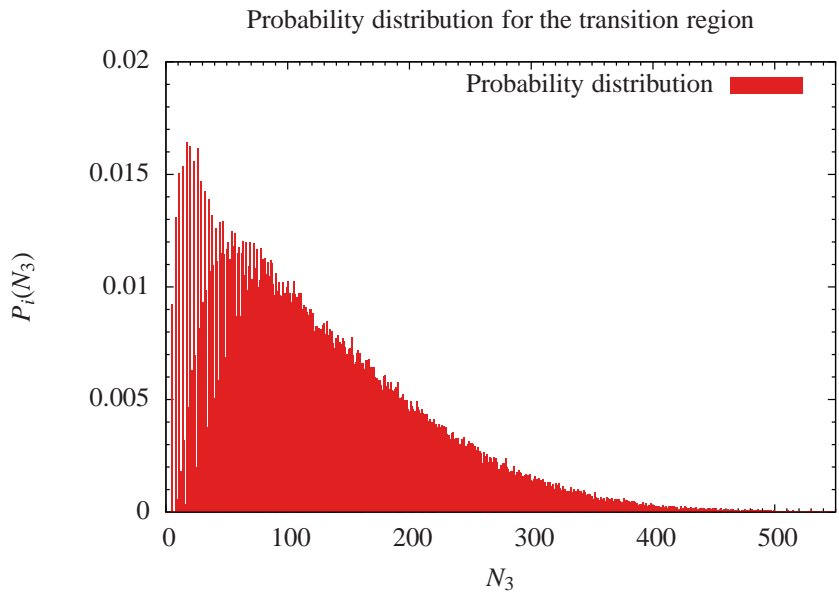}
\caption{{\footnotesize The probability distribution $P(N_3)$
from the transition region near the end of the blob ($i=18$).
For small $N_3$ the distribution splits into three families (top).
For large $N_3$ the split disappears, but the distribution
is highly asymmetric (bottom, no colour-coding for the three families). }}
\label{artifacts}
\end{figure}

Inside the stalk the situation is very different and
we observe pronounced finite-size effects. This is
to be expected since we are basically at the cutoff scale.
The distribution of three-volumes splits into three separate families.
Inside each family the discrete three-volumes differ by 3,
such that we have three sets of possible $N_3$-values, $\{5, 8, 11, \dots\}$,
$\{6,9,12,\dots\}$ and $\{7,10,13,\dots\}$.
When we start to move from the stalk and into the bulk, some memory
of this will remain for small $N_3(i)$, as shown in Fig.\ \ref{artifacts}.
The bulk distribution of Fig.\ \ref{cos3} extends over
approximately 42 lattice spacings in the time direction. For distances up to
six lattice spacings from the ``boundary'' between the stalk and bulk
(on both sides) one can observe phenomena like those shown in
Fig.\ \ref{artifacts}, which of course are most pronounced close to the
boundary. The figure shows the distribution $P_{17}(N_3)$
of three-volume at five time steps from the ``boundary'' between the
stalk and the bulk (see Fig.\ \ref{cos3} for the position of bulk
and stalk).
For small $N_3$ one observes the split into three distributions, which for $N_3 > 100$
merge into one. However, even then the picture
is quite different
from the Gaussian distribution far away from the
stalk, shown in Fig.\ \ref{gauss}.
When we consider $P_i(N_3)$ for $|i| < 16$,
the chance of finding a $N_3$ sufficiently small to
observe the finite-size effects just mentioned
is essentially zero. From
Fig.\ \ref{figuu_a} it is clear that the real test of the power law
requires small $N_3$, but we cannot use $N_3 < 100$ for
the reasons just stated. A fair way to state our result is that the data are
{\it consistent} with a $N_3^{-5/3}$ law, as seen from Fig.\ \ref{figuu_a}.
On the other hand, if one performs a fit to {\it determine} the power 5/3, the error bars
will be quite large.

\begin{figure}[t]
\centerline{\scalebox{1.00}{\rotatebox{0}{\includegraphics{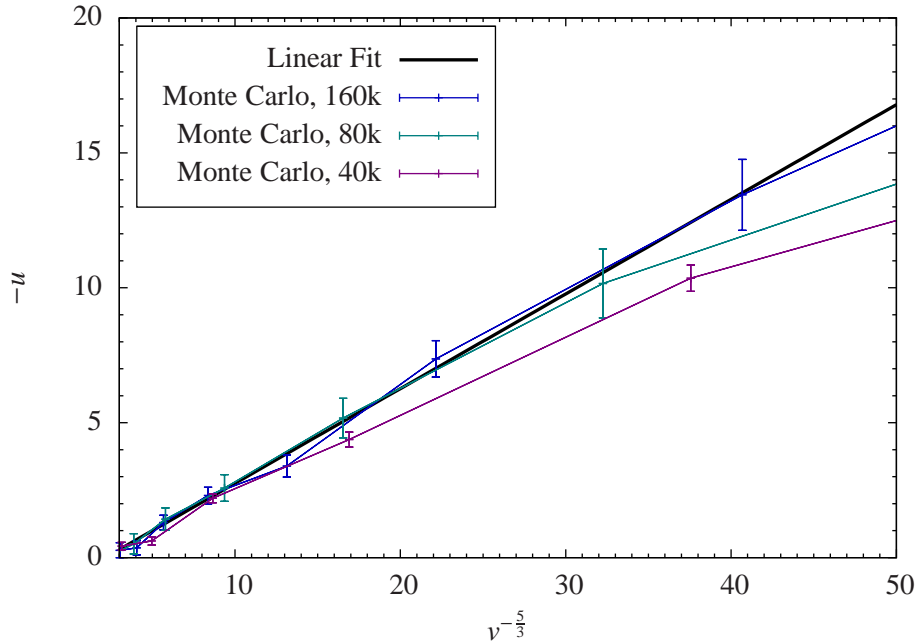}}}}
\caption{\label{figuu_b} {\footnotesize The dimensionless
second derivative $u= N_4^{5/4} U''(N_3)$
plotted against $\n^{-5/3}$, where $\n = N_3/N_4^{3/4}$ is the
dimensionless spatial volume, for $N_4^{(4,1)}= 40.000$,
80.000 and 160.000, $\kp_0 = 2.2$ and $\Del = 0.6$. One expects a universal
straight line near the origin (i.e. for large volumes) if
the power law $U(N_3) \propto N_3^{1/3}$ is correct.}}
\end{figure}

Apart from obtaining the correct power law $ N_3^{-5/3}$ for the potential
for a given spacetime volume $N_4$, it is equally
important that the coefficient
in front of this term be independent of $N_4$.
This seems to be the case as is shown in Fig.\ \ref{figuu_b}, where
we have plotted the measured potentials in terms of reduced, dimensionless
variables which make the comparison between measurements for
different $N_4$ easier.

In summary, we conclude that the data allow us to reconstruct the action
\rf{n7b} with good precision.
Quantifying the short-distance artifacts for one quantity
gives us a good idea of the scale at which they occur for a
given four-volume, but still leaves us with the difficulty of disentangling
them from genuine quantum-gravity signatures in this and other
observables. For example, the non-Gaussian character of the distribution of
fluctuations around the sphere \rf{n1} at small $N_3(i)$
observed above could indicate new short-distance physics, say,
along the lines suggested in the asymptotic safety scenario
\cite{laureu}. We presently do not know how to relate the
deviation from the four-sphere at a small scale factor
described there to our explicit small-$N_3(i)$ ``observations'', but it would
clearly be interesting to do so.

\subsection{Relation to Ho\v rava-Lifshitz gravity (part II)}

As described above our data in phase C fits well to \rf{n7b},
the discretized version of the minisuperspace action \rf{n5}. 
However, there is a residual ambiguity in the interpretation of the time coordinate,
as it appears in the identification \rf{n3}, which takes the form of an overall,
finite scaling between time and spatial directions.
As we have emphasized a number of
times, due to the entropic nature of the effective action,
there is no compelling reason to take the microscopic geometric
length assignments literally. 
We have chosen the time ``coordinate'' $t$ such that we
obtain a round four-sphere. However, the shape of the universe clearly
changes in terms of the number of
lattice spacings in time direction relative to
those in the spatial directions
when we vary the bare coupling constants $\kp_0$ and $\Del$.
Although this change is qualitatively in agreement with
the change of $\ta$ as a function of $\kp_0$ and $\Del$ (recall
eqs.\ \rf{act4dis} and \rf{actshort} which define $\ta$ and $\Del$, as well
as Fig.\ \ref{figalfa}), there
is no detailed quantitative agreement, as we have already mentioned.
Thus, rather than choosing time to be
consistent with an $S^4$-geometry, an alternative interpretation at this
stage is to assume
that the effective action is changing. Since the CDT set-up allows us to 
refer to its distinguished time slicing, a deformation \`a la Ho\v rava-Lifshitz
suggests itself. A corresponding effective Euclidean continuum action, 
including the measure, and expressed
in terms of standard metric variables could be of the form
\beql{horava}
S_H = \frac{1}{16\pi G} \int \d^3x\ \d t \; N \sqrt{g}
\Big((K_{ij}K^{ij}-\lam K^2) + (-\g R^{(3)} +2\La + V(g_{ij})\Big),
\eeq
where $K_{ij}$ denotes the extrinsic curvature, $g_{ij}$ the three-metric
on the spatial slices, $R^{(3)}$ the corresponding three-dimensional scalar
curvature, $N$ the lapse function, and finally $V$ a ``potential''
which in Ho\v rava's continuum
formulation would contain higher orders of spatial derivatives, potentially
rendering $S_H$ renormalizable. In our case we are not committed
to any particular choice of potential $V(g_{ij})$, since we are
not imposing renormalizability of the theory in any conventional
sense. An effective $V(g_{ij})$ could be generated
by entropy, i.e.\ by the measure, and may not relate
to any discussion of the theory being renormalizable.
The kinetic term depending on the extrinsic
curvature is the most general such term which is at most second order in
time derivatives and consistent with spatial diffeomorphism invariance.
The parameter $\lambda$ appears in the (generalized) DeWitt metric, which
defines an ultralocal metric on the classical space of all
three-metrics.\footnote{The value of $\lambda$ governs the signature of
the generalized DeWitt metric
$$
G_\lambda^{ijkl}=\frac{1}{2}\sqrt{\det g} (g^{ik} g^{jl}+g^{il}g^{jk}-
2\lambda g^{ij} g^{kl}),
$$ which is positive definite for $\lambda <1/3$, indefinite
for $\lambda =1/3$ and negative definite for $\lambda >1/3$. The role of $\lambda$
in three-dimensional CDT quantum gravity has been analyzed in detail in \cite{cdtlambda}.}
The parameter $\gamma$ can be related to a relative scaling between
time and spatial directions.
When $\lam =\g =1$ and $V=0$ we recover the standard (Euclidean)
Einstein-Hilbert action.

Making a simple minisuperspace ansatz with compact spherical slices,
which assumes homogeneity and isotropy of the spatial three-metric $g_{ij}$,
and fixing the lapse to $N=1$, the Euclidean action (\ref{horava}) becomes
a function of the scale factor $a(t)$
(see also \cite{elias,brandenberger,calcagni}), that is,
\beql{mini}
S_{mini} = \frac{2 \pi^2}{16\pi G} \int \d t \; a(t)^3 \Big( 3(1-3\lambda)\
\frac{\dot{a}^2}{a^2} -\gamma\ \frac{6}{a^2} +2 \Lambda+ \tV(a)\Big).
\eeq
The first three terms in the parentheses define the
IR limit, while the potential term $\tV(a)$ contains
inverse powers of the scale factor $a$ coming from possible
higher-order spatial derivative terms.

Our reconstruction of  the effective action
from the computer data, as described above,
is compatible with the functional form of the minisuperspace action
\rf{mini}. Importantly, we so far have not been able
to determine the constant $\tilde{k}_2$ in front of the potential term in
\rf{n7bb} with any precision,
which would enable us to fix
the ratio $(1-3\lam)/2\g$ appearing in \rf{mini}.
Also, as we have already discussed, it is presently not possible to determine
$\tV(a)$, which could be important for small values of the scale factor.
Once we have developed a better computer algorithm which allows us to
approach the B-C phase transition line, testing such Ho\v rava-Lifshitz 
scenarios will hopefully become easier.

\subsection{Some preliminary conclusions}

Let us emphasize a remarkable aspect of our results.
Our starting point was the Regge action for CDT,
as described in Sec.\ \ref{latticemodel} above.
However, the effective action we have generated dynamically
by performing the nonperturbative
sum over histories is only indirectly related
to this ``bare'' action. Likewise, the coupling constant $k_1$ which
appears in front of the effective action, and which we view as related to the
gravitational coupling constant $G$ by eq.\ \rf{n7c}, has no obvious direct
relation to the ``bare'' coupling $\kp_0$
appearing in the Regge action \rf{actshort}.
Nevertheless the leading terms in the
effective action for the scale factor are
precisely the ones presented in \rf{n5} or, more generally,
in the effective Ho\v rava-type action \rf{mini}.
The fact that a kinetic term with a second-order derivative appears as the
leading term in an effective action is perhaps less surprising,
but that it has precisely the form \rf{n5} or \rf{mini}, including
the term $N_3(i)^{1/3}$, is
remarkable and very encouraging for the entire
CDT quantization program. Until now our results are compatible with 
the Einstein-Hilbert action,
but better data are required to discriminate between the actions \rf{n5} and
\rf{mini}. In general, solutions
to the more general action \rf{mini} will be stretched spheres
\cite{bh}, which have conical singularities at the end points
(in our case the locations where the stalk starts). However,
these singularities in the curvature will not be visible in the
observable $V_3(t)$.

\section{Fluctuations around de Sitter space\label{fluctuations}}

We have shown that the action \rf{n7b}
gives an excellent description of the measured shape
$\bN_3(i)$ of the extended universe. 
Assuming that the
three-volume fluctuations around $\bN_3(i)$ are sufficiently small
so that a quadratic approximation is valid, we have also shown that
we can use the measured fluctuations
to reconstruct the discretized version \rf{n7b} of
the minisuperspace action \rf{n5},
where $k_1$ and $\tk_2$ are independent of the total four-volume $N_4$ used
in the simulations. This certainly provides strong evidence that both the
minisuperspace description of the dynamical
behaviour of the (expectation value of
the) three-volume, and the semiclassical
quadratic truncation for the description of
the quantum fluctuations in the three-volume are essentially correct.

In the following we will test in more detail how well the
actions \rf{n5} and \rf{n7b}
describe the data encoded in the covariance matrix $\hC$.
\begin{figure}[t]
\centerline{{\scalebox{1.0}{\rotatebox{0}{\includegraphics{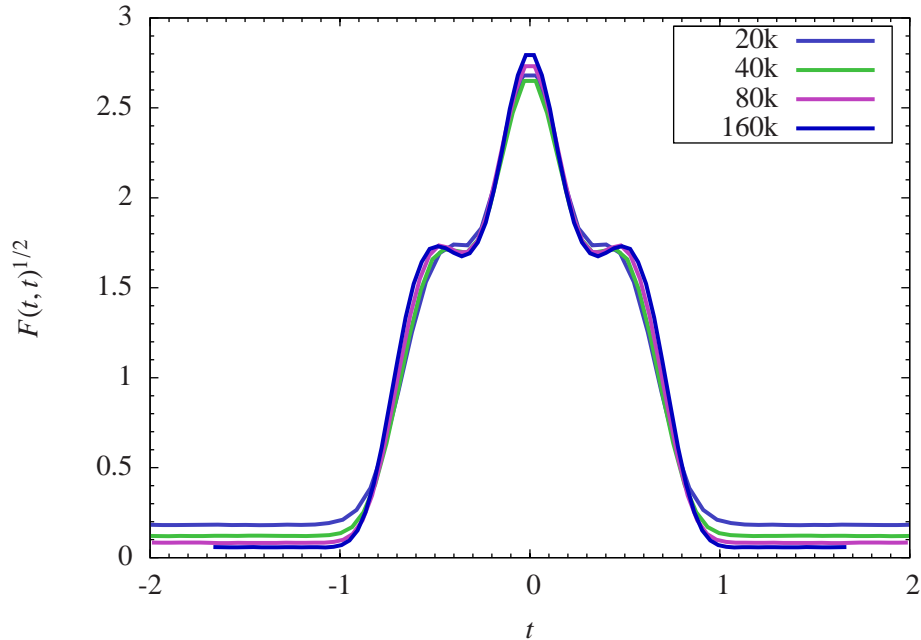}}}}}
\caption{\label{fluctu}{\footnotesize Analyzing the quantum
fluctuations of Fig.\ \ref{cos3}:
diagonal entries $F(t,t)^{1/2}$ of the universal scaling function
$F$ from \rf{n7f}, for $N_4^{(4,1)}=$ 20.000,
40.000, 80.000 and 160.000.}}
\end{figure}
The correlation function was defined in the previous section by
\beq\label{3h.1}
C_{N_4}(i,i') = \la \delta N_3(i) \delta N_3 (i')\ra,
~~~~\del N_3(i) \equiv N_3(i) -\bN_3(i),
\eeq
where we have included an additional subscript $N_4$ to emphasize that
$N_4$ is kept constant in a given simulation.
The first observation extracted from
the Monte Carlo simulations is that under a change in the four-volume
$C_{N_4}(i,i')$ scales as\footnote{We stress again that
the form \rf{n7f} is only valid in that part of the universe whose
spatial extension is considerably
larger than the minimal $S^3$ constructed from five tetrahedra.
(The spatial volume
of the stalk typically fluctuates between five and fifteen tetrahedra.)}
\beq
C_{N_4}(i,i') =
N_4 \; F\Big({i}/N^{1/4}_4,{i'}/N_4^{1/4}\Big), \label{n7f}
\eeq
where $F$ is a universal scaling function.
This is illustrated by Fig.\ \ref{fluctu} for the rescaled version
of the diagonal part $C_{N_4}^{1/2}(i,i)$,
corresponding precisely to the
quantum fluctuations $\la (\del N_3(i))^2\ra^{1/2}$
of Fig.\ \ref{cos3}. While the height of the curve in Fig.\ \ref{cos3}
will grow as $N_4^{3/4}$, the superimposed fluctuations
will only grow as $N_4^{1/2}$. We conclude that {\it for fixed bare
coupling constants} the relative fluctuations
will go to zero in the infinite-volume limit .

From the way the factor $\sqrt{N_4}$ appears as an overall scale in
eq.\ \rf{n7bb} it is clear that to the extent
a quadratic expansion around the effective background
geometry is valid one will have a scaling
\beq\label{sc1}
\la \del N_3(i) \del N_3(i')\ra =
N_4^{3/2} \la \del n_3(t_i) \del n_3(t_{i'})\ra =
N_4 F(t_i,t_{i'}),
\eeq
where $t_i = i/N_4^{1/4}$. This implies that \rf{n7f} provides additional
evidence for the validity of the quadratic approximation
and the fact that our choice of action (\ref{n7b}),
with $k_1$ independent of $N_4$, is indeed consistent.

To demonstrate in detail that the full
function $F(t,t')$ and not only its diagonal part is described by the
effective actions \rf{n5}, \rf{n7b}, let us for convenience adopt a
continuum language and compute its expected behaviour.
Expanding \rf{n5} around the classical solution
according to $V_3(t) = V_3^{cl}(t) + x(t)$,
the quadratic fluctuations are given by
\bea
\la x(t) x(t')\ra &\! =\! &
\int \cD x(s)\; x(t)x(t')\;
e^{ -\frac{1}{2}\int\!\!\int d s d s' x(s) M(s,s') x(s')} \nonumber\\
&\! =\! &  M^{-1}(t,t'),\label{n7a}
\eea
where $\cD x(s)$ is the normalized measure and
the quadratic form $M(t,t')$ is determined by expanding the
effective action $S$ to second order in $x(t)$,
\beq\label{n8}
S(V_3) = S(V_3^{cl}) + \frac{1}{18\pi G}\frac{B}{V_4}  \int \d t \; x(t) \hH
 x(t).
\eeq
In expression \rf{n8}, $\hH$ denotes the Hermitian operator
\beq\label{n9}
\hH= -\frac{\d }{\d t} \frac{1}{\cos^3 (t/B)}\frac{\d }{\d t} -
\frac{4}{B^2\cos^5(t/B)},
\eeq
which must be diagonalized
under the constraint that $\int dt \sqrt{g_{tt}}\; x(t) =0$, since
$V_4$ is kept constant.

Let $e^{(n)}(t)$ be the eigenfunctions of the quadratic form
given by \rf{n8} with the volume constraint enforced\footnote{One simple
way to find the eigenvalues and eigenfunctions approximately, including
the constraint, is to discretize the differential operator, imposing
that the (discretized) eigenfunctions vanish at the boundaries
$t= \pm B \pi/2$ and finally adding the
constraint as a term $\xi \Big(\int dt\,
x(t)\Big)^2$ to the action, where the coefficient $\xi$ is taken large. The
differential operator then becomes an ordinary matrix whose eigenvalues and
eigenvectors can be found numerically. Stability with respect
to subdivision and choice of $\xi$ is easily checked.},
ordered according to increasing eigenvalues $\lam_n$.
As we will discuss shortly, the lowest eigenvalue is $\lam_1 =0$,
associated with translational invariance in time direction,
and should be left out
when we invert $M(t,t')$, because we precisely fix the centre of volume
when making our measurements. Its dynamics is therefore not accounted for
in the correlator $C(t,t')$.

\begin{figure}[t]
\centerline{{\scalebox{0.65}{\rotatebox{0}{\includegraphics{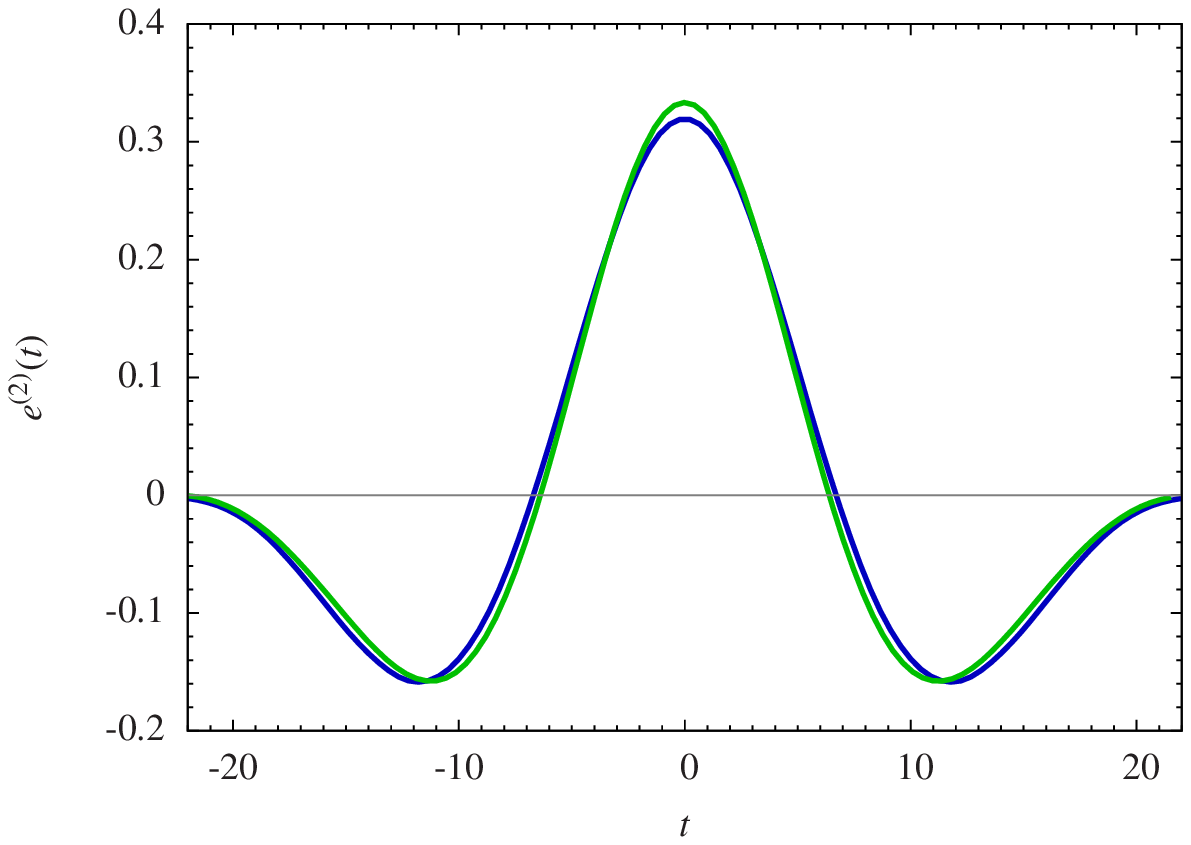}}}}
{\scalebox{0.65}{\rotatebox{0}{\includegraphics{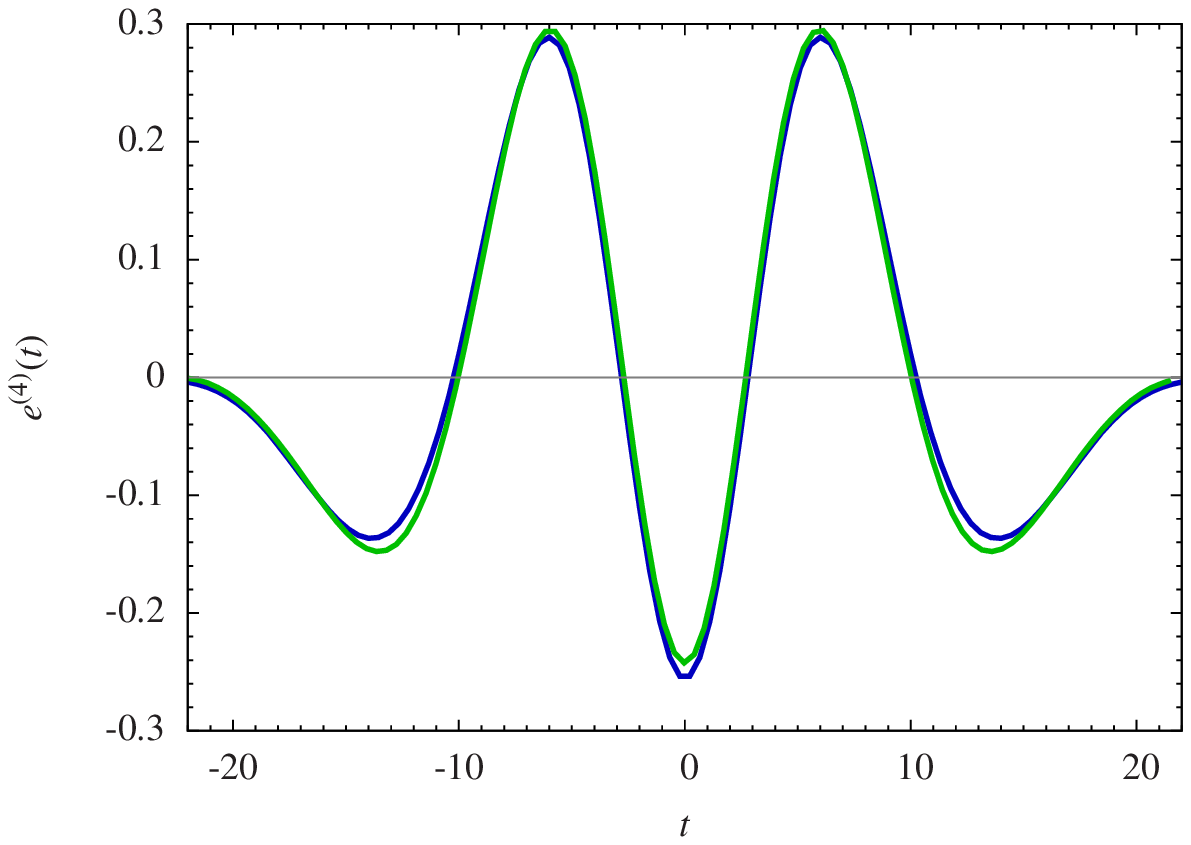}}}}}
\caption{\label{fig8}{\footnotesize 
Comparing the two highest even eigenvectors of the
covariance matrix $C(t,t')$ measured directly
(green curves) with the two lowest even
eigenvectors of $M^{-1}(t,t')$, calculated semiclassically (blue curves).}}
\end{figure}
If this cosmological continuum model were to give the correct
description of the computer-generated universe, the matrix
\beq\label{n12}
M^{-1}(t,t') = \sum_{n=2}^\infty \frac{e^{(n)}(t)e^{(n)}(t')}{\lam_n}.
\eeq
should be proportional to the measured correlator $C(t,t')$.
Fig.\ \ref{fig8} shows the eigenfunctions $e^{(2)}(t)$ and $e^{(4)}(t)$
(with two and four zeros respectively),
calculated from $\hH$ with the constraint $\int dt \sqrt{g_{tt}}\; x(t) =0$
imposed. Simultaneously
we show the corresponding eigenfunctions
calculated from the data, i.e.\ from the matrix $C(t,t')$, which
correspond to the (normalizable) eigenfunctions with the highest and third-highest
eigenvalues. The agreement is very good,
in particular when taking into consideration that
no parameter has been adjusted in the action (we simply take
$B=s_0 N_4^{1/4}\Del t$ in \rf{n2} and \rf{n8}, which for $N_4 = 362.000$
gives $B=14.47a_t$, where $a_t$ denotes the time distance between successive
slices).

The reader may wonder why the first eigenfunction exhibited
has two zeros. As one would expect, the ground state eigenfunction $e^{(0)}(t)$
of the Hamiltonian \rf{n9}, corresponding to the lowest eigenvalue,
has no zeros, but does not satisfy the volume
constraint $\int dt \sqrt{g_{tt}}\; x(t) =0$.
The eigenfunction $e^{(1)}(t)$ of $\hH$ with next-lowest
eigenvalue has one zero and is given by the simple analytic function
\beq\label{e1}
e^{(1)}(t) = \frac{4}{\sqrt{\pi B}}  \sin \Big(\frac{t}{B}\Big) \,
\cos^2 \Big(\frac{t}{B}\Big) = c^{-1} \frac{d V_3^{cl} (t)}{dt},
\eeq
where $c$ is a constant.
One realizes immediately that $e^{(1)}$ is the translational zero mode of the
classical solution $V_3^{cl}(t)$ ($\propto \cos^3 t/B$). Since the action is
invariant under time translations we have
\beq\label{cl1}
S(V_3^{cl}(t+\Del t)) = S(V_3^{cl}(t)),
\eeq
and since $V_3^{cl}(t)$ is a solution to the
classical equations of motion we find to second order (using the
definition \rf{e1})
\beq\label{cl2}
S(V_3^{cl}(t+\Del t)) = S(V_3^{cl}(t)) +
\frac{c^2 (\Del t)^2 }{18\pi G}\frac{B}{V_4}
\int dt \;e^{(1)}(t) \hH e^{(1)}(t),
\eeq
consistent with $e^{(1)}(t)$ having eigenvalue zero.

It is clear from Fig.\ \ref{fig8} that some of the eigenfunctions of $\hH$
(with the volume constraint imposed) agree very well with the
measured eigenfunctions.
All even eigenfunctions (those symmetric with respect to reflection about the
symmetry axis located at the centre of volume) turn out to agree very well.
The odd eigenfunctions of $\hH$ agree less well with the
eigenfunctions calculated from the measured $C(t,t')$.
The reason seems to be that we have not managed to eliminate the motion
of the centre of volume completely from our measurements. As already mentioned
above, there is an inherent ambiguity in fixing the centre of volume, which
turns out to be sufficient to reintroduce the zero mode in the data.
Suppose we had by mistake misplaced the centre of volume by a small distance
$\Del t$.
This would introduce a modification
\beq\label{mod}
\Del V_3 = \frac{dV_3^{cl}(t)}{dt} \; \Del t
\eeq
proportional to the zero mode of the potential $V_3^{cl}(t)$.
It follows that the zero mode can re-enter whenever we have an ambiguity
in the position of the centre of volume.
In fact, we have found that the first odd eigenfunction
extracted from the data can be perfectly described by
a linear combination of $e^{(1)}(t)$ and $e^{(3)}(t)$.
It may be surprising at first that an ambiguity of one lattice
spacing can introduce a significant mixing. However, if we translate
$\Del V_3$ from eq.\ \rf{mod} to ``discretized'' dimensionless units using
$V_3(i) \sim N_4^{3/4} \cos (i/N_4^{1/4})$,
we find that $\Del V_3 \sim \sqrt{N_4}$,
which because of $\la (\del N_3(i))^2\ra \sim N_4$ is of
the same order of magnitude as the fluctuations
themselves. In our case, this
apparently does affect the odd eigenfunctions.

\section{The size of the universe}\label{size}

Let us now return to equation \rf{n7c},
\beq\label{n7cc}
{G} = \frac{a^2}{k_1} \frac{\sqrt{\tC_4}\; \ts_0^2}{3\sqrt{6} },
\eeq
which relates the parameter $k_1$ extracted from
the Monte Carlo simulations to
Newton's constant in units of the cutoff $a$, $G/a^2$.
For the bare coupling constants
$(\kp_0,\Del)= (2.2,0.6)$ we have high-statistics measurements
for $N_4$ ranging from 45.500 to 362.000 four-simplices
(equivalently, $N_4^{(4,1)}$
ranging from 20.000 to 160.000 four-simplices). The choice of
$\Del$ determines the asymmetry parameter $\ta$, and the
choice of $(\kp_0,\Del)$ determines the ratio $\xi$
between $N_4^{(3,2)}$ and $N_4^{(4,1)}$. This in turn determines
the ``effective'' four-volume $\tC_4$ of an average four-simplex, which
also appears in \rf{n7cc}. The number $\ts_0$ in \rf{n7cc}
is determined directly from the time extension $T_{\rm univ}$
of the extended universe according to
\beq\label{width}
T_{\rm univ}=\pi\; \ts_0 \Big(N_4^{(4,1)}\Big)^{1/4}.
\eeq
Finally, from our measurements we have determined $k_1= 0.038$.
Taking everything together according to  \rf{n7cc},
we obtain $G\approx 0.23 a^2$, or
$\ell_{Pl}\approx 0.48 a$, where  $\ell_{Pl} = \sqrt{G}$ is the Planck length.

From the identification of the volume of the four-sphere,
$V_4 =  8\pi^2 r^4/{3} = \tC_4 N_4^{(4,1)} a^4$,
we obtain that $r=3.1 a$. In other words,
{\it the linear size $\pi R$ of the quantum de Sitter universes
studied here lies in the range of 12-21 Planck lengths for $N_4$ in the
range mentioned above and for the bare
coupling constants chosen as $(\kp_0,\Del)=(2.2,0.6)$}.

Our dynamically generated universes are therefore not very big, and the
quantum fluctuations around their average shape are large as is apparent from
Fig.\ \ref{cos3}. It is rather
surprising that the semiclassical minisuperspace formulation is applicable
for universes of such a small size,
a fact that should be welcome news to anyone
performing semiclassical calculations to describe the
behaviour of the early universe.
However, in a certain sense our lattices are still coarse compared
to the Planck scale $\ell_{Pl}$ because the Planck length is
roughly half a lattice
spacing. If we are after a theory of quantum gravity valid on all scales,
we are in particular interested in
uncovering phenomena associated with Planck-scale
physics. In order to collect data free from
unphysical short-distance lattice artifacts at this
scale, we would ideally like to work with a
lattice spacing much smaller than the Planck length,
while still being able to set by hand the physical volume of the
universe studied on the computer.

The way to achieve this, under the assumption that the coupling
constant  $G$ of formula \rf{n7cc} is indeed a true measure
of the gravitational
coupling constant, is as follows.
We are free to vary the discrete four-volume $N_4$ and the bare coupling
constants $(\kp_0, \Del)$ of the Regge action. Assuming for the
moment that the semiclassical minisuperspace action is valid,
the effective coupling constant $k_1$ in front of it
will be a function of the bare coupling constants $(\kp_0, \Del)$,
and can in principle be determined as described above for the case
$(\kp_0, \Del)=(2.2,0.6)$. If we adjusted the bare coupling
constants such that in the limit as $N_4 \to \infty$ both
\beq\label{n100}
V_4 \sim N_4 a^4~~~{\rm and}~~~G\sim a^2/k_1(\kp_0,\Del)
\eeq
remained constant (i.e.\ $k_1(\kp_0,\Del) \sim 1/\sqrt{N_4}$),
we would eventually reach a region where the lattice spacing $a$ was
significantly smaller than the Planck length, in which event the lattice
could be used to approximate spacetime structures of Planckian size
and we could initiate a genuine study of the sub-Planckian regime.
\begin{figure}[t]
\centerline{{\scalebox{0.6}{\rotatebox{0}{\includegraphics{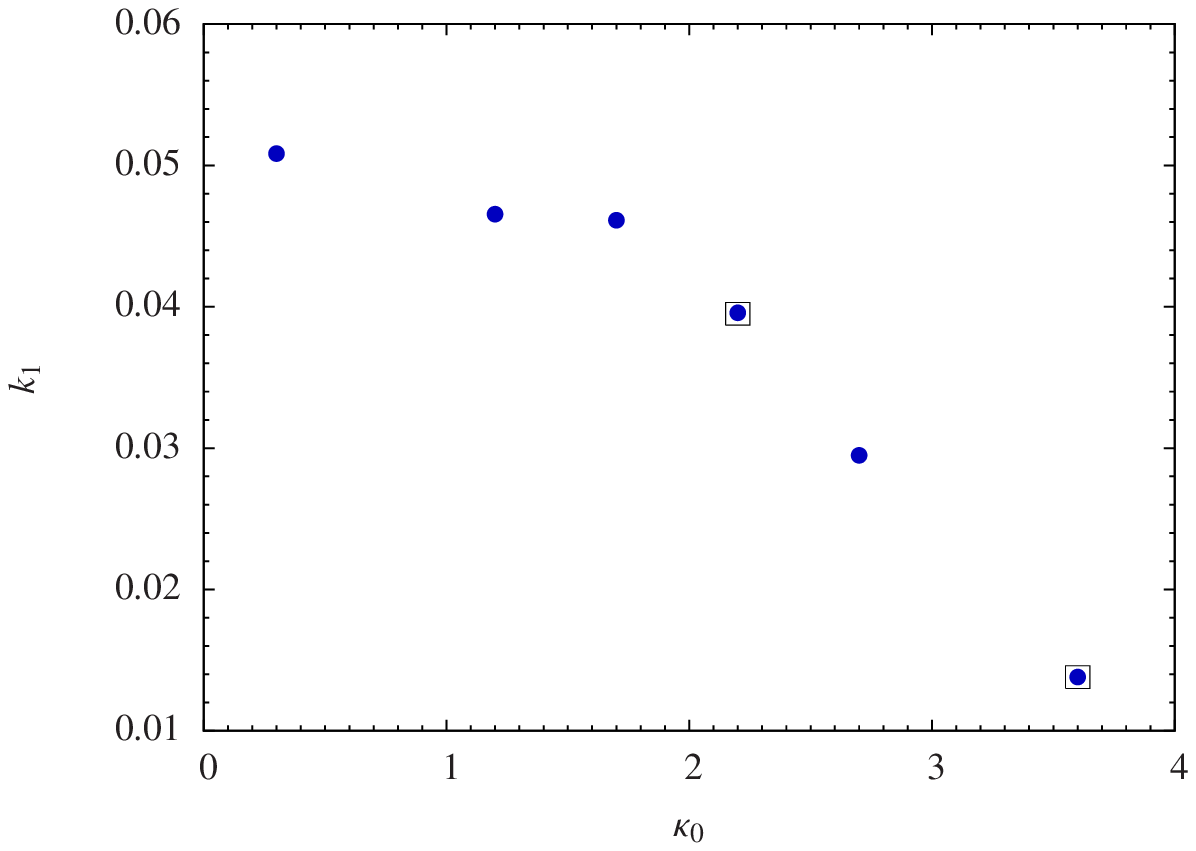}}}}
{\scalebox{0.6}{\rotatebox{0}{\includegraphics{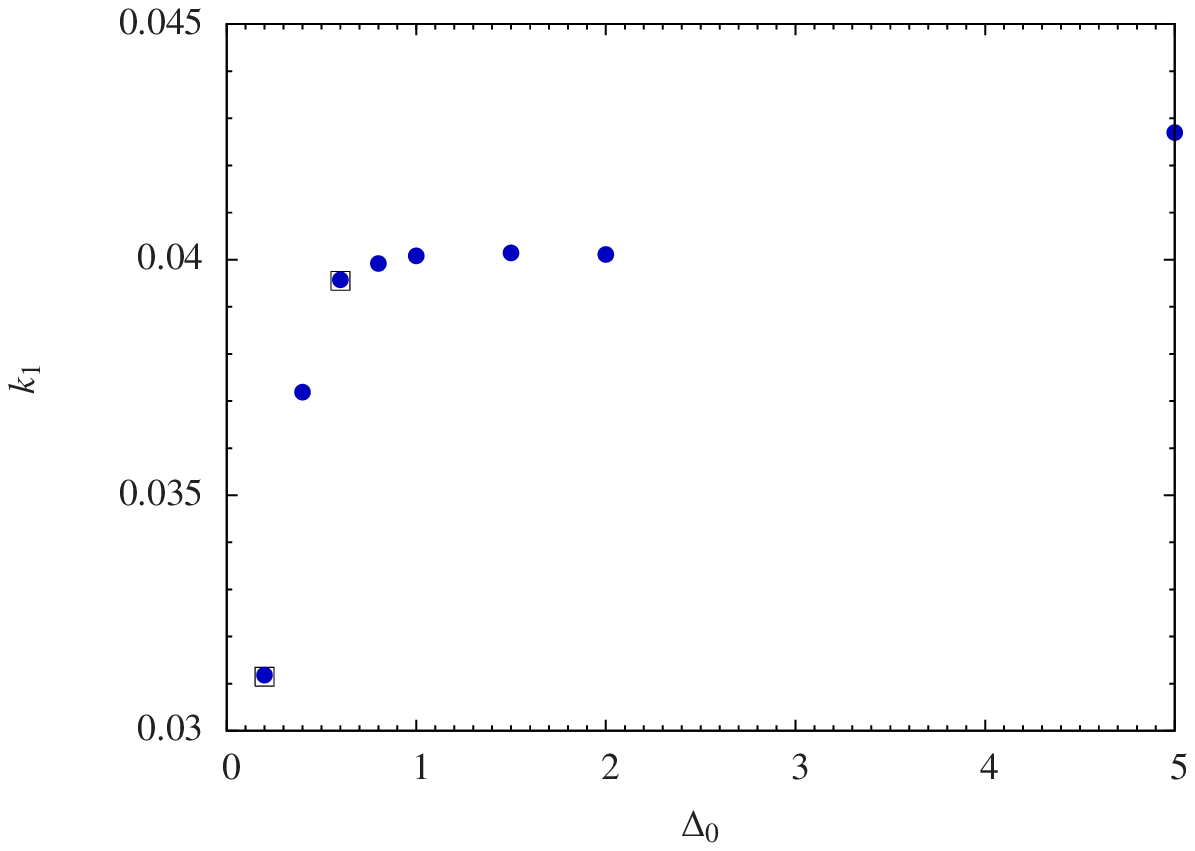}}}}}
\caption{\label{fig12}{\footnotesize The measured effective
coupling constant $k_1$ as function
of the bare $\kp_0$ (left, $\Del = 0.6$ fixed)
and the asymmetry $\Del$ (right,
$\kp_0=2.2$ fixed). The marked point near the middle of the data points sampled
is the point $(\kp_0,\Del)=(2.2,0.6)$
where most measurements in the remainder of the
paper were taken. The other marked points are those
closest to the two phase
transitions lines, to the A-C phase transition line on the left figure,
and the B-C phase transition line on right figure.}}
\end{figure}
Since we have no control over the effective coupling constant
$k_1$, the first obvious question which arises is whether we can at all
adjust the bare coupling constants in such a way that at large
scales we still see a four-dimensional universe, with
$k_1$ going to zero at the same time.
The answer seems to be in the affirmative, as we will now go on to explain.

Fig.\ \ref{fig12} shows the results of extracting
$k_1$ for a range of bare coupling constants for which we still
observe an extended universe. In the top figure
$\Del = 0.6$ is kept constant while $\kp_0$ is varied.
For $\kp_0$ sufficiently large
we eventually reach a point where the A-C  transition takes place (the point
in the square in the bottom right-hand corner is the measurement closest to
the transition we have looked at).
For even larger values of $\kp_0$, beyond this transition,
the universe disintegrates
into a number of small universes,
in a CDT-analogue of the branched-polymer phase of
Euclidean quantum gravity, as described above. The plot shows that
the effective coupling constant $k_1$ becomes smaller and
possibly goes to zero as the phase transition point is approached, although
our current data do not yet allow us to conclude that $k_1$
does indeed vanish at the transition point.

Conversely, the bottom figure of Fig.\ \ref{fig12}
shows the effect of varying  $\Del$, while keeping $\kp_0=2.2$ fixed.
As $\Del$ is decreased from 0.6, we
eventually hit the B-C phase transition, separating the physical phase of
extended universes from the CDT-equivalent of the
crumpled phase of Euclidean quantum gravity, where the
entire universe will be concentrated within a few time steps.
Strictly speaking we are not able to go much closer than to $\Del=0.3$,
as already explained. It seems that
the effective coupling constant $k_1$ starts to decrease
in value when $\Del$ is decreasing from 0.6, but since we cannot
get very close to the phase boundary, we cannot in any
convincing way say that $k_1 \to 0$ along the B-C phase boundary.

To extract the coupling constant $G$
from \rf{n7cc} we not only have to take into account
the change in $k_1$, but also
that in $\ts_0$ (the width of the distribution $N_3(i)$) and
in the effective four-volume $\tC_4$ as a
function of the bare coupling constants.
Combining these changes, we arrive at
a slightly different picture. Approaching the B-C boundary
the gravitational coupling constant $G$ does not vary much,
despite the fact that $1/k_1$ increases. This is a consequence of $\ts_0$
decreasing considerably, as can be seen from Fig.\ \ref{cos3_b}.
On the other hand, when we approach the A-C
the effective gravitational coupling constant $G$ increases, more or
less like $1/k_1$, where the behaviour of $k_1$ is shown in
Fig.\ \ref{fig12} (left). This implies that
the Planck length $\ell_{Pl} = \sqrt{G}$ increases
from approximately $0.48 a$ to $0.83 a$ when $\kp_0$ changes from
2.2 to 3.6. On the basis of these arguments it seems that we have
to go very close to the phase boundaries in order to
penetrate into the sub-Planckian regime.

One interesting issue under investigation is whether and to what extent
the simple minisuperspace description remains
valid as we go to shorter scales.
This raises the interesting possibility of being able to test explicitly the
scaling violations of $G$ predicted by renormalization group methods
in the context of asymptotic safety \cite{reuteretc}. We will discuss
this  in detail in Sec.\ \ref{renormalization}.

\section{The fractal  dimensions}
\label{spectraldim}

\subsection{The spectral dimension}

One way to obtain information about the geometry of our quantum universe
is by studying a diffusion process on the underlying geometric ensemble.
We will use this technique to determine the
so-called {\it spectral dimension} $D_S$ of the ensemble of geometries.

As a start let us discuss briefly diffusion on a $d$-dimensional
manifold with a fixed,
smooth Riemannian metric $g_{ab}(\xi)$. The diffusion equation has the form
\beq\label{ja2}
\frac{\prt}{\prt \sg} \, K_g(\xi,\xi_0;\sg) = \Del_g K_g (\xi,\xi_0;\sg),
\eeq
where $\sg$ is a fictitious diffusion time, $\Del_g$ the Laplace
operator of the metric $g_{ab}(\xi)$ and $K_g(\xi,\xi_0;\sg)$
the probability density of diffusion from $\xi_0$ to $\xi$ in
diffusion time $\sg$. We will consider diffusion processes which
initially are peaked at some point $\xi_0$,
\beq\label{ja3}
K_g(\xi,\xi_0;\sg\equ 0) = \frac{1}{\sqrt{\det g(\xi)}}\, \del^d(\xi-\xi_0).
\eeq
For the special case of a flat Euclidean metric, we have
\beq\label{flat1}
K_g(\xi,\xi_0;\sg) = \frac{\e^{-d_g^2(\xi,\xi_0)/4\sg}}{(4\pi \sg)^{d/2}},
\qquad g_{ab}(\xi)\equ \delta_{ab}.
\eeq
For general curved spaces $K_g$ has the well-known
asymptotic expansion
\beq\label{ja4}
K_g(\xi,\xi_0;\sg) \sim \frac{\e^{-d_g^2(\xi,\xi_0)/4\sg}}{\sg^{d/2}}
\sum_{r\equ 0}^\infty a_r(\xi,\xi_0)\, \sg^r
\eeq
for small $\sg$,
where $d_g(\xi,\xi_0)$ denotes the geodesic distance
between $\xi$ and $\xi_0$.
Note the appearance of the power $\sg^{-d/2}$ in this relation, reflecting
the dimension $d$ of the manifold, just like in the formula \rf{flat1} for
flat space. This happens because small values of $\sg$ correspond to short
distances and for any given smooth metric short distances imply
approximate flatness.

A quantity that is easier to measure in numerical simulations is the average
{\it return probability} $P_{g}(\sg)$, which possesses an analogous expansion
for small $\sg$,
\beq\label{ja5}
P_{g}(\sg) \equiv \frac{1}{V} \int  d^d\xi \sqrt{\det g(\xi)} \;
K_g(\xi,\xi;\sg) \sim
\frac{1}{\sg^{d/2}} \sum_{r\equ 0}^\infty A_r \sg^r,
\eeq
where $V$ is the spacetime volume $V\equ\int d^d\xi \sqrt{\det g(\xi)}$ and
the expansion coefficients $A_r$ are given by
\beq\label{adef}
A_r = \frac{1}{V}\int  d^d\xi \sqrt{\det g(\xi)} \;a_r(\xi,\xi).
\eeq
For an infinite flat space,
we have $P_g(\sg)\equ  1/(4\pi\sg )^{d/2}$ and thus can
extract the dimension $d$ by taking the logarithmic derivative,
\beq\label{ja5a}
-2\ \frac{d \log P_g(\sg)}{d\log \sg} = d,
\eeq
independent of $\sg$. For non-flat spaces and/or finite volume $V$,
one can still use
eq.\ \rf{ja5a} to extract the dimension, but there will be
corrections for sufficiently large $\sg$.
For finite volume in particular, $P_g(\sg)$ goes to $1/V$ for
$\sg \gg V^{2/d}$ since the zero mode of the Laplacian $-\Del_g$
will dominate the diffusion in this region.
For a given diffusion time $\sg$ the behaviour of $P_g(\sg)$ is determined by
eigenvalues $\lambda_n$ of $-\Del_g$ with $\lambda_n \le 1/\sg$,
and the contribution from higher eigenvalues is exponentially suppressed.
Like in the flat case, where diffusion over a time $\sg$ probes the geometry
at a linear scale $\sqrt{\sg}$, large $\sg$ corresponds to large distances away
from the origin $\xi_0$ of the diffusion process,
and small $\sg$ to short distances.

The construction above can be illustrated by the simplest example of
diffusion in one dimension.
The solution to the diffusion equation on the real axis,
starting at the origin,  is
\beq\label{jb1}
K(\xi,\sg) = \frac{\e^{-\xi^2/4\sg}}{\sqrt{4\pi \sg}},~~~~
K(k,\sg) = \e^{-k^2 \sg},
\eeq
where $K(k,\sg)$ denotes the Fourier transform of $K(\xi,\sg)$.
The eigenvalues of the Laplace operator are of course just given by $k^2$.
In order to illustrate the finite-volume effect, let us compactify
the real line to a circle of length $L$. The return probability is now given by
\beq\label{jb2}
P_L(\sg)= \frac{1}{L}\sum_{n=-\infty}^\infty \e^{-k_n^2 \sg} =
\frac{1}{\sqrt{4\pi \sg}}
\sum_{m=-\infty}^{\infty} \e^{-L^2m^2/4\sg},~~~~k_n=\frac{2\pi n}{L},
\eeq
where in the second step we have performed a Poisson resummation to
highlight the $\sg^{-1/2}$-dependence for small $\sg$. In line with the
discussion above, the associated spectral dimension $D_S(\sg)$
is constant (and equal to one) up to
$\sg$-values of the order $L^2/4\pi^2$, and then goes to zero monotonically.

In applying this set-up to four-dimensional quantum gravity
in a path integral formulation,
we are interested in measuring the expectation value of the return
probability $P_g(\sg)$. Since $P_g(\sg)$ defined in \rf{ja5}
is invariant under reparametrizations, it
makes sense to take its quantum average over all geometries of
a given spacetime volume $V_4$,
\beq\label{ja7}
P_{V_4}(\sg) =
\frac{1}{\ Z(V_4)} \int\!\! \cD [g_{ab}] \; e^{-{S_E}(g_{ab})}
\del\left(\int d^4x \sqrt{ g}-V_4\right) \, P_g(\sg),
\eeq
in accordance with the definition of expectation values \rf{constantV3}.
Since the small-$\sg$ behaviour of $P_g(\sg)$ is the same for each smooth
geometry, it might seem obvious that the
same is true for their integral $P_{V_4}(\sg)$,
but this need not be so. Firstly, the scale $\sg$ in \rf{ja7} is held fixed,
independent of the geometry $g_{ab}$, while the expansion \rf{ja5} contains
reference to higher powers of the curvature of $g_{ab}$.
Secondly, one should keep in mind that a typical geometry which contributes
to the path integral -- although continuous -- is unlikely to be smooth.
This does not invalidate our treatment, since diffusion processes can be
meaningfully defined on much more general objects than smooth
manifolds. For example, the return probability for diffusion on
fractal structures is well studied in statistical physics and takes
the form
\beq\label{ja6}
P_N(\sg)= \sg^{-D_S/2} \; F\Big(\frac{\sg}{N^{2/D_S}}\Big),
\eeq
where $N$ is the ``volume'' associated with the fractal
structure and $D_S$ the so-called {\it spectral dimension},
which is not necessarily an integer.
An example of fractal structures are branched polymers
which generically have $D_S\equ 4/3$ \cite{thordur-john,anrw}.
Extensive numerical simulations \cite{2dspectral,aa}
have shown that in two-dimensional quantum gravity
the only effect of integrating over geometries is to replace
the asymptotic expansion \rf{ja5}, which contains
reference to powers of the curvature related to
a specific metric, by the simpler form \rf{ja6}.

Our next task is to define diffusion on the class of metric spaces
under consideration, the piecewise linear structures
defined by the causal triangulations $T$.
At this point we will cheat a little. When translating eq.\ \rf{ja2}
into a form suitable for our piecewise linear geometries
we will assume $\ta\equ 1$, although this is strictly speaking
not the case. The motivation is that the expression for $\ta\neq 1$ is
more complicated, and that following the diffusion on large
triangulations consumes a lot of computer time. Since we are presently
only trying to determine $D_S$ which is a reasonably universal exponent,
we expect it to be independent of the exact discretized implementation of
the Laplacian $\Del_g$ on the piecewise linear geometry.

We start from an initial probability distribution
\beq
K_T(i,i_0;\sg\equ 0) = \delta_{i,i_0},
\eeq
which vanishes everywhere except at a randomly chosen (4,1)-simplex $i_0$,
and define the diffusion process by the evolution rule
\beq
K_T(j,i_0;\sg+1) = \frac{1}{5}\sum_{k\to j} K_T(k,i_0;\sg).
\label{evo4d}
\eeq
These equations are the simplicial analogues of \rf{ja3} and \rf{ja2},
with the triangulation
(together with its Euclideanized edge-length assignments)
playing the role of $g_{ab}$, and
$k\to j$ denoting the five nearest neighbours of the four-simplex $j$.
Clearly \rf{evo4d} is only the discretized Laplacian if the length
assignments of all links is 1 (in lattice units), i.e.\ $\ta\equ 1$, as
just discussed. In this process, the total probability
\beq
\sum_j K_T(j,i_0;\sg) =1
\eeq
is conserved. The return probability to a simplex $i_0$ is then
defined as $P_T(i_0;\sg)\equ  K_T(i_0,i_0;\sg)$ and the
quantum average as
\beq\label{ja7a}
P_{N_4}(\sg)= \frac{1}{ Z(N_4)} \sum_{T_{N_4}}
e^{-{S}_E(T_{N_4})}\;
\frac{1}{N_4}\sum_{i_0 \in T_{N_4}} K_{T_{N_4}}(i_0,i_0;\sg),
\eeq
where $T_{N_4}$ denotes a triangulation with $N_4$ four-simplices,
and ${S}_E(T_{N_4})$ and $ Z(N_4)$ are the obvious
simplicial analogues of the continuum quantities at fixed four-volume.
Assuming that the return probability behaves according to \rf{ja6},
with $N\equ N_4$, we can extract the value of the fractal dimension $D_S$
by measuring the logarithmic derivative as in \rf{ja5a} above, as long
as the diffusion time is not much larger than $N_4^{2/D_S}$,
\beq\label{ja1}
D_S(\sg) = -2 \;\frac{d\log P_{N_4}(\sg)}{d\log \sg}+
\mbox{finite-size corrections}.
\eeq

From the experience with numerical simulations of two-dimensional 
Euclidean quantum
gravity in terms of dynamical triangulations \cite{ajw,2dspectral,aa}, we
expect some irregularities in the behaviour of the return probability for
the smallest $\sg$, i.e. close to the cutoff scale. Typically, the
behaviour of $P_N(\sg)$ for odd and even diffusion steps $\sg$ will
be quite different for small $\sg$ and merge
only for $\sg \approx 20-30$. After the merger, the curve enters
a long and stable regime where the right-hand side of \rf{ja1} is
independent of $\sg$, before finite-size effects start to dominate
which force it to go to zero.

The origin of the odd-even asymmetry can again be illustrated by the
simple case of diffusion on a  circle, whose
solution is the discretized version of the solution \rf{jb2}. In this
case the asymmetry of the return probability between odd and even time steps
is extreme: if we use the one-dimensional version
of the simple evolution equation \rf{evo4d}, we obtain
\beq\label{jb3}
P_L(\sg)= \left\{
\begin{array}{cl}
0 & ~~\mbox{for $\sg$ {odd,}} \\
~&~\\
\displaystyle{\frac{1}{2^\sg}}\;
\begin{pmatrix} \sg  \\ {\sg}/{2} \end{pmatrix}
 &~~\mbox{ for $\sg$ even,}
\end{array}
\right.
\eeq
as long as $\sg < L/2$, where $L$ is the discrete volume of the circle
(i.e. the number of its edges).
It is of course possible to eliminate this asymmetry by using an
``improved" discretized diffusion equation, but in the case of
higher-dimensional random geometries like those used in four-dimensional
causal dynamical triangulations this is not really necessary.
\begin{figure}[t]
\centerline{\scalebox{1.0}{\rotatebox{0}{\includegraphics{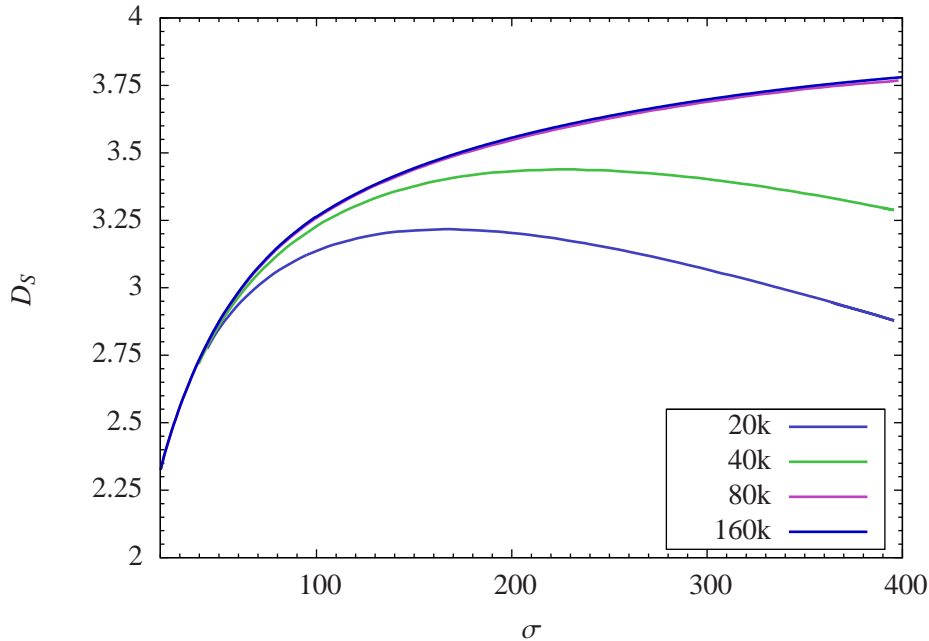}}}}
\caption[phased]{{\footnotesize The spectral dimension $D_S(\sg)$ of causal
dynamical triangulations as a function of the diffusion time $\sg$, which
is a direct measure of the distance scale probed. The measurements
were taken at volumes $ N_4^{(4,1)}\equ 20$k (bottom curve), 40k and
80k (top curve), and for $\kappa_0\equ 2.2$, $\Delta\equ 0.6$ and
$N\equ 80$.}}
\label{new4d2a}
\end{figure}

The results of measuring the spacetime spectral dimension $D_S$
were first reported in \cite{spectral-d} and discussed in detail in
\cite{blp}. We work with system sizes of
up to $ N_4^{(4,1)}\equ 80$k with $\kappa_0\equ 2.2$, $\Delta\equ 0.6$ and
$N\equ 80$. Since we are interested in the bulk properties of quantum
spacetime and since the volume is not distributed evenly in the
time direction (cf. Fig.\ \ref{unipink}),
we always start the diffusion process from a four-simplex
adjacent to the slice of maximal three-volume.
When this is done the
variations in the curve $D_S(\sg)$ for different
generic choices of triangulation $T$ and starting simplices $i_0$ are small.
The data curves presented in Fig.\ \ref{new4d2a} were
obtained by averaging over 400 different diffusion processes
performed on independent configurations.
We have omitted error bars from Fig.\ \ref{new4d2a} to illustrate
how the curves converge to a shape that represents $D_S(\sg)$ in
the infinite-volume
limit, and which is given by the envelope of the data curves for finite volume.
For the two lower volumes, $ N_4^{(4,1)}\equ 20$k and $ N_4^{(4,1)}\equ 40$k,
there still are clear finite-volume effects for large diffusion times $\sg$.

\begin{figure}[t]
\centerline{\scalebox{1.0}{\rotatebox{0}{\includegraphics{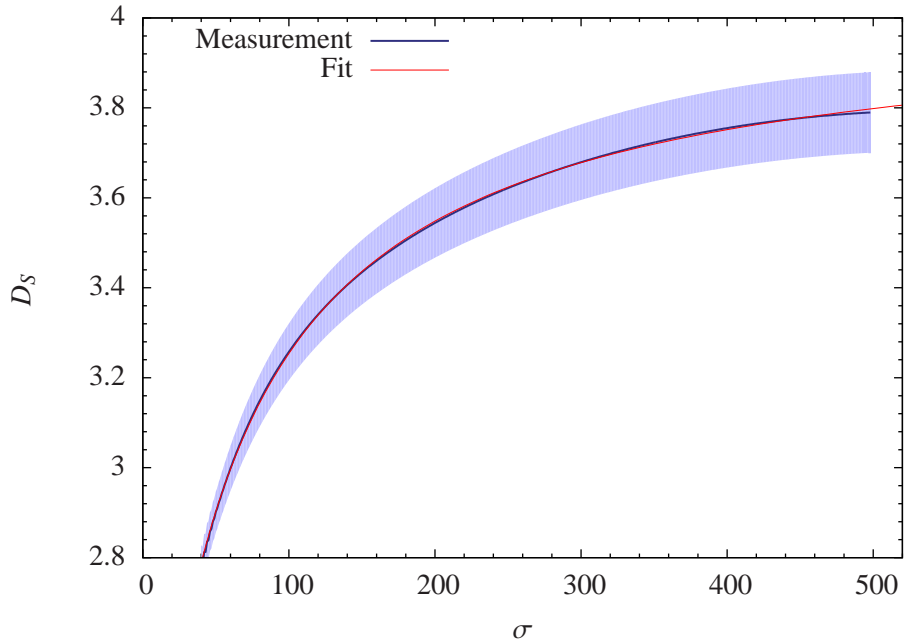}}}}
\caption[phased]{{\footnotesize The spectral dimension $D_S$ of the universe as
function of the diffusion time $\sg$, measured for
$\kappa_0\equ 2.2$, $\Delta\equ 0.6$ and $N\equ 80$, and a spacetime volume
$N_4\equ 181$k. The averaged measurements lie along the central curve,
together with a superimposed best fit
$D_S(\sg) = 4.02\mi 119/(54\plu\sg)$. The two outer
curves represent error bars.}}
\label{d4s2.2b4}
\end{figure}

By contrast, the top curve -- corresponding to the maximal volume
$ N_4^{(4,1)}\equ 80$k -- continues to rise for increasing $\sg$, which makes
it plausible that we can ignore any finite-size effects and that it is a good
representative of the infinite-volume limit in the
$\sg$-range considered.\footnote{In both Figs.\ \ref{new4d2a} and
\ref{d4s2.2b4} we have only plotted the region where the curves
for odd and even $\sg$ coincide, in order to exclude short-distance
lattice artifacts. The two curves merge at about $\sg\equ 40$.
Since the diffusion distance grows as $\sqrt{\sg}$, a
return diffusion time of 40 corresponds to just a few steps
away from the initial four-simplex.}
We will therefore concentrate on analyzing the
shape of this curve, which is presented separately in Fig.\ \ref{d4s2.2b4},
now with error bars included. (More precisely, the two outer curves
represent the envelopes to the tops and bottoms of the error bars.) The
error grows linearly with $\sg$, due to the occurrence of the
log$\, \sg$ in \rf{ja1}.

The remarkable feature of the curve $D_S(\sg)$ is its slow approach to the
asymptotic value of $D_S(\sg)$ for large $\sg$.
This type of behaviour has never been
observed previously in systems of random geometry (see e.g.\
\cite{ajw,2dspectral,diffusion}), and again
underlines that causal dynamical triangulations in four dimensions
behave qualitatively differently, and that the quantum geometry produced
is in general richer. The new phenomenon we observe here is a
{\it scale dependence of the spectral dimension}, which has emerged
dynamically. This is to be contrasted with fractal structures which
show a self-similar behaviour at all scales.

A best three-parameter fit which
asymptotically approaches a constant is of the form
\beq\label{ja9}
D_S(\sg) =  a -\frac{b}{\sg+c} = 4.02-\frac{119}{54+\sg}.
\eeq
The constants $a$, $b$ and $c$ have been determined by using the
full data range $\sg \in [40,400]$
and the curve shape agrees well with the measurements, as can be seen from
Fig.\ \ref{d4s2.2b4}.
Integrating \rf{ja9} we obtain
\beq\label{ja8a}
P(\sg) \sim \frac{1}{\sg^{a/2} (1+c/\sg)^{b/2c}},
\eeq
from which we deduce the limiting cases
\beq\label{ja8b}
P(\sg) \sim \left\{
\begin{array}{cl}
\displaystyle{{\sg^{-a/2}}} &~~\mbox{for large $\sg$,}\\
~&~\\
\displaystyle{{\sg^{-(a-b/c)/2}}} &~~ \mbox{for small $\sg$}.
\end{array}
\right.
\eeq

We conclude that the quantum geometry generated by
causal dynamical triangulations has a scale-dependent
spectral dimension which increases continuously from
$a\mi b/c$ to $a$ with increasing distance.
Substituting the values for $a$, $b$ and $c$ obtained from the
fit \rf{ja9}, and taking into account their variation as we vary the
$\sg$-range $[\sg_{\rm min}, \sg_{\rm max}]$ and use different
weightings for the errors, we obtained
the asymptotic values \cite{spectral-d,blp}
\beq\label{ja10}
D_S(\sg\equ \infty) = 4.02 \pm 0.1
\eeq
for the ``long-distance spectral dimension" and
\beq\label{ja11}
D_S(\sg\equ 0)= 1.80 \pm 0.25
\eeq
for the ``short-distance spectral dimension".

A dynamically generated
scale-dependent dimension with this behaviour
signals the existence of an effective ultraviolet cutoff for
theories of gravity, brought about by the
(highly nonperturbative) behaviour of the
quantum-geometric degrees of freedom on the very smallest scale.
Of course, one should not lose sight of the fact that this is a numerical
result, based on data and fits. It would be desirable to have a
model which exhibits a scale-dependent spectral dimension and can
be understood analytically, in order to illuminate the mechanism at
work in CDT quantum gravity. An example of
such a model has been constructed recently \cite{wheater2}.

After we observed the dynamical reduction (to $\sim\! 2$) of the spectral
dimension at short distances, there has been a considerable amount of 
theoretical work showing that $D_S=2$ in the UV regime comes about naturally 
also in the asymptotic safety scenario \cite{laureu}, in Ho\v rava-Lifshitz gravity \cite{horspec} 
and even more generally \cite{carlip}. Comparative studies of the behaviour of
the spectral dimension away from the asymptotic UV and IR regimes have been
conducted in \cite{curvefit}.
In the next section we will discuss how one can imagine making
contact with these other theories of quantum gravity and also
why the UV spectral dimension in these theories is equal to 2.

Finally, it should be mentioned that the observed short-distance
value $D_S(\sg\equ 0)= 1.80 \pm 0.25$ agrees within measuring accuracy
with old measurements
of the spectral dimension in Euclidean DT quantum gravity in
the so-called ``crinkled'' phase \cite{crinkled1}. This
phase was ``discovered'' when a number of gauge fields were coupled to
Euclidean four-dimensional gravity using the formalism of dynamical triangulations.
The effect of the gauge fields is to change the
typical geometry from branched-polymer type
with Hausdorff dimension 2 and spectral dimension 4/3
to a different type of geometry with Hausdorff dimension close to 4 and
a spectral dimension $1.7\pm 0.2$. It was shown subsequently that
one can obtain the same geometric results by forgetting
about the $U(1)$-gauge fields altogether and instead assigning an additional  weight
$\prod_{t\in T}o(t)^\b$ to each triangulation $T$, where $t$ denotes a
triangle in $T$ and $o(t)$ is the order of $t$, i.e.\ the number
of four-simplices to which $t$ belongs. For suitable values of
$\b$ one could reproduce most of the geometric features resulting
from the addition of $U(1)$-gauge fields \cite{crinkled2}. Since the Regge curvature is
located at the triangles and -- when using identical building blocks
like in DT -- only depends on the order of $t$, it was natural to view
the weight as an effective higher-order curvature term in the action,
resulting from integrating out the gauge fields. However, this otherwise
appealing interpretation was somewhat undermined by the observation
that precisely this weight factor arises when changing the assignment
of the $U(1)$-gauge field from the links to the dual links \cite{crinkled3},
suggesting that one is dealing merely with discretization ambiguities, which
are unlikely to leave an imprint in the continuum limit. The conclusion at the
time was that the crinkled phase, just like
the other phases in pure, four-dimensional Euclidean DT, is a lattice artifact.
Nevertheless, the crinkled phase was reconsidered recently in
\cite{crinkled4}, where the spectral dimension was measured again and
$D_S(\sg\equ 0)$ found to be in agreement with the old results obtained
in \cite{crinkled1,crinkled2}. At the same time it
was noticed that it displays a scale-dependence similar to that
observed in four-dimensional CDT. If other physically desirable features can
be shown to be present in this phase, it may mean that there is a truly isotropic
model which belongs to the same universality class as CDT. This is an interesting
suggestion which deserves further study.

In closing, let us remark that --
apart from allowing us to define and measure the spectral
dimension -- the diffusion equation is an immensely useful tool for
studying the geometry of quantum space. A lot of
interesting work has been reported, both in three- \cite{bh}
and four-dimensional CDT \cite{rounds4}, related to the more detailed
shape of the computer-generated quantum universes.

\subsection{The Hausdorff dimension}

The diffusion equation probes certain aspects of the underlying
spacetime, which -- important for our purposes -- extend into the 
quantum regime, as we have seen above. 
For smooth manifolds the spectral dimension defined from
the diffusion equation agrees with the ordinary, topological dimension
of the manifold and its so-called Hausdorff dimension. 
For fractal geometries, like those encountered
in the path integral, the Hausdorff dimension can differ
from the spectral dimension, while there may be no ``ordinary'' dimension.
Here we will define and measure the Hausdorff dimension both  for
a spatial slice and for the entire spacetime between two spatial
slices. Let us start with a spatial slice of a
given configuration of spacetime as it appears in the
path integral. Due to the presence of a time foliation
we have well-defined spatial slices with the topology
of three-spheres at  discretized times $t_n$, constructed by gluing
together tetrahedra.

Our algorithm for determining the Hausdorff dimension $d_h$ of the spatial
slices is as follows. From a given time slice  of discrete volume $N_3$,
we randomly pick a tetrahedron $i_0$. From $i_0$, we move out by
one step and obtain $n(1)$ tetrahedra at distance 1 from $i_0$
(that is, its nearest neighbours). Moving out by a further step,
there will be $n(2)$ tetrahedra
at distance 2, and so on until all tetrahedra have been visited.
The numbers $n(r)$ recorded for each (discrete) distance $r$ sum up to
\beq
\sum_{r\equ 0}^{r_{max}} n(r)=N_3.
\label{v3av}
\eeq
Finally, we measure the average linear extension
\beq
\langle r \rangle = \frac{1}{N_3}\sum_r r n(r)
\label{extenav}
\eeq
of the spatial slice.
For a slice of volume $N_3$, this process is repeated $N_3/50 +1$ times
with different randomly chosen initial tetrahedra. The expectation value
$\langle r \rangle$ is then averaged over these
measurements. In this way, we obtain for every slice a data pair
\beq
\{ \langle\langle r \rangle\rangle, N_3\},
\label{pair}
\eeq
representing one ``bare" measurement.

\begin{figure}[ht]
\vspace{-3cm}
\psfrag{log2}{\bf{\Large $\log\langle r\rangle$}}
\psfrag{log1}{\Large\bf $\log\langle N_3\rangle$}
\centerline{\scalebox{0.6}{\rotatebox{0}{\includegraphics{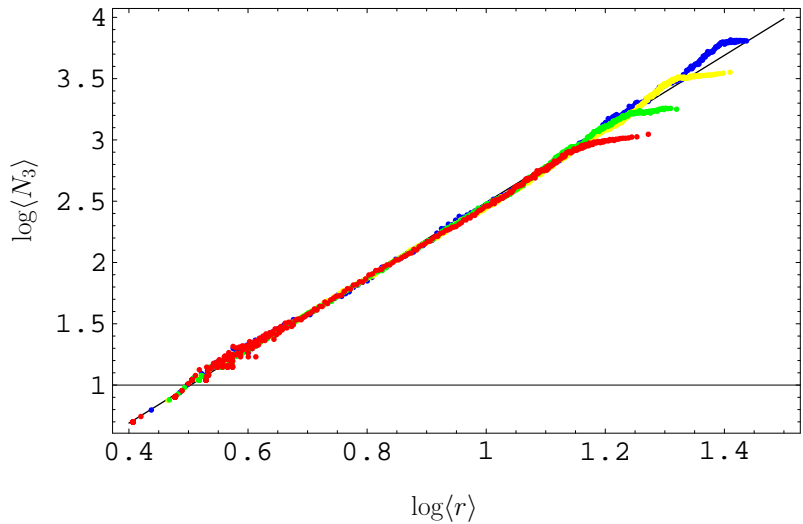}}}}
\vspace{-4.0cm}
\caption[phased]{{\footnotesize Log-log plot
of the average linear geodesic size $\langle r\rangle_{N_3}$ versus the
three-volume $N_3$, measured for $\Delta\equ 0.4$ and $\kp_0=2.2$.
The straight line corresponds to a Hausdorff
dimension $d_h\equ 3$. Similar measurements for $\Delta\equ 0.5$ and
$0.6$ and $\kp_0=2.2$  yield virtually indistinguishable results.
}}
\label{haus2.2b4}
\end{figure}

This process is performed for all  spatial slices of  the
given geometry. We have typically done this for about 1000
different spacetime geometries.
The final results are sorted by their
$ \langle\langle r \rangle\rangle$-value
(in practice a continuous variable)
and averaged over a sequence of 100 consecutive
points. The data points are
then displayed on a log-log plot. In the presence of finite-size scaling,
we expect them to follow a curve
\beq
\langle N_3\rangle (r) \propto \langle r\rangle^{d_h},
\label{3dscal}
\eeq
defining a spatial Hausdorff dimension $d_h$.
Our results, which show universal behaviour and scaling,
are presented in Fig.\ \ref{haus2.2b4}.

\begin{figure}[ht]
\vspace{-3cm}
\psfrag{log1}{\bf{\Large $\log\langle r \rangle$}}
\psfrag{log2}{\Large\bf $\log\langle N\rangle$}
\centerline{\scalebox{0.6}{\rotatebox{0}{\includegraphics{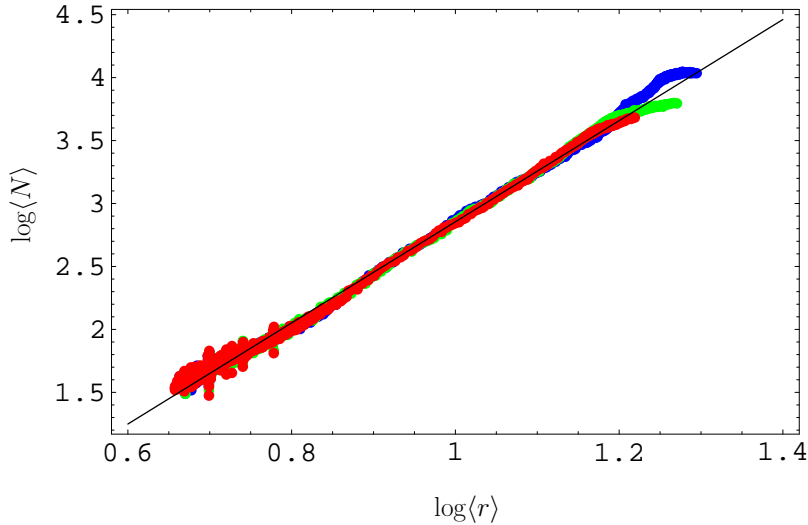}}}}
\vspace{-4.5cm}
\caption[phased]{{\footnotesize Log-log plot
of the average linear geodesic size $\langle r\rangle_{N}$ versus the
volume $N$ of a thick slice. The straight line corresponds to a Hausdorff
dimension $d_H\equ 4$.
}}
\label{Haus2.2b6.ps}
\end{figure}

Another substructure of spacetime whose geometry can be investigated
straightforwardly are what we shall call ``thick" spatial slices, namely, the
sandwiches of geometry contained in between two adjacent spatial slices
of integer times $t_n$ and $t_{n+1}$, with topology $I\times S^3$.
Such thick spatial slices  are made up
of four-simplices of types (4,1) and (3,2) and their time-reversed counterparts
(1,4) and (2,3).
To determine the Hausdorff dimension $d_H$ of the thick slices,
we pick a random
four-simplex from such a slice.
We proceed then exactly as we did when measuring the
Hausdorff dimension $d_h$ of the spatial slices at integer-$t_n$.
Starting from the
randomly chosen initial four-simplex, we move out,
one step at a time, to the nearest neighbours at distance $r$, from the
previous set of building blocks at distance $r-1$, except
that we never allow the crossing of the time slices $t_n$ and
$t_{n+1}$. At each step, we keep track of the number $n(r)$ of building blocks,
whose sum is equal to the total slice volume $N$,
\beq
\sum_{r\equ 0}^{r_{max}} n(r) = N,
\label{vav}
\eeq
which enables us to compute the average linear size
\beq
\langle r \rangle = \frac{1}{N}\sum_r r n(r)
\eeq
of the thick slice. For a given slice of volume $N$,
this process is then repeated
$N_4/100+1$ times with randomly chosen initial building blocks. The average
size $\langle r\rangle$ is again averaged over this set and the data pair
$\{\langle\langle r \rangle\rangle, N\}$ stored. Using the same
procedure as described following formula (\ref{pair}) above, we extract the
Hausdorff dimension $d_H$ from the relation
\beq
N_4 \propto \langle r \rangle^{d_H}.
\label{hausagain}
\eeq
The results are presented in Fig.\ \ref{Haus2.2b6.ps} in the form of a
log-log plot.
The straight line corresponds to $d_H\equ 4$, and the best fitted value is
$d_H \equ 4.01\pm 0.05$. Measurements were
performed on spacetimes with total size $\tilde N_4=$20, 40 and 80k, at
$\kappa_0\equ 2.2$, $\Delta\equ 0.6$. -- We conclude that the
transition from a genuine constant-time slice to a thick slice adds one
extra dimension to the Hausdorff dimension -- the
thickened slice already ``feels''
some aspects of the four-dimensionality of the full spacetime.
Note that this would not be true for a thick slice of a classical, regular
lattice, whose Hausdorff dimension would be the same as that of a
``thin" slice. The likely mechanism for the dimensional increase in
the triangulations is the appearance of ``short-cuts" between two
tetrahedral building blocks, say, once one moves slightly away
from a slice of integer-$t_n$. Finally, it is also true
that if we simply measure the Hausdorff dimension without any
restriction to stay within a ``thick'' spatial slice we
still obtain $d_H=4$.

\section{Making contact with asymptotic safety}\label{renormalization}

As we discussed earlier, it is technically challenging to get close to the
B-C phase transition line, which is needed if we want to achieve a
higher resolution in the UV regime, such that the lattice spacing
is much smaller than the Planck
length. Also, we do not know yet whether in such a limit
we have an isotropic scaling behaviour in space and time, 
like in the asymptotic safety approach,
or need to invoke a more general, anisotropic scenario as outlined above.
For the time being, let us assume that the end point
$P_0$ of the B-C transition line in the phase diagram of
Fig.\ \ref{phasediagram}
corresponds to an isotropic phase transition point.
How can one make contact with the gravitational
renormalization group treatment?
The standard way would be to ``measure'' observables (by lattice Monte
Carlo simulations), like a mass
in QCD or the string tension in Yang-Mills theory. For definiteness, let us
consider the string tension $\sigma$, which
has mass dimension two. The measurements, for some choice $g_0$ of
the bare coupling constant, will give us a number $\sigma(g_0)$.
We now write
\beq\label{ny40}
\sigma(g_0) = \sigma_R\, a^2(g_0),
\eeq
where $\sigma_R$ is the dimensionful, physical string tension and $a(g_0)$ describes
the dependence of the lattice spacing $a$ on the bare coupling constant $g_0$.
Similarly, one can try to measure the mass of the glueball, which
on the lattice will be some number $m_0(g_0)$. The relation analogous to \rf{ny40} is
\beq\label{ny40a}
m_0(g_0)= m_R\, a(g_0),
\eeq
where $m_R$ is the physical glueball mass.
Being able to write down relations like this for
all observables, where $a(g_0)$ is determined by the renormalization
group equation
\beq\label{ny41}
a\, \frac{\d g_0}{\d a} = -\b (g_0),
\eeq
allows us to define a continuum theory at a fixed point $g_0^*$ where
$\b(g_0)=0$, since there we can take $a(g_0) \to 0$ when $g_0 \to g_0^*$.
In the case of QCD or Yang-Mills theory the fixed point is the Gaussian
fixed point $g_0^*=0$, but in the more general setting of asymptotic safety
it will be non-Gaussian.

Assume now that we have a fixed point for gravity. The gravitational coupling
constant is dimensionful, and we can write for the bare coupling
constant
\beq\label{ny42}
 G(a) = a^2 \hG(a),~~~~ a\,  \frac{\d \hG}{\d a} = -\b(\hG),~~~~
\b(\hG) = 2\hG -c \hG^3 +\cdots\ .
\eeq
The IR fixed point $\hG =0$ corresponds to $G$ constant while the
putative non-Gaussian fixed point corresponds to $\hG\to \hG^*$,
i.e.\ $G(a) \to
\hG^* a^2$. In our case it is tempting to identify our dimensionless
constant $k_1$ with $1/\hG$, up to the constant of proportionality
given in \rf{n7c}. Close to the UV fixed point we have
\beq\label{ny43}
\hG(a) = \hG^* - K a^\tc,~~~k_1 = k_1^* + K a^\tc,~~~~~~\tc=-\b'(\hG^*).
\eeq
Usually one relates the lattice spacing near the fixed point
to the bare coupling constants with the help of some correlation
length $\xi$. Assume that $\xi$ diverges according to
\beq\label{ny44}
\xi(g_0) = \frac{c}{|g_0 -g_0^*|^\n}
\eeq
in the limit as we tune the bare coupling constant $g_0 \to g_0^*$.
This correlation length is associated with a field correlator and
usually some physical mass $m_{ph}$ (like the glue-ball mass
referred to in \rf{ny40a}) by means of
\beq\label{ny45}
\frac{|n_1-n_2|}{\xi(g_0)} = m_{ph} (a |n_1-n_2|) = m_{ph}|x_1-x_2|,
\eeq
where $|n_1-n_2|$ is a discrete lattice distance and
$|x_1-x_2|$ a physical distance. Requiring the physical
quantities $|x_1-x_2|$ and
$m_{ph}$ to remain constant as $a \to 0$ then fixes $a$ as a function
of the bare coupling constant,
\beq\label{ny46}
a = \frac{1}{c \, m_{ph}} \; |g_0-g_0^*|^\n.
\eeq
Eq.\ \rf{ny46} is only valid close to the fixed point and
should be compared to the renormalization group equation \rf{ny41},
from which we deduce that $\n = 1/|\b'(g_0^*)|$.

In the gravitational case at hand we do not (yet) have
observables which would allow
us to define meaningful correlation lengths.
At any rate, it is by no means a settled issue
how to {\it define} such a concept in a theory where one integrates
over all geometries, and where the length is itself a function
of geometry, as already emphasized in the Introduction
(see \cite{correlationlength} for related discussions).
Instead, we construct from our computer-generated ``data''
an effective action, where all degrees of freedom,
apart from the scale factor, have been integrated out.
We impose the constraint that the data are fitted
to a universe of total lattice four-volume $N_4$.
Measurements are performed at
different, fixed values of $N_4$, all the while maintaining the
relation\footnote{In principle, we should be taking
into account the different volumes of the two types of
four-simplices, which depend on $\Del$,
but we will ignore these details to streamline the presentation.}
\beq\label{ny47}
 V_4 = N_4 a^4.
\eeq
We then ``remove the regulator" by investigating the limit
$N_4\rightarrow\infty$.
In ordinary lattice field theory, we have two options for changing
$N_4$; either we keep $a$ fixed, and therefore change $V_4$, or
we keep $V_4$ fixed and change $a$. Let us illustrate the
difference in terms of a scalar field on a lattice.
Its dimensionless action can be written as
\beq\label{ny48}
S= \sum_{i} \Big( \sum_\m (\phi(i+\m)-\phi(i))^2 + m_0^2  \phi^2(i)\Big),
\eeq
where $i$ labels discrete lattice points and $\m$ unit
vectors in the different lattice directions. The correlation length is
approximately $1/m_0$ lattice spacings. Holding $a$ fixed and
increasing $N_4$ is not going to
change the correlation length in any significant way if $N_4$ is sufficiently
large. Thus the interpretation for fixed $a$ is straightforward:
the physical volume $V_4$ is simply increased and finite-size
effects will become smaller and smaller.
However, we can also insist on an interpretation where $V_4$ is kept fixed,
$N_4$ is increased
and $a$ decreased accordingly. In this case, the lattice becomes finer
and finer with increasing $N_4$. But now the physical interpretation of
\rf{ny48} will change with increasing $N_4$, even if no bare coupling
constant is changed, and the correlation length is still approximately
$1/m_0$ lattice spacings. Since the {\it physical} lattice length
$a$ decreases proportional to $1/N_4^{1/4}$ the {\it physical} correlation
length is going to zero, and the physical mass to infinity. This can
be made explicit in \rf{ny48} by introducing the lattice
spacing $a$,
\beq\label{ny49}
S= \frac{1}{a^2} \sum_i a^4 \left( \sum_\m
\Big( \frac{(\phi(i+\m)-\phi(i))^2}{a^2}\Big)
+ \frac{m_0^2}{a^2}  \phi^2(i)\right).
\eeq
The physical mass is $m_{ph}=m_0/a$ and goes to infinity
unless we adjust $m_0$. The factor $1/a^2$ in front of
the sum can be reabsorbed in a redefinition of $\phi$ if desired.

In our case it is natural to consider $V_4$ as fixed if we want
to make contact with the continuum framework of asymptotic safety,
since this will allow us to vary $a$. However, like in
the ordinary field theory case just discussed, it requires
the fine-tuning of some coupling constants to consider $V_4$ as
fixed when $N_4\to \infty$. In the
free field case we had to fine-tune the bare mass $m_0$ to
zero as $N_4 \to \infty$. In the gravity case the need for
a fine-tuning of some kind is also clear if we consider Fig.\ \ref{cos3}.
The bars indicated in the figure refer to a specific spacetime
four-volume which we label by $N_4$, without any reference to $a$.
For a given choice of bare coupling constants the size of the bars reflects
real fluctuations of the three-volumes of our quantum universe. When $N_4$
increases, but the bare coupling constants are kept fixed, we saw that
the relative size of the fluctuations went to zero. This is an indication
that we should view $V_4$ as going to infinity when $N_4 \to \infty$.
It is natural to assume that if we want to have the physics associated
with the continuum volume $V_4$ (approximately) invariant when
$N_4\to \infty$, the {\it relative} fluctuations 
$\del V_3(t)/V_3(t)$ of $V_3(t)$ must be constant as a function of $N_4$. This is only
possible by changing the bare coupling constants. As discussed
above, it is not a foregone conclusion that this can be achieved. 
For instance, we may have to
enlarge the coupling constant space to also include the fluctuations of
$N_4$ (and thus only talk about $\la V_4 \ra$, and not $V_4$), i.e.\ reintroduce
the cosmological constant. 

Let us assume it is
possible to change the bare coupling constants in such a way that we can
stay on a trajectory where $V_4$ can be considered fixed when
$N_4 \to \infty$. Choosing the right starting point, such a
trajectory may lead us to a fixed point, say,
the point $P_0$ in our phase diagram. We can also imagine a
scenario where different choices of starting point lead us to
different points along the B-C transition line, if it is a
second-order phase transition line.
If $P_0$ is a UV fixed point, eq.\ \rf{ny43} makes it clear what to
expect. Using \rf{ny47}, we can convert it into an equation
involving $N_4$ and suitable for application in CDT simulations, namely,
\beq\label{ny50}
k_1(N_4)=k_1^c -\tilde{K} N_4^{-\tc/4}.
\eeq
When we measured $k_1(N_4)$ deep inside phase $C$ (at the
point ($\kp_0,\Del)=(2.2,0.6)$),
we did not find any $N_4$-dependence of $k_1$. However,
according to the insights just presented,
we should observe such a dependence
close to a UV fixed point when we follow
a path in the coupling constant space where the continuum four-volume $V_4$ can
be considered constant. An explicit
verification of such a relation will have to await more reliable
computer simulations close to the phase transition lines.

How could such an analysis proceed more concretely in
CDT quantum gravity? Recall the ``na\"ive''
renormalization conditions \rf{ny17a} and \rf{ny17d},
introduced mainly to illustrate how a renormalization procedure could
lead to finite renormalized cosmological and gravitational constants,
both with a semiclassical
interpretation. Close to the UV fixed point, we know that
$G$ will not be constant when we change scale, but $\hat G$ will.
Writing $G(a)= a^2 \hat G^*$, eqs.\ \rf{ny17a} and \rf{ny17d} are changed to
\beq\label{ny51}
\kp_4-\kp_4^c = \frac{\La}{\hat G^*} \,a^2,~~~~k_1(\kp_0^c) = \frac{1}{\hat G^*}.
\eeq
The first of these relations now looks two-dimensional (cf. eq.\ \rf{ny4})!
Nevertheless, the expectation value of the four-volume still satisfies the
correct relation
\beq\label{ny52}
\la V_4\ra = \la N_4\ra \;a^4 \propto \frac{1}{\La^2},
\eeq
as follows from \rf{ny17e}. The second one of the relations in \rf{ny51}
tells us that it is not per se a problem if $k_1$ does not
go to zero along the B-C phase boundary, something we did not find 
strongly supported by the data (see the discussion in Sec.\ \ref{size}).

Formulas like \rf{ny51} and \rf{ny52} illustrate that one way to get
a handle on the relation $V_4 = a^4 N_4$ is by invoking the cosmological
constant. This is natural since the four-volume is the variable
conjugate to the cosmological constant. Unfortunately, as already
mentioned a number of times, it is not very practical from the point of view 
of the numerical
simulations and would add another coupling constant to the discussion, thus 
complicating the numerics.
It is important to try to find an independent way to
ensure one stays on trajectories $V_4 = const.$.

Finally, a UV fixed point of $\hG$ implies that $G(a)$ scales
``anomalously'' like $a^2$. This anomalous scaling near the fixed point
is the reason why the spectral dimension in both the asymptotic safety
and the Ho\v rava-Lifshiz scenario is 2 and not 4
\cite{laureu,horspec}.

\section{Conclusion}
\label{discussion}

In this report, we have described an attempt to formulate a nonperturbative
theory of four-dimensional quantum gravity using the path integral. More specifically,
we have collected and
reviewed results obtained by using {\it Causal Dynamical Triangulations}. 

For gravitational theories, 
the path-integral prescription to ``sum over all geometries" leaves 
open what is meant by ``all geometries".
Our choice follows closely that of the path integral in
ordinary quantum mechanics. There, one starts by summing over a set of 
piecewise linear paths and -- after taking a suitable continuum limit --
ends up summing over all continuous paths. 
For the case of gravity, this motivated us to construct the path integral
by summing over a class of piecewise linear geometries, which have the additional
virtue that their geometry can be defined without the use of coordinates. 
This enabled us to define a genuinely geometric path integral with a lattice cutoff.
The entire construction is well defined for Lorentzian signature, and the
setting accommodates in a natural way the use of different lattice spacings in the
time and spatial directions, although in the first place one would not expect the choice 
to matter from a continuum perspective. 

Being able to formulate a path integral over Lorentzian geometries in this way is
excellent, but needs
to be followed by evaluating it if we are going to extract physical consequences from it.
The nonperturbative nature of the path integral makes this difficult in the Lorentzian
regime, certainly in four dimensions, which has our main interest. What comes to
our help at this stage is a remarkable feature of the theory, namely, that any
piecewise linear geometry contributing to the path integral
allows for a rotation to Euclidean signature in such
a way that the usually formal relation between the Lorentzian-
and the Euclidean-signature action is valid. After performing this ``Wick rotation",
we can try to evaluate the path integral in the Euclidean
sector, where we have the usual tools from Euclidean
quantum field theory available, in particular, Monte Carlo simulations.
By varying the bare coupling constants of
the regularized lattice theory we mapped out the phase diagram of this
``auxiliary" Euclidean theory. Somewhat surprisingly, it closely resembled
a so-called Lifshitz phase diagram with a tricritical point; even more
surprisingly, the characterization of the different phases in terms of an
order parameter we loosely called ``average geometry" had some similarity 
with the phases of a Lifshitz diagram. While the qualitative features of the phase 
diagram had been known to us for some time, it was the advent of a new class of
potential quantum gravity theories, the so-called Ho\v rava-Lifshitz theories,
which made us realize that an anisotropic scaling of time vs. space may play a
role in the interpretation of the observed phase structure.

Not only does our original path integral have Lorentzian signature, but we
also use its Lorentzian character to impose a notion of causality {\it on
individual path-integral histories}, as first suggested by Teitelboim.
All path-integral configurations are required to have a product topology, of a given
three-dimensional space and the time direction. The absence of topology changes
and associated causality-violating branching points are in practice
implemented by building up the histories in triangulated steps of unit proper time.
The original motivation behind this choice was the
search for a class of geometries distinctly different from
the full set of Euclidean geometries of a fixed topology, which up to this point had led only
to physically meaningless results in a purely Euclidean implementation
of dynamical triangulations. 
The presence of a time foliation and causality constraints in the new, causal
formulation enabled us -- despite the randomness of the geometric structure --
to prove that the transfer matrix is (site)-reflection positive
and thus most likely allows for a unitary theory.

It is interesting that several of the concepts just outlined have
counterparts in the new continuum theories suggested by Ho\v rava, which were
arrived at by a rather different route. The latter may throw a 
new light on the role of ``asymmetry" between space and time.
The possibility to introduce a {\it finite} relative scaling between space and time
distances is of course present in any lattice theory, once
one has identified a time direction. Given the sparseness
of critical points it was therefore natural to assume that also in CDT quantum
gravity spacetime isotropy -- to the extent
it is a meaningful property of the eventual, physical theory and its (Euclidean) 
ground state or solutions  -- can be restored at a UV critical point. So far, our
analysis of the emergent de Sitter space in four dimensions is not in conflict with 
this interpretation, but in light of Ho\v rava's results one should certainly take
the possibility of anisotropic\footnote{both in the sense of a finite relative
scaling for macroscopic geometry and of a different scaling {\it dimension}
in the UV}, unitary limits into consideration. 
Our lattice formulation appears to be the perfect setting to investigate this issue.
The lattice implementation of Ho\v rava-Lifshitz gravity was
initiated recently (in three spacetime dimensions) in \cite{california}. 
Based on the new evidence that the B-C transition line is second order, a major 
effort is currently under way to understand the nature of the
theory when approaching this line: do we find evidence of an anisotropy between space
and time, with different scaling dimensions, and does the exponent
of anisotropy change along the line? Given that in the theory of critical
phenomena end points are usually associated with a transition of higher order
than those of generic points along the
line, do they play a special role? 

Most efforts until now have been focused on understanding
the quantum universes well inside phase C, produced by the
computer simulations. We have observed a universe whose average
geometry can be identified with that of a round four-sphere, with relatively 
small quantum fluctuations of the scale factor. We view this as
the infrared limit of our theory. The behaviour of the scale factor,
including quantum fluctuations, is described accurately
by a minisuperspace model \`a la Hartle and Hawking, which {\it assumes
homogeneity and isotropy from the outset}, as reflected in the
truncated ansatz 
\beq\label{hi}
ds^2 = dt^2 +a^2(t) d\Om_3^2,
\eeq
for the spacetime metric, with
$a(t)$ the scale factor of the universe, and $d \Om_3^2$ the
line element of the three-sphere $S^3$. Quantizing such a reduced model leads to
what one usually calls quantum cosmology, which entails a
quantum-mechanical, as opposed to a quantum field-theoretical
description of the universe. 
In our model no such assumption is put in by hand. Instead, we
integrate out all degrees of freedom except the scale factor {\it at the
level of the quantum theory}.
Nevertheless, to the precision we have been able to measure, the dynamics of this scale
factor, including its fluctuations, is governed by the same minisuperspace
action \rf{n5}. Let us reiterate that this result is not a triviality of the
kind ``of course, the infrared sector has to agree with general
relativity since you are using the Einstein-Hilbert action''. It is true that
$S^4$ is a solution
of the equations of motion for the Euclidean Einstein-Hilbert action with
positive cosmological constant,
but it is only a saddle point. There is absolutely no reason why we
should see this solution in our computer simulations if the classical
action was dominant. What we have found is a truly nonperturbative result, originating
from a subtle interplay between entropic contributions from the measure
and the Einstein-Hilbert action. This is related to the observation that
the CDT model can be viewed as an ``entropic'' theory, due to the
geometric nature of the Einstein action on the piecewise linear
geometries used. 
Being able to count the number of CDT configurations
for a given four-volume and a given number of vertices would give
us an exact, analytic handle on the partition function
of full quantum gravity! As we have argued in Sec.\ \ref{entro}, the partition function 
plays the role of a generating function for the number of triangulations.

The analysis of the infrared properties of the quantum universe allowed
us to relate the coupling constant appearing in the effective
action for the scale factor to the lattice spacing, thereby obtaining
an estimate of the Planck scale in terms of the unit lattice length $a$.
Well inside phase C we established that one Planck length corresponds
to roughly half a lattice unit, which means that despite their small size
(the largest universe being about 20 Planck lengths across), in this part of
phase space we are not yet probing the ``sub-Planckian'' regime. On the
other hand, we are clearly dealing with {\it quantum} universes, since even
the most macroscopic observable, the total universe volume (a.k.a. the scale factor),
undergoes sizeable quantum fluctuations, as illustrated by Fig.\ \ref{cos3}.
Although this is clearly an interesting and relevant regime to understand
the quantum structure of spacetime (recall that we are operating with a
``lattice window" which is merely of size $\sim\!\! 10^4$), one obvious aim is
to move to other regions in the
bare coupling constant space where the Planck length is larger than
a single lattice spacing, and study the UV physics there. For example, this
may enable us to quantitatively study quantum deviations from the standard
Friedmann behaviour for small scale factors (close to the big bang) \cite{maitra}.
These regions of coupling constant space appear
to be located close to the transition lines of
the phase diagram. Of course, if we then still want to work with universes of
size $20 \ell_{\rm Pl}$, much larger lattices will be required. 
We are encountering the usual challenges
of studying quantum field theories on a lattice!

The identification of the gravitational coupling constant is so far
based on the reconstruction of the effective action from measuring
the scale factor. It is important to get an independent
verification of its interpretation as the coupling constant
of gravity. Since gravity -- at least on sufficiently large scales -- is defined
as the universal attraction between ``matter'', 
one obvious way of trying this is by coupling
matter to the theory. It is technically straightforward to do this
in the DT formalism, and therefore also in CDT.
In the context of Euclidean DT quantum gravity, combined theories of gravity 
and various matter field theories have
been studied extensively, in both three and four dimensions \cite{4dDTmatter,jaj,ty}.
Surprisingly little new insight was gained from comparing the
phase diagram with matter to that
without matter\footnote{An exception is
\cite{ty}, which claimed that the addition of {\it many} $U(1)$-fields
would cure the problem of finding only a first-order phase transition in four-dimensional DT.
However, the claim is in disagreement with the results of \cite{jaj}. The disagreement
has never been resolved, nor has the claim been substantiated further.}. At least perturbatively
this may be understood from the fact that adding matter
to the action will from a gravitational point of view mainly lead to
a shift in the bare gravitational coupling constant. 

Again in
the context of Euclidean DT attempts have been made to study the
converse effect, namely, how gravity influences the matter
sector \cite{bs1}, by trying to measure how certain well-chosen
matter correlators change, compared to those of a flat or de Sitter spacetime.
A preliminary investigation along the same lines in CDT has revealed that
both our universe and the value of the gravitational
coupling constant $G$ (extracted as explained above) are too small to
make the effect reported in \cite{bs1} observable. Alternatively,
one could try to measure the attraction between two
``test particles''. As emphasized already in the Introduction, it is highly nontrivial
to find a theoretically well-defined way of implementing this in the full,
nonperturbative theory.  
Merely considering the length scales involved (a strongly quantum-fluctuating
universe of less than $20 \ell_{\rm Pl}$ across) makes it clear that there
is considerable scope for being misled by na\"ive classical
considerations. This does not yet account for the fact that we are trying to
extract properties of the causal, Lorentzian theory from making measurements
in the Euclidean sector. Before embarking on substantial simulations and
measurements, we should better have a
precise idea of what we are looking for. 

Somewhat simpler conceptually is the situation of just a single
test particle, whose presence will modify
the average ``background'' geometry away from the
de Sitter geometry observed for pure gravity. One point
of reference one would expect to be relevant here is the
de Sitter-Schwarzschild solution with Euclidean signature,
a connection which has recently been addressed analytically \cite{klr}.
The translation of this analysis into a concrete measuring prescription 
for the combined system ``quantum universe plus particle" is work in
progress. -- As we have seen throughout this report, this is just one of many
interesting works in progress to further illuminate the nature of quantum gravity,
both addressing conceptual challenges and dealing with their numerical
implementation.

\section*{Acknowledgments}
\noindent We thank K.\ Anagnostopoulos, D.\ Benedetti, T.\ Budd, B.\ Dittrich, J.\ Gizbert-Studnicki,
S.\ Jordan,  I.\ Khavkine, C.F.\ Kristjansen, R.L.\ Maitra, I.\ Pushkina, P.\ Reska, T.\ Trzesniewski,
G.\ Vernizzi, Y.\ Watabiki, W.\ Westra, F.\ Zamponi and S. Zohren for collaboration
on various parts of the ongoing CDT program. J.A. thanks Utrecht
University and the Perimeter Institute for hospitality and financial support.
J.A. and A.G. acknowledge financial
support by the Danish Research Council (FNU) through the grant
``Quantum gravity and the role of black holes''.
J.J.\ acknowledges partial support of the Polish Ministry of
Science and Higher Education under grant 182/N-QGG/2008/0.
A.G.\ acknowledges a partial support by the Polish Ministry of 
Science grant N N202 229137 (2009-2012).

\section*{Appendix 1: Lorentzian angles}

\begin{figure}[t]
\centerline{\scalebox{0.55}{\rotatebox{0}{\includegraphics{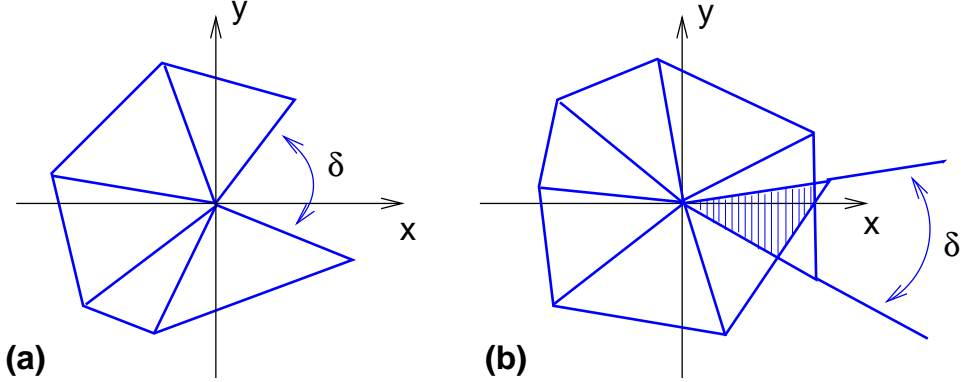}}}}
\caption[euangle]{{\footnotesize Positive (a) and negative (b)
Euclidean deficit angles $\delta$.}}
\label{euangle}
\end{figure}

In this appendix we will describe some properties of
Lorentzian angles (or ``boosts'') which appear in the Regge
scalar curvature as rotations about spacelike bones (that is,
spacelike links in three and spacelike triangles in four dimensions). This summarizes
the treatment and conventions of \cite{sorkin}. The familiar form of
a contribution of a single bone $b$ to the total curvature and therefore
the action is
\begin{equation}
\Delta_{b} S={\rm volume}(b)\delta_{b},\;\;\;\;
\delta_{b} =2\pi -\sum_{i}\Theta_{bi},
\end{equation}
where the volume of the bone is by definition real and positive and
$\delta_{b}$ is the deficit angle around $b$. For the case of
Euclidean angles this is conveniently illustrated by a
two-dimensional example, where the bone is simply a point with volume
1. A positive deficit angle $\delta$ (Fig.\ \ref{euangle}a)
implies a positive Gaussian
curvature and therefore a positive contribution to the action,
whereas an ``excess'' angle $\delta$ contributes negatively
(Fig.\ \ref{euangle}b).
\begin{figure}[t]
\centerline{\scalebox{0.55}{\rotatebox{0}{\includegraphics{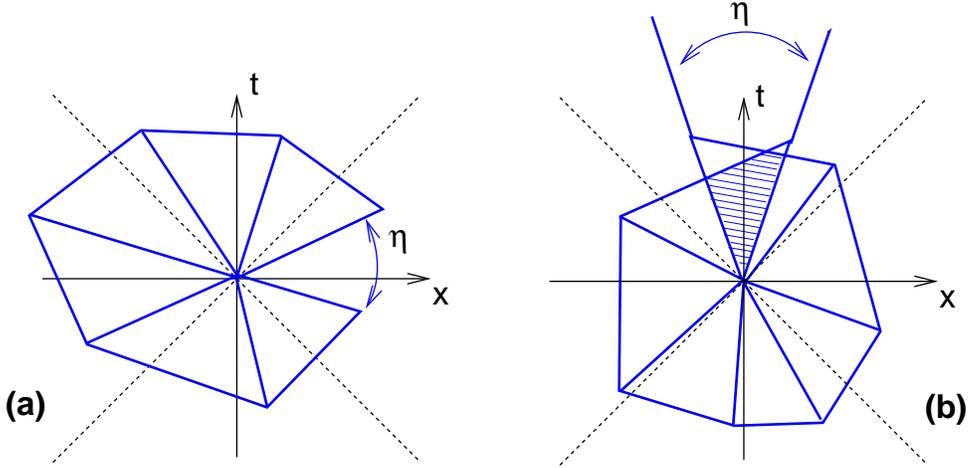}}}}
\caption[lorangle]{{\footnotesize 
Positive spacelike (a) and timelike (b) Lorentzian
deficit angles $\eta$.}}
\label{lorangle}
\end{figure}
Lorentzian angles are also defined to be additive, in such a way that
a complete rotation around a spacelike bone gives $2\pi$ in the flat
case. However, note that the angles are in general {\it complex}, and
can become arbitrarily large in the vicinity of the light cone.
In their contribution to the action, we have to distinguish between two cases.
If the Lorentzian deficit angle $\delta\equiv\eta$ is spacelike
(Fig.\ \ref{lorangle}a), it contributes
as $\Delta_{b} S={\rm volume}(b)\eta_{b}$, just
like in the Euclidean case. By contrast, if it is timelike
(Fig.\ \ref{lorangle}b),
the deficit angle contributes with the opposite sign, that is,
as $\Delta_{b} S=-{\rm volume}(b)\eta_{b}$. Therefore, both a
spacelike defect and a timelike excess increase the action, where as
a timelike defect or a spacelike excess decrease it.

The deficit angles in Secs 2.1 and 2.2 were calculated using
\begin{equation}
\cos\Theta =\frac{\langle \vec v_{1},\vec v_{2}\rangle }{
\langle \vec v_{1},\vec v_{1}\rangle^{\frac{1}{2}}
\langle \vec v_{2},\vec v_{2}\rangle^{\frac{1}{2}} },\;\;\;\;
\sin\Theta =\frac{\sqrt{
\langle \vec v_{1},\vec v_{1}\rangle
\langle \vec v_{2},\vec v_{2}\rangle -
\langle \vec v_{1},\vec v_{2}\rangle^{2} }}{
\langle \vec v_{1},\vec v_{1}\rangle^{\frac{1}{2}}
\langle \vec v_{2},\vec v_{2}\rangle^{\frac{1}{2}} },
\end{equation}
for pairs of vectors $\vec v_{1},\vec v_{2}$, and the flat
Minkowskian scalar product $\langle \cdot,\cdot \rangle$.
By definition, the square roots are positive imaginary for
negative arguments.

\section*{Appendix 2: Link-reflection positivity in two dimensions}

In this appendix we will demonstrate the link-reflection positivity of
the discrete model of two-dimensional Lorentzian quantum gravity
introduced in \cite{al}. Recall that link reflection is the
reflection at a plane of half-integer $t$. We choose it to lie at
$t=1/2$ and fix the boundary spatial lengths at the initial time $t=-T+1$
and the final time $t=T$ to $l_{-T+1}=l_{T}=l$. In order to
describe a two-dimensional Lorentzian universe with these boundary
conditions, we must not only specify the geometry of the spatial
sections (simply given by $l_{t}$, $-T+1\leq t\leq T$), but the
connectivity of the entire two-dimensional triangulation.

A convenient way of
parametrizing the connectivity that is symmetric with respect to
incoming and outgoing triangles at any given slice of constant $t$
is as follows. For any spatial slice at some integer time $t$,
each of the $l_t$ spatial edges forms the base of one incoming
and one outgoing triangle. The geometry of the adjoining sandwiches
is determined by how these triangles are glued together
pairwise along their timelike edges. These gluing patterns
correspond to distinct ordered partitions of the $l_t$ triangles
(either above or below $t$) into $k$ groups, $1\leq k\leq l_t$.
We denote the partitions collectively by $m(t)=\{ m_r(t),\ r=1,\ldots ,k\}$
for the incoming triangles and by
$n(t)=\{ n_r(t),\ r=1,\ldots ,k'\}$ for the outgoing triangles.
The constraints on these variables are obviously
$\sum_{r=1}^k m_r(t) =\sum_{r=1}^{k'} n_r(t) =l_t$ and a matching
condition for adjacent slices, namely, $k'(t)=k(t+1)$.\footnote{
This parametrization is closely related to the description of
two-dimensional Lorentzian gravity in terms of ``half-edges'' \cite{lottietal}.}
In terms of these variables, the (unmarked) one-step propagator
is given by
\bea
G_g(l_1,l_2)&=& g^{l_1+l_2} \sum_{k\geq 1}\; \frac{1}{k}
\sum_{ {n_r,m_r\geq 1,\ r=1,\ldots,k \atop
\sum_{q=1}^k n_q =l_1,\ \sum_{p=1}^k m_p =l_2 } } 1 \nonumber\\
&=& g^{l_1+l_2} \sum_{k\geq 1}\; \frac{1}{k} \
{l_1-1 \choose l_1-k } {l_2-1 \choose l_2-k},
\label{1step}
\eea
where $g={\rm e}^{-\lambda}$ depends on the two-dimensional
cosmological constant $\lambda$. It is obvious from (\ref{1step}) that
the propagator depends symmetrically on the local variables $m$ and $n$.
The partition function for the entire system is
\begin{equation}
Z_{2d}(g)=G_g(l_{-T+1},l_{-T+2}) \prod_{t=-T+2}^{T-1}\,
\sum_{l_t\geq 1} l_t\ G_{g}(l_t,l_{t+1}),
\end{equation}
in accordance with (\ref{iter}).

Following the discussion of link reflection in Sec.\ 6, we consider
now functions $F(\{ m,n\} )$ that depend only on the geometry above
the reflecting plane at $t=1/2$, i.e. on $m(t)$, $n(t)$, $t\geq 1$.
The reflection (\ref{lreflect}) acts on the partition data according
to $\theta_l (m(t))=n(-t+1)$ and $\theta_l(n(t))=m(-t+1)$.
Without loss of generality, we can assume that $F$ depends
symmetrically on $m(t)$ and $n(t)$, $\forall t$. Computing the
expectation value (\ref{linkref}), we obtain
\bea
\lefteqn{\langle (\theta_l F)F\rangle =}\nonumber\\
&& Z_{2d}^{-1} \sum_{l_0,l_1 \geq 1}
l_0 l_1 G_g(l_0,l_1)  \Bigl( G_g(l_{-T+1},l_{-T+2})
\prod_{t=-T+2}^{-1} \sum_{l_t\geq 1}
l_t G_g(l_t,l_{t+1}) (\theta_l F) \Bigr)\nonumber\\
&& \times\Bigl( \prod_{t=1}^{T-2} \sum_{l_{t+1}\geq 1}
G_g(l_t,l_{t+1}) l_{t+1}
G_g(l_{T-1},l_T) F \Bigr)\nonumber
\eea
i.e.
\bea
\langle (\theta_l F)F\rangle& =&
Z_{2d}^{-1} \sum_{l_0,l_1 \geq 1} l_0 l_1 \sum_{k\geq 1}\
\frac{1}{k}\ g^{l_0 +l_1}{l_0-1 \choose l_0-k}{l_1-1 \choose l_1-k }\
\overline{{\bf F}(l_0)} {\bf F}(l_1)\nonumber\\
&=&Z_{2d}^{-1} \sum_{k\geq 1}\ \frac{1}{k}\
\overline{{\cal F}(k)}{\cal F}(k).
\label{alongeq}
\eea
Since the last expression is a sum of positive terms, we have
hereby shown that two-dimensional Lorentzian gravity is
link-reflection positive.









\end{document}